\documentclass[12pt]{article}

\usepackage[utf8]{inputenc}
\usepackage{geometry}
\usepackage{setspace}
\usepackage{float}
\usepackage{array}
\usepackage{booktabs}
\usepackage{longtable}
\usepackage{pdflscape}

\usepackage{amsmath}
\usepackage{amsthm}
\usepackage{amssymb}
\usepackage{amsfonts}
\usepackage{bm}
\usepackage{siunitx}
\DeclareMathOperator{\sgn}{sgn}

\usepackage{graphicx}
\usepackage[bf]{caption}
\usepackage{subcaption}
\usepackage{tikz}
\usetikzlibrary{arrows.meta}
\usetikzlibrary{patterns}
\usetikzlibrary{decorations}
\usetikzlibrary{decorations.pathmorphing}
\usetikzlibrary{decorations.pathreplacing}
\usetikzlibrary{decorations.text}
\usetikzlibrary{intersections}
\usetikzlibrary{calc}
\usetikzlibrary{positioning}
\usepackage{pgfplots}
\pgfplotsset{compat=1.16}
\usepackage{standalone}

\usepackage{etoolbox}
\usepackage{datetime}
\usepackage{comment}

\usepackage[bottom]{footmisc}

\usepackage[style=authoryear-comp,backend=biber,natbib=true,giveninits=true,
            uniquename=false,uniquelist=false,maxbibnames=99,maxcitenames=2,
            dashed=false,doi=false,isbn=false,url=false,eprint=false,date=year]{biblatex}
\DeclareNameAlias{sortname}{family-given}
\DeclareNameAlias{default}{family-given}
\renewbibmacro{in:}{\ifentrytype{article}{}{\printtext{\bibstring{in}\intitlepunct}}}
\DeclareFieldFormat[article]{volume}{#1}
\DeclareFieldFormat[article]{number}{\mkbibparens{#1}}
\DeclareFieldFormat[article]{pages}{#1}
\renewbibmacro*{volume+number+eid}{%
  \printfield{volume}\setunit*{\addspace}\printfield{number}\setunit{\addcomma\space}\printfield{eid}}
\renewbibmacro*{note+pages}{%
  \printfield{note}\setunit{\ifentrytype{article}{\addcolon\space}{\bibpagespunct}}\printfield{pages}\newunit}
\AtBeginDocument{\let\cite\citet}
\usepackage{xcolor}
\usepackage{hyperref}

\renewcommand{\arraystretch}{1.2}
\newdateformat{monthyeardate}{\monthname[\THEMONTH] \THEYEAR}

\definecolor{custom_color}{RGB}{221, 0,0}
\definecolor{CoolRed}{HTML}{CC3333}
\definecolor{CoolBlue}{HTML}{4682B4}
\definecolor{CoolGreen}{HTML}{43AA8B}

\hypersetup{
    colorlinks=true,
    linkcolor={red},
    filecolor=blue,
    citecolor={blue},
    urlcolor={blue},
}

\makeatletter
\patchcmd{\@maketitle}{\LARGE}{\Large}{}{}
\makeatother

\newcommand{\J}{\mathcal{J}}

\newcommand{\CPset}{\mathcal{P}}

\newtheorem{assumption}{Assumption}

\newtheorem{result}{Proposition}

\newcommand{\sM}{\mathsf{M}}

\newcommand{\sP}{\mathsf{P}}
\newcommand{\sMP}{\mathsf{MP}}
\newcommand{\sU}{\mathsf{U}}
\newcommand{\AM}{\mathcal{A}^{\sM}}
\newcommand{\AMP}{\mathcal{A}^{\sMP}}

\newcounter{resultvariantcnt}
\newenvironment{resultvariant}[3][]{
    \stepcounter{resultvariantcnt}
    \renewcommand{\theresult}{\ref*{#2}#3}
    
    \addtocounter{result}{-1}
    \begin{result}[#1]
}{
    \end{result}
}

\title{\textbf{Perks Under Pressure? Demand Shocks and Firm Provision of Workplace Amenities}%
    \thanks{
        \textit{Acknowledgments}:
        This project has received funding from the European Union’s Horizon Europe Research and Innovation Programme under Grant Agreement No. 101043127. We thank Erik Hurst, Patrick Kline, Thomas Le Barbanchon, Barbara Petrongolo, Claudio Michelacci, Lukas Rodrian, Raffaele Saggio, and participants at the 2026 CEPR-Banca d'Italia Labor Workshop, the 2026 ELSE Workshop at Bocconi University, Berlin Applied Micro Seminar, Catholic University of Milan, Frisch Centre for Economic Research, and Statistics Norway for helpful comments.
    }

    \vspace{2mm}
}

\author{
  \begin{tabular}{cccc}
    Manudeep Bhuller\thanks{University of Oslo, Statistics Norway, CEPR, IZA, CESifo. E-mail: \href{mailto:manudeep.bhuller@econ.uio.no}{manudeep.bhuller@econ.uio.no}.} \qquad
    &
    Lukas Delgado-Prieto\thanks{University of Oslo, RFBerlin. E-mail: \href{mailto:laprieto@econ.uio.no}{laprieto@econ.uio.no}.}  \qquad
    &
    Samuel \v Skoda\thanks{University of Oslo. E-mail: \href{mailto:samuel.skoda@econ.uio.no}{samuel.skoda@econ.uio.no}.}
  \end{tabular}
}
\date{}

\begin{document}

\maketitle

\begin{center}
    First Version: October 2026 \\
    This Version: November 2026
\end{center}

\begin{abstract}
Do firms adjust workplace amenities in response to demand shocks? We develop a monopsony framework in which firms jointly set wages and amenities and show that responses to demand shocks depend crucially on the structure of amenity provision costs. Using Norwegian data that combine matched employer-employee records, firms’ cost accounts, and non-pecuniary job attributes drawn from job postings, we exploit three distinct sources of firm-level demand shocks: abrupt changes in value added, public procurement contract wins, and export demand shifts. We document multi-margin adjustments in amenities alongside wages and employment, and discuss implications for the measurement of employer market power.
\end{abstract}

\vfill

\vspace{-0.1cm}


\thispagestyle{empty}
\clearpage

\onehalfspacing

\setcounter{page}{1}

\providecommand{\sym}[1]{\ifmmode^{#1}\else\(^{#1}\)\fi}
\newcommand{\coeffnote}{Standard errors in parentheses. ${*} p<0.10, {**} p<0.05, {***} p<0.01.$}

\newcommand{\coefemploymentraw}{0.160}
\newcommand{\coefwagestayersraw}{0.034}
\newcommand{\coefamenmultiplierwlraw}{-0.00121}
\newcommand{\coeftotalcompperworkerraw}{0.032}

\newcommand{\coefvalueadded}{0.247\sym{***}}
\newcommand{\sevalueadded}{(0.019)}
\newcommand{\coefsevalueadded}{0.247\sym{***} (0.019)}
\newcommand{\pctvalueadded}{28.0\%}
\newcommand{\meanvalueadded}{15.57}
\newcommand{\obsvalueadded}{554,967}
\newcommand{\firmsvalueadded}{43,880}

\newcommand{\coefoperatingrevenues}{0.200\sym{***}}
\newcommand{\seoperatingrevenues}{(0.015)}
\newcommand{\coefseoperatingrevenues}{0.200\sym{***} (0.015)}
\newcommand{\pctoperatingrevenues}{22.2\%}
\newcommand{\meanoperatingrevenues}{16.51}
\newcommand{\obsoperatingrevenues}{554,967}
\newcommand{\firmsoperatingrevenues}{43,880}

\newcommand{\coefemployment}{0.160\sym{***}}
\newcommand{\seemployment}{(0.016)}
\newcommand{\coefseemployment}{0.160\sym{***} (0.016)}
\newcommand{\pctemployment}{17.3\%}
\newcommand{\meanemployment}{2.38}
\newcommand{\obsemployment}{554,967}
\newcommand{\firmsemployment}{43,880}

\newcommand{\coefempfte}{0.176\sym{***}}
\newcommand{\seempfte}{(0.015)}
\newcommand{\coefseempfte}{0.176\sym{***} (0.015)}
\newcommand{\pctempfte}{19.2\%}
\newcommand{\meanempfte}{2.03}
\newcommand{\obsempfte}{554,967}
\newcommand{\firmsempfte}{43,880}

\newcommand{\coefincumbents}{0.040\sym{***}}
\newcommand{\seincumbents}{(0.004)}
\newcommand{\coefseincumbents}{0.040\sym{***} (0.004)}
\newcommand{\pctincumbents}{4.1\%}
\newcommand{\meanincumbents}{2.29}
\newcommand{\obsincumbents}{554,967}
\newcommand{\firmsincumbents}{43,880}

\newcommand{\coefwagestayers}{0.034\sym{***}}
\newcommand{\sewagestayers}{(0.003)}
\newcommand{\coefsewagestayers}{0.034\sym{***} (0.003)}
\newcommand{\pctwagestayers}{3.5\%}
\newcommand{\meanwagestayers}{5.59}
\newcommand{\obswagestayers}{554,967}
\newcommand{\firmswagestayers}{43,880}

\newcommand{\coefpayinkindstayers}{0.0195\sym{***}}
\newcommand{\sepayinkindstayers}{(0.0075)}
\newcommand{\coefsepayinkindstayers}{0.0195\sym{***} (0.0075)}
\newcommand{\pctpayinkindstayers}{2.0\%}
\newcommand{\meanpayinkindstayers}{6.39}
\newcommand{\obspayinkindstayers}{387,194}
\newcommand{\firmspayinkindstayers}{33,749}

\newcommand{\coefwagehires}{0.028\sym{***}}
\newcommand{\sewagehires}{(0.004)}
\newcommand{\coefsewagehires}{0.028\sym{***} (0.004)}
\newcommand{\pctwagehires}{2.9\%}
\newcommand{\meanwagehires}{5.29}
\newcommand{\obswagehires}{432,607}
\newcommand{\firmswagehires}{40,248}

\newcommand{\coefwageincumbents}{0.029\sym{***}}
\newcommand{\sewageincumbents}{(0.003)}
\newcommand{\coefsewageincumbents}{0.029\sym{***} (0.003)}
\newcommand{\pctwageincumbents}{3.0\%}
\newcommand{\meanwageincumbents}{5.49}
\newcommand{\obswageincumbents}{554,967}
\newcommand{\firmswageincumbents}{43,880}

\newcommand{\coeftotalcompstayers}{0.037\sym{***}}
\newcommand{\setotalcompstayers}{(0.003)}
\newcommand{\coefsetotalcompstayers}{0.037\sym{***} (0.003)}
\newcommand{\pcttotalcompstayers}{3.8\%}
\newcommand{\meantotalcompstayers}{10.57}
\newcommand{\obstotalcompstayers}{534,224}
\newcommand{\firmstotalcompstayers}{42,336}

\newcommand{\coeftotalcomphires}{0.045\sym{***}}
\newcommand{\setotalcomphires}{(0.010)}
\newcommand{\coefsetotalcomphires}{0.045\sym{***} (0.010)}
\newcommand{\pcttotalcomphires}{4.6\%}
\newcommand{\meantotalcomphires}{9.90}
\newcommand{\obstotalcomphires}{405,087}
\newcommand{\firmstotalcomphires}{39,974}

\newcommand{\coefmwagestayers}{0.038\sym{***}}
\newcommand{\semwagestayers}{(0.003)}
\newcommand{\coefsemwagestayers}{0.038\sym{***} (0.003)}
\newcommand{\pctmwagestayers}{3.8\%}
\newcommand{\meanmwagestayers}{10.55}
\newcommand{\obsmwagestayers}{534,155}
\newcommand{\firmsmwagestayers}{42,336}

\newcommand{\coefmwagehires}{0.046\sym{***}}
\newcommand{\semwagehires}{(0.011)}
\newcommand{\coefsemwagehires}{0.046\sym{***} (0.011)}
\newcommand{\pctmwagehires}{4.7\%}
\newcommand{\meanmwagehires}{9.89}
\newcommand{\obsmwagehires}{404,953}
\newcommand{\firmsmwagehires}{39,973}

\newcommand{\coefwagecost}{0.193\sym{***}}
\newcommand{\sewagecost}{(0.015)}
\newcommand{\coefsewagecost}{0.193\sym{***} (0.015)}
\newcommand{\pctwagecost}{21.3\%}
\newcommand{\meanwagecost}{15.16}
\newcommand{\obswagecost}{551,109}
\newcommand{\firmswagecost}{43,634}

\newcommand{\coefwagecostperworker}{0.034\sym{***}}
\newcommand{\sewagecostperworker}{(0.003)}
\newcommand{\coefsewagecostperworker}{0.034\sym{***} (0.003)}
\newcommand{\pctwagecostperworker}{3.4\%}
\newcommand{\meanwagecostperworker}{12.78}
\newcommand{\obswagecostperworker}{551,109}
\newcommand{\firmswagecostperworker}{43,634}

\newcommand{\coeftotalnonwagecost}{0.187\sym{***}}
\newcommand{\setotalnonwagecost}{(0.015)}
\newcommand{\coefsetotalnonwagecost}{0.187\sym{***} (0.015)}
\newcommand{\pcttotalnonwagecost}{20.6\%}
\newcommand{\meantotalnonwagecost}{12.47}
\newcommand{\obstotalnonwagecost}{546,639}
\newcommand{\firmstotalnonwagecost}{43,435}

\newcommand{\coeftotalnonwageperworker}{0.028\sym{***}}
\newcommand{\setotalnonwageperworker}{(0.005)}
\newcommand{\coefsetotalnonwageperworker}{0.028\sym{***} (0.005)}
\newcommand{\pcttotalnonwageperworker}{2.8\%}
\newcommand{\meantotalnonwageperworker}{10.09}
\newcommand{\obstotalnonwageperworker}{546,639}
\newcommand{\firmstotalnonwageperworker}{43,435}

\newcommand{\coefnonwagecost}{0.191\sym{***}}
\newcommand{\senonwagecost}{(0.011)}
\newcommand{\coefsenonwagecost}{0.191\sym{***} (0.011)}
\newcommand{\pctnonwagecost}{21.0\%}
\newcommand{\meannonwagecost}{12.14}
\newcommand{\obsnonwagecost}{543,821}
\newcommand{\firmsnonwagecost}{43,409}

\newcommand{\coefnonwageperworker}{0.031\sym{***}}
\newcommand{\senonwageperworker}{(0.009)}
\newcommand{\coefsenonwageperworker}{0.031\sym{***} (0.009)}
\newcommand{\pctnonwageperworker}{3.1\%}
\newcommand{\meannonwageperworker}{9.76}
\newcommand{\obsnonwageperworker}{543,821}
\newcommand{\firmsnonwageperworker}{43,409}

\newcommand{\coefpenspremcost}{0.173\sym{***}}
\newcommand{\sepenspremcost}{(0.023)}
\newcommand{\coefsepenspremcost}{0.173\sym{***} (0.023)}
\newcommand{\pctpenspremcost}{18.9\%}
\newcommand{\meanpenspremcost}{11.62}
\newcommand{\obspenspremcost}{302,876}
\newcommand{\firmspenspremcost}{28,976}

\newcommand{\coefpenspremperworker}{0.017\sym{*}}
\newcommand{\sepenspremperworker}{(0.010)}
\newcommand{\coefsepenspremperworker}{0.017\sym{*} (0.010)}
\newcommand{\pctpenspremperworker}{1.7\%}
\newcommand{\meanpenspremperworker}{9.10}
\newcommand{\obspenspremperworker}{302,876}
\newcommand{\firmspenspremperworker}{28,976}

\newcommand{\coefrentalcost}{0.110\sym{***}}
\newcommand{\serentalcost}{(0.012)}
\newcommand{\coefserentalcost}{0.110\sym{***} (0.012)}
\newcommand{\pctrentalcost}{11.6\%}
\newcommand{\meanrentalcost}{12.84}
\newcommand{\obsrentalcost}{493,805}
\newcommand{\firmsrentalcost}{40,620}

\newcommand{\coefexternallabor}{0.195\sym{***}}
\newcommand{\seexternallabor}{(0.019)}
\newcommand{\coefseexternallabor}{0.195\sym{***} (0.019)}
\newcommand{\pctexternallabor}{21.5\%}
\newcommand{\meanexternallabor}{12.79}
\newcommand{\obsexternallabor}{257,143}
\newcommand{\firmsexternallabor}{25,364}

\newcommand{\coefamenmultiplierwl}{-0.00121\sym{***}}
\newcommand{\seamenmultiplierwl}{(0.00038)}
\newcommand{\coefseamenmultiplierwl}{-0.00121\sym{***} (0.00038)}
\newcommand{\pctamenmultiplierwl}{$-$0.12\%}
\newcommand{\meanamenmultiplierwl}{0.09}
\newcommand{\obsamenmultiplierwl}{551,159}
\newcommand{\firmsamenmultiplierwl}{43,638}

\newcommand{\coeftotalnonwagecomp}{0.147\sym{***}}
\newcommand{\setotalnonwagecomp}{(0.019)}
\newcommand{\coefsetotalnonwagecomp}{0.147\sym{***} (0.019)}
\newcommand{\pcttotalnonwagecomp}{15.8\%}
\newcommand{\meantotalnonwagecomp}{8.80}
\newcommand{\obstotalnonwagecomp}{459,416}
\newcommand{\firmstotalnonwagecomp}{39,055}

\newcommand{\coeftotalothercomp}{0.151\sym{***}}
\newcommand{\setotalothercomp}{(0.021)}
\newcommand{\coefsetotalothercomp}{0.151\sym{***} (0.021)}
\newcommand{\pcttotalothercomp}{16.3\%}
\newcommand{\meantotalothercomp}{8.36}
\newcommand{\obstotalothercomp}{134,725}
\newcommand{\firmstotalothercomp}{15,334}

\newcommand{\coeftranstravelperworker}{0.012}
\newcommand{\setranstravelperworker}{(0.016)}
\newcommand{\coefsetranstravelperworker}{0.012 (0.016)}
\newcommand{\pcttranstravelperworker}{1.2\%}
\newcommand{\meantranstravelperworker}{7.11}
\newcommand{\obstranstravelperworker}{126,509}
\newcommand{\firmstranstravelperworker}{14,605}

\newcommand{\coefcarperworker}{0.018\sym{**}}
\newcommand{\secarperworker}{(0.008)}
\newcommand{\coefsecarperworker}{0.018\sym{**} (0.008)}
\newcommand{\pctcarperworker}{1.8\%}
\newcommand{\meancarperworker}{8.06}
\newcommand{\obscarperworker}{276,402}
\newcommand{\firmscarperworker}{27,223}

\newcommand{\coefelectrocommsperworker}{-0.002}
\newcommand{\seelectrocommsperworker}{(0.002)}
\newcommand{\coefseelectrocommsperworker}{-0.002 (0.002)}
\newcommand{\pctelectrocommsperworker}{$-$0.18\%}
\newcommand{\meanelectrocommsperworker}{5.95}
\newcommand{\obselectrocommsperworker}{403,610}
\newcommand{\firmselectrocommsperworker}{35,108}

\newcommand{\coeftotalcompperworker}{0.032\sym{***}}
\newcommand{\setotalcompperworker}{(0.003)}
\newcommand{\coefsetotalcompperworker}{0.032\sym{***} (0.003)}
\newcommand{\pcttotalcompperworker}{3.3\%}
\newcommand{\meantotalcompperworker}{12.87}
\newcommand{\obstotalcompperworker}{551,289}
\newcommand{\firmstotalcompperworker}{43,641}

\newcommand{\coefhires}{0.190\sym{***}}
\newcommand{\sehires}{(0.023)}
\newcommand{\coefsehires}{0.190\sym{***} (0.023)}
\newcommand{\pcthires}{20.9\%}
\newcommand{\meanhires}{0.15}
\newcommand{\obshires}{554,967}
\newcommand{\firmshires}{43,880}

\newcommand{\coefretentions}{-0.053\sym{***}}
\newcommand{\seretentions}{(0.006)}
\newcommand{\coefseretentions}{-0.053\sym{***} (0.006)}
\newcommand{\pctretentions}{$-$5.1\%}
\newcommand{\meanretentions}{0.88}
\newcommand{\obsretentions}{554,967}
\newcommand{\firmsretentions}{43,880}

\newcommand{\coefsharenonwagecost}{-0.000}
\newcommand{\sesharenonwagecost}{(0.000)}
\newcommand{\coefsesharenonwagecost}{-0.000 (0.000)}
\newcommand{\ppsharenonwagecost}{-0.0}
\newcommand{\meansharenonwagecost}{0.06}
\newcommand{\obssharenonwagecost}{551,339}
\newcommand{\firmssharenonwagecost}{43,643}

\newcommand{\coefsharenonwagewage}{-0.000}
\newcommand{\sesharenonwagewage}{(0.000)}
\newcommand{\coefsesharenonwagewage}{-0.000 (0.000)}
\newcommand{\ppsharenonwagewage}{-0.0}
\newcommand{\meansharenonwagewage}{0.07}
\newcommand{\obssharenonwagewage}{551,159}
\newcommand{\firmssharenonwagewage}{43,638}

\newcommand{\coefprobnonwagecost}{0.008\sym{***}}
\newcommand{\seprobnonwagecost}{(0.003)}
\newcommand{\coefseprobnonwagecost}{0.008\sym{***} (0.003)}
\newcommand{\ppprobnonwagecost}{0.8}
\newcommand{\meanprobnonwagecost}{0.91}
\newcommand{\obsprobnonwagecost}{551,339}
\newcommand{\firmsprobnonwagecost}{43,643}

\newcommand{\coefprobexternallabor}{0.006\sym{***}}
\newcommand{\seprobexternallabor}{(0.002)}
\newcommand{\coefseprobexternallabor}{0.006\sym{***} (0.002)}
\newcommand{\ppprobexternallabor}{0.6}
\newcommand{\meanprobexternallabor}{0.46}
\newcommand{\obsprobexternallabor}{551,339}
\newcommand{\firmsprobexternallabor}{43,643}

\newcommand{\coeffulltimestayers}{0.011\sym{***}}
\newcommand{\sefulltimestayers}{(0.001)}
\newcommand{\coefsefulltimestayers}{0.011\sym{***} (0.001)}
\newcommand{\ppfulltimestayers}{1.1}
\newcommand{\meanfulltimestayers}{0.91}
\newcommand{\obsfulltimestayers}{534,367}
\newcommand{\firmsfulltimestayers}{42,336}

\newcommand{\coeffulltimehires}{0.025\sym{***}}
\newcommand{\sefulltimehires}{(0.005)}
\newcommand{\coefsefulltimehires}{0.025\sym{***} (0.005)}
\newcommand{\ppfulltimehires}{2.5}
\newcommand{\meanfulltimehires}{0.60}
\newcommand{\obsfulltimehires}{405,415}
\newcommand{\firmsfulltimehires}{39,977}

\newcommand{\coeffixedstayers}{0.001}
\newcommand{\sefixedstayers}{(0.002)}
\newcommand{\coefsefixedstayers}{0.001 (0.002)}
\newcommand{\ppfixedstayers}{0.1}
\newcommand{\meanfixedstayers}{0.67}
\newcommand{\obsfixedstayers}{534,367}
\newcommand{\firmsfixedstayers}{42,336}

\newcommand{\coeffixedhires}{-0.005}
\newcommand{\sefixedhires}{(0.004)}
\newcommand{\coefsefixedhires}{-0.005 (0.004)}
\newcommand{\ppfixedhires}{$-$0.5}
\newcommand{\meanfixedhires}{0.41}
\newcommand{\obsfixedhires}{405,415}
\newcommand{\firmsfixedhires}{39,977}

\newcommand{\coefbonusstayers}{0.004\sym{**}}
\newcommand{\sebonusstayers}{(0.002)}
\newcommand{\coefsebonusstayers}{0.004\sym{**} (0.002)}
\newcommand{\ppbonusstayers}{0.4}
\newcommand{\meanbonusstayers}{0.04}
\newcommand{\obsbonusstayers}{534,367}
\newcommand{\firmsbonusstayers}{42,336}

\newcommand{\coefbonushires}{0.003}
\newcommand{\sebonushires}{(0.002)}
\newcommand{\coefsebonushires}{0.003 (0.002)}
\newcommand{\ppbonushires}{0.3}
\newcommand{\meanbonushires}{0.03}
\newcommand{\obsbonushires}{405,415}
\newcommand{\firmsbonushires}{39,977}

\newcommand{\coefinkindstayers}{0.008\sym{***}}
\newcommand{\seinkindstayers}{(0.002)}
\newcommand{\coefseinkindstayers}{0.008\sym{***} (0.002)}
\newcommand{\ppinkindstayers}{0.8}
\newcommand{\meaninkindstayers}{0.51}
\newcommand{\obsinkindstayers}{534,367}
\newcommand{\firmsinkindstayers}{42,336}

\newcommand{\coefinkindhires}{0.014\sym{***}}
\newcommand{\seinkindhires}{(0.005)}
\newcommand{\coefseinkindhires}{0.014\sym{***} (0.005)}
\newcommand{\ppinkindhires}{1.4}
\newcommand{\meaninkindhires}{0.29}
\newcommand{\obsinkindhires}{405,415}
\newcommand{\firmsinkindhires}{39,977}

\newcommand{\coefjobads}{0.026\sym{***}}
\newcommand{\sejobads}{(0.007)}
\newcommand{\coefsejobads}{0.026\sym{***} (0.007)}
\newcommand{\ppjobads}{2.6}
\newcommand{\meanjobads}{0.04}
\newcommand{\obsjobads}{554,967}
\newcommand{\firmsjobads}{43,880}

\newcommand{\coefseparations}{-0.005\sym{***}}
\newcommand{\seseparations}{(0.001)}
\newcommand{\coefseseparations}{-0.005\sym{***} (0.001)}
\newcommand{\ppseparations}{$-$0.5}
\newcommand{\meanseparations}{0.05}
\newcommand{\obsseparations}{476,749}
\newcommand{\firmsseparations}{43,878}

\newcommand{\coefgoodfinance}{0.005}
\newcommand{\segoodfinance}{(0.004)}
\newcommand{\coefsegoodfinance}{0.005 (0.004)}
\newcommand{\ppgoodfinance}{0.5}
\newcommand{\meangoodfinance}{0.13}
\newcommand{\obsgoodfinance}{150,461}
\newcommand{\firmsgoodfinance}{21,114}

\newcommand{\coefgoodcareer}{0.023\sym{***}}
\newcommand{\segoodcareer}{(0.006)}
\newcommand{\coefsegoodcareer}{0.023\sym{***} (0.006)}
\newcommand{\ppgoodcareer}{2.3}
\newcommand{\meangoodcareer}{0.25}
\newcommand{\obsgoodcareer}{150,461}
\newcommand{\firmsgoodcareer}{21,114}

\newcommand{\coefgoodworkplace}{0.022\sym{***}}
\newcommand{\segoodworkplace}{(0.007)}
\newcommand{\coefsegoodworkplace}{0.022\sym{***} (0.007)}
\newcommand{\ppgoodworkplace}{2.2}
\newcommand{\meangoodworkplace}{0.29}
\newcommand{\obsgoodworkplace}{150,461}
\newcommand{\firmsgoodworkplace}{21,114}

\newcommand{\coefaddedperks}{0.011}
\newcommand{\seaddedperks}{(0.010)}
\newcommand{\coefseaddedperks}{0.011 (0.010)}
\newcommand{\ppaddedperks}{1.1}
\newcommand{\meanaddedperks}{0.20}
\newcommand{\obsaddedperks}{150,461}
\newcommand{\firmsaddedperks}{21,114}

\newcommand{\coefcareer}{0.004}
\newcommand{\secareer}{(0.005)}
\newcommand{\coefsecareer}{0.004 (0.005)}
\newcommand{\ppcareer}{0.4}
\newcommand{\meancareer}{0.08}
\newcommand{\obscareer}{150,461}
\newcommand{\firmscareer}{21,114}

\newcommand{\coeftraining}{0.024\sym{***}}
\newcommand{\setraining}{(0.005)}
\newcommand{\coefsetraining}{0.024\sym{***} (0.005)}
\newcommand{\pptraining}{2.4}
\newcommand{\meantraining}{0.24}
\newcommand{\obstraining}{150,461}
\newcommand{\firmstraining}{21,114}

\newcommand{\coefconvenient}{-0.017\sym{***}}
\newcommand{\seconvenient}{(0.006)}
\newcommand{\coefseconvenient}{-0.017\sym{***} (0.006)}
\newcommand{\ppconvenient}{$-$1.7}
\newcommand{\meanconvenient}{0.36}
\newcommand{\obsconvenient}{150,461}
\newcommand{\firmsconvenient}{21,114}

\newcommand{\coefhomeoffice}{-0.0002}
\newcommand{\sehomeoffice}{(0.0008)}
\newcommand{\coefsehomeoffice}{-0.0002 (0.0008)}
\newcommand{\pphomeoffice}{-0.0}
\newcommand{\meanhomeoffice}{0.00}
\newcommand{\obshomeoffice}{150,461}
\newcommand{\firmshomeoffice}{21,114}

\newcommand{\coefextenvalueadded}{0.260\sym{***}}
\newcommand{\seextenvalueadded}{(0.017)}
\newcommand{\coefseextenvalueadded}{0.260\sym{***} (0.017)}
\newcommand{\pctextenvalueadded}{29.6\%}
\newcommand{\meanextenvalueadded}{15.50}
\newcommand{\obsextenvalueadded}{3,817,030}
\newcommand{\firmsextenvalueadded}{77,385}

\newcommand{\coefextenemployment}{0.161\sym{***}}
\newcommand{\seextenemployment}{(0.012)}
\newcommand{\coefseextenemployment}{0.161\sym{***} (0.012)}
\newcommand{\pctextenemployment}{17.5\%}
\newcommand{\meanextenemployment}{2.31}
\newcommand{\obsextenemployment}{3,817,030}
\newcommand{\firmsextenemployment}{77,385}

\newcommand{\coefextenwagestayers}{0.044\sym{***}}
\newcommand{\seextenwagestayers}{(0.003)}
\newcommand{\coefseextenwagestayers}{0.044\sym{***} (0.003)}
\newcommand{\pctextenwagestayers}{4.5\%}
\newcommand{\meanextenwagestayers}{5.48}
\newcommand{\obsextenwagestayers}{3,817,030}
\newcommand{\firmsextenwagestayers}{77,385}

\newcommand{\coefextentotalcompperworker}{0.036\sym{***}}
\newcommand{\seextentotalcompperworker}{(0.002)}
\newcommand{\coefseextentotalcompperworker}{0.036\sym{***} (0.002)}
\newcommand{\pctextentotalcompperworker}{3.7\%}
\newcommand{\meanextentotalcompperworker}{12.85}
\newcommand{\obsextentotalcompperworker}{3,715,051}
\newcommand{\firmsextentotalcompperworker}{77,050}

\newcommand{\coefextenamenmultiplierwl}{-0.00047\sym{*}}
\newcommand{\seextenamenmultiplierwl}{(0.00024)}
\newcommand{\coefseextenamenmultiplierwl}{-0.00047\sym{*} (0.00024)}
\newcommand{\pctextenamenmultiplierwl}{$-$0.05\%}
\newcommand{\meanextenamenmultiplierwl}{0.09}
\newcommand{\obsextenamenmultiplierwl}{3,714,409}
\newcommand{\firmsextenamenmultiplierwl}{77,045}

\newcommand{\coefextenwagecost}{0.355\sym{***}}
\newcommand{\seextenwagecost}{(0.025)}
\newcommand{\coefseextenwagecost}{0.355\sym{***} (0.025)}
\newcommand{\pctextenwagecost}{42.7\%}
\newcommand{\meanextenwagecost}{17.30}
\newcommand{\obsextenwagecost}{3,817,030}
\newcommand{\firmsextenwagecost}{77,385}

\newcommand{\coefextenwagecostperworker}{0.194\sym{***}}
\newcommand{\seextenwagecostperworker}{(0.013)}
\newcommand{\coefseextenwagecostperworker}{0.194\sym{***} (0.013)}
\newcommand{\pctextenwagecostperworker}{21.5\%}
\newcommand{\meanextenwagecostperworker}{14.98}
\newcommand{\obsextenwagecostperworker}{3,817,030}
\newcommand{\firmsextenwagecostperworker}{77,385}

\newcommand{\coefextennonwagecost}{0.204\sym{***}}
\newcommand{\seextennonwagecost}{(0.011)}
\newcommand{\coefseextennonwagecost}{0.204\sym{***} (0.011)}
\newcommand{\pctextennonwagecost}{22.7\%}
\newcommand{\meanextennonwagecost}{12.10}
\newcommand{\obsextennonwagecost}{3,669,504}
\newcommand{\firmsextennonwagecost}{76,834}

\newcommand{\coefextennonwageperworker}{0.042\sym{***}}
\newcommand{\seextennonwageperworker}{(0.004)}
\newcommand{\coefseextennonwageperworker}{0.042\sym{***} (0.004)}
\newcommand{\pctextennonwageperworker}{4.3\%}
\newcommand{\meanextennonwageperworker}{9.76}
\newcommand{\obsextennonwageperworker}{3,669,504}
\newcommand{\firmsextennonwageperworker}{76,834}

\newcommand{\coefextenpenspremcost}{0.182\sym{***}}
\newcommand{\seextenpenspremcost}{(0.015)}
\newcommand{\coefseextenpenspremcost}{0.182\sym{***} (0.015)}
\newcommand{\pctextenpenspremcost}{20.0\%}
\newcommand{\meanextenpenspremcost}{11.48}
\newcommand{\obsextenpenspremcost}{1,848,444}
\newcommand{\firmsextenpenspremcost}{55,016}

\newcommand{\coefextenpenspremperworker}{0.019\sym{***}}
\newcommand{\seextenpenspremperworker}{(0.004)}
\newcommand{\coefseextenpenspremperworker}{0.019\sym{***} (0.004)}
\newcommand{\pctextenpenspremperworker}{1.9\%}
\newcommand{\meanextenpenspremperworker}{8.97}
\newcommand{\obsextenpenspremperworker}{1,848,444}
\newcommand{\firmsextenpenspremperworker}{55,016}

\newcommand{\coefextengoodfinance}{0.004\sym{**}}
\newcommand{\seextengoodfinance}{(0.002)}
\newcommand{\coefseextengoodfinance}{0.004\sym{**} (0.002)}
\newcommand{\ppextengoodfinance}{0.4}
\newcommand{\meanextengoodfinance}{0.11}
\newcommand{\obsextengoodfinance}{909,697}
\newcommand{\firmsextengoodfinance}{35,426}

\newcommand{\coefextengoodcareer}{0.007\sym{***}}
\newcommand{\seextengoodcareer}{(0.002)}
\newcommand{\coefseextengoodcareer}{0.007\sym{***} (0.002)}
\newcommand{\ppextengoodcareer}{0.7}
\newcommand{\meanextengoodcareer}{0.23}
\newcommand{\obsextengoodcareer}{909,697}
\newcommand{\firmsextengoodcareer}{35,426}

\newcommand{\coefextengoodworkplace}{0.010\sym{***}}
\newcommand{\seextengoodworkplace}{(0.003)}
\newcommand{\coefseextengoodworkplace}{0.010\sym{***} (0.003)}
\newcommand{\ppextengoodworkplace}{1.0}
\newcommand{\meanextengoodworkplace}{0.26}
\newcommand{\obsextengoodworkplace}{909,697}
\newcommand{\firmsextengoodworkplace}{35,426}

\newcommand{\coefextenaddedperks}{0.004\sym{*}}
\newcommand{\seextenaddedperks}{(0.003)}
\newcommand{\coefseextenaddedperks}{0.004\sym{*} (0.003)}
\newcommand{\ppextenaddedperks}{0.4}
\newcommand{\meanextenaddedperks}{0.17}
\newcommand{\obsextenaddedperks}{909,697}
\newcommand{\firmsextenaddedperks}{35,426}

\newcommand{\coefextencareer}{0.004\sym{**}}
\newcommand{\seextencareer}{(0.001)}
\newcommand{\coefseextencareer}{0.004\sym{**} (0.001)}
\newcommand{\ppextencareer}{0.4}
\newcommand{\meanextencareer}{0.07}
\newcommand{\obsextencareer}{909,697}
\newcommand{\firmsextencareer}{35,426}

\newcommand{\coefextentraining}{0.007\sym{***}}
\newcommand{\seextentraining}{(0.002)}
\newcommand{\coefseextentraining}{0.007\sym{***} (0.002)}
\newcommand{\ppextentraining}{0.7}
\newcommand{\meanextentraining}{0.22}
\newcommand{\obsextentraining}{909,697}
\newcommand{\firmsextentraining}{35,426}

\newcommand{\coefextenhomeoffice}{-0.0001}
\newcommand{\seextenhomeoffice}{(0.0001)}
\newcommand{\coefseextenhomeoffice}{-0.0001 (0.0001)}
\newcommand{\ppextenhomeoffice}{-0.0}
\newcommand{\meanextenhomeoffice}{0.00}
\newcommand{\obsextenhomeoffice}{909,697}
\newcommand{\firmsextenhomeoffice}{35,426}

\newcommand{\coefextenpermanent}{0.004\sym{**}}
\newcommand{\seextenpermanent}{(0.002)}
\newcommand{\coefseextenpermanent}{0.004\sym{**} (0.002)}
\newcommand{\ppextenpermanent}{0.4}
\newcommand{\meanextenpermanent}{0.84}
\newcommand{\obsextenpermanent}{909,697}
\newcommand{\firmsextenpermanent}{35,426}

\newcommand{\coefextenconvenient}{-0.011\sym{***}}
\newcommand{\seextenconvenient}{(0.002)}
\newcommand{\coefseextenconvenient}{-0.011\sym{***} (0.002)}
\newcommand{\ppextenconvenient}{$-$1.1}
\newcommand{\meanextenconvenient}{0.24}
\newcommand{\obsextenconvenient}{909,697}
\newcommand{\firmsextenconvenient}{35,426}

\newcommand{\coefextenregular}{0.009\sym{***}}
\newcommand{\seextenregular}{(0.002)}
\newcommand{\coefseextenregular}{0.009\sym{***} (0.002)}
\newcommand{\ppextenregular}{0.9}
\newcommand{\meanextenregular}{0.63}
\newcommand{\obsextenregular}{909,697}
\newcommand{\firmsextenregular}{35,426}

\newcommand{\coefextenpayinkindstayers}{0.0634\sym{***}}
\newcommand{\seextenpayinkindstayers}{(0.0067)}
\newcommand{\coefseextenpayinkindstayers}{0.0634\sym{***} (0.0067)}
\newcommand{\pctextenpayinkindstayers}{6.6\%}
\newcommand{\meanextenpayinkindstayers}{8.63}
\newcommand{\obsextenpayinkindstayers}{3,233,844}
\newcommand{\firmsextenpayinkindstayers}{70,655}

\newcommand{\coefextenshareinkindstayers}{0.013\sym{***}}
\newcommand{\seextenshareinkindstayers}{(0.001)}
\newcommand{\coefseextenshareinkindstayers}{0.013\sym{***} (0.001)}
\newcommand{\ppextenshareinkindstayers}{1.3}
\newcommand{\meanextenshareinkindstayers}{0.63}
\newcommand{\obsextenshareinkindstayers}{3,817,030}
\newcommand{\firmsextenshareinkindstayers}{77,385}

\newcommand{\elasw}{3.63}
\newcommand{\elasal}{3.67}
\newcommand{\elasA}{-0.003}
\newcommand{\Abar}{1.09}
\newcommand{\markdownAM}{0.786}
\newcommand{\markdownAMtextbook}{0.784}
\newcommand{\markdownM}{0.718}
\newcommand{\correctionpp}{0.2}
\newcommand{\deepeningpp}{6.8}
\newcommand{\elasrs}{0.171}

\newcommand{\elaswredu}{4.68}
\newcommand{\elasalredu}{4.85}
\newcommand{\elasAredu}{-0.008}
\newcommand{\Abarredu}{1.09}
\newcommand{\markdownAMredu}{0.829}
\newcommand{\markdownAMtextbookredu}{0.824}
\newcommand{\markdownMredu}{0.758}
\newcommand{\correctionppredu}{0.5}
\newcommand{\deepeningppredu}{7.1}
\newcommand{\elasrsredu}{0.138}

\newcommand{\coefexptotalexports}{0.534\sym{***}}
\newcommand{\seexptotalexports}{(0.136)}
\newcommand{\coefseexptotalexports}{0.534\sym{***} (0.136)}
\newcommand{\pctexptotalexports}{70.6\%}
\newcommand{\meanexptotalexports}{14.61}
\newcommand{\obsexptotalexports}{32,568}
\newcommand{\firmsexptotalexports}{3,867}

\newcommand{\coefexpwagestayers}{0.074\sym{***}}
\newcommand{\seexpwagestayers}{(0.016)}
\newcommand{\coefseexpwagestayers}{0.074\sym{***} (0.016)}
\newcommand{\pctexpwagestayers}{7.7\%}
\newcommand{\meanexpwagestayers}{5.44}
\newcommand{\obsexpwagestayers}{36,759}
\newcommand{\firmsexpwagestayers}{3,957}

\newcommand{\coefexpemployment}{0.138\sym{***}}
\newcommand{\seexpemployment}{(0.035)}
\newcommand{\coefseexpemployment}{0.138\sym{***} (0.035)}
\newcommand{\pctexpemployment}{14.8\%}
\newcommand{\meanexpemployment}{2.84}
\newcommand{\obsexpemployment}{36,759}
\newcommand{\firmsexpemployment}{3,957}

\newcommand{\coefexpvalueadded}{0.293\sym{***}}
\newcommand{\seexpvalueadded}{(0.052)}
\newcommand{\coefseexpvalueadded}{0.293\sym{***} (0.052)}
\newcommand{\pctexpvalueadded}{34.0\%}
\newcommand{\meanexpvalueadded}{16.18}
\newcommand{\obsexpvalueadded}{36,759}
\newcommand{\firmsexpvalueadded}{3,957}

\newcommand{\coefexptotalcompperworker}{0.080\sym{***}}
\newcommand{\seexptotalcompperworker}{(0.024)}
\newcommand{\coefseexptotalcompperworker}{0.080\sym{***} (0.024)}
\newcommand{\pctexptotalcompperworker}{8.3\%}
\newcommand{\meanexptotalcompperworker}{13.08}
\newcommand{\obsexptotalcompperworker}{33,475}
\newcommand{\firmsexptotalcompperworker}{3,933}

\newcommand{\coefexpamenmultiplierwl}{-0.00542}
\newcommand{\seexpamenmultiplierwl}{(0.00589)}
\newcommand{\coefseexpamenmultiplierwl}{-0.00542 (0.00589)}
\newcommand{\pctexpamenmultiplierwl}{$-$0.54\%}
\newcommand{\meanexpamenmultiplierwl}{0.14}
\newcommand{\obsexpamenmultiplierwl}{33,463}
\newcommand{\firmsexpamenmultiplierwl}{3,933}

\newcommand{\elaswexp}{1.86}
\newcommand{\elasalexp}{2.01}
\newcommand{\elasAexp}{-0.039}
\newcommand{\markdownAMexp}{0.667}
\newcommand{\markdownAMtextbookexp}{0.651}
\newcommand{\correctionppexp}{1.7}
\newcommand{\coefexpemploymentraw}{0.138}
\newcommand{\coefexpwagestayersraw}{0.074}
\newcommand{\coefexpamenmultiplierwlraw}{-0.00542}
\newcommand{\Abarexp}{1.15}
\newcommand{\markdownMexp}{0.580}
\newcommand{\deepeningppexp}{8.7}
\newcommand{\elasrsexp}{0.254}

\newcommand{\coefexpintvalueadded}{0.304\sym{***}}
\newcommand{\seexpintvalueadded}{(0.018)}
\newcommand{\coefseexpintvalueadded}{0.304\sym{***} (0.018)}
\newcommand{\pctexpintvalueadded}{35.5\%}
\newcommand{\meanexpintvalueadded}{16.18}
\newcommand{\obsexpintvalueadded}{36,759}
\newcommand{\firmsexpintvalueadded}{3,957}

\newcommand{\coefexpintemployment}{0.147\sym{***}}
\newcommand{\seexpintemployment}{(0.015)}
\newcommand{\coefseexpintemployment}{0.147\sym{***} (0.015)}
\newcommand{\pctexpintemployment}{15.9\%}
\newcommand{\meanexpintemployment}{2.84}
\newcommand{\obsexpintemployment}{36,759}
\newcommand{\firmsexpintemployment}{3,957}

\newcommand{\coefexpintwagestayers}{0.049\sym{***}}
\newcommand{\seexpintwagestayers}{(0.006)}
\newcommand{\coefseexpintwagestayers}{0.049\sym{***} (0.006)}
\newcommand{\pctexpintwagestayers}{5.0\%}
\newcommand{\meanexpintwagestayers}{5.44}
\newcommand{\obsexpintwagestayers}{36,759}
\newcommand{\firmsexpintwagestayers}{3,957}

\newcommand{\coefexpintamenmultiplierwl}{-0.00113}
\newcommand{\seexpintamenmultiplierwl}{(0.00131)}
\newcommand{\coefseexpintamenmultiplierwl}{-0.00113 (0.00131)}
\newcommand{\pctexpintamenmultiplierwl}{$-$0.11\%}
\newcommand{\meanexpintamenmultiplierwl}{0.14}
\newcommand{\obsexpintamenmultiplierwl}{33,463}
\newcommand{\firmsexpintamenmultiplierwl}{3,933}

\newcommand{\elaswexpint}{3.02}
\newcommand{\elasalexpint}{3.09}
\newcommand{\elasAexpint}{-0.008}
\newcommand{\Abarexpint}{1.15}
\newcommand{\markdownAMexpint}{0.756}
\newcommand{\markdownAMtextbookexpint}{0.751}
\newcommand{\markdownMexpint}{0.657}
\newcommand{\correctionppexpint}{0.4}
\newcommand{\deepeningppexpint}{9.9}
\newcommand{\elasrsexpint}{0.160}

\newcommand{\coefprocwagestayers}{0.012\sym{***}}
\newcommand{\seprocwagestayers}{(0.003)}
\newcommand{\coefseprocwagestayers}{0.012\sym{***} (0.003)}
\newcommand{\pctprocwagestayers}{1.3\%}
\newcommand{\meanprocwagestayers}{5.70}
\newcommand{\obsprocwagestayers}{375,883}
\newcommand{\firmsprocwagestayers}{18,243}

\newcommand{\coefprocemployment}{0.079\sym{***}}
\newcommand{\seprocemployment}{(0.007)}
\newcommand{\coefseprocemployment}{0.079\sym{***} (0.007)}
\newcommand{\pctprocemployment}{8.2\%}
\newcommand{\meanprocemployment}{3.17}
\newcommand{\obsprocemployment}{375,883}
\newcommand{\firmsprocemployment}{18,243}

\newcommand{\coefprocvalueadded}{0.101\sym{***}}
\newcommand{\seprocvalueadded}{(0.010)}
\newcommand{\coefseprocvalueadded}{0.101\sym{***} (0.010)}
\newcommand{\pctprocvalueadded}{10.7\%}
\newcommand{\meanprocvalueadded}{16.65}
\newcommand{\obsprocvalueadded}{375,883}
\newcommand{\firmsprocvalueadded}{18,243}

\newcommand{\coefproctotalcompperworker}{0.003}
\newcommand{\seproctotalcompperworker}{(0.005)}
\newcommand{\coefseproctotalcompperworker}{0.003 (0.005)}
\newcommand{\pctproctotalcompperworker}{0.35\%}
\newcommand{\meanproctotalcompperworker}{13.14}
\newcommand{\obsproctotalcompperworker}{372,834}
\newcommand{\firmsproctotalcompperworker}{18,154}

\newcommand{\coefprocamenmultiplierwl}{-0.00150}
\newcommand{\seprocamenmultiplierwl}{(0.00115)}
\newcommand{\coefseprocamenmultiplierwl}{-0.00150 (0.00115)}
\newcommand{\pctprocamenmultiplierwl}{$-$0.15\%}
\newcommand{\meanprocamenmultiplierwl}{0.11}
\newcommand{\obsprocamenmultiplierwl}{372,770}
\newcommand{\firmsprocamenmultiplierwl}{18,151}

\newcommand{\elaswproc}{6.32}
\newcommand{\elasalproc}{7.19}
\newcommand{\elasAproc}{-0.019}
\newcommand{\Abarproc}{1.12}
\newcommand{\markdownAMproc}{0.878}
\newcommand{\markdownAMtextbookproc}{0.863}
\newcommand{\markdownMproc}{0.786}
\newcommand{\correctionppproc}{1.4}
\newcommand{\deepeningppproc}{9.1}
\newcommand{\elasrsproc}{0.123}

\newcommand{\coefprocintvalueadded}{0.301\sym{***}}
\newcommand{\seprocintvalueadded}{(0.026)}
\newcommand{\coefseprocintvalueadded}{0.301\sym{***} (0.026)}
\newcommand{\pctprocintvalueadded}{35.1\%}
\newcommand{\meanprocintvalueadded}{16.65}
\newcommand{\obsprocintvalueadded}{375,883}
\newcommand{\firmsprocintvalueadded}{18,243}

\newcommand{\coefprocintemployment}{0.208\sym{***}}
\newcommand{\seprocintemployment}{(0.019)}
\newcommand{\coefseprocintemployment}{0.208\sym{***} (0.019)}
\newcommand{\pctprocintemployment}{23.1\%}
\newcommand{\meanprocintemployment}{3.17}
\newcommand{\obsprocintemployment}{375,883}
\newcommand{\firmsprocintemployment}{18,243}

\newcommand{\coefprocintwagestayers}{0.034\sym{***}}
\newcommand{\seprocintwagestayers}{(0.002)}
\newcommand{\coefseprocintwagestayers}{0.034\sym{***} (0.002)}
\newcommand{\pctprocintwagestayers}{3.5\%}
\newcommand{\meanprocintwagestayers}{5.70}
\newcommand{\obsprocintwagestayers}{375,883}
\newcommand{\firmsprocintwagestayers}{18,243}

\newcommand{\coefprocintamenmultiplierwl}{0.00126\sym{**}}
\newcommand{\seprocintamenmultiplierwl}{(0.00055)}
\newcommand{\coefseprocintamenmultiplierwl}{0.00126\sym{**} (0.00055)}
\newcommand{\pctprocintamenmultiplierwl}{0.13\%}
\newcommand{\meanprocintamenmultiplierwl}{0.11}
\newcommand{\obsprocintamenmultiplierwl}{372,770}
\newcommand{\firmsprocintamenmultiplierwl}{18,151}

\newcommand{\elaswprocint}{6.09}
\newcommand{\elasalprocint}{5.87}
\newcommand{\elasAprocint}{0.006}
\newcommand{\Abarprocint}{1.12}
\newcommand{\markdownAMprocint}{0.854}
\newcommand{\markdownAMtextbookprocint}{0.859}
\newcommand{\markdownMprocint}{0.765}
\newcommand{\correctionppprocint}{-0.4}
\newcommand{\deepeningppprocint}{8.9}
\newcommand{\elasrsprocint}{0.113}

\newcommand{\shareexpagri}{1.1}
\newcommand{\shareexpmining}{2.3}
\newcommand{\shareexpmanuf}{41.5}
\newcommand{\shareexpelec}{0.1}
\newcommand{\shareexpwater}{0.9}
\newcommand{\shareexpconstr}{3.1}
\newcommand{\shareexptrade}{37.6}
\newcommand{\shareexptransp}{4.2}
\newcommand{\shareexpinfo}{2.5}
\newcommand{\shareexprealestate}{0.4}
\newcommand{\shareexpprof}{3.8}
\newcommand{\shareexprest}{2.7}
\newcommand{\shareintagri}{1.4}
\newcommand{\shareintmining}{0.5}
\newcommand{\shareintmanuf}{9.6}
\newcommand{\shareintelec}{0.4}
\newcommand{\shareintwater}{0.4}
\newcommand{\shareintconstr}{17.5}
\newcommand{\shareinttrade}{28.0}
\newcommand{\shareinttransp}{4.8}
\newcommand{\shareintaccom}{5.0}
\newcommand{\shareintinfo}{4.5}
\newcommand{\shareintfinance}{0.0}
\newcommand{\shareintrealestate}{3.1}
\newcommand{\shareintprof}{10.6}
\newcommand{\shareintrest}{14.0}
\newcommand{\shareprocagri}{1.0}
\newcommand{\shareprocmining}{0.9}
\newcommand{\shareprocmanuf}{9.9}
\newcommand{\shareprocelec}{0.7}
\newcommand{\shareprocwater}{1.1}
\newcommand{\shareprocconstr}{26.1}
\newcommand{\shareproctrade}{19.8}
\newcommand{\shareproctransp}{5.2}
\newcommand{\shareprocaccom}{1.7}
\newcommand{\shareprocinfo}{8.1}
\newcommand{\shareprocrealestate}{0.6}
\newcommand{\shareprocprof}{12.8}
\newcommand{\shareprocrest}{12.2}

\newcommand{\paEtaHat}{5.5}
\newcommand{\paEtaHatTwo}{5.51}
\newcommand{\paEtaLo}{4.6}
\newcommand{\paEtaHi}{7.1}
\newcommand{\paRhoHat}{0.060}
\newcommand{\paRhoLo}{0.036}
\newcommand{\paRhoHi}{0.091}
\newcommand{\paEtaAcc}{4.2}
\newcommand{\paRhoAcc}{0.067}
\newcommand{\paEtaAccTwo}{4.22}

\newcommand{\paAmBot}{1.08}
\newcommand{\paAmTop}{1.14}
\newcommand{\paAdBot}{1.50}
\newcommand{\paAdTop}{1.55}
\newcommand{\paThetaPct}{38}
\newcommand{\paPhiBot}{-0.38}
\newcommand{\paPhiTop}{0.05}
\newcommand{\paExclBot}{seven}
\newcommand{\paAMBot}{0.64}
\newcommand{\paAMTop}{0.88}
\newcommand{\paAMBotPct}{64}
\newcommand{\paAMTopPct}{88}
\newcommand{\paMBot}{0.42}
\newcommand{\paMTop}{0.57}
\newcommand{\paGapLo}{22}
\newcommand{\paGapHi}{31}
\newcommand{\paGapMLo}{6}
\newcommand{\paGapMHi}{11}
\newcommand{\paCondLo}{0.84}
\newcommand{\paCondHi}{0.88}
\newcommand{\paCondPctLo}{12}
\newcommand{\paCondPctHi}{16}

\newcommand{\paEtaHatEmp}{6.1}
\newcommand{\paEtaHatEmpTwo}{6.13}
\newcommand{\paEtaLoEmp}{4.9}
\newcommand{\paEtaHiEmp}{15}
\newcommand{\paRhoHatEmp}{0.047}
\newcommand{\paRhoLoEmp}{0.015}
\newcommand{\paRhoHiEmp}{0.085}
\newcommand{\paEtaAccEmpTwo}{4.66}
\newcommand{\paRhoAccEmp}{0.043}
\newcommand{\paPhiBotEmp}{-0.36}
\newcommand{\paPhiTopEmp}{0.04}
\newcommand{\paAMBotEmp}{0.66}
\newcommand{\paAMTopEmp}{0.89}
\newcommand{\paMBotEmp}{0.43}
\newcommand{\paMTopEmp}{0.58}

\newcommand{\paAttrVal}{0.421}
\newcommand{\paAttrPct}{42}
\newcommand{\paValPerm}{0.210}
\newcommand{\paValReg}{0.099}
\newcommand{\paWtpColl}{0.094}
\newcommand{\paWtpHome}{0.065}
\newcommand{\paWtpIA}{0.173}
\newcommand{\paAblFlex}{0.080}
\newcommand{\paAblShift}{$-$0.171}
\newcommand{\paNuReg}{0.171}
\newcommand{\paNuFlex}{0.251}
\newcommand{\paPermPP}{0.4}

\newcommand{\paTabAMM}{0.786}
\newcommand{\paTabAM}{0.788}
\newcommand{\paTabMM}{0.718}
\newcommand{\paTabM}{0.520}
\newcommand{\paTabGapMM}{6.8}
\newcommand{\paTabGap}{26.8}
\newcommand{\paTabGapMMInt}{7}
\newcommand{\paTabGapInt}{27}
\newcommand{\paTabCorrM}{0.2}
\newcommand{\paTabCorr}{0.4}
\newcommand{\paTabMacc}{0.720}
\newcommand{\paTabGapAcc}{6.8}

\newcommand{\paTabAMMRedu}{0.829}
\newcommand{\paTabAMRedu}{0.830}
\newcommand{\paTabMMRedu}{0.758}
\newcommand{\paTabMRedu}{0.547}
\newcommand{\paTabGapMMRedu}{7.1}
\newcommand{\paTabGapRedu}{28.2}
\newcommand{\paTabGapMMIntRedu}{7}
\newcommand{\paTabGapIntRedu}{28}
\newcommand{\paTabCorrMRedu}{0.5}
\newcommand{\paTabCorrRedu}{0.6}
\newcommand{\paTabMaccRedu}{0.758}
\newcommand{\paTabGapAccRedu}{7.1}

\newcommand{\paTabAMMProcint}{0.854}
\newcommand{\paTabAMProcint}{0.857}
\newcommand{\paTabMMProcint}{0.765}
\newcommand{\paTabMProcint}{0.558}
\newcommand{\paTabGapMMProcint}{8.9}
\newcommand{\paTabGapProcint}{30.0}
\newcommand{\paTabGapMMIntProcint}{9}
\newcommand{\paTabGapIntProcint}{30}
\newcommand{\paTabCorrMProcint}{-0.4}
\newcommand{\paTabCorrProcint}{-0.2}
\newcommand{\paTabMaccProcint}{0.768}
\newcommand{\paTabGapAccProcint}{8.9}

\newcommand{\paTabAMMProc}{0.878}
\newcommand{\paTabAMProc}{0.875}
\newcommand{\paTabMMProc}{0.786}
\newcommand{\paTabMProc}{0.569}
\newcommand{\paTabGapMMProc}{9.1}
\newcommand{\paTabGapProc}{30.6}
\newcommand{\paTabGapMMIntProc}{9}
\newcommand{\paTabGapIntProc}{31}
\newcommand{\paTabCorrMProc}{1.4}
\newcommand{\paTabCorrProc}{1.2}
\newcommand{\paTabMaccProc}{0.784}
\newcommand{\paTabGapAccProc}{9.1}

\newcommand{\paTabAMMExpint}{0.756}
\newcommand{\paTabAMExpint}{0.757}
\newcommand{\paTabMMExpint}{0.657}
\newcommand{\paTabMExpint}{0.482}
\newcommand{\paTabGapMMExpint}{9.9}
\newcommand{\paTabGapExpint}{27.5}
\newcommand{\paTabGapMMIntExpint}{10}
\newcommand{\paTabGapIntExpint}{28}
\newcommand{\paTabCorrMExpint}{0.4}
\newcommand{\paTabCorrExpint}{0.6}
\newcommand{\paTabMaccExpint}{0.658}
\newcommand{\paTabGapAccExpint}{9.9}

\newcommand{\paTabAMMExp}{0.667}
\newcommand{\paTabAMExp}{0.666}
\newcommand{\paTabMMExp}{0.580}
\newcommand{\paTabMExp}{0.424}
\newcommand{\paTabGapMMExp}{8.7}
\newcommand{\paTabGapExp}{24.2}
\newcommand{\paTabGapMMIntExp}{9}
\newcommand{\paTabGapIntExp}{24}
\newcommand{\paTabCorrMExp}{1.7}
\newcommand{\paTabCorrExp}{1.6}
\newcommand{\paTabMaccExp}{0.579}
\newcommand{\paTabGapAccExp}{8.7}

\newcommand{\paTabEMP}{-0.006}
\newcommand{\paTabAMP}{1.52}
\newcommand{\paPhiTab}{-0.09}
\newcommand{\paPhiAvg}{-0.05}

\newcommand{\paTaxAMDrop}{5.3}
\newcommand{\paTaxPhiMax}{0.03}
\newcommand{\paTaxMBot}{0.42}
\newcommand{\paTaxMTop}{0.57}
\newcommand{\paPayMBot}{0.39}
\newcommand{\paPayMTop}{0.52}
\newcommand{\paTaxElasDropPct}{12}
\newcommand{\paTaxWiden}{6.5}
\newcommand{\paTaxReval}{4.4}
\newcommand{\paTaxElasPart}{2.1}
\newcommand{\paTaxAMWithout}{0.79}
\newcommand{\paTaxAMWith}{0.73}
\newcommand{\paTaxEpsMid}{3.8}

\section{Introduction}
\label{sec:introduction}

\begin{refsection}
Firms attract and retain workers through wages and non-wage workplace amenities, such as health insurance, flexible scheduling, on-the-job training, and office space, which together determine the value of a job. A large empirical literature estimates the pass-through of firm-level demand shocks to wages \citep{card2018firms,kline2019profits,lamadon2022imperfect}, but how workplace amenities adjust to such shocks remains largely unexplored. This omission matters because, if amenities adjust endogenously alongside wages, both the measurement of employer market power and the welfare interpretation of wage differentials may change.\footnote{See, for instance, the discussion in the recent survey by \cite{caldwell2026jep}.} If part of the labor supply response to firm-level demand shocks operates through adjustments in workplace amenities, then conventional approaches aimed at quantifying wage markdowns may not accurately characterize employer market power. At the same time, whether or not growing firms adjust amenities alongside wages has direct welfare implications.

The endogenous provision of workplace amenities raises three questions that motivate our analysis. First, how do firms choose the optimal mix of wages and amenities in the presence of employer market power? For an exogenously given job value, a firm that offers better amenities can attract the same workers at a lower wage. With employer market power, however, the firm endogenously chooses the job value through the amenity-wage bundle it offers. Second, what determines the adjustments in wages and amenities in response to firm-level demand shocks? Idiosyncratic demand shocks are particularly informative about firms' adjustments because they change the hiring incentives without shifting the relative prices of amenities. With employer market power, growing firms may adjust both wages and amenities to attract workers, but such adjustments may not be proportional, creating a wedge between the pass-through of shocks to wages and to total compensation. Third, what do conventional estimates of the firm-level labor supply elasticity capture when amenities adjust alongside wages, and how can we quantify welfare-relevant measures of employer market power?

To answer these questions, we develop a monopsony framework in which firms not only have wage-setting power but also determine the provision of workplace amenities. Higher compensation---higher wages or better amenities---attracts more workers, so firms face an upward-sloping labor supply curve, as in the classical monopsony model \citep{manning2003real,manning2021monopsony}. The mix of wages and amenities offered in the compensation bundle is determined by workers' valuation of amenities and the amenity provision costs facing the firm. Amenities are multi-dimensional and may differ in the structure of their costs, which the firm pays either through foregone output or directly in money. For instance, some amenities have costs that are proportional to firm size (e.g., pension premiums), others have fixed costs (e.g., on-site facilities), and yet others exhibit congestion or coordination costs (e.g., schedule flexibility).

The heterogeneity in amenity cost structures yields rich comparative statics on how firms respond to shocks.  Crucially, firm-level demand shocks affect amenity provision through two channels in our framework: (i) a taste-cost ratio that pins amenity adjustment to the wage pass-through, and (ii) a scale adjustment that alters the amenity provision cost. Amenities with constant per-worker costs track wages; amenities with economies of scale adjust more than wages; and amenities that become costlier to provide at scale adjust less—or even move opposite to wages if provision costs are sufficiently convex. When provision costs depend on firm size, per-worker \textit{amenity costs} may not track the \textit{amenity level} the firm provides, leading to distinct predictions in our framework about amenity costs and provision. The size-dependent cost structure shapes the distribution of employer market power across firms.

In our framework, the ratio of the employment response to the wage response to a demand shock provides an \textit{effective} labor supply elasticity that incorporates the endogenous amenity responses. The welfare-relevant measure of employer market power is the \textit{total compensation markdown}---the ratio of total compensation to the marginal product---rather than the conventional wage markdown. The gap between the two markdowns is governed by an \textit{amenity cost multiplier}, which captures how total compensation costs scale relative to the wage bill. We implement our framework using Norwegian data, provide new evidence on adjustments in amenities and quantify the distribution of total compensation markdowns.

We address several methodological challenges in our empirical investigation. First, amenities are multi-dimensional and hard to measure: most data sources capture only a single margin of adjustment, yet one may expect firms to re-optimize across all margins simultaneously \citep{mas2025non}. We address this concern using Norwegian matched employer-employee data that contain (i) detailed information on contract types and in-kind compensation, which we link to (ii) administrative records on firms’ non-wage personnel costs and (iii) non-pecuniary job attributes from publicly posted job advertisements \citep{bhuller2024ads}. Second, identification requires exogenous variation in labor demand at the firm level. We exploit three distinct firm-level demand shifters: (i) an \emph{internal} shock design that uses abrupt within-firm changes in value added, as in \citet{lamadon2022imperfect}; (ii) a \emph{procurement} shock based on firms winning their first-ever public procurement contract, in the spirit of \citet{kroft2025imperfect} and \citet{volpe2024job}; and (iii) a shift-share \emph{export} demand shock that combines firm-level pre-shock exposure to country-product export markets with shifts in international import demand around the 2007--2009 Great Recession, following \citet{garin2024export} and \citet{bhuller2025wage}. The comprehensive coverage of workplace amenities in our linked dataset and the availability of firm-level demand shocks allow us to provide novel evidence on adjustments in workplace amenities along multiple margins. Third, identifying total compensation markdowns requires jointly evaluating firms’ amenity responses alongside their wage and employment adjustments. We develop a structural identification and estimation approach that recovers (i) the degree of substitutability between wages and amenities in workers' preferences and (ii) the dispersion of idiosyncratic tastes, which is the underlying source of employer market power in our model. We then show how these structural preference parameters, together with estimates of the effective labor supply and amenity cost multiplier elasticities, identify the distribution of total compensation markdowns across firms.

Our empirical evidence shows that firm-level demand shocks generate sizable responses across the multiple margins of non-wage compensation that we observe. Firms' total spending on non-wage compensation moves in the same direction as wage costs, with pension premiums and other non-wage personnel costs rising by a similar magnitude as the wage bill. By contrast, per-worker non-wage spending responses differ across components. Firms expand coverage of in-kind benefits and raise benefit amounts, offer more full-time contracts, and tilt postings toward jobs with regular work schedules, permanent contracts, on-the-job training and better workplace quality, while advertised schedule flexibility declines.

Quantifying adjustments in wages and employment across the firm distribution in response to demand shocks, we find effective labor supply elasticities ranging from 2 to 8, with employment responding more than wages. Using firms’ adjustments in non-wage personnel spending and in the observed set of valued workplace amenities, we estimate the amenity cost multiplier elasticities to be significantly negative at the bottom of the firm productivity distribution. Taken together, our evidence suggests that wages absorb a larger share of demand shocks than non-wage costs, especially among below-median firms.

To analyze how firms' endogenous amenity provision shapes employer market power, we incorporate firms’ responses in amenity provision and non-wage spending, in addition to their wage and employment adjustments, into our structural framework. Using this approach, we recover an elasticity of substitution between wages and amenities in worker preferences that is statistically greater than one but economically close to it, and a structural labor supply elasticity of $\paEtaHat$, suggesting comparatively modest dispersion of idiosyncratic tastes. Total compensation markdowns rise from roughly $\paAMBotPct$ percent of the marginal product in the bottom decile of the firm distribution to $\paAMTopPct$ percent in the top decile, while wage markdowns are substantially lower. We estimate the amenity cost structure to be convex for most firms.

An implication of our findings is that approaches inferring markdowns from the ratio of \textit{observed} wages to marginal products (e.g., the production function approach) risk attributing a large share of the compensation that workers receive in the form of non-wage benefits to employer rent extraction. The measurement challenge is less severe for elasticity-based approaches that rely on the \textit{estimated} pass-through of demand shocks: the markdowns implied by the average labor supply elasticity are close to the total compensation markdowns we estimate for the average firm. This suggests that firms’ responses to demand shocks remain informative about employer market power even when amenity provision is endogenous.

\paragraph{Related literature and contribution.} Our paper makes four key contributions. First, we develop a tractable monopsony framework in which firms jointly set wages and multi-dimensional amenities with heterogeneous, size-dependent provision costs. Second, we show that standard IV estimands based on firm-level demand shocks identify the distribution of effective labor supply elasticities---which incorporate the endogenous amenity responses---rather than the structural dispersion of idiosyncratic tastes, and provide conditions under which the distribution of welfare-relevant total compensation markdowns can be identified from the data. Third, we provide direct evidence on the pass-through of firm-level demand shocks across multiple margins of workplace amenities. Finally, we develop a structural approach that allows us to quantify the degree of substitutability between wages and amenities in worker preferences and the distribution of total compensation markdowns across firms.

Our paper relates to a nascent literature that characterizes how firms determine the optimal amenity-wage bundle in the presence of employer market power \citep[e.g.,][]{dube2022power}. We depart from existing work by allowing for multi-dimensional amenities with heterogeneous cost structures and by studying the implications of endogenous amenity provision for the measurement of employer market power.\footnote{Notably, \citet{mas2025non} considers the case of public-good amenities, which correspond to what we refer to as fixed amenities, and shows in a competitive setting that such amenities create convexity in the firm's isoprofit curve, making larger firms more likely to provide them. Some recent studies let specific non-pecuniary job attributes affect output, e.g., work hours in \citet{lachowska2026work}, managerial delegation in \citet{diporto2026careers}, and schedule flexibility in \citet{adams2023flexible}, which we refer to as production-relevant amenities. Other  studies focus on how labor market institutions shape the amenity-wage bundle under monopsony, including minimum wages \citep{dube2022power}, union bargaining power \citep{lagos2025union}, and the enforcement of mandated workplace standards \citep{colonna2025}. We allow for multi-dimensional amenities with heterogeneous cost structures and study firm adjustments to firm-level demand shocks.} The joint determination of heterogeneous-cost amenities matters because amenities with different cost structures interact through the employment margin: concave-cost amenities amplify the effective labor supply elasticity, while convex-cost amenities dampen it. These interactions have consequences for the measurement of employer market power that would be missed in a single-amenity model. Notably, the composition of the cost structure determines whether the effective labor supply elasticity exceeds or falls short of the structural taste elasticity, which is the underlying source of employer market power. Although it endogenizes the provision of multi-dimensional amenities with heterogeneous costs under employer market power, our framework remains tractable.

Our work also relates to an empirical literature measuring workplace amenities. Most studies rely on either broad survey categories of benefits or single amenity dimensions, such as safety \citep{lee2019randomized} or scheduling flexibility \citep{mas2017valuing}. \citet{mas2025non} highlights the difficulty of capturing the full scope of non-wage job attributes, and \citet{woodbury1983substitution} and \citet{woodbury1991tax} emphasize that the substitution between wages and benefits depends on the relative prices of many distinct margins, yet most empirical work observes only aggregate fringe-benefit expenditure or a single category. We tackle this concern by combining two complementary data sources from Norway. From the administrative employment registers and firm cost accounts, we construct measures of employment conditions (e.g., contract types) and non-wage personnel costs. From the near-universe of posted job ads, we measure workplace amenities as they appear to prospective job seekers, using the text-analysis approach of \citet{bhuller2024ads}. This comprehensive combination of amenity measures is, to our knowledge, novel and directly addresses the concern that single-dimension studies understate both the firm's adjustment and the worker's overall valuation of the job.

Our paper further relates to the literature on the pass-through of firm shocks to workers. The pass-through of firm-level shocks to wages is a central prediction of monopsony models \citep{kline2025labor} and has been tested extensively \citep{card2018firms, kline2019profits,lamadon2022imperfect}. Some existing evidence suggests that high-paying or more productive firms also offer better amenities  \citep{sockin2022show,roussille2025bidding,caldwell2025firm}, and that total compensation gaps between education groups exceed wage gaps \citep{maestas2023value}. However, there is limited direct evidence on how firms adjust amenity provision in response to firm-level demand shocks.\footnote{To the best of our knowledge, only two studies have considered such responses to firm-level demand shocks. \cite{kroft2025imperfect} study firms winning government procurement auctions and report finding no evidence that this affects workers' safety in the US construction industry (see their Supplemental
Appendix H). \cite{charles2022demand} find that output-price increases in US metal mining raise firm-level injury rates and safety-regulation violations, consistent with firms reallocating effort from safety toward production.} Our framework has further implications for measuring employer market power. We show that the standard approach of regressing log employment on log wages, instrumented with a demand shifter, identifies an effective labor supply elasticity that incorporates the endogenous amenity response, because firms adjust wages and amenities simultaneously after the shock and both affect labor supply. This reinterpretation is distinct from standard endogeneity concerns: even with an exogenous instrument, the IV estimand identifies a different object when amenities adjust. We provide conditions under which the welfare-relevant total compensation markdowns can be recovered from the estimates of effective labor supply elasticities and amenity cost multiplier elasticities, and quantify these markdowns in our setting.

Finally, our paper relates to recent work on why employer market power differs across firms. Recent studies focus on the role of preference heterogeneity and worker-firm sorting: wage-sensitive workers may sort into high-paying firms \citep{volpe2024job}, while richer workers may also care less about wages and sort into larger firms operating in more concentrated markets \citep{berger2026labor}. Alternatively, productive firms may pay above workers' outside options and thus face less elastic labor supply \citep{seegmiller2025valuing}, while workers with a comparative advantage at their firm have less to gain from leaving \citep{bils2025robinson}. Our framework instead points to a distinct mechanism in which differences in market power across firms reflect differences in amenity cost structures.

The remainder of the paper proceeds as follows. Section~\ref{sec:model} develops the monopsony framework with endogenous, multi-dimensional amenities, derives the equilibrium markdown decomposition, and characterizes firm responses to demand shocks. Section~\ref{sec:data_empirical} describes the Norwegian administrative data sources, our measurement of workplace amenities and non-wage compensation, the construction of the three firm-level demand shocks, and the event-study research design. Section~\ref{sec:demand_shocks} presents our empirical evidence on multi-margin amenity responses, while Section~\ref{sec:discussion} quantifies the implications for the measurement of employer market power. Section~\ref{sec:conclusion} concludes with a discussion and avenues for future work.

\section{Framework}
\label{sec:model}

We provide a monopsony model where firms jointly determine wages and a bundle of amenities. We characterize the equilibrium under endogenous amenity provision and clarify how much of the marginal product accrues to workers depending on the amenity cost structure. Finally, we use the model to derive predictions on firm responses to demand shocks.

\subsection{Workers}
\label{subsec:workers}

A population of $N$ workers with utility $U_{ij} = V_j \cdot \xi_{ij}$ chooses among a set of firms $\mathcal{J}$, where $V_j$ is the common utility component and $\xi_{ij}$ are idiosyncratic tastes. The common component is represented by constant elasticity of substitution (CES) preferences over wages $w_j$ and an $n$-dimensional vector of amenities $\mathbf{a}_j = (a_{j1}, a_{j2}, \ldots, a_{jn}) \in \mathbb{R}_+^n$.\footnote{For brevity, we abstract from taxation here, but our theoretical results extend easily to the case of a proportional tax on labor earnings. In Appendix \ref{app:taxation}, we also consider the case of progressive taxation.} Each worker $i$ thus evaluates the common wage-amenity compensation bundle offered by firm $j$ as follows:
\begin{equation*}
V_j \equiv V_j(w_j, \mathbf{a}_j) = \Bigl(w_j^\rho + \sum_{k=1}^n \gamma_k\, a_{jk}^\rho\Bigr)^{1/\rho}, \qquad \rho < 1,
\end{equation*}
where preference weights $\gamma_k > 0$ govern how strongly workers value each amenity $k$ relative to wages, while $\rho < 1$ determines the elasticity of substitution ($1/(1-\rho)>0$) between wages and amenities.\footnote{As preference weights $\gamma_k > 0$ $\forall k$, each component of $\mathbf{a}_j$ enters workers' utility with a non-negative value. Nevertheless, we can accommodate the presence of workplace disamenities (e.g., inconvenient hours or level of injury risk), as a disamenity can be redefined as the deviation from a baseline level of the amenity.} Wages and amenities can be substitutes $(\rho > 0)$ or complements $(\rho < 0)$ in the compensation bundle, while in the limiting case $\rho\to0$ the common utility index becomes log-additive in wages and amenities (i.e., Cobb--Douglas preferences).%

The idiosyncratic tastes are independent draws $\xi_{ij}\sim\text{Fréchet}(\eta,\bar{a}_j)$, where the shape parameter $\eta>0$ drives the dispersion of the taste shocks, and the firm-specific scale parameter $\bar{a}_j>0$ captures exogenous differences in firm attractiveness (e.g., location, brand, or industry appeal). Given this structure, we can derive the labor supply facing each firm as:

\begin{equation}
L_j = \frac{N\,\bar{a}_j}{\mathbb{V}^\eta}\,V_j^\eta, \qquad \mathbb{V}^\eta \equiv \sum_{l \in \mathcal{J}} \bar{a}_l V_l^\eta,
\label{eq:L_j}
\end{equation}
where $\eta$ is the structural labor supply elasticity with respect to the job value $V_j$ and $\mathbb{V}$ is an aggregate attractiveness index over all firms.  The formal derivation is in Appendix~\ref{app:microfound}.

\subsection{Firms}
\label{subsec:firms}

The firm's output depends on a homogeneous labor input, $Y(L_j) = \psi_j\, f(L_j)$, with $f'>0$ and $f''<0$ and firm-specific exogenous productivity $\psi_j > 0$. The output price is exogenous and normalized to one. Each firm $j$ is atomistic and takes $\bar{a}_j$ and $\mathbb{V}$ as given, and chooses wages and amenities $(w_j, \mathbf{a}_j)$ to maximize profit subject to its labor-supply curve~\eqref{eq:L_j}:

\begin{equation}
\max_{w_j,\mathbf{a}_j}\;\; \pi_j \;=\; \psi_j\,f(L_j) - w_j L_j - \mathbf{C}(\mathbf{a}_j, L_j), \qquad \text{s.t.}\quad L_j = \frac{N\,\bar{a}_j}{\mathbb{V}^\eta}\,V_j^\eta.
\label{eq:profit}
\end{equation}

We assume that $f(L_j)=L_j^\alpha$ with $\alpha\in(0,1)$. The amenity bundle $\mathbf{a}_j$ has a provision cost $\mathbf{C}(\mathbf{a}_j, L_j)\equiv\sum_{k=1}^n C_{jk}(a_{jk}, L_j)$. We let the cost $C_{jk}(a_{jk}, L_j)$ of amenity $k$ be linear in the level $a_{jk}$ of amenity provision, multiplied by an exogenous firm-specific unit cost factor $c_{jk}>0$, and a weakly increasing \emph{scaling function} $h_k(L_j) > 0$ governing how the per-worker amenity cost depends on firm size, such that:
\begin{equation}
C_{jk}(a_{jk}, L_j) \;=\; c_{jk}\,h_k(L_j)\,a_{jk}.
\label{eq:cost_k}
\end{equation}
Our formulation of provision costs allows for \textit{monetized amenities}, for which the firm faces a monetary cost, as well as \textit{production-relevant amenities}, which are paid through foregone output.\footnote{In Appendix Table~\ref{tab:literature}, we provide an overview of existing studies of imperfect competition with endogenous amenities, clarifying what assumptions are maintained in these studies regarding the amenity provision cost structure, and how each case compares to the cost structure specified in our framework.} The costs of monetized amenities appear in the firm's accounts, where $h_k(L_j)$ can be a function of size (e.g., capturing bulk discounts). The costs of production-relevant amenities are measured in the output that the firm forgoes  (e.g., due to the coordination or supervision that the provision of flexible hours or work from home require, or the hours a trainee spends off the line). For such amenities, we let the costs scale with the firm's output, $h_k(L_j)=g_k(f(L_j))$ for a function $g_k$, which need not be concave in $L_j$ even though $f$ is.
We parameterize the size dependence through the \emph{size-elasticity of amenity cost}
\begin{equation*}
\varepsilon_L^{h_k} \;\equiv\; \frac{L_j\,h_k'(L_j)}{h_k(L_j)},
\end{equation*}
which measures how the per-unit cost of amenity $k$ responds to employment size. We assume that $\varepsilon_L^{h_k}\geq0$ for all $k$, so that adding workers never lowers the total provision cost given $a_{jk}$.

To fix ideas, consider the iso-elastic case $h_k(L_j) = L_j^\delta$ for some $\delta \geq 0$, giving $\varepsilon_L^{h_k} = \delta$. Four canonical regimes span the range from pure ``public-good'' amenities with fixed costs to amenities with convex costs: (i) \emph{fixed costs} ($\delta = 0$, $h_k = 1$), where per-unit provision cost is independent of headcount (e.g., an on-site gym or a childcare facility); (ii) \emph{concave-in-$L_j$ costs} ($0 < \delta < 1$, $h_k = L_j^\delta$), for instance, capturing provision costs for monetized amenities with economies of scale in purchasing or contracting (e.g., bulk discounts on software licenses, negotiated rates for meal plans, group transportation passes, or employee assistance programs), and production-relevant amenities with costs that are proportional to output ($\delta=\alpha$); (iii)  \emph{per-worker costs} ($\delta = 1$, $h_k = L_j$), where provision costs scale linearly with employment (e.g., pension contributions, or individually paid travel and food compensation for staff); (iv) \emph{convex-in-$L_j$ costs} ($\delta > 1$, $h_k = L_j^\delta$), where provision becomes more costly per worker as the firm grows, due to coordination or congestion (e.g., flexible scheduling that involves pairwise negotiation over shift swaps and meeting alignment, or shared-resource amenities with capacity constraints, such as recreation areas). These regimes span a range of cost structures with different implications for per-worker amenity costs $C_{jk}/L_j$. Notably, per-worker costs decline (increase) with firm size for concave-cost (convex-cost) amenities.

\subsection{Firm Optimum and Equilibrium}
\label{subsec:foc}

We characterize the firm's solution to~\eqref{eq:profit} in Appendix~\ref{app:firm_problem}, which provides $n+1$ first-order conditions for wages $w^*_j$ and amenities $\mathbf{a}^*_j$. By combining the solutions for wages and each amenity, we derive the following $n$ MRS conditions:

\begin{equation}
\underbrace{\gamma_k \left(\frac{a_{jk}^*}{w_j^*}\right)^{\!\rho-1}}_{\text{worker's relative valuation}}
\;=\;
\underbrace{\frac{c_{jk}\, h_k(L_j)}{L_j}}_{\text{cost per worker}},
\qquad \forall\, k \in \{1, \dots, n\}.
\label{eq:MRS}
\end{equation}
The left side is the worker's marginal utility of amenity $k$ relative to wages. The right side is the firm's per-worker cost of a marginal increase in $a_{jk}$. Rearranging~\eqref{eq:MRS}, we get

\begin{equation}
\frac{a_{jk}^*}{w_j^*} \;=\; \left[\; \underbrace{\left(\frac{\gamma_k}{c_{jk}}\right)}_{\text{taste--cost ratio}} \cdot\; \underbrace{\left(\frac{L_j}{h_k(L_j)}\right)}_{\text{cost-scaling factor}} \;\right]^{\!\frac{1}{1-\rho}},
\qquad \forall\, k \in \{1, \dots, n\},
\label{eq:astar_ratio}
\end{equation}
where the optimal amenity-wage ratio is pinned down by two forces: (i) The \emph{taste--cost ratio} $(\gamma_k/c_{jk})$ captures the worker-firm dimension: higher worker valuation $\gamma_k$ raises the ratio while higher firm-specific unit cost $c_{jk}$ lowers it, so cross-firm variation in $c_{jk}$ generates variation in amenity provision at any given wage level. (ii) The \emph{cost-scaling factor} $(L_j/h_k(L_j))$ either rises, is invariant, or falls with firm size, as $\varepsilon_L^{h_k} \lessgtr 1$, capturing how amenity provision cost scales with employment. In all cases, the level of amenity provision $a_{jk}^*$ is proportional to wages $w_{j}^*$, scaled by a size-dependent factor that depends on $L_j$ but not on $w_j$.\footnote{For notational ease, we drop the superscript ``$^*$'' that denotes the firm's optimal solutions in the following discussion of the model's equilibrium, unless we find it necessary to include it to avoid confusion.}

The total compensation paid by the firm consists of the wage bill and total amenity costs, $TC_j(w_j, \mathbf{a}_j,L_j) \equiv w_j L_j + \sum_k C_{jk}(a_{jk}, L_j)$. We define the \emph{amenity cost multiplier} as the ratio of total compensation to the wage bill,

\begin{equation}
\mathcal{A}(L_j) \;\equiv\; \frac{TC_j(w_j,\mathbf{a}_j, L_j)}{w_j L_j} \;=\; 1 + \sum_k \omega_k(L_j),
\label{eq:amultiplier}
\end{equation}
so total compensation costs the firm $\mathcal{A}(L_j)$ for every unit of the wage bill, counting provision costs paid in money and in output alike. Notably, while both equalities follow from the definitions of $\mathcal{A}(L_j)$ and $\omega_k$, what is specific to our model is that the shares depend on $L_j$ alone, where $\omega_k(L_j) \equiv C_{jk}/(w_j L_j)$ is amenity $k$'s cost relative to the wage bill. Although $TC_j$, $w_j L_j$, and $C_{jk}$ all individually depend on $w_j$, their ratios do not: substituting the optimal $a_{jk}^*$ from~\eqref{eq:astar_ratio} cancels the wage, leaving them as a function of $L_j$ alone, hence the notation $\mathcal{A}(L_j)$ and $\omega_k(L_j)$.\footnote{The wage-invariance of $\mathcal{A}(L_j)$ and $\omega_k(L_j)$ relies on two features: (i) CES preferences are homothetic, so the worker's MRS depends only on the ratios $a_{jk}/w_j$; (ii) amenity provision costs are linear in $a_{jk}$ in \eqref{eq:cost_k}.} Substituting~\eqref{eq:astar_ratio} into $\omega_k$ gives
\begin{equation}
\omega_k(L_j) \;=\; \left(\frac{\gamma_k}{c_{jk}^\rho}\right)^{\frac{1}{1-\rho}} \left(\frac{L_j}{h_k(L_j)}\right)^{\frac{\rho}{1-\rho}} \;>\; 0.
\label{eq:omega_explicit}
\end{equation}
Two special cases are worth noting. First, under Cobb--Douglas preferences $(\rho=0)$, we get $\omega_k = \gamma_k$ for all $k$, implying that amenity cost shares simply reflect workers' amenity tastes. Second, if costs scale per-worker $(h_k(L_j)=L_j)$ for all $k$, then $\omega_k = (\gamma_k/c_{jk}^\rho)^{1/(1-\rho)}$, so cost shares depend on tastes and unit provision costs. In both cases, $\mathcal{A}(L_j)$ is independent of $L_j$, so the total amenity provision cost is a constant ``markup'' on the wage.

\label{subsec:equilibrium}The MRS conditions \eqref{eq:astar_ratio} pin down each amenity as a wage-proportional function of $L_j$. Multiplying the MRS condition~\eqref{eq:MRS} by $a_{jk}/w_j$ gives $\gamma_k\,a_{jk}^\rho = \omega_k(L_j)\,w_j^\rho$, so the wages and amenities contribute $w_j^\rho$ and $\sum_k\omega_k(L_j)\,w_j^\rho$ to $V_j^\rho$. The CES aggregator therefore collapses to a wage-proportional job value scaled by the amenity multiplier $\mathcal{A}(L_j) = 1+\sum_k\omega_k(L_j)$,
\begin{equation}
V_j \;=\; \mathcal{A}(L_j)^{1/\rho}\,w_j.
\label{eq:V_S}
\end{equation}
Inserting~\eqref{eq:V_S} into the labor-supply curve~\eqref{eq:L_j} eliminates the $n$ amenity choices and yields the effective labor supply facing the firm.\footnote{At $\rho=0$, \eqref{eq:V_S} becomes $V_j\propto w_j\prod_k\bigl[L_j/h_k(L_j)\bigr]^{\gamma_k/(1+\sum_m\gamma_m)}$, where the constant depends only on the parameters $\gamma_k$ and $c_{jk}$. The labor supply equation~\eqref{eq:eq_supply} also holds at $\rho=0$ as a limit: the factor $\mathcal{A}(L_j)^{\eta/\rho}$ diverges, but so does every term of $\mathbb{V}^{\eta}$ by the same factor, and their ratio is finite.} Together with the first-order condition for wages, we characterize the equilibrium of our model by two equations with two unknowns $(w_j, L_j)$:
\begin{align}
\text{(Wage Equation)}\qquad w_j \;&=\; \frac{\alpha\psi_j\,L_j^{\alpha-1}}{1 + \mathcal{A}(L_j)/\eta + \sum_k \omega_k(L_j)\,\varepsilon_L^{h_k}}, \label{eq:eq_demand}\\[4pt]
\text{(Labor Supply)}\qquad L_j \;&=\; (N\bar{a}_j/\mathbb{V}^\eta)\,w_j^\eta\,\mathcal{A}(L_j)^{\eta/\rho}. \label{eq:eq_supply}
\end{align}

Solving~\eqref{eq:eq_demand}--\eqref{eq:eq_supply}, we get the equilibrium wage and employment in terms of firm primitives,
\begin{equation}
w^*_j \;=\; w^*\!\bigl(\psi_j,\,\bar{a}_j,\,\mathbf{c}_j;\,\mathbb{V},\mathbf{h},\bm{\gamma},\,\alpha,\eta,\rho\bigr), \qquad L^*_j \;=\; L^*\!\bigl(\psi_j,\,\bar{a}_j,\,\mathbf{c}_j;\,\mathbb{V},\mathbf{h},\bm{\gamma},\,\alpha,\eta,\rho\bigr),
\label{eq:star_w_L}
\end{equation}
where $(\mathbf{c}_j,\mathbf{h},\bm{\gamma})$ are vectors of firm-specific unit costs, scale-dependent cost functions, and tastes, respectively, associated with the $n$-dimensional vector of endogenous amenities $\mathbf{a}_j$. The equilibrium amenities $\mathbf{a}^*_j$ follow from evaluating~\eqref{eq:astar_ratio} at $(w^*_j, L^*_j)$, and $\mathcal{A}^*=\mathcal{A}(L^*_j)$. An interior equilibrium exists and is unique when labor supply slopes upward and the wage equation (labor demand) slopes downward (see Appendix~\ref{app:existence} for the formal conditions).

The wage equation \eqref{eq:eq_demand} expresses the wage as a fraction of the marginal revenue product (MRP) $\alpha\psi_j L_j^{\alpha-1}$. With production-relevant amenities, this is the marginal product at gross productivity $\psi_j$; the output the amenities absorb is a cost in~\eqref{eq:profit} and sits on the compensation side, inside $\mathcal{A}(L_j)$, valued at the worker's marginal rate of substitution, by~\eqref{eq:MRS}. The markdowns below therefore compare everything workers receive, in money and in kind, with the gross marginal product. The denominator comprises the direct wage cost (equal to 1), and two additional amenity-related terms: (i) $\mathcal{A}(L_j)/\eta$ is the infra-marginal compensation cost (raising $w_j$ to attract one more worker also raises every amenity proportionally, since $a_{jk}\propto w_j$), and (ii) $\omega_k(L_j)\varepsilon_L^{h_k}$ is the marginal amenity-provision cost (hiring one more worker raises $h_k(L_j)$ at fixed amenity levels). Both channels widen the wedge between wages and the MRP.\footnote{In Appendix Section \ref{app:value_taker}, we compare the composition of the wage-amenity bundle offered by the monopsonist to the bundle provided under a \emph{value-taking} conduct. Proposition \ref{res:underprovision} shows that the monopsonist will strictly under-provide employment, wages, and job value as compared to the value-taking benchmark, while amenities will be under-provided unless the provision costs are strongly convex.}

\paragraph{Wage-elasticity of the effective labor supply curve.} The trade-off that monopsonistic firms face between employment and wages is summarized by the wage-elasticity of the effective labor supply curve. Implicitly differentiating the labor supply equation, $L_j = (N\bar{a}_j/\mathbb{V}^\eta) V_j^\eta$ with $\mathbf{a}^*_j$ substituted out, with respect to the wage yields (Appendix~\ref{app:wage_elasticity}):
\begin{equation}
\varepsilon_w^L(L_j) \;\equiv\; \frac{d\ln L_j}{d\ln w_j}\bigg|_{\text{supply}} \;=\; \frac{\eta}{1 - \dfrac{\eta}{1-\rho}\,\Phi(L_j)}, \qquad \Phi(L_j) \;\equiv\; \sum_k \sigma_k(L_j)\,(1-\varepsilon_L^{h_k}),
\label{eq:eps_w_def}
\end{equation}
where $\sigma_k(L_j) \equiv C_{jk}/TC_j = \omega_k(L_j)/\mathcal{A}(L_j)$ is amenity $k$'s share of total compensation.  The \textit{cost-curvature index} $\Phi(L_j)$ is a $\sigma_k$-weighted average of $(1-\varepsilon_L^{h_k})$: positive when concave-cost amenities dominate, negative when convex-cost amenities dominate, and zero at per-worker costs on aggregate. The effective labor supply elasticity $\varepsilon_w^L(L_j)$ differs from the structural taste dispersion parameter $\eta$ as the amenity bundle also adjusts when $w_j$ changes and workers respond to the complete change in the wage-amenity compensation bundle. Specifically, if $\Phi(L_j)>0$, the amenity adjustment amplifies the supply response ($\varepsilon_w^L(L_j) > \eta$), while if $\Phi(L_j)<0$, the amenity adjustment dampens the supply response ($\varepsilon_w^L(L_j)< \eta$).\footnote{Note that if $\eta\,\Phi(L_j) > 1-\rho$, the denominator of~\eqref{eq:eps_w_def} would turn negative and the labor supply curve would bend backwards. We throughout maintain the assumption that $\eta\,\Phi(L_j)<1-\rho$ (Condition~(i) of Appendix~\ref{app:existence}), which effectively places an upper bound on the degree of concavity of the cost-curvature index. In Section~\ref{sec:discussion}, we show that the effective labor supply elasticity estimates are indeed strictly positive.}

In recent work, \citet{kline2025labor} points out that common iso-elastic supply specifications used in the monopsony literature impose a zero super-elasticity. Notably, as $\Phi(L_j)$ itself depends on firm size, $\varepsilon_w^L(L_j)$ is not constant along the supply curve. The super-elasticity is generally non-zero in our framework, and it depends crucially on the cost structure \eqref{eq:super_elasticity}.

\paragraph{Size-elasticity of the amenity cost multiplier.} How amenity provision costs vary with firm size is captured by the size-elasticity of the amenity cost multiplier. Differentiating the  multiplier $\mathcal{A}(L_j)=1+\sum_k\omega_k(L_j)$ along the labor supply curve (Appendix~\ref{app:wage_elasticity}), gives its size-elasticity as the $\sigma_k$-weighted counterpart of the cost-curvature index,

\begin{equation}
\varepsilon_L^\mathcal{A}(L_j) \;\equiv\; \frac{d\ln\mathcal{A}(L_j)}{d\ln L_j} \;=\; \sum_k\sigma_k(L_j)\,\frac{\rho\,}{1-\rho}(1-\varepsilon_L^{h_k}) \;=\; \frac{\rho}{1-\rho}\Phi(L_j).
\label{eq:epsLA}
\end{equation}
It captures how amenity costs relative to the wage bill change with employment: the multiplier rises with firm size when $\rho$ and $\Phi(L_j)$ share sign, falls when they oppose, and is invariant ($\varepsilon_L^\mathcal{A}=0$) under per-worker costs on aggregate ($\Phi(L_j)=0$) or under Cobb--Douglas preferences ($\rho=0$). Notably, if $\rho=0$, then the amenity cost multiplier is constant ($\varepsilon_L^\mathcal{A}=0$) for \emph{any} cost structure, since fixed cost shares keep amenity costs relative to the wage bill constant even as amenity \emph{levels} move.  Although $\rho=0$ implies $\varepsilon_L^\mathcal{A}=0$, the cost-curvature index $\Phi(L_j)$ still shifts the supply elasticity~\eqref{eq:eps_w_def}, as workers respond to the amenity levels that are provided with the wage.

\subsection{Markdowns and Amenity Cost Structure}
\label{subsec:defs}

How much of the marginal product of labor accrues to workers in equilibrium? We derive the total compensation markdown from the firm's optimum, where its marginal cost equals its marginal product. Differentiating the compensation bill $L_j\mathcal{A}(L_j)w_j$ along the supply curve,
\begin{align*}
\mathrm{MC}(L_j) \;&\equiv\; \frac{d[L_j\,\mathcal{A}(L_j)\,w_j]}{dL_j} \;=\; \mathcal{A}(L_j)\,w_j + L_j\,\frac{d[\mathcal{A}(L_j)w_j]}{dL_j},
\end{align*}
where substituting the \emph{total compensation supply elasticity} $\varepsilon_{\mathcal{A}w}^L$, the response of $L_j$ to total compensation $\mathcal{A}(L_j)w_j$ along the supply curve, gives:
\begin{align*}
\mathrm{MC}(L_j) &=\; \mathcal{A}(L_j)\,w_j\,\biggl(1 + \underbrace{\frac{d\ln[\mathcal{A}(L_j)w_j]}{d\ln L_j}}_{\equiv\, (\varepsilon_{\mathcal{A}w}^L)^{-1}}\biggr) \;=\; \mathcal{A}(L_j)\,w_j\,\biggl(1 + \underbrace{\frac{d\ln w_j}{d\ln L_j}}_{=\, (\varepsilon_w^L)^{-1}} + \underbrace{\frac{d\ln\mathcal{A}(L_j)}{d\ln L_j}}_{=\, \varepsilon_L^\mathcal{A}}\biggr) \\
&=\; \mathcal{A}(L_j)\,w_j\,\biggl(1 + \frac{1}{\varepsilon_{\mathcal{A}w}^L}\biggr)  \;=\; \mathcal{A}(L_j)\,w_j\,\biggl(1 + \frac{1}{\varepsilon_w^L} + \varepsilon_L^\mathcal{A}\biggr).
\end{align*}
Equating $\mathrm{MC}(L_j) = \mathrm{MRP}(L_j)$, rearranging and defining $\mathcal{M}(L_j) \equiv w_j/\mathrm{MRP}(L_j)$, yields:
\begin{equation}
\mathcal{A}(L_j) \mathcal{M}(L_j) \;=\; \frac{\varepsilon_{\mathcal{A}w}^L}{1+\varepsilon_{\mathcal{A}w}^L} \;=\; \frac{\varepsilon_w^L}{(\varepsilon_w^L+1) + \varepsilon_w^L\,\varepsilon_L^\mathcal{A}} \;=\; \frac{\eta}{\eta + 1 - \eta\,\Phi(L_j)},
\label{eq:foc_AwMRP}
\end{equation}
where we refer to $\mathcal{A}(L_j) \mathcal{M}(L_j)$ as the \emph{total compensation markdown}. Equation~\eqref{eq:foc_AwMRP} provides three equivalent expressions for this markdown. The first expression is the canonical inverse-elasticity monopsony markdown formula applied to $\varepsilon_{\mathcal{A}w}^L$.  The second decomposes it through $\varepsilon_w^L$ and $\varepsilon_L^\mathcal{A}$; the wedge $\varepsilon_w^L\,\varepsilon_L^\mathcal{A}$ is the amenity-feedback correction relative to the canonical markdown formula $\varepsilon_w^L/(1+\varepsilon_w^L)$.  The third expresses it in terms of the Fr{\'e}chet-shape parameter $\eta$, which represents the structural elasticity (reflecting taste dispersion) in the canonical monopsony model, and $\Phi(L_j)$, which captures the amenity cost structure.\footnote{To get the third expression, we plug in the closed forms of $\varepsilon_w^L$ and $\varepsilon_L^\mathcal{A}$ from~\eqref{eq:eps_w_def} and~\eqref{eq:epsLA}, or equivalently $\varepsilon_{\mathcal{A}w}^L = \eta/[1-\eta\,\Phi(L_j)]$ obtained via the identity $1/\varepsilon_{\mathcal{A}w}^L = 1/\varepsilon_w^L + \varepsilon_L^\mathcal{A}$. The same substitution recovers~\eqref{eq:MC_mono}.}

Dividing~\eqref{eq:foc_AwMRP} by $\mathcal{A}(L_j)$, we can further derive the canonical \emph{wage markdown} $\mathcal{M}(L_j) \equiv w_j/\mathrm{MRP}(L_j)$ as the total compensation markdown deflated by the amenity multiplier:

\begin{equation}
\mathcal{M}(L_j) \;=\; \underbrace{\frac{1}{\mathcal{A}(L_j)}}_{\substack{\text{amenity}\\\text{deflator}}}\cdot\frac{\eta}{\eta + 1 - \underbrace{\eta\,\Phi(L_j)}_{\substack{\text{cost}\\\text{curvature}}}}.
\label{eq:M_formula}
\end{equation}

Three observations follow.  First, the equilibrium total compensation markdown $\mathcal{A}^*\mathcal{M}^*$ deviates from the canonical $\eta/(\eta+1)$ benchmark in the direction of $\Phi(L_j)$: the total compensation markdown is above the canonical markdown when $\Phi(L_j)>0$ (concave costs), below when $\Phi(L_j)<0$ (convex costs), and at the benchmark when $\Phi(L_j)=0$ (per-worker costs).  Second, $\mathcal{M}(L_j)$ sits below $\mathcal{A}(L_j)\mathcal{M}(L_j)$ by the factor $\mathcal{A}(L_j)>1$. This wedge reflects relabeling between wages and amenities, not between workers and the firm.  Third, the first expression in~\eqref{eq:foc_AwMRP} is the canonical inverse-elasticity formula applied to the \emph{total compensation} elasticity $\varepsilon_{\mathcal{A}w}^L$. It reduces to the wage-based formula when the two elasticities coincide, that is, when the amenity cost multiplier is invariant ($\varepsilon_L^\mathcal{A}=0$, since $1/\varepsilon_{\mathcal{A}w}^L = 1/\varepsilon_w^L + \varepsilon_L^\mathcal{A}$). By~\eqref{eq:epsLA} this holds under Cobb--Douglas preferences ($\rho=0$) or aggregate per-worker costs ($\Phi(L_j)=0$). In either case, $\mathcal{A}^*=\bar{\mathcal{A}}$, and thus we have  $\mathcal{A}^*\mathcal{M}^*=\bar{\mathcal{A}}\mathcal{M}^*=\varepsilon_w^L/(\varepsilon_w^L+1)$, so wage-based elasticity measures can recover the total compensation markdown even though wages are only part of the compensation. We summarize these observations in Proposition~\ref{res:markdown_amenities}.

\begin{result}[Markdowns and Amenity Cost Structure]
\label{res:markdown_amenities}
\textit{With any valued endogenous amenity: \textbf{(i)} The total compensation markdown deviates from the canonical benchmark in the direction of the cost-curvature index: $\mathcal{A}(L_j)\mathcal{M}(L_j) \gtrless \eta/(\eta+1)$ iff $\Phi(L_j) \gtrless 0$, with equality at the per-worker benchmark $\Phi(L_j)=0$. \textbf{(ii)} The wage markdown $\mathcal{M}(L_j)$ is strictly below the total compensation markdown $\mathcal{A}(L_j)\mathcal{M}(L_j)$ by the factor $\mathcal{A}(L_j) > 1$. \textbf{(iii)} The canonical formula recovers the total compensation markdown exactly, $\mathcal{A}^*\mathcal{M}^*=\varepsilon_w^L/(\varepsilon_w^L+1)$, in two special cases: (a) Cobb--Douglas preferences ($\rho=0$), and (b) per-worker amenity costs on aggregate ($\Phi(L_j)=0$). In both, the amenity cost multiplier is constant, $\mathcal{A}^*=\bar{\mathcal{A}}$, so the wage markdown is $\mathcal{M}^*=\mathcal{A}^*\mathcal{M}^*/\bar{\mathcal{A}}$. In case (b), in addition, $\varepsilon_w^L=\eta$.}
\end{result}

\begin{figure}[t!]
    \centering
    \caption{Markdowns across Alternative Amenity-Cost Regimes.}
    \label{fig:monopsony_panels}
    \hspace*{-4.5cm}%
    \begin{subfigure}{0.5\textwidth}
        \centering
        \captionsetup{margin={3cm,-3.5cm}}
        \resizebox{!}{0.765\textwidth}{\begin{tikzpicture}[
    xscale=5.8, yscale=3.7,
    declare function={
        AwPw(\x)=sqrt(\x);
        wPw(\x)=0.667*sqrt(\x);
        mcpw(\x)=1.5*sqrt(\x);
        mrpf(\x)=0.825/sqrt(\x);
    }
]
    \path[use as bounding box] (-0.55,-0.18) rectangle (1.85,1.82);

    \draw[->,thick] (0,0) -- (1.45,0) node[below] {$L$};
    \draw[->,thick] (0,0) -- (0,1.72) node[left] {$w$};

    \draw[thick,CoolRed!60,dashed,domain=0.23:1.30,smooth,variable=\x]
        plot (\x,{0.6667*mrpf(\x)});
    \node[color=CoolRed!80,fill=white,inner sep=1pt] at (0.22,1.25)
        {$\tfrac{\eta}{\eta+1}\,MRP(L)$};

    \draw[thick,color=CoolGreen,domain=0.23:1.30,smooth,variable=\x]
        plot (\x,{mrpf(\x)});
    \node[color=CoolGreen,fill=white,inner sep=1pt] at (0.22,1.70) {$MRP(L)$};

    \draw[thick,color=black,domain=0.15:1.22,smooth,variable=\x]
        plot (\x,{AwPw(\x)}) node[right,fill=white,inner sep=1pt] {$w(L)$};
    \draw[thick,color=orange,domain=0.15:1.30,smooth,variable=\x]
        plot (\x,{mcpw(\x)}) node[right] {$MC(L)$};

    \draw[dashed,gray!60] (0.550,0) node[below,font=\scriptsize,text=black] {$L^{*}$} -- (0.550,1.1127);
    \draw[dashed,gray!60] (0,1.1127) node[left,font=\scriptsize,text=black] {$MRP^{*}$} -- (0.550,1.1127);
    \draw[dashed,gray!60] (0,0.7416) node[left,font=\scriptsize,text=black] {$w^{*}$} -- (0.550,0.7416);

    \node[circle,inner sep=1pt,fill=black] at (0.550,1.1127) {};
    \node[circle,inner sep=1pt,fill=CoolRed] at (0.550,0.7416) {};

\end{tikzpicture}}
        \caption{Canonical Monopsony}
        \label{fig:monopsony_classic}
    \end{subfigure}\hspace{1.5cm}%
    \begin{subfigure}{0.5\textwidth}
        \centering
        \captionsetup{margin={3cm,-3.5cm}}
        \resizebox{!}{0.765\textwidth}{\begin{tikzpicture}[
    xscale=5.8, yscale=3.7,
    declare function={
        AwPw(\x)=sqrt(\x);
        wPw(\x)=0.667*sqrt(\x);
        mcpw(\x)=1.5*sqrt(\x);
        mrpf(\x)=0.825/sqrt(\x);
    }
]
    \path[use as bounding box] (-0.55,-0.18) rectangle (1.85,1.82);

    \draw[->,thick] (0,0) -- (1.45,0) node[below] {$L$};
    \draw[->,thick] (0,0) -- (0,1.72) node[left] {$w$};

    \draw[thick,CoolRed!60,dashed,domain=0.23:1.30,smooth,variable=\x]
        plot (\x,{0.6667*mrpf(\x)});
    \node[color=CoolRed!80,fill=white,inner sep=1pt] at (0.22,1.25)
        {$\tfrac{\eta}{\eta+1}\,MRP(L)$};

    \draw[thick,color=CoolGreen,domain=0.23:1.30,smooth,variable=\x]
        plot (\x,{mrpf(\x)});
    \node[color=CoolGreen,fill=white,inner sep=1pt] at (0.22,1.70) {$MRP(L)$};

    \draw[thick,color=black,domain=0.15:1.22,smooth,variable=\x]
        plot (\x,{AwPw(\x)}) node[right,fill=white,inner sep=1pt] {$\mathcal{A}(L)\,w(L)$};
    \draw[thick,color=CoolBlue,domain=0.15:1.22,smooth,variable=\x]
        plot (\x,{wPw(\x)}) node[right,yshift=-10pt] {$w(L)$};
    \draw[thick,color=orange,domain=0.15:1.30,smooth,variable=\x]
        plot (\x,{mcpw(\x)}) node[right] {$MC(L)$};

    \draw[dashed,gray!60] (0.550,0) node[below,font=\scriptsize,text=black] {$L^{*}$} -- (0.550,1.1127);
    \draw[dashed,gray!60] (0,1.1127) node[left,font=\scriptsize,text=black] {$MRP^{*}$} -- (0.550,1.1127);
    \draw[dashed,gray!60] (0,0.4948) node[left,font=\scriptsize,text=black] {$w^{*}$} -- (0.550,0.4948);
    \draw[dashed,gray!60] (0,0.7416) node[left,font=\scriptsize,text=black] {$\mathcal{A}^{*}w^{*}$} -- (0.550,0.7416);

    \node[circle,inner sep=1pt,fill=black] at (0.550,1.1127) {};
    \node[circle,inner sep=1pt,fill=CoolRed] at (0.550,0.7416) {};
    \node[circle,inner sep=1pt,fill=black] at (0.550,0.4948) {};

\end{tikzpicture}}
        \caption{Per-Worker Amenity Costs ($\Phi(L_j)=0$)}
        \label{fig:monopsony_perworker}
    \end{subfigure}
    \par\medskip
    \hspace*{-4.5cm}%
    \begin{subfigure}{0.5\textwidth}
        \centering
        \captionsetup{margin={3cm,-3.5cm}}
        \resizebox{!}{0.765\textwidth}{\begin{tikzpicture}[
    xscale=5.8, yscale=3.7,
    declare function={
        AwSh(\x)=0.8360*pow(\x,0.2);
        wSh(\x)=0.5578*pow(\x,0.2);
        mcsh(\x)=1.0032*pow(\x,0.2);
        mrpf(\x)=0.825/sqrt(\x);
    }
]
    \path[use as bounding box] (-0.55,-0.18) rectangle (1.85,1.82);

    \draw[->,thick] (0,0) -- (1.45,0) node[below] {$L$};
    \draw[->,thick] (0,0) -- (0,1.72) node[left] {$w$};

    \draw[thick,CoolRed!60,dashed,domain=0.23:1.30,smooth,variable=\x]
        plot (\x,{0.6667*mrpf(\x)});
    \node[color=CoolRed!80,fill=white,inner sep=1pt] at (0.22,1.25)
        {$\tfrac{\eta}{\eta+1}\,MRP(L)$};

    \draw[thick,color=CoolGreen,domain=0.23:1.30,smooth,variable=\x]
        plot (\x,{mrpf(\x)});
    \node[color=CoolGreen,fill=white,inner sep=1pt] at (0.22,1.70) {$MRP(L)$};

    \draw[thick,color=black,domain=0.15:1.22,smooth,variable=\x]
        plot (\x,{AwSh(\x)}) node[right,fill=white,inner sep=1pt] {$\mathcal{A}(L)\,w(L)$};
    \draw[thick,color=CoolBlue,domain=0.15:1.22,smooth,variable=\x]
        plot (\x,{wSh(\x)}) node[right] {$w(L)$};
    \draw[thick,color=orange,domain=0.15:1.22,smooth,variable=\x]
        plot (\x,{mcsh(\x)}) node[right] {$MC(L)$};

    \draw[dashed,gray!60] (0.756,0) node[below,font=\scriptsize,text=black] {$L^{*}$} -- (0.756,0.9488);
    \draw[dashed,gray!60] (0,0.9488) node[left,font=\scriptsize,text=black] {$MRP^{*}$} -- (0.756,0.9488);
    \draw[dashed,gray!60] (0,0.7906) node[left,font=\scriptsize,text=black] {$\mathcal{A}^{*}w^{*}$} -- (0.756,0.7906);
    \draw[dashed,gray!60] (0,0.5275) node[left,font=\scriptsize,text=black] {$w^{*}$} -- (0.756,0.5275);

    \node[circle,inner sep=1pt,fill=black] at (0.756,0.9488) {};
    \node[circle,inner sep=1pt,fill=black] at (0.756,0.7906) {};
    \node[circle,inner sep=1pt,fill=black] at (0.756,0.5275) {};
    \node[circle,inner sep=1pt,fill=CoolRed] at (0.756,0.6325) {};

\end{tikzpicture}}
        \caption{Concave Amenity Costs ($\Phi(L_j)>0$)}
        \label{fig:monopsony_concave}
    \end{subfigure}\hspace{1.5cm}%
    \begin{subfigure}{0.5\textwidth}
        \centering
        \captionsetup{margin={3cm,-3.5cm}}
        \resizebox{!}{0.765\textwidth}{\begin{tikzpicture}[
    xscale=5.8, yscale=3.7,
    declare function={
        AwWd(\x)=1.1961*pow(\x,0.8);
        wWd(\x)=0.7981*pow(\x,0.8);
        mcwd(\x)=2.1530*pow(\x,0.8);
        mrpf(\x)=0.825/sqrt(\x);
    }
]
    \path[use as bounding box] (-0.55,-0.18) rectangle (1.85,1.82);

    \draw[->,thick] (0,0) -- (1.45,0) node[below] {$L$};
    \draw[->,thick] (0,0) -- (0,1.72) node[left] {$w$};

    \draw[thick,CoolRed!60,dashed,domain=0.23:1.30,smooth,variable=\x]
        plot (\x,{0.6667*mrpf(\x)});
    \node[color=CoolRed!80,fill=white,inner sep=1pt] at (0.22,1.25)
        {$\tfrac{\eta}{\eta+1}\,MRP(L)$};

    \draw[thick,color=CoolGreen,domain=0.23:1.30,smooth,variable=\x]
        plot (\x,{mrpf(\x)});
    \node[color=CoolGreen,fill=white,inner sep=1pt] at (0.22,1.70) {$MRP(L)$};

    \draw[thick,color=black,domain=0.15:1.22,smooth,variable=\x]
        plot (\x,{AwWd(\x)}) node[right,fill=white,inner sep=1pt] {$\mathcal{A}(L)\,w(L)$};
    \draw[thick,color=CoolBlue,domain=0.15:1.22,smooth,variable=\x]
        plot (\x,{wWd(\x)}) node[right] {$w(L)$};
    \draw[thick,color=orange,domain=0.15:0.755,smooth,variable=\x]
        plot (\x,{mcwd(\x)});
    \node[color=orange,fill=white,inner sep=1pt] at (0.76,1.70) {$MC(L)$};

    \draw[dashed,gray!60] (0.478,0) node[below,font=\scriptsize,text=black] {$L^{*}$} -- (0.478,1.193);
    \draw[dashed,gray!60] (0,1.193) node[left,font=\scriptsize,text=black] {$MRP^{*}$} -- (0.478,1.193);
    \draw[dashed,gray!60] (0,0.663) node[left,font=\scriptsize,text=black] {$\mathcal{A}^{*}w^{*}$} -- (0.478,0.663);
    \draw[dashed,gray!60] (0,0.442) node[left,font=\scriptsize,text=black] {$w^{*}$} -- (0.478,0.442);

    \node[circle,inner sep=1pt,fill=black] at (0.478,1.193) {};
    \node[circle,inner sep=1pt,fill=black] at (0.478,0.663) {};
    \node[circle,inner sep=1pt,fill=black] at (0.478,0.442) {};
    \node[circle,inner sep=1pt,fill=CoolRed] at (0.478,0.7956) {};

\end{tikzpicture}}
        \caption{Convex Amenity Costs ($\Phi(L_j)<0$)}
        \label{fig:monopsony_convex}
    \end{subfigure}
    \par\medskip
    \begin{minipage}{\textwidth}
    \scriptsize
    \emph{Notes:} Each panel plots the firm's choice in $(L_j, w_j)$ space, with the equilibrium $(L_j^*, w_j^*)$ marked by black dashed lines.  Panel (a) is the benchmark without endogenous amenities: the wedge $1-\mathcal{M}^*$ separates $w_j^*$ from $\mathrm{MRP}^*$, and the markdown equals the canonical formula $\eta/(\eta+1)$.  Panels (b)--(d) allow endogenous amenities, so $\mathrm{MRP}^*$ splits three ways: wage $w_j^*$, amenity cost $(\mathcal{A}^* - 1)\,w_j^*$, and firm surplus $(1-\mathcal{A}^*\mathcal{M}^*)\,\mathrm{MRP}^*$.  At per-worker costs (b), $\mathcal{A}(L_j)$ is constant in $L_j$, and the total compensation markdown $\mathcal{A}^*\mathcal{M}^*$ equals $\eta/(\eta+1)$.  Concave costs (c) raise $\mathcal{A}^*\mathcal{M}^*$ above the benchmark, narrowing firm surplus; convex costs (d) lower it, widening firm surplus. The illustration sets $\rho=0$, so $\mathcal{A}(L_j)$ is constant in $L_j$ ($\varepsilon_L^\mathcal{A}=0$). Different cost technologies generate different $\Phi(L_j)$ values, which shift the equilibrium supply elasticity $\varepsilon_w^L$ and the firm-surplus wedge $1-\mathcal{A}^*\mathcal{M}^*$.
    \end{minipage}
\end{figure}
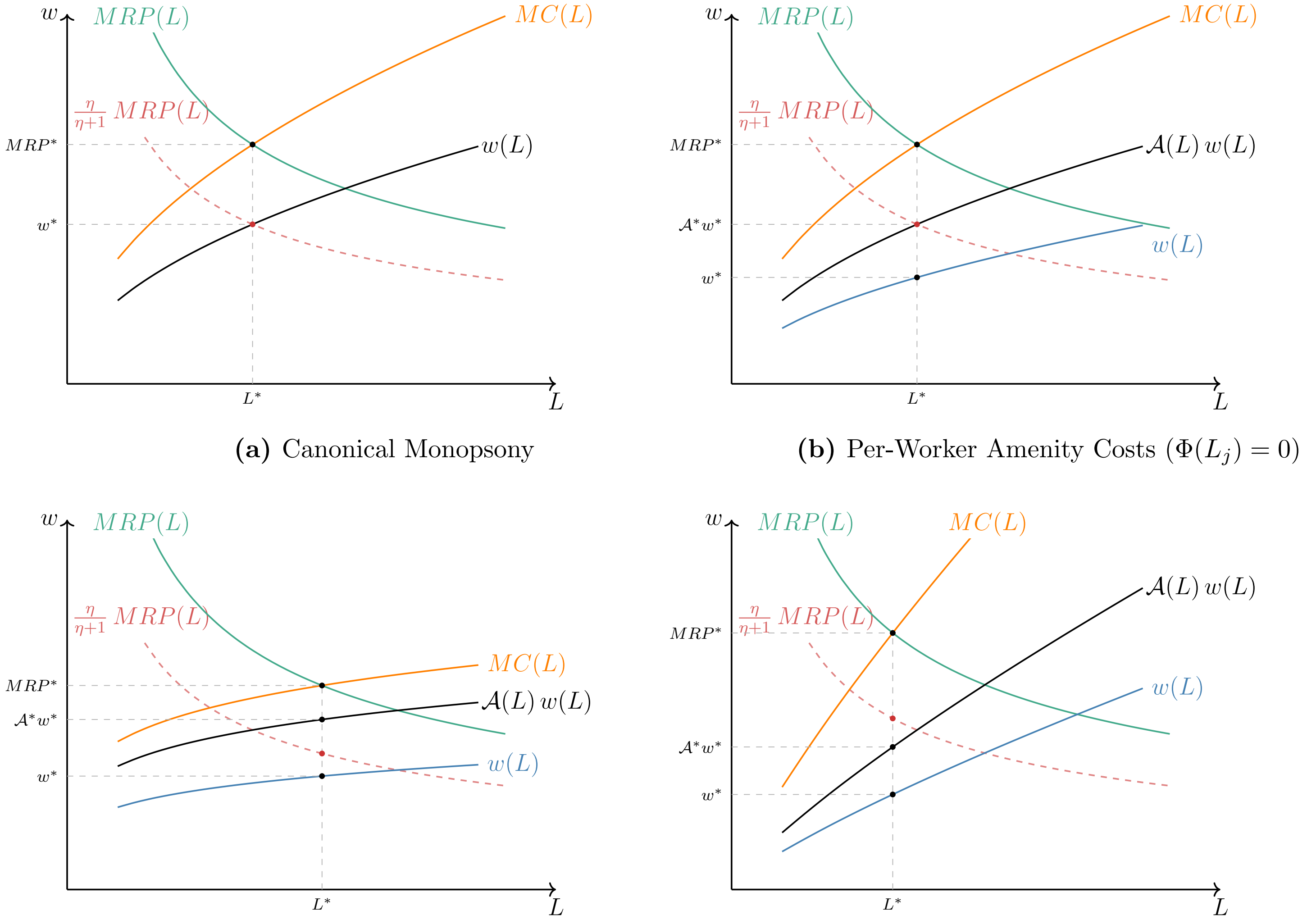

Figure~\ref{fig:monopsony_panels} maps the markdown decomposition in the $(L_j,w_j)$ space. In canonical monopsony without endogenous amenities (panel a), $\mathrm{MC}(L_j)$ crosses $\mathrm{MRP}(L_j)$ at an employment level below the value-taking benchmark, and the wage markdown wedge $1-\mathcal{M}^*$ separates the wage from the marginal product. With endogenous amenities (panels b–d), total compensation $\mathcal{A}(L_j)w_j$ replaces $w_j$ as the firm's effective price of labor and the total compensation markdown $\mathcal{A}^*\mathcal{M}^*$ is the welfare-relevant object. At the optimum, the marginal revenue product $\mathrm{MRP}^*$ splits into three parts: wage $w_j^*$, amenity cost $(\mathcal{A}^*-1)w_j^*$, and firm surplus $(1-\mathcal{A}^*\mathcal{M}^*)\,\mathrm{MRP}^*$. The wedge between the wage and the marginal product, $1-\mathcal{M}^*$, widens as amenities absorb part of what would otherwise be wages. The three cost regimes share a common mechanism: hiring one more worker triggers a re-optimization of amenity provision, which feeds back into the supply curve.  Whether the feedback amplifies or dampens supply, and hence whether $\mathcal{A}^*\mathcal{M}^*$ exceeds or falls short of the canonical $\eta/(\eta+1)$, depends on how average amenity cost per worker scales with firm size.\footnote{The illustration shows the stylized case with $\rho=0$, so $\mathcal{A}(L_j)$ is constant within each panel and the panels differ only through $\Phi(L_j)$, which moves the supply elasticity $\varepsilon_w^L$ and the firm-surplus wedge together.}

We now review each cost regime in turn: (i) With per-worker costs ($\Phi(L_j)=0$), amenity costs relative to the wage bill are scale-invariant, so there is a \textit{neutral feedback} on labor supply: $\varepsilon_w^L = \eta$. We have $\mathrm{MC} = \mathcal{A}(L_j)w_j(1+1/\eta)$, which mirrors the canonical $\mathrm{MC} = w_j(1+1/\eta)$, so $\mathcal{A}^*\mathcal{M}^* = \eta/(\eta+1)$. Panel (b) therefore reproduces panel (a) with the price of labor relabeled from $w$ to $\mathcal{A}(L_j)w$, and the same equilibrium and value-taking employment.
    (ii) Under concave amenity costs ($\Phi(L_j)>0$), amenity costs per worker are diluted by firm size, as fixed costs spread across more workers, so we get \textit{positive feedback} on labor supply. The firm provides \emph{more} amenity per worker as it grows. This attracts workers at any wage, which flattens $\mathcal{A}(L_j)w$ and lifts $\varepsilon_w^L$ above $\eta$. The total compensation markdown $\mathcal{A}^*\mathcal{M}^*$ rises above $\eta/(\eta+1)$ and the wedge at the firm's optimum narrows. Equilibrium employment is higher than at per-worker costs (panel c).  (iii) Under convex amenity costs ($\Phi(L_j)<0$), each additional worker is more expensive to equip, so amenities fall relative to the wage as the firm grows, and we get \textit{negative feedback} on labor supply: attracting the next worker requires a larger wage hike, which steepens $\mathcal{A}(L_j)w$ and pulls $\varepsilon_w^L$ below $\eta$. The total compensation markdown $\mathcal{A}^*\mathcal{M}^*$ falls below $\eta/(\eta+1)$ and the wedge at the firm's optimum widens, so the intuition that ``amenities reduce markdowns'' reverses, and employment is lower (panel d).

\paragraph{Size-elasticity of the wage markdowns.} How wage markdowns adjust with firm size is captured by the elasticity $\varepsilon_L^\mathcal{M}(L_j) \;\equiv\; \frac{d\ln\mathcal{M}(L_j)}{d\ln L_j}$. We derive the functional form of this elasticity in Appendix~\ref{app:wage_eq_slope}. Notably, under iso-elastic costs and substitutes ($\rho>0$), $\varepsilon_L^\mathcal{M}<0$ if the firm faces concave costs, so markdowns deepen as firms grow. Conversely, $\varepsilon_L^\mathcal{M}>0$ if the firm faces convex costs, so markdowns are more compressed in larger firms. The wage markdown's size-elasticity will partly determine how firm size mediates the wage and employment responses to productivity shocks in our comparative statics in Section~\ref{subsec:cs_psi}.

\paragraph{Conduct versus cost structure.}  Having defined the total compensation markdown as the gap between total compensation and marginal product in~\eqref{eq:foc_AwMRP}, we ask how much of this gap reflects employer market power, and how much the structure of amenity provision costs. We can express the total compensation markdown as the product of two factors:

\begin{equation}
\label{eq:AM_decomp}
\mathcal{A}(L_j)\,\mathcal{M}(L_j)\;=\;\underbrace{\frac{1}{1-\Phi(L_j)}}_{\text{cost curvature}}\;\times\;\underbrace{\frac{1-\Phi(L_j)}{1-\Phi(L_j)+1/\eta}}_{\text{conduct}}.
\end{equation}
This decomposition serves as an \textit{accounting identity} featuring two terms that can be interpreted as capturing the cost curvature wedge and the conduct wedge, respectively. The first factor is the share of its marginal product that a firm would pay if it took the job value as exogenously given (Appendix~\ref{app:value_taker}). It differs from one because an extra worker changes what the amenities cost for the inframarginal workers: the firm pays less than the marginal product when they become costlier ($\Phi(L_j)<0$) and more when they become cheaper ($\Phi(L_j)>0$). The second factor captures the additional gap under monopsony, where an extra worker also requires a higher job value for every worker. Attributing the total compensation markdown as a whole to employer market power would conflate cost curvature with conduct. In Section~\ref{sec:discussion}, we thus decompose the cost curvature and conduct components.\footnote{We interpret the decomposition in \eqref{eq:AM_decomp} as an accounting exercise as it is valid only at the monopsonist's employment level $L^*_j$. A value-taking firm, whose marginal cost~\eqref{eq:MC_vt} lacks the conduct term $\mathcal{A}(L_j)w(L_j)/\eta$ of the monopsonist's marginal cost~\eqref{eq:MC_mono}, would expand to $L^c_j>L^*_j$ (Proposition~\ref{res:underprovision}), where its compensation is $1/(1-\Phi(L^c_j))$ of its marginal product and the cost-curvature wedge differs from the one at $L^*_j$.}

\subsection{Productivity Shocks}
\label{subsec:cs_psi}

We now characterize how the firm's wages, employment, and amenity provision respond to a positive firm-specific productivity shock $d\psi_j>0$. In our framework, the comparative statics for a negative productivity shock, $d\psi_j<0$, are symmetric, and the predictions further carry over to the same-signed firm-level product-demand shocks.\footnote{We normalized product prices $p_j=1$ in Section~\ref{subsec:firms}. Under firm-level price heterogeneity $p_j$, the comparative statics of a product price change $dp_j$ are identical to those of a productivity change $d\psi_j$.} The productivity shock enters the equilibrium conditions of our model only through the wage equation~\eqref{eq:eq_demand}. Total log-differentiation of the wage equation~\eqref{eq:eq_demand} and labor supply~\eqref{eq:eq_supply} yields
\begin{align*}
\text{Wage Equation:}&\quad d\ln w_j \;=\; d\ln\psi_j \;-\; \bigl[(1-\alpha) - \varepsilon_L^\mathcal{M} \bigr] d\ln L_j,\\
\text{Labor Supply:}&\quad d\ln L_j \;=\; \varepsilon_w^L\,d\ln w_j,
\end{align*}
where the bracketed term is the wage-equation slope, $\varepsilon_L^\mathcal{M}$ is the wage markdown's size-elasticity from~\eqref{eq:chi_def}, and $\varepsilon_w^L$ is the labor supply elasticity from~\eqref{eq:eps_w_def}. Substituting gives

\begin{equation*}
\frac{d\ln w_j}{d\ln\psi_j} \;=\; \frac{1}{1 + \varepsilon_w^L\!\left(1-\alpha - \varepsilon_L^\mathcal{M}\right)} \;>\; 0, \qquad \frac{d\ln L_j}{d\ln\psi_j} \;=\; \varepsilon_w^L\cdot\frac{d\ln w_j}{d\ln\psi_j} \;>\; 0,
\end{equation*}
where the denominator is positive under Conditions~(i) and~(ii) of Appendix~\ref{app:existence}, so both wages and employment rise. The ratio of these two responses to a productivity shifter is the natural empirical analog of the \textit{local} labor supply elasticity, $d\ln L_j/d\ln w_j = \varepsilon_w^L(L_j)$, and underlies the instrumental variable (IV) strategy in \citet{lamadon2022imperfect}.\footnote{As $\varepsilon_w^L(L_j)$ varies with firm size when the super-elasticity~\eqref{eq:super_elasticity} is non-zero, an IV regression that pools across firms produces a weighted average of local elasticities rather than a single structural target parameter. The IV estimand collapses to a constant target parameter in two special cases: (a) Cobb--Douglas preferences ($\rho=0$) combined with iso-elastic costs $\varepsilon_L^{h_k} = \delta \geq 0$ $\forall\, k$, reducing the cost-curvature index to a constant ($\Phi(L_j) = \bar{\Phi}$); (b) per-worker amenity costs scale on aggregate ($\Phi(L_j) = 0$), where $\varepsilon_w^L = \eta$ exactly.}

How amenities respond is more nuanced and depends on what is measured: the amenity level $a_{jk}$ and the per-worker amenity cost $C_{jk}/L_j$ have distinct comparative statics, and the two need not move together. Since $a_{jk}^*$ is wage-proportional with a size-dependent factor, log-differentiating \eqref{eq:astar_ratio} with respect to $\psi_j$ and using $d\ln(L_j/h_k(L_j))/d\ln L_j = 1-\varepsilon_L^{h_k}$ decomposes the level response into a direct wage channel and an indirect firm-size channel:

\begin{equation}
\frac{d\ln a_{jk}}{d\ln\psi_j} \;=\; \Bigg[\,\underbrace{\vphantom{\frac{\varepsilon_w^L}{1-\rho}}1}_{\text{wage channel}} \;+\; \underbrace{\frac{1}{1-\rho}(1-\varepsilon_L^{h_k})\varepsilon_w^L}_{\text{firm-size channel}}\,\Bigg]\,\frac{d\ln w_j}{d\ln\psi_j}.
\label{eq:dlna}
\end{equation}
The wage channel captures that wages and amenities must always move together according to the MRS condition \eqref{eq:astar_ratio} when the firm-size channel is absent. The sign of the firm-size channel depends on $1-\varepsilon_L^{h_k}$: positive at concave costs ($\varepsilon_L^{h_k}<1$), zero at per-worker costs ($\varepsilon_L^{h_k}=1$), negative at convex costs ($\varepsilon_L^{h_k}>1$), while its strength depends on $\varepsilon_w^L(L_j)$ and $\rho$.\footnote{The bracket in~\eqref{eq:dlna} equals $\varepsilon_w^L$ times the supply-curve gradient~\eqref{eq:path_gradient} from Proposition~\ref{res:underprovision}: the shock moves the firm up its supply curve, so the amenity level rises precisely when this gradient is positive.}

Per-worker cost of amenity $k$ is $C_{jk}/L_j = \omega_k(L_j)\,w_j$, the wage scaled by $\omega_k$ from \eqref{eq:omega_explicit}, which is amenity $k$'s cost relative to the wage bill. Log-differentiating, $d\ln(C_{jk}/L_j) = d\ln\omega_k(L_j) + d\ln w_j$, and substituting $d\ln\omega_k(L_j)/d\ln L_j = \rho(1-\varepsilon_L^{h_k})/(1-\rho)$ from \eqref{eq:dlnomega}, together with $d\ln L_j = \varepsilon_w^L\,d\ln w_j$ gives

\begin{equation}
\frac{d\ln(C_{jk}/L_j)}{d\ln\psi_j} \;=\; \Bigg[\,\underbrace{\vphantom{\frac{\varepsilon_w^L}{1-\rho}}1}_{\text{wage channel}} \;+\; \underbrace{\frac{\rho}{1-\rho}(1-\varepsilon_L^{h_k})\varepsilon_w^L}_{\text{share channel}}\,\Bigg]\,\frac{d\ln w_j}{d\ln\psi_j}.
\end{equation}
The structure of cost responses mirrors~\eqref{eq:dlna} except the second channel carries an extra factor $\rho$.  Under Cobb--Douglas preferences ($\rho=0$) or per-worker amenity cost structure ($\varepsilon_L^{h_k}=1$), the share channel vanishes and the adjustment in per-worker amenity costs tracks wages exactly.  Otherwise, the share channel's sign depends on $\rho(1-\varepsilon_L^{h_k})$: it adds to the wage channel under substitutes with concave costs ($\rho>0$, $\varepsilon_L^{h_k}<1$) or complements with convex costs ($\rho<0$, $\varepsilon_L^{h_k}>1$), and subtracts under the opposite pairings. Thus, per-worker costs fall only when the share channel is negative and outweighs the wage channel.

Taken together, we note that the sign of the amenity level response depends on whether the cost structure is sufficiently convex, while the per-worker cost response depends on both the degree of cost convexity \emph{and} the wage-amenity substitutability or complementarity in preferences (the sign of $\rho$).  The two responses can therefore move in opposite directions: substitutes with strongly convex costs can deliver a falling amenity level alongside rising per-worker costs, while complements with strongly concave costs can deliver the reverse.

The bundle aggregates these individual responses through the amenity multiplier $\mathcal{A}(L_j)$. Using the expression for the multiplier size-elasticity $\varepsilon_L^\mathcal{A}$ from \eqref{eq:epsLA} and $d\ln L_j = \varepsilon_w^L\,d\ln w_j$,
\begin{equation}
\frac{d\ln\mathcal{A}(L_j)}{d\ln\psi_j} \;=\; \varepsilon_L^\mathcal{A}\cdot\varepsilon_w^L\cdot\frac{d\ln w_j}{d\ln\psi_j}.
\end{equation}
Thus, the sign of the amenity cost multiplier response is governed by the sign of the multiplier size-elasticity, as the last two terms are positive. Analogous to the share channel for per-worker amenity costs, the sign of $\varepsilon_L^\mathcal{A}$ depends jointly on the preference structure $\rho$ and the cost-curvature index $\Phi(L_j)$. Crucially, it rises only when $\sgn(\rho)=\sgn(\Phi(L_j))$. Collecting the sign conditions for the three objects yields the following comparative statics.

\begin{result}[Productivity and Endogenous Amenity Provision]
\label{res:demand_shocks}
\textit{Following a positive productivity shock $d\psi_j>0$: \textbf{(i)} The amenity level $a^*_{jk}$ rises if and only if $\varepsilon_L^{h_k} - 1 \;<\; \frac{1-\rho}{\varepsilon_w^L}$. \textbf{(ii)} Per-worker amenity cost $C^*_{jk}/L^*_j$ rises if and only if $\rho\,(\varepsilon_L^{h_k}-1) \;<\; \frac{1-\rho}{\varepsilon_w^L}$, which fails if $\rho$ and $\varepsilon_L^{h_k}-1$ share sign and their product exceeds the threshold: substitutes with strongly convex costs, or complements with strongly concave costs. \textbf{(iii)} The amenity cost multiplier $\mathcal{A}^*$ rises if and only if $\rho$ and $\Phi(L_j)$ share sign, and is invariant at $\rho=0$ or $\Phi(L_j)=0$.}
\end{result}

Empirically, the level response of an amenity relative to the wage response isolates the sign of its cost curvature, whereas per-worker amenity cost and the amenity cost multiplier combine preference patterns with cost curvature.\footnote{We illustrate Proposition~\ref{res:demand_shocks} numerically in Appendix~\ref{app:psi_figures} for a single amenity under alternative cost structures and preferences. We trace wages, the amenity cost multiplier, compensation shares and markdowns across the productivity distribution. The amenity cost multiplier is constant under Cobb--Douglas preferences (Appendix Figure~\ref{fig:markdown_psi_rho0}). Concave costs raise it under substitutes and lower it under complements, and convex costs do the reverse (Appendix Figures~\ref{fig:markdown_psi_rho_pos} and~\ref{fig:markdown_psi_rho_neg}). Total compensation markdowns and total compensation shares vary across the productivity distribution unless the cost-curvature index is constant ($\Phi(L_j)=\bar\Phi$). Wage markdowns vary unless, in addition, preferences are Cobb--Douglas or costs are per-worker ($\Phi(L_j)=0$).} Their direction thus identifies the sign of the interaction between these elements rather than either element in isolation. Interpreting amenity cost responses thus requires having knowledge about the preference parameter $\rho$. In Section \ref{sec:data_empirical} below, we discuss the model's identification and describe our data sources.

\section{Data and Identification}
\label{sec:data_empirical}

We describe our data sources, including measures of workplace amenities and firm-level demand shocks, and present our identification strategy, including the conditions under which the shocks identify effective supply elasticities and the structural parameters capturing idiosyncratic taste dispersion $\eta$ and amenity-wage substitutability $\rho$. Our analysis combines several Norwegian administrative data sources that capture standard firm- and worker-level outcomes, along with non-pecuniary job attributes recorded in employment registers, detailed firm cost accounts that capture non-wage compensation, and job ads that provide information on amenities offered to prospective hires. We focus on private-sector firms throughout. Our data sources span the period 2002--2022. Some of our analyses use the period 2015--2022, where we have more detailed information on contract types.

\subsection{Measurement of Amenities and Non-Wage Compensation}
\label{subsec:data_amenities}

We use three distinct data sources for the measurement of workplace amenities and non-wage compensation: (i) monthly matched employer--employee data, which provide worker- and employer-level information on contract types and in-kind compensation, (ii) annual firm-level cost accounts, which provide information on non-wage compensation, and (iii) data on publicly posted job vacancies, including the full-text portion of postings, with information on advertised workplace amenities and other job attributes at the ad level.

We have access to population-wide, matched employer--employee records linking all workers to their employers, from which we construct annual firm-level employment measures and annual worker- and firm-level hourly wages for the period 2002--2022. We also use more detailed monthly records (A-meldingen), which provide worker-employer-level information on contract types: from January 2015, we can define indicators for full-time or part-time contracts and fixed-hours or variable-hours contracts. For our analyses, we construct annual firm-level measures of shares of full-time and fixed-hours (non-variable) contracts.\footnote{To abstract from seasonal patterns, our reference month with A-meldingen is November.} The same data source also provides information on the monetary values of compensation received as divisible in-kind benefits in addition to wages.\footnote{These benefits include access to a free car, covered electronic communication (e.g., cellphone), covered lodging or food, employee loans, other personal benefits (e.g., covered membership at a fitness center), and travel allowances, all of which the employer reports to the authorities as taxable in-kind compensation.} We use this information to construct annual firm-level measures of in-kind compensation per worker and of the share of workers receiving in-kind compensation.

Using the information drawn from the matched employer--employee records---contract types and in-kind compensation---we can construct annual firm-level measures, respectively, across all workers employed in the firm, for stayers who remained in the same job across multiple years, and for incumbents who worked in a firm for some years and later changed to a different employer in our data. To avoid concerns about compositional changes in the workforce following demand shocks, we primarily focus on the stayers sample---workers who remained employed at the same firm throughout the event window.\footnote{We applied initial restrictions to the dataset, keeping workers with a non-missing firm identifier and at least $52 \times 4$ annual hours of work, and retained the highest-wage contract per worker-firm-year.}

We also have access to annual firm-level cost accounts, which allow us to decompose total labor costs into wage bill and non-wage personnel costs. These data are available for the period 2002--2022. Non-wage personnel costs consist of two components: employer-sponsored pension contributions above the regulated minimum and other non-wage personnel costs (e.g., employee perks, gifts, gym access, company cabin, travel and food allowances). Notably, taxable non-wage personnel costs that are divisible across workers are partly reported in the employment records and contained in our measures of in-kind compensation at the worker-firm level. However, firm-level non-wage personnel costs include additional components beyond divisible in-kind compensation, such as employer-sponsored pension contributions and non-divisible employee perks. Having observable firm-level non-wage labor costs allows us to quantify measures of compensation costs beyond hourly wages.\footnote{The annual firm-level cost accounts also provide information on payroll taxes, covering public pensions, sickness leave payments, unemployment insurance, etc. However, as payroll taxes are statutorily determined and levied proportionally to wage costs, there is no scope for firm-level adjustments in these (relative to wage costs) in response to shocks. We thus do not include payroll taxes in our measures of non-wage compensation.} Using firms' employment counts, we also construct per-worker measures for non-wage labor cost variables.

As an additional source of information on workplace amenities, we also use the near-universe of publicly posted job advertisements for the period 2002--2022. Specifically, we use information on workplace attributes that workers may value and that are mentioned in job ads by employers but not necessarily reflected in firms’ non-wage personnel costs. Such attributes could be salient aspects of employment conditions that are valuable to workers (e.g., the possibility to work from home or to have flexible working hours) but not captured in our measures of contract types drawn from the matched employer--employee records, or reflect more subjective aspects of the workplace (e.g., a good social environment), as communicated by the employers in job postings. Recently, \citet{bhuller2024ads} developed a text-analysis methodology to extract such information about workplace amenities from the full text portions of job postings. We use their approach to assign each job ad with mentions of specific workplace amenities and further use information on the employer posting each ad to construct annual firm-level measures of the share of postings listing each workplace amenity for our analysis. Our data sources include a register of job ads maintained by the Norwegian Labour and Welfare Administration (NAV) and a collection of job ads posted on the major private online job platform FINN.\footnote{As detailed in \citet{bhuller2024ads}, the employment agency collects information about vacancies from several sources, including online job boards and newspapers, as well as direct job opening reports from employers. Under the Labor Market Act §7, Norwegian employers must report publicly posted job openings to the agency, which maintains a comprehensive database of such ads to provide job-seekers with current information on suitable job opportunities. We accessed similar information from the FINN job platform for our analysis. Both sources provide employer identifiers, which allow us to construct firm-level measures of offered workplace amenities and to link job ad data to our remaining data sources. As in \citet{bhuller2024ads}, we exclude job ads posted by recruitment or temporary employment agencies, as we cannot link such ads to the actual firm or establishment where the job is posted.} Each job posting is categorized according to mentions of amenity attributes that we classify into eight different categories: the presence of on-the-job training, the presence of financial benefits (such as pensions, insurance, or mortgage support), workplace quality (such as interesting tasks, remote work, or social environment), additional perks (such as gym, meals, parking, or company car), regular work schedules, flexible working hours, permanent jobs, and good career opportunities. These measures capture amenities as they appear to prospective hires in job ads, which can be thought of as providing valuable signals about what workers may expect the firm to offer.\footnote{Notably, mentions of amenities detected in job ads are predictive of revealed-preference measures of job quality, such as the employer value of \cite{sorkin2018ranking} and the poaching index of \cite{bagger2019empirical}, and validated against realized job attributes and choices from a survey experiment \citep{bhuller2024ads}.}

Appendix Table \ref{tab:variable_construction} provides an overview of labor cost components and amenity measures we use, while Appendix Figure~\ref{fig:descr_contracts_jobads} shows the prevalence of non-wage compensation and non-pecuniary job attributes across the firm value-added distribution. High value-added firms have higher hourly wage costs as well as higher hourly non-wage compensation, and most non-pecuniary job attributes vary systematically across the firm distribution. In our data, the ratio of non-wage compensation to the wage bill varies around 10\% across firms, with high value-added firms having a higher share of compensation on pension premiums.

To summarize, while the firm accounts data provide precise information on the costs of monetized amenities, allowing us to directly measure how firms allocate resources across margins of non-wage compensation, information on non-wage job attributes from matched employer--employee data and job postings allows us to capture measures of amenities that are not necessarily visible in firms' non-wage personnel costs. We expect non-wage job attributes such as flexible hours, regular work schedules, training, workplace quality, permanent jobs, and full-time contracts to be paid largely in output terms. Notably, the measures of non-pecuniary job attributes and non-wage compensation derived from high-quality administrative registers have limited scope for measurement error. Moreover, having information on both wages and per-worker measures of non-wage personnel costs allows us to link our empirical analysis to the theoretical insights developed above. Since different amenities likely scale their provision costs with firm size differently---pension premiums may scale per worker, while on-site facilities involve fixed costs---disaggregating components allows us to test the model's predictions across amenities with potentially different cost structures.

As discussed in Section \ref{subsec:cs_psi}, our model makes distinct predictions about how demand shocks affect the level of amenity provision and per-worker amenity costs. Having measures of both allows drawing sharper inference. Job ads complement the tangible measures of contract types and non-wage benefits monetized in firms accounts by capturing more subjective, harder-to-quantify attributes, allowing us to track adjustments across a comprehensive amenity bundle. Together, these sources provide the most comprehensive multi-margin coverage of amenities we are aware of. This helps address the concern in \citet{mas2025non} that studies focusing on a single amenity dimension understate the firm's potential total adjustment.

\subsection{Measurement of Demand Shocks}

We use three distinct measures of firm-level demand shocks. First, we use annual firm-level balance sheets to construct events of abrupt changes in firm value-added, as in \cite{lamadon2022imperfect}. Such events capture both sudden firm-level changes in product demand and productivity over time, after purging common changes within the firm's local labor market. Second, we use information on public procurement contracts, exploiting the fact that contract winners gain substantial revenue streams that in turn affect labor demand, as in \cite{kroft2025imperfect} and \cite{volpe2024job}. Third, we use changes in international trade patterns following the Great Recession to construct firm-level exposure to export-demand shocks, as in \cite{garin2024export} and \cite{bhuller2025wage}. We next discuss each of these research designs in more detail. We refer to the first design as using an ``internal'' demand shock, while the remaining two designs rely on demand shocks that are ``external'' to the firm.

\paragraph{Internal shock.} Annual firm-level balance sheets provide information on firm revenues and value-added. We use the annual growth in value added from one year to the next to construct the internal shock, following \cite{lamadon2022imperfect}. Throughout, we index each event by its cohort $c$, the year when a firm is exposed to the shock. Let $\Delta VA_{cj}=\frac{VA_{cj,0}-VA_{cj,-1}}{VA_{cj,-1}}$ denote value-added growth for firm $j$ in cohort $c$ between event time $-1$ and $0$, and let $\Delta VA^{Median}_c$ denote the median value-added growth within the firm's two-digit industry in the same cohort. We define the exposure to an internal demand shock as follows:
\[
    Z^{Int}_{cj} =
        \begin{cases}
        1, & \text{if } \Delta VA_{cj} > \Delta VA^{Median}_c, \\
        0, & \text{if } \Delta VA_{cj} \leq \Delta VA^{Median}_c.
        \end{cases}
\]

\noindent We implement the internal shock design on different samples; Appendix Tables~\ref{tab:main_outcomes_sample_shocks}, \ref{tab:main_outcomes_sample} and \ref{tab:jobads_exten} provide summary statistics for each of these samples. The first is the extended sample, comprising cohorts from 2005--2019, which includes \firmsextenvalueadded \ firms with complete information on exposure to the internal shock, along with linked employer--employee data, detailed firm cost accounts over a 7-year window, and a minimum of two stayers. For around half of these firms, we can further link measures of amenities from job ads, resulting in $\firmsextentraining$ firms in the job postings sample. The third is the contracts sample, covering only the 2018--2019 cohorts, where there are \firmsvalueadded \ firms that can be linked to detailed matched employer--employee data with information on contract types and in-kind compensation, and at least two stayers.

\paragraph{Procurement shock.} We use administrative records on all public procurement contracts awarded in Norway, from which we identify firms that win a government contract for the first time. Public procurement has been used as a source of variation in firm-level demand in \cite{kroft2025imperfect} for U.S.\ construction firms. \cite{volpe2024job} adapts this strategy to Norwegian firms, and we follow a similar approach. We restrict the sample to two-digit sectors in which at least five firms won their first procurement contract between 2012 and 2022, and use the universe of first-time winners in the 2012--2019 cohorts. We define exposure to the procurement shock as a firm $j$ winning a government contract for the first time as:
\[
    Z^{Proc}_{cj} =
        \begin{cases}
        1, & \text{if firm } j \text{ wins its first procurement contract in cohort } c, \\
        0, & \text{if firm } j \text{ never wins a procurement contract.}
        \end{cases}
\]
The control group consists of never-treated firms within the same two-digit sectors, and was selected through a propensity score procedure.\footnote{We estimate a logit model for winning using data from one period before the firm wins, on pre-treatment employment, wages, value added, revenues, one-digit sector fixed effects, and commuting zone fixed effects, and restrict the control group to never-treated firms with a predicted winning probability above 1\%.} Appendix Table~\ref{tab:main_outcomes_sample_shocks} provides summary statistics for the sample used for this shock. For the 2012--2019 cohorts, we use \firmsprocvalueadded \ firms in the procurement shock design, for which we can link matched employer--employee data with detailed firm cost accounts over a 7-year window, and have at least two stayers.

\paragraph{Export shock.} We use changes in the patterns of international trade that took place following the Great Recession to construct a firm-level export demand shock exposure. This strategy follows \cite{garin2024export} for Portugal and \cite{bhuller2025wage} for Norway. We link the matched employer--employee data to firm-level customs records that contain exported values by 8-digit HS product code and destination country. We combine this with publicly available bilateral trade flow data from the \textit{BACI-CEPII} dataset \citep{gaulier2010baci}, disaggregated at the HS6 product and country-pair level, to construct the country-product demand shifts that form the export shock design.

Following \cite{garin2024export} and \cite{bhuller2025wage}, we use a shift-share strategy that interacts firms' pre-period export exposure across country-product pairs with changes in international import demand around the 2007--2009 Great Recession, $Z^{Ext}_{cj} = \sum_{p\in\CPset} s_{jp} \Delta_{p}$, where the exposure shares $s_{jp} = \text{Exp}_{jp}^{2005\text{--}07}/\left(\sum_{p\in\CPset}\text{Exp}_{jp}^{2005\text{--}07}\right)$ are constructed from firm $j$'s real exported value to country-product $p$ over 2005--2007. The symmetric growth rate is defined as $\Delta_{p} = (M_{p}^{\text{post}} - M_{p}^{\text{pre}}) / [(M_{p}^{\text{post}} + M_{p}^{\text{pre}})/2]$, where $M_{p}$ denotes the total value of non-domestic imports (i.e., leaving out imports from Norway) of country-product $p$ in the pre (2006--07) and post (2009--10) periods \citep{bhuller2025wage}.

For the export shock, we focus on a single cohort that experienced changes in export demand around 2007--2009 and link information from matched employer--employee data and detailed firm cost accounts over a 10-year window from 2002 to 2011. Notably, the time periods and available data differ substantially from those in the remaining analysis, as the information on contract types and divisible in-kind compensation is not available for the period used in the export-shock design. Appendix Table~\ref{tab:main_outcomes_sample_shocks} provides summary statistics for the main sample of \firmsexpvalueadded \ firms used in the export shock design.\footnote{As in \cite{bhuller2025wage}, we restrict the sample to exporting firms whose exports accounted for at least 1\% of revenues in 2005--2007, to capture firms in which exports constitute a meaningful economic activity.}

In Appendix Table~\ref{tab:industry_shares_shocks}, we report the industry composition of each shock sample (including both treatment and control firms). The export shock is dominated by manufacturing (\shareexpmanuf\%) and trade (\shareexptrade\%), reflecting the concentration of exporting activity in goods-producing and wholesale sectors. The procurement shock sample is instead tilted toward construction (\shareprocconstr\%), trade (\shareproctrade\%), and service sectors, consistent with the types of firms that typically win public contracts. The internal shock sample is more diversified, with trade (\shareinttrade\%) and construction (\shareintconstr\%) as the largest sectors. These differences in industry composition across designs should be kept in mind when comparing the shock responses.

\subsection{Estimation of Shock Responses}

For each shock, we estimate dynamic firm-level responses from event-study models as follows:

\begin{equation}
    \label{eq:dynamic_did}
    \ln Y_{jt}  =
    \sum_{\tau \in \mathcal{T}} \beta_{Y,\tau} \ Z_{cj} \ \mathbf{1}\{\tau = t - c\}
    + \delta^Y_{g(j)t} + \delta^Y_{\mathrm{CZ}(j)t} + \gamma^Y_j + v^Y_{jt},
\end{equation}
where $Y_{jt}$ represents different outcomes, including employment $L_{jt}$ and mean wages $w_{jt}$, besides value added $VA_{jt}$ and measures of amenities or non-wage compensation as discussed above, while $Z_{cj} \in \{Z^{Int}_{cj}, Z^{Proc}_{cj}, Z^{Ext}_{cj}\}$ denotes the relevant shock measure used in each shock design. The relative time-to-event year is indexed by $\tau$, so $\mathbf{1}\{\tau = t - c\}$ is an indicator for event time relative to the shock year $c$, and $\mathcal{T}$ is the set of relative years included. The term $\gamma^Y_j$ denotes firm fixed effects, which control for constant firm-level parameters, such as baseline firm productivity. The terms $\delta^Y_{g(j)t}$ and $\delta^Y_{\mathrm{CZ}(j)t}$ denote industry-by-year and commuting-zone-by-year fixed effects, where $g(j)$ is the two-digit industry of firm $j$ and $\mathrm{CZ}(j)$ is the commuting zone where it is located \citep{Bhuller2009}.\footnote{For multi-establishment firms, we use the location of the establishment with most employees.} We refer to the commuting zone-industry pair as the firm's local labor market $\ell(j)=\{g(j),\mathrm{CZ}(j)\}$. These terms isolate the effects from market-level demand shocks, which could otherwise confound the results. In the stacked designs of the internal and procurement shocks, all fixed effects are cohort-specific.

All event-study figures report coefficients with 95\% confidence intervals. For the internal shock, the baseline period is the event time $-2$, and we estimate the model using a balanced $\mathcal{T}=[-3,+3]$ event window, with standard errors clustered at the industry and commuting zone level. For the procurement shock, we use the same event window, but the baseline period is event time $-1$, and we cluster standard errors at the firm level. For these two designs, we restrict to firms observed in all event years that have at least two stayers. For the export shock, we use a single cohort window $\mathcal{T}=[2002,2011]$, with a baseline period of 2006, and stayers are defined as workers present at the firm for at least 8 years within the estimation window, and we restrict to firms with at least two stayers. Standard errors in this design are clustered at the firm level. Lastly, as a summary statistic, we report the post-shock average effect as $\bar{\beta}_Y = (\hat{\beta}_{Y,1} + \hat{\beta}_{Y,2} + \hat{\beta}_{Y,3})/3$ for the internal and procurement shock, and as $\bar{\beta}_Y = (\hat{\beta}_{Y,2010} + \hat{\beta}_{Y,2011})/2$ for the export shock. For outcomes in logarithms, we report percentage changes, which we compute as $100 \times (e^{\bar{\beta_Y}} - 1)$.

\subsection{Identification}
\label{subsec:identification_structural}

The dynamic firm-level responses from the event-study models \eqref{eq:dynamic_did} provide reduced-form estimates of the impacts of each demand shock on employment, wages, and amenity measures. A causal interpretation of these estimates as well as the identification of the wage-elasticity of effective labor supply $\varepsilon_w^L(L_j)$ and the size-elasticity of amenity cost multiplier $\varepsilon_L^\mathcal{A}(L_j)$ based on the shock responses requires additional assumptions. Next, we provide these identification conditions and discuss the connections to our theoretical model in Section \ref{sec:model}.

To simplify the presentation, we consider two periods, before ($t=0$) and after ($t=1$) a demand shock $Z_{j}$. We denote our data using first differences, $\Delta \ln X = \ln X_1 - \ln X_0$ for some variable $X$. Our data are then $\{\Delta\ln L_{j}, \Delta\ln w_{j}, \Delta\ln a_{jk}, \Delta\ln {VA}_{j}, Z_j\}_{j\in\J}$. We present these conditions in terms of a binary shock indicator $Z_{j}$, applying to the internal and procurement shocks, whereas analogous conditions are needed for the continuous export shock.

\begin{assumption}[Relevance]
\label{assu:relevance}
Firms exposed to the shock experience differential changes in value added within their local labor market $\ell(j)$,
\[
\mathbb{E}[\Delta \ln {VA}_{j}\mid Z_{j}=1,\ell(j)]
\neq
\mathbb{E}[\Delta \ln {VA}_{j}\mid Z_{j}=0,\ell(j)],
\]
where $Z_{j} \in \{Z^{Int}_{j}, Z^{Proc}_{j}, Z^{Ext}_{j}\}$. For the internal shock, this relevance condition is satisfied by construction. For the procurement and export shocks, it is an empirical requirement. In the context of our theoretical model in Section \ref{sec:model}, this assumption implies that each shock $Z_{j}$ must be correlated with changes in the structural productivity parameter $\Delta \psi_{j}$, which determines the location of the firm's labor demand curve.
\end{assumption}

\begin{assumption}[Exogeneity]
\label{assu:exogeneity}
Let $Y_{j}(1)$ and $Y_{j}(0)$ denote the potential outcomes with and without the shock, respectively. Then, absent the shock, the treated and control firms would have followed parallel trends, within their local labor market $\ell(j)$:
\[
\mathbb{E}[\Delta \ln Y_{j}(0)\mid Z_{j}=1,\ell(j)]
=
\mathbb{E}[\Delta \ln Y_{j}(0)\mid Z_{j}=0,\ell(j)].
\]

This assumption states that the residual change $\Delta v^Y_{j}$ is uncorrelated with the shock $Z_{j}$. In the context of our theoretical model in Section \ref{sec:model}, each shock $Z_{j}$ must be uncorrelated with changes in other firm-level determinants of wages and employment in \eqref{eq:star_w_L} and amenity provision in \eqref{eq:astar_ratio}, i.e., exogenous amenities $\Delta \bar{a}_{j}$ and unit costs of amenities $\Delta \mathbf{c}_{j}$.
\end{assumption}

We provide evidence supporting Assumption \ref{assu:relevance}, documenting statistically significant expansion of firms following each of the three demand shocks (Appendix Figure \ref{fig:value_added_shocks}). Averaged across the post-event periods, exposure to the internal demand shock increases firm value added by $\pctextenvalueadded$, while a procurement win increases it by $\pctprocvalueadded$ and a one-unit increase in export demand exposure increases it by $\pctexpvalueadded$. We also empirically assess parallel trends using pre-shock event-study coefficients, embedded in Assumption \ref{assu:exogeneity}. Across outcomes and all three shock designs, we find no evidence of pre-trends, suggesting that the parallel trends assumption is plausible. We thus interpret post-shock event-study estimates as the causal reduced-form impacts of demand shocks on firm employment, wages, and amenity provision.

\paragraph{Elasticity of effective labor supply.}
The wage-elasticity of effective labor supply $\varepsilon_w^L(L_j)$ is a key object in our analysis. As discussed in Section \ref{subsec:cs_psi}, $\varepsilon_w^L(L_j)$ varies with firm size when the super-elasticity~\eqref{eq:super_elasticity} is non-zero. Heuristically, the ratios of employment responses $\bar{\beta}_L(L_j)$ to wage responses $\bar{\beta}_w(L_j)$, the post-shock averages of the coefficients in~\eqref{eq:dynamic_did} measured across the firm distribution, provide the natural empirical analogs for $\varepsilon_w^L(L_j)$. This object can be estimated equivalently in an IV approach, where the demand shifter $Z_j$ serves as an instrument for the wage change $\Delta \ln w_j$ across the firm distribution. In an IV approach, a relevance condition analogous to Assumption \ref{assu:relevance} should be satisfied, requiring $\bar{\beta}_w(L_j)\ne0$ across the firm distribution. Assumption \ref{assu:exogeneity} clarifies which amenity-related components may adjust without challenging the identification of $\varepsilon_w^L(L_j)$; in an IV approach, the analogous conditions can be stated as exclusion restrictions. Under these assumptions, we can identify the wage-elasticity of effective labor supply $\varepsilon_w^L(L_j)$.

\paragraph{Elasticity of amenity cost multiplier.}
The size-elasticity of the amenity cost multiplier $\varepsilon_L^\mathcal{A}(L_j)$ is another key object in our analysis. Its identification, and that of the structural preference parameters ($\eta,\rho$), requires additional assumptions. The multiplier $\mathcal{A}(L_j)$ in~\eqref{eq:amultiplier} contains all endogenous amenities that the firm provides, both monetized and production-relevant, whereas the firm cost accounts record only the monetized amenities.
    To make progress, we partition endogenous amenities into three disjoint sets: $\sM$, the monetized amenities, including all non-wage costs as well as financial benefits and perks listed in job ads; $\sP$, the production-relevant contract types and other attributes listed in job ads; and $\sU$, the amenities observed in neither source. We construct an augmented multiplier that embeds the observed production-relevant amenities, as Assumption \ref{ass:sep} formalizes.

\begin{assumption}[Partition]
\label{ass:sep}
The sets partition the amenities, $\sM\cup\sP\cup\sU=\{1,\dots,n\}$ and $\sM\cap\sP=\sM\cap\sU=\sP\cap\sU=\emptyset$. The firm accounts contain provision costs of all amenities in $\sM$ and of no other amenities, and the non-wage job attributes recorded in job ads data contain all amenities in $\sP$. Hence, the monetized amenity cost multiplier is $\AM=1+\sum_{k\in\sM}\omega_{jk}$, and the augmented multiplier is $\AMP=\AM+\sum_{k\in\sP}\omega_{jk}$.
\end{assumption}

Notably, the monetized amenity cost multiplier $\AM$ is the ratio of \emph{monetized compensation} $w_jL_j+\sum_{k\in\sM}C_{jk}$ to the wage bill, and the augmented multiplier $\AMP$ is the ratio of \emph{augmented compensation} $w_jL_j+\sum_{k\in\sM\cup\sP}C_{jk}$ to the wage bill. Meanwhile, the amenity multiplier $\mathcal{A}$ in~\eqref{eq:amultiplier} equals $\AMP+\sum_{k\in\sU}\omega_{jk}$. As the augmented multiplier $\AMP$ leaves out the unobserved amenities in $\sU$, its shock responses equal those of $\mathcal{A}$ only under Assumption \ref{ass:prop} stated below.

\begin{assumption}[Proportionality]
\label{ass:prop}
Firms exposed to the shock experience on average the same change in $\ln(\mathcal{A}_j/\AMP_j)$ as other firms within their local labor market $\ell(j)$,
\[
\begin{aligned}
&\mathbb{E}[\Delta\ln\AMP_j\mid Z_{j}=1,\ell(j)]-\mathbb{E}[\Delta\ln\AMP_j\mid Z_{j}=0,\ell(j)]\\
&\qquad=\mathbb{E}[\Delta\ln\mathcal{A}_j\mid Z_{j}=1,\ell(j)]-\mathbb{E}[\Delta\ln\mathcal{A}_j\mid Z_{j}=0,\ell(j)].
\end{aligned}
\]
In the context of our theoretical model in Section \ref{sec:model}, the amenity cost multiplier $\mathcal{A}$ and the augmented multiplier $\AMP$ have the same size-elasticity, $\varepsilon_L^{\mathcal{A}}(L_j)=\varepsilon_L^{\AMP}(L_j)$, so the shock leaves the unobserved amenities’ share of total compensation unchanged, on average.
\end{assumption}

As the ratio of multipliers $\mathcal{A}_j/\AMP_j$ equals total compensation $w_j L_j + \sum_{k\in\sM\cup\sP\cup\sU} C_{jk}$ divided by augmented compensation $w_jL_j+\sum_{k\in\sM\cup\sP}C_{jk}$, Assumption~\ref{ass:prop} implies that shock-exposed firms experience the same average log change in total compensation relative to augmented compensation as other firms within their local labor market. Under  Assumption~\ref{ass:prop}, $\varepsilon_L^{\mathcal{A}}=\varepsilon_L^{\AMP}$, hence the total compensation markdown $\mathcal{AM}$ in~\eqref{eq:foc_AwMRP} is identified as long as both $\varepsilon_w^L$ and $\varepsilon_L^{\AMP}$ are identified. Note, however, that Assumption~\ref{ass:prop} does not impose $\mathcal{A}=\AMP$, which requires the stronger assumption of \textit{no unobserved endogenous amenities}. Thus, even though Assumptions~\ref{assu:relevance}-\ref{ass:prop} provide the identification of $\mathcal{AM}$, identifying its two components: the amenity multiplier $\mathcal{A}$ and the wage markdown $\mathcal{M}$, requires a stronger assumption.

In our empirical analysis, the multiplier shock responses we estimate are derived from shock responses in the measured amenities in $\sM$ and $\sP$. The monetized amenity cost multiplier $\AM$ is the ratio of the wage bill plus non-wage personnel costs to the wage bill. The ratios of its shock responses $\bar{\beta}_{\AM}(L_j)$ to employment responses $\bar{\beta}_L(L_j)$, measured across the firm distribution, provide the natural empirical analogs for its size-elasticity, $\varepsilon_L^{\AM}(L_j)$. In an IV approach, where the demand shifter $Z_j$ serves as an instrument for the employment change $\Delta\ln L_j$, the relevance condition analogous to Assumption \ref{assu:relevance} requires $\bar{\beta}_L(L_j)\neq0$.

Moving from $\varepsilon_L^{\AM}(L_j)$ to the size-elasticity of the augmented multiplier $\varepsilon_L^{\AMP}(L_j)$ requires the cost shares of the amenities in $\sP$. From~\eqref{eq:MRS}, the firm provides each amenity until its per-worker cost equals workers' valuation. Multiplying both sides of~\eqref{eq:MRS} by $a_{jk}/w_j$ gives $\omega_{jk}=C_{jk}/(w_jL_j)=\gamma_k(a_{jk}/w_j)^{\rho}$, so workers' taste $\gamma_k$ for attribute $k$ together with $\rho$ gives $k$'s cost share. Appendix~\ref{app:structural_estimation} describes its measurement. Along the shock, $\gamma_k$ and $\rho$ are fixed, so each attribute's cost share moves by $d\ln\omega_{jk}=\rho\,(d\ln a_{jk}-d\ln w_j)$, and differentiating $\AMP=\AM+\sum_{k\in\sP}\omega_{jk}$ with respect to $\ln L_j$ gives the amenity multiplier's size-elasticity,
\begin{equation}
\varepsilon_L^{\mathcal{A}}=\varepsilon_L^{\AMP}=\sigma^{\sM}_j\,\varepsilon_L^{\AM}+\rho\sum_{k\in\sP}\sigma^{\sP}_{jk}\Bigl(\frac{d\ln a_{jk}}{d\ln L_j}-\frac{1}{\varepsilon_w^L}\Bigr).
\label{eq:AP_resp}
\end{equation}
The first equality follows from Assumption~\ref{ass:prop}. The second writes $\varepsilon_L^{\AMP}$ as the sum of two terms. The first term is the elasticity of the monetized amenity cost multiplier, weighted by the share of monetized compensation in augmented compensation, $\sigma^{\sM}_j=\AM_j/\AMP_j$. The second term adds the response of each attribute's cost share, weighted by the share of attribute $k\in\sP$ in augmented compensation, $\sigma^{\sP}_{jk}=\omega_{jk}/\AMP_j$. The weights sum to one, $\sigma^{\sM}_j+\sum_{k\in\sP}\sigma^{\sP}_{jk}=1$. Each cost share moves with the attribute's provision relative to the wage, where $d\ln a_{jk}/d\ln L_j$ is the ratio of the attribute's response to the employment response. Since the elasticity $\varepsilon_L^{\AMP}$ depends on $\rho$, both directly and through the measured shares $\sigma^{\sM}_j$ and $\sigma^{\sP}_{jk}$, it is only jointly identified with the structural parameters, to which we turn next.

\paragraph{Structural parameters.}
In our framework, a productivity shock $d\psi_j$ traces out, simultaneously, the effective labor supply elasticity $\varepsilon_w^L(L_j)$ and the amenity cost multiplier elasticity $\varepsilon_L^\mathcal{A}(L_j)$. Combining~\eqref{eq:eps_w_def} and~\eqref{eq:epsLA} gives the structural relationship between them,
\begin{equation}
\frac{1}{\varepsilon_w^L(L_j)} \;=\; \frac{1}{\eta} \;-\; \frac{1}{\rho}\varepsilon_L^\mathcal{A}(L_j).
\label{eq:eos_link}
\end{equation}
As $\eta$ and $\rho$ are parameters that are assumed to be common to all firms while $\Phi(L_j)$ varies with size, the pair $(\varepsilon_w^L(L_j),\varepsilon_L^\mathcal{A}(L_j))$ traced across the distribution of firms identifies both parameters: across firms, $1/\varepsilon_w^L$ moves with $\varepsilon_L^\mathcal{A}$ at slope $-1/\rho$ and intercept $1/\eta$.\footnote{In our model, $\varepsilon_L^{\mathcal{A}}(L_j)$ close to zero reflects a cost structure close to per-worker on aggregate ($\Phi(L_j) \approx 0$) or preferences close to Cobb--Douglas ($\rho \approx 0$). A zero response can also reflect lumpy adjustment: if changing an amenity involves a fixed cost, the firm leaves it at its pre-shock level unless the shock is large enough.} Under Assumption~\ref{ass:prop}, $\varepsilon_L^{\mathcal{A}}=\varepsilon_L^{\AMP}$, and thus by substituting~\eqref{eq:AP_resp} into~\eqref{eq:eos_link}, we can jointly identify $(\eta,\rho)$ from the responses of employment $L_j$, wages $w_j$, the monetized amenity cost multiplier $\AM_j$ and the production-relevant attributes $a_{jk}$, and their valuations $\gamma_k$ for ${k\in\sP}$. With $(\eta,\rho)$ in hand, the amenity cost multiplier elasticity $\varepsilon_L^\mathcal{A}$ follows from~\eqref{eq:AP_resp} and the cost-curvature index follows from~\eqref{eq:eps_w_def}, $\Phi(L_j)=(1-\rho)(1/\eta-1/\varepsilon_w^L)$. The total compensation markdown~\eqref{eq:foc_AwMRP} requires only the two elasticities $(\varepsilon_w^L,\varepsilon_L^\mathcal{A})$ and follows once $(\eta,\rho)$ are identified.\footnote{Proposition \ref{res:markdown_amenities} shows that, unless $\rho=0$ or $\Phi(L_j)=0$, the distribution of $\varepsilon_w^L(L_j)$ alone does not identify the total compensation markdown $\mathcal{A}(L_j)\mathcal{M}(L_j)$. From \eqref{eq:foc_AwMRP}, this requires both the labor supply elasticity $\varepsilon_w^L(L_j)$ and the amenity cost multiplier elasticity $\varepsilon_L^\mathcal{A}(L_j)$.} We obtain the variation across firms that the identification requires by estimating the shock responses separately within deciles of the firm distribution and fitting $(\eta,\rho)$ to the resulting elasticities across deciles. Appendix~\ref{app:identification} describes this non-linear estimation, and considers cases in which this variation is absent, i.e., $\Phi(L_j)=0$ or $\rho=0$, where $(\eta,\rho)$ are at most partially identified.

\section{Evidence on Workplace Amenity Provision}
\label{sec:demand_shocks}

We now provide causal evidence from the three event-study designs on adjustments in employment, wages and amenities in response to firm-level demand shocks.

\subsection{Impacts on Wages and Employment}

A key focus of the recent literature quantifying employer market power is how firm-level demand shocks affect employment and wages, which jointly inform the elasticity of firm-level labor supply in canonical monopsony models \citep{lamadon2022imperfect,garin2024export}. Before turning to our evidence on firm responses across various margins of non-wage compensation, we show the employment and wage responses induced by each shock, along with estimates of the implied firm-level labor supply elasticities.

\begin{figure}
\centering
\caption{The Effects of Demand Shocks on Employment and Wages.}
\label{fig:emp_wage_shocks}

\begin{subfigure}[t]{0.48\textwidth}
    \includegraphics[width=\linewidth]{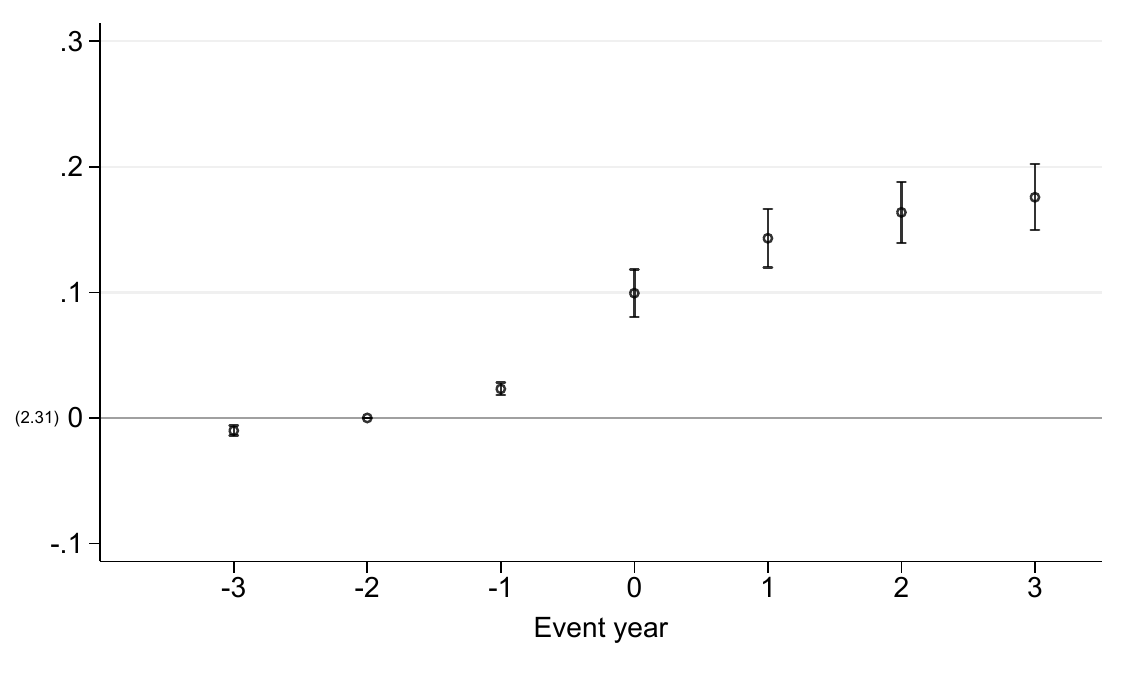}
    \caption{Internal Shock: Employment}
\end{subfigure}\hfill
\begin{subfigure}[t]{0.48\textwidth}
    \includegraphics[width=\linewidth]{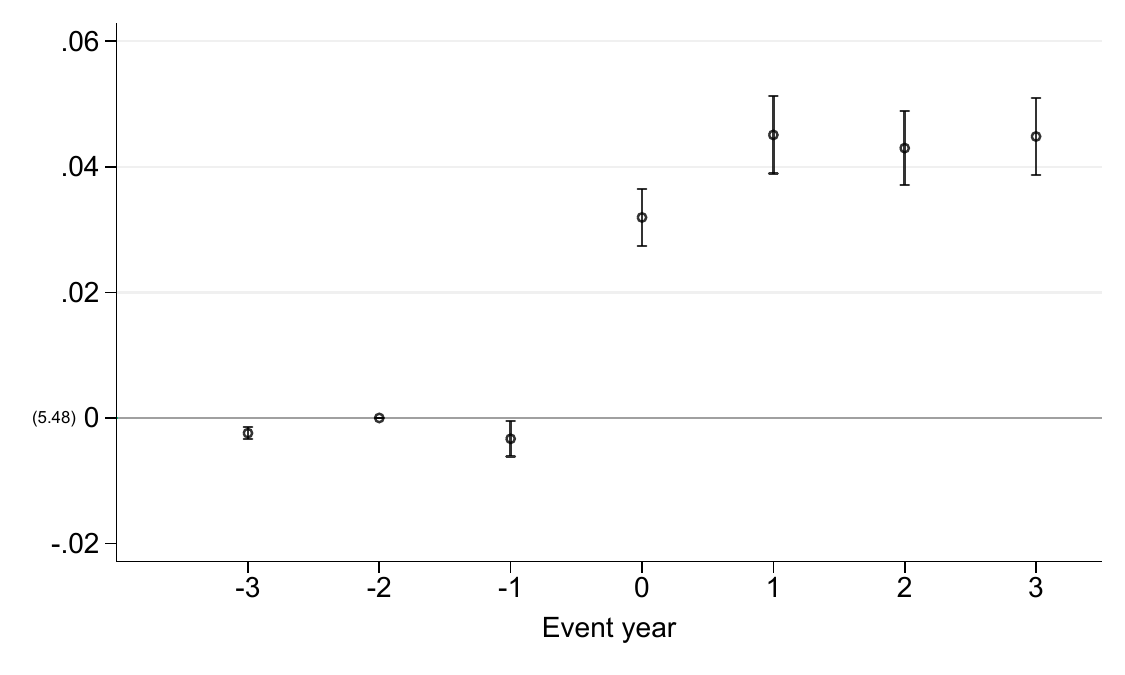}
    \caption{Internal Shock: Wages}
\end{subfigure}

\vspace{0.8em}

\begin{subfigure}[t]{0.48\textwidth}
    \includegraphics[width=\linewidth]{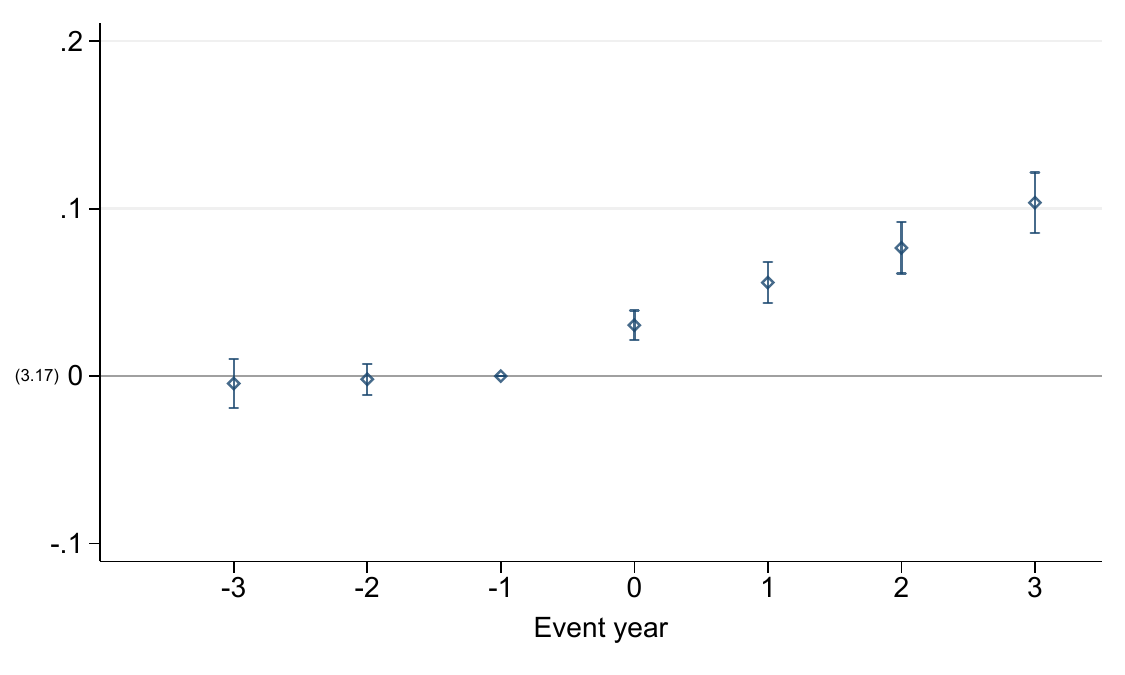}
    \caption{Procurement Shock: Employment}
\end{subfigure}\hfill
\begin{subfigure}[t]{0.48\textwidth}
    \includegraphics[width=\linewidth]{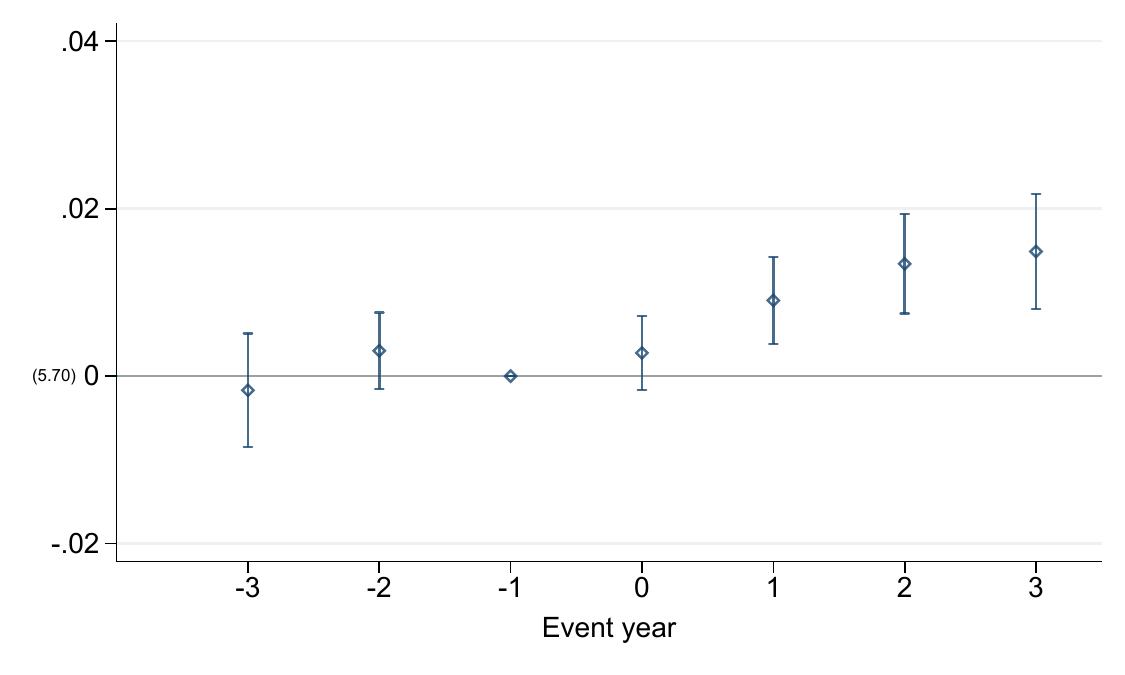}
    \caption{Procurement Shock: Wages}
\end{subfigure}

\vspace{0.8em}

\begin{subfigure}[t]{0.48\textwidth}
    \includegraphics[width=\linewidth]{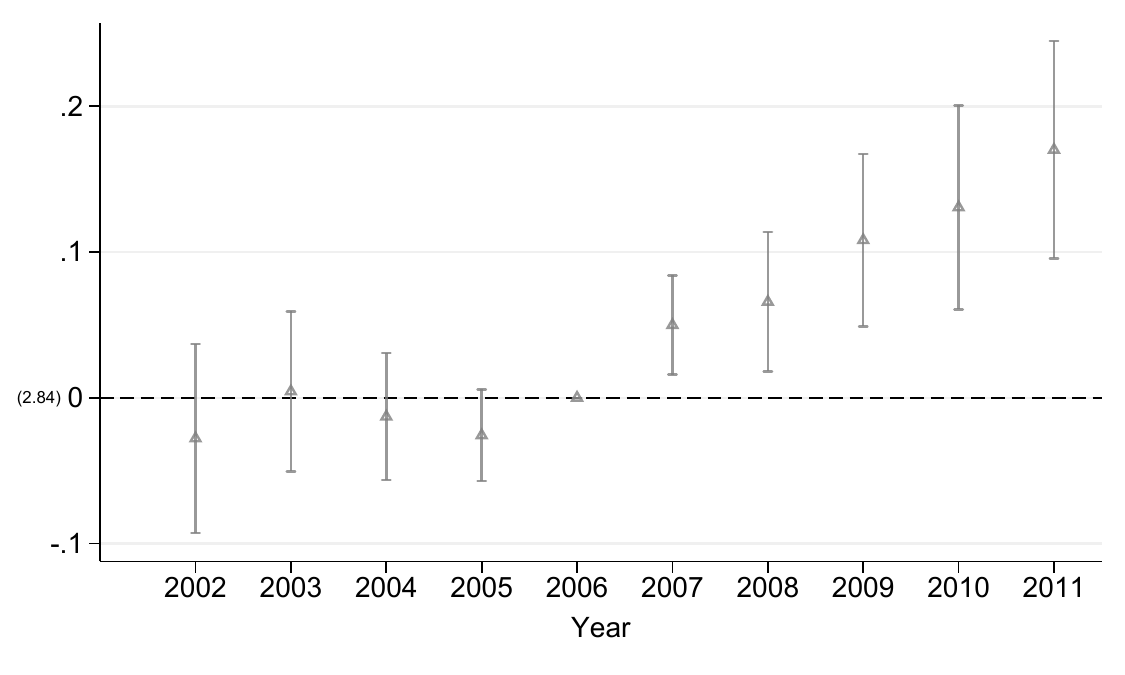}
    \caption{Export Shock: Employment}
\end{subfigure}\hfill
\begin{subfigure}[t]{0.48\textwidth}
    \includegraphics[width=\linewidth]{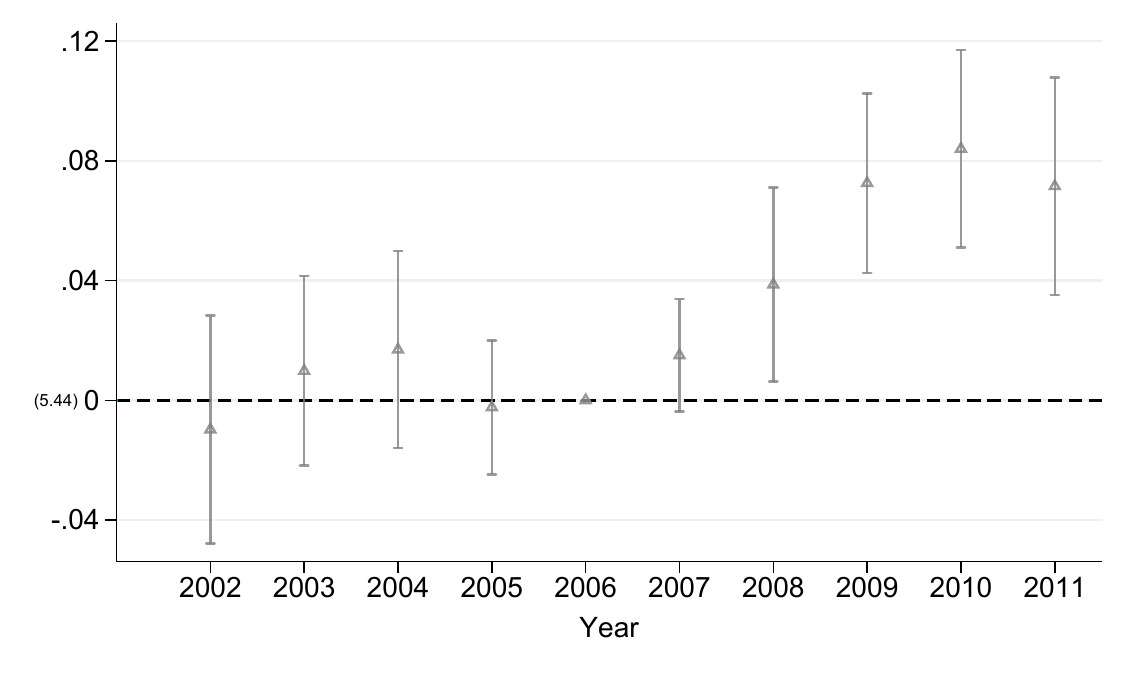}
    \caption{Export Shock: Wages}
\end{subfigure}

\begin{minipage}{\textwidth}
\scriptsize
\emph{Notes:} The main datasets used to construct outcomes are matched employer--employee data. Coefficients are event-time estimates with 95\% confidence intervals from the dynamic specification in Eq.~(\ref{eq:dynamic_did}). All outcomes are in logs. Wage outcomes are expressed in hourly terms and constructed from the stayers sample. Panels~(a)--(b): internal shock for cohorts 2005--2019 with standard errors clustered at the industry and commuting zone level. Panels~(c)--(d): procurement shock for cohorts 2012--2019 with standard errors clustered at the firm level. For these two shocks, we restrict to firms observed in all years of the balanced $[-3,+3]$ event window and having at least two stayers. Panels~(e)--(f): export shock with year 2006 as the baseline period and standard errors clustered at the firm level. Stayers in this shock are defined as workers who are present at the firm for at least 8 years during the estimation window. The export sample is restricted to firms with an export-to-revenue ratio of at least 1\% in 2005--2007.
\end{minipage}
\end{figure}

Figure~\ref{fig:emp_wage_shocks} reports employment and wage responses across the three demand shocks.\footnote{As visualized across panels~(a)--(c) of Appendix Figure~\ref{fig:value_added_shocks}, all three shocks induce salient increases in firm value added, but the magnitudes differ. When interpreting the evidence on reduced-form impacts presented here, it is important to note these differences in relative shock intensities.} Averaged across the post-event period, we find that employment expands by \pctextenemployment\ and stayers' wages increase by \pctextenwagestayers\ using the internal shock (panels~a--b). These reduced-form impacts provide an implied effective labor supply elasticity of $\varepsilon^L_w=\bar{\beta}_L/\bar{\beta}_w \approx \elasw$ and an implied wage pass-through of $\varepsilon^w_{VA}=\bar{\beta}_w/\bar{\beta}_{VA} \approx \elasrs$. For the procurement shock (panels~c--d), we find that employment expands by \pctprocemployment\ and stayers' wages increase by \pctprocwagestayers, while for the export shock (panels~e--f), employment expands by \pctexpemployment\ and stayers' wages increase by \pctexpwagestayers. Appendix Table~\ref{tab:main_outcomes_postavg_shocks} provides detailed point estimates across the three shocks.

While the magnitudes of employment and wage impacts differ across the three shocks, partly reflecting differences in shock intensities (Appendix Figure~\ref{fig:value_added_shocks}) and partly differences in firm composition (Appendix Table~\ref{tab:industry_shares_shocks}), we find that firms respond significantly to each shock on both the employment and wage margins. The qualitative pattern is nevertheless consistent across all three designs: employment responds more than wages, and the implied elasticities are broadly similar despite differences in the sources of variation exploited.\footnote{In Appendix Figure~\ref{fig:bothshocks_emp_wage}, panels~(a)-(b), we provide additional evidence on firm-level employment and wage responses using the internal shock design, focusing on the extended sample (2005–2019 cohorts) and the contracts sample (2018–2019 cohorts). We find very similar employment responses, with minor differences across samples in the wage effects. Further, we re-estimate the internal shock specification for firms in the procurement-shock sample (panels~(c)-(d)) and the export-shock sample (panels~(e)-(f)), for comparison to the external shock estimates. Once sample differences are accounted for, we find similar employment and wage responses across the designs, suggesting that the different wage responses in the export- or procurement-shock sample reflect the composition of firms across samples rather than the design of the shock itself.}

Using the internal shock design, in Appendix Figure~\ref{fig:app_emp_wages}, we further document that our findings are robust across alternative definitions of firm-level employment outcomes, such as using full-time equivalent employment. Our main evidence on wage impacts focuses on stayers who remained employed with the same firm; however, we also report the wage responses for hires, as well as the wage and employment responses for all incumbent workers by following their future labor market trajectories.

\subsection{Impacts on Non-Wage Compensation and Amenity Provision}
\label{subsec:jobads_contracts}

Having established how firm-level demand shocks affect employment and wage margins, the remaining analysis examines whether firms also adjust workplace amenity provision in response to such shocks. In existing studies, this margin is typically assumed fixed or omitted, despite being an important component of the compensation bundle. We now use detailed information on categories of non-wage compensation to study this margin of response.

\begin{figure}[h!]
\centering
\caption{The Effects of Demand Shock on Wage and Non-Wage Compensation.}
\label{fig:nonwage_types}

\begin{subfigure}[t]{0.48\textwidth}
    \includegraphics[width=\linewidth]{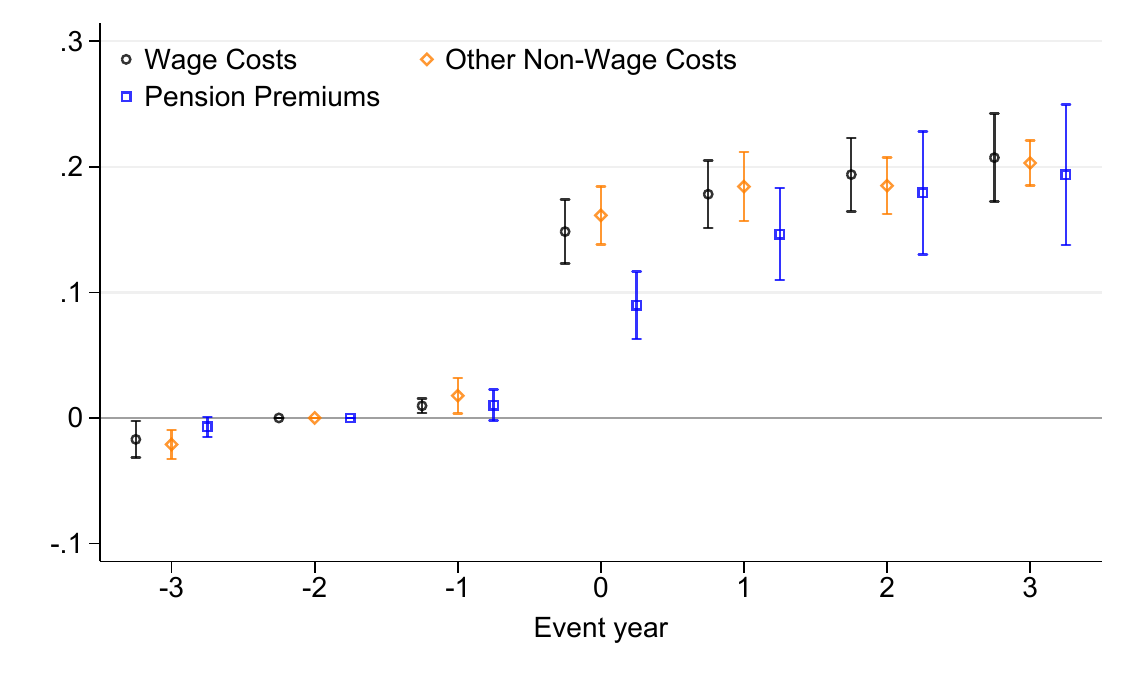}
    \caption{Total Compensation}
\end{subfigure}\hfill
\begin{subfigure}[t]{0.48\textwidth}
    \includegraphics[width=\linewidth]{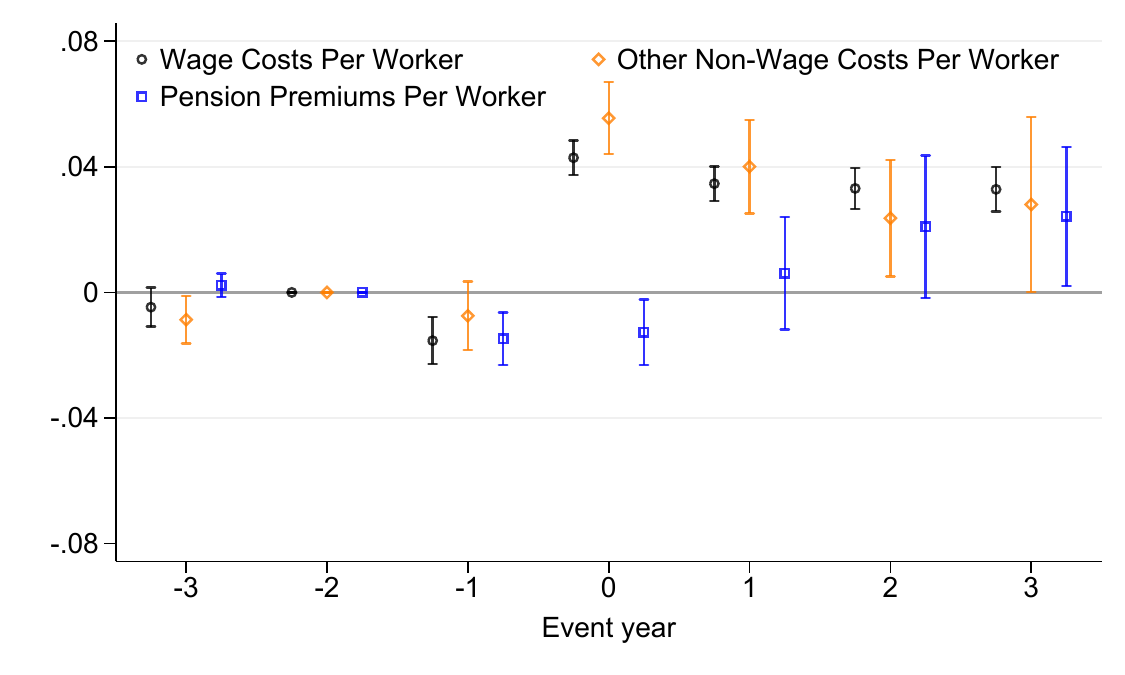}
    \caption{Per-Worker Compensation}
\end{subfigure}
\begin{minipage}{\textwidth}
\scriptsize
\emph{Notes:} Outcomes are constructed from firm cost accounts and matched employer--employee data. Coefficients are event-time estimates with 95\% confidence intervals from the dynamic specification in Eq.~(\ref{eq:dynamic_did}). All outcomes are in logs. Panel~(a) reports total compensation costs by type, and panel~(b) reports per-worker costs. We use the internal shock design for the contracts sample covering the 2018--2019 cohorts, restricted to firms observed throughout the balanced $[-3,+3]$ event window with at least two stayers. Standard errors are clustered at the industry and commuting zone level.
\end{minipage}
\end{figure}

\paragraph{Non-wage compensation.} Figure~\ref{fig:nonwage_types} reports the responses of wage and
non-wage components of compensation, measured using firms' cost accounts, to
demand shocks. Panel~(a) reports total personnel costs and panel~(b) reports
per-worker costs, both using the internal shock design (contracts sample), highlighting heterogeneity across non-wage cost components. Focusing on total personnel costs (panel a), we find that wage costs rise by \pctwagecost, pension premiums by
\pctpenspremcost, and other non-wage personnel costs by \pctnonwagecost. Turning to per-worker personnel costs (panel b), we find that wage costs increase by $\pctwagecostperworker$, pension premiums per worker by \pctpenspremperworker, and other non-wage personnel costs by \pctnonwageperworker.\footnote{Appendix Figure~\ref{fig:external_nonwage_types} shows the corresponding responses for the procurement shock (panels~a--b) and the export shock (panels~c--d). Wage costs rise under both shocks, in total and per worker. Total non-wage costs also rise, by less than wage costs. The per-worker responses of the non-wage components are imprecisely estimated, with confidence intervals that include zero.}

One interpretation for the heterogeneous responses across non-wage components, consistent with Proposition~\ref{res:demand_shocks}-\textit{\textbf{(ii)}}, is that per-worker non-wage cost responses vary with the structure of amenity provision costs. When wages and amenities enter as substitutes ($\rho>0$) in worker preferences, components whose provision cost scales less than proportionally with employment exhibit positive per-worker responses, while those with sufficiently convex cost structures may decline even as total spending rises. As discussed in Section~\ref{subsec:cs_psi}, to disentangle whether these differences reflect cost curvature or the structure of preferences, we need to observe shock responses at the level of amenity provision, which we turn to next.

\begin{figure}[H]
\centering
\caption{The Effects of Demand Shock on Advertised Job Attributes.}
\label{fig:contract_types_jobads}
\begin{subfigure}[t]{0.48\textwidth}
    \centering
    \includegraphics[width=\linewidth]{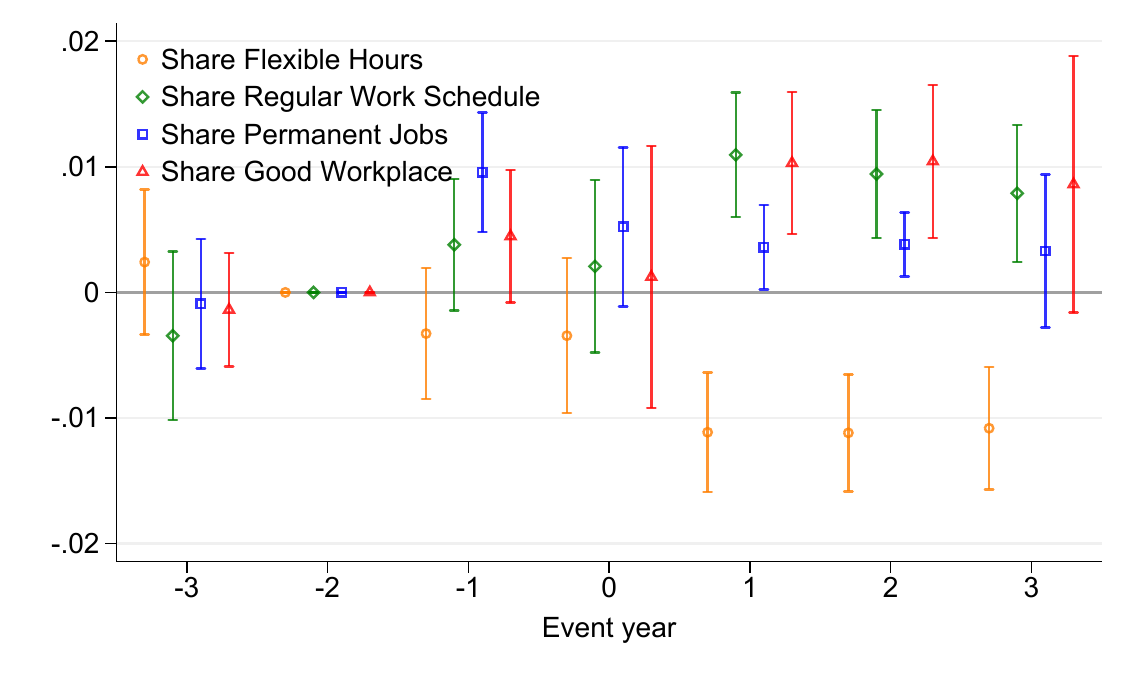}
    \caption{Flexibility and Workplace Conditions}
\end{subfigure}\hfill
\begin{subfigure}[t]{0.48\textwidth}
    \includegraphics[width=\linewidth]{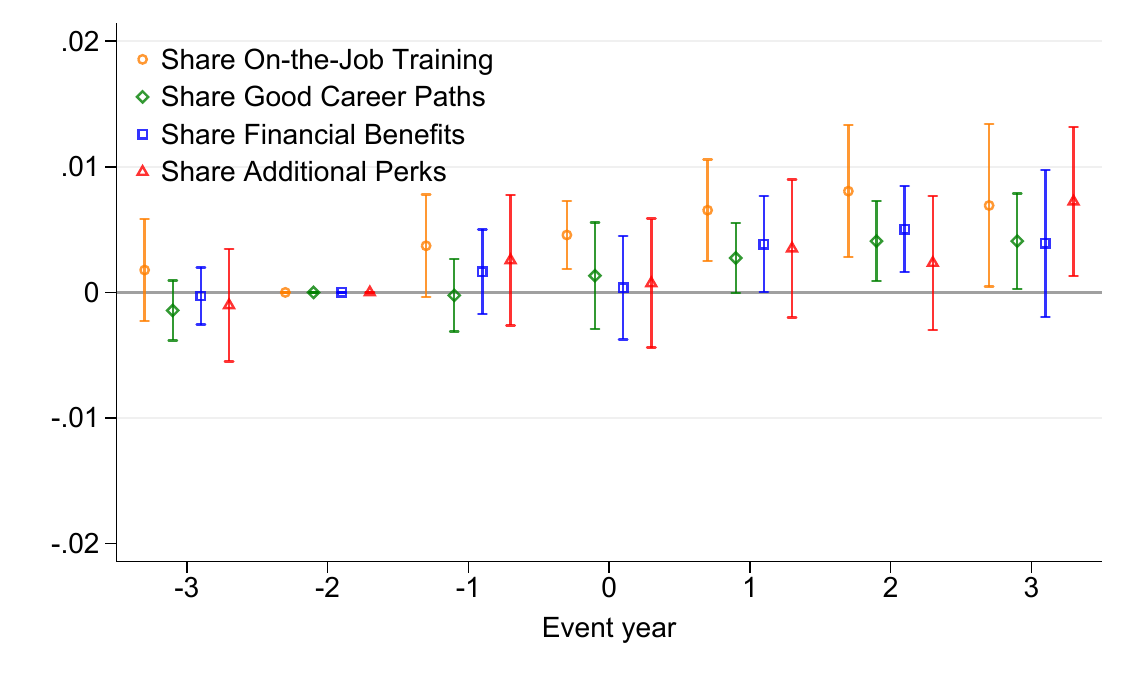}
    \caption{Training, Career Paths and Other Perks}
\end{subfigure}

\begin{minipage}{\textwidth}
\scriptsize
\emph{Notes:} Outcomes are constructed from job ads data (FINN/NAV) linked to the extended sample. Coefficients are event-time estimates with 95\% confidence intervals from the dynamic specification in Eq.~(\ref{eq:dynamic_did}). Outcomes measure the shares of posted job ads mentioning each listed workplace amenity. We use the internal shock design for the 2005--2019 cohorts, restricted to firms observed in all years of the balanced $[-3,+3]$ event window, with at least two stayers throughout the full estimation period. Job-ad outcomes are measured for posting firms. Standard errors are clustered at the industry and commuting zone level.
\end{minipage}
\end{figure}

\paragraph{Advertised job attributes.} We next examine non-wage job attributes advertised by posting firms in Figure~\ref{fig:contract_types_jobads}. Each outcome measures the share of a firm's posted job ads that mention a specific amenity or a category of related amenities. Appendix Table~\ref{tab:jobads_exten} reports post-shock averages, pre-shock means, and number of observations for all advertised-amenity outcomes. First, in panel~(a), we consider mentions of attributes related to the flexibility of working hours and workplace conditions. We find an overall increase in mentions of good workplace quality by \ppextengoodworkplace\ pp. However, mentions of flexible hours respond negatively, with an estimated change of \ppextenconvenient\ pp.\footnote{By Proposition~\ref{res:demand_shocks}, an amenity declines after a positive shock when its provision cost is sufficiently convex in firm size. For instance, firms may face convex coordination costs in the provision of flexible work hours.} In contrast, the share of ads offering regular work schedules increases by \ppextenregular\ pp and the share offering permanent jobs by \ppextenpermanent\ pp, relative to pre-shock mean shares of \meanextenregular\ and \meanextenpermanent, respectively. Next, in panel~(b), we consider mentions of attributes related to on-the-job training, good career paths, financial benefits (e.g., pensions, insurance) and additional perks. The mentions of good career paths increase by \ppextencareer\ pp, on-the-job training by \ppextentraining\ pp, additional perks by \ppextenaddedperks\ pp, and financial benefits by \ppextengoodfinance\ pp following the shock. These increases are statistically significant, suggesting that firms enhance the non-wage bundle offered to prospective hires following the demand shock.

\begin{figure}[H]
\centering
\caption{The Effects of Demand Shock on Contract Types and In-Kind Compensation.}
\label{fig:inkind_margins}
\begin{subfigure}[t]{0.48\textwidth}
    \centering
    \includegraphics[width=\linewidth]{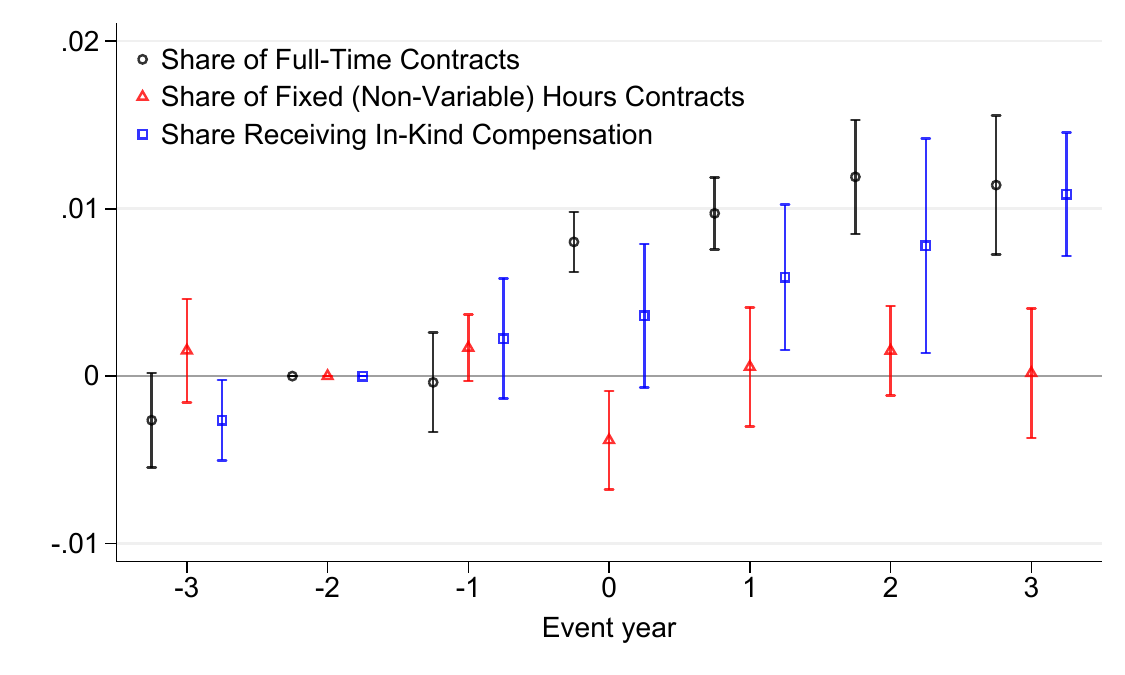}
    \caption{Contract Types}
\end{subfigure}\hfill
\begin{subfigure}[t]{0.48\textwidth}
    \includegraphics[width=\linewidth]{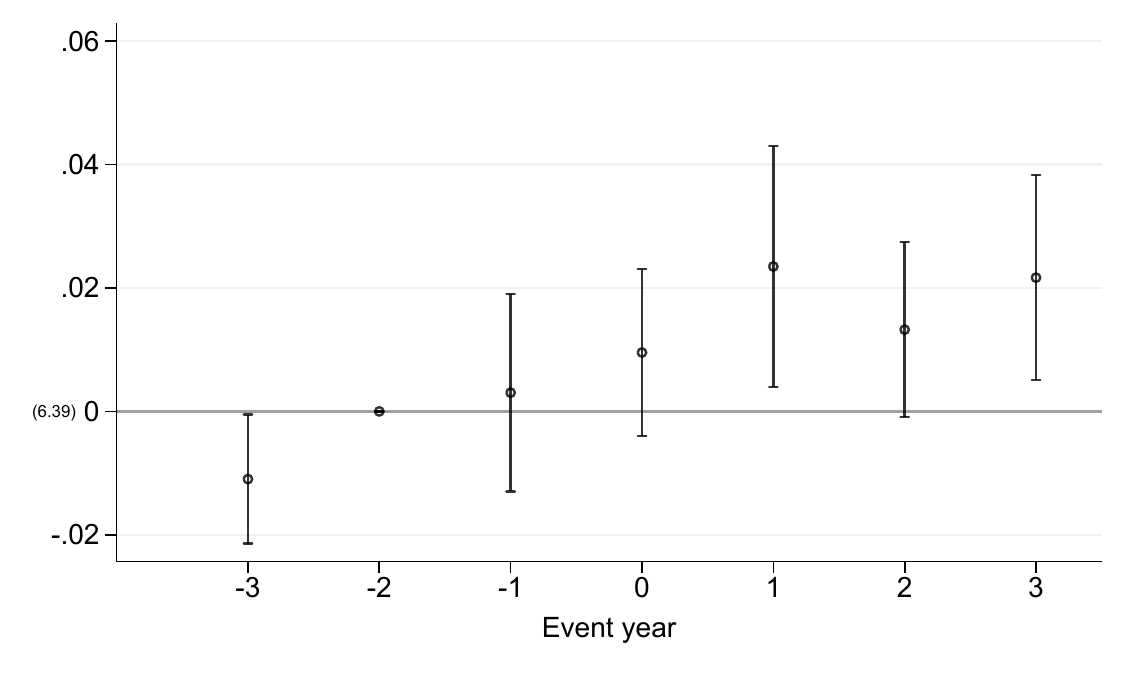}
    \caption{In-Kind Compensation}
\end{subfigure}

\begin{minipage}{\textwidth}
\scriptsize
\emph{Notes:} Outcomes are constructed from matched employer--employee data (A-meldingen). Coefficients are event-time estimates with 95\% confidence intervals from the dynamic specification in Eq.~(\ref{eq:dynamic_did}). Panel~(a) measures the shares of stayers in full-time (above 20 hours of work per week) and fixed (non-variable) hours contracts, and the share of stayers receiving in-kind compensation as part of their pay contract. Panel~(b) measures the log of in-kind compensation amount received among stayers receiving in-kind compensation as part of their pay contract. We use the internal shock design for the contracts sample covering the 2018–2019 cohorts, restricted to firms observed in all years of the balanced $[-3,+3]$ event window and with at least two stayers throughout the full estimation period. Standard errors are clustered at the industry and commuting zone level.
\end{minipage}
\end{figure}

\paragraph{Contract types and in-kind compensation.}
We now turn to contract types and in-kind compensation from matched employer--employee data. We use the internal shock design on the contracts sample covering the 2018--2019 cohorts. Figure~\ref{fig:inkind_margins} documents adjustments in contract types and in-kind compensation among stayers. In panel~(a), full-time shares increase by $\ppfulltimestayers$ pp, while the share of fixed (non-variable) hours contracts remains largely unchanged. The share of stayers receiving any form of in-kind compensation increases by \ppinkindstayers\ pp, and panel~(b) shows that in-kind pay conditional on receiving any increases by \pctpayinkindstayers. Both margins respond positively to a positive shock, indicating that firms provide in-kind benefits to more workers and also increase the in-kind benefit amounts.

\section{Implications for Employer Market Power}
\label{sec:discussion}

Having documented the multi-margin adjustments to firm-level demand shocks, we now turn to the implications of our findings for measuring employer market power. Following the identification approach described in Section~\ref{subsec:identification_structural}, we start by estimating the demand shock responses across the firm distribution. Using the distributional responses in wages, employment, non-wage costs and non-wage job attributes, we estimate the structural parameters $(\eta,\rho)$. Finally, we recover the implied distributions of the cost-curvature index and the total compensation markdowns, which we decompose into cost curvature and conduct.

\begin{figure}[h!]
\centering
\caption{The Distribution of Effective Labor Supply Elasticities.}
\label{fig:elasticities_deciles_labor}
\begin{subfigure}[t]{0.48\textwidth}
    \includegraphics[width=\linewidth]{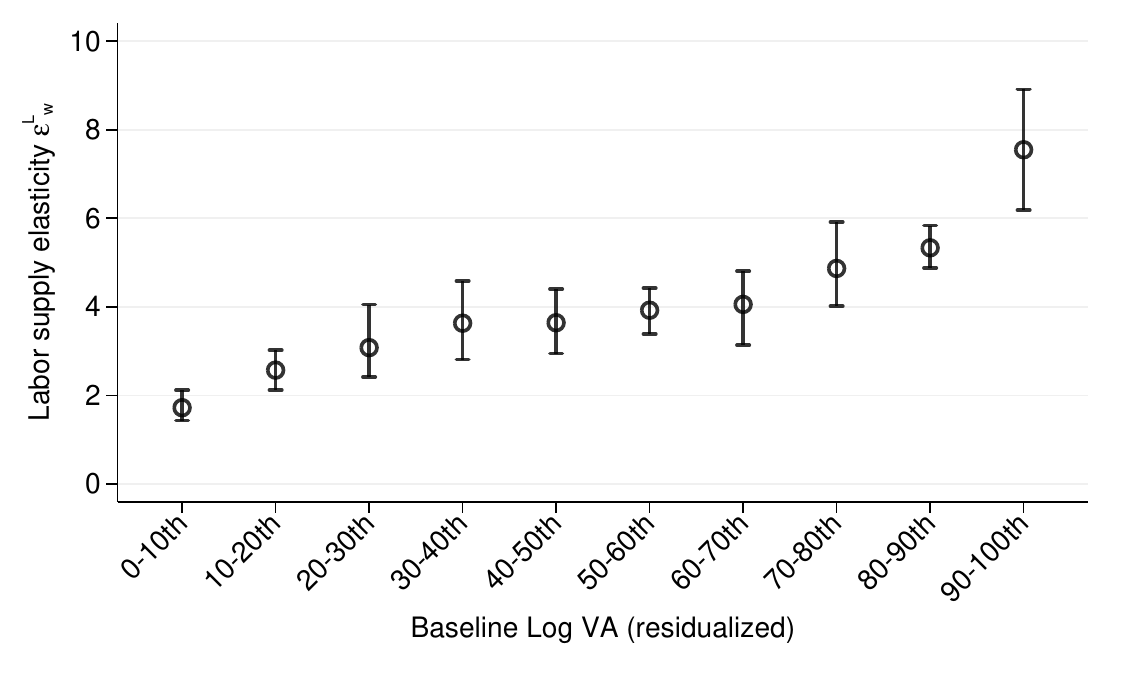}
    \caption{Ranked by Baseline Value Added}
\end{subfigure}\hfill
\begin{subfigure}[t]{0.48\textwidth}
    \includegraphics[width=\linewidth]{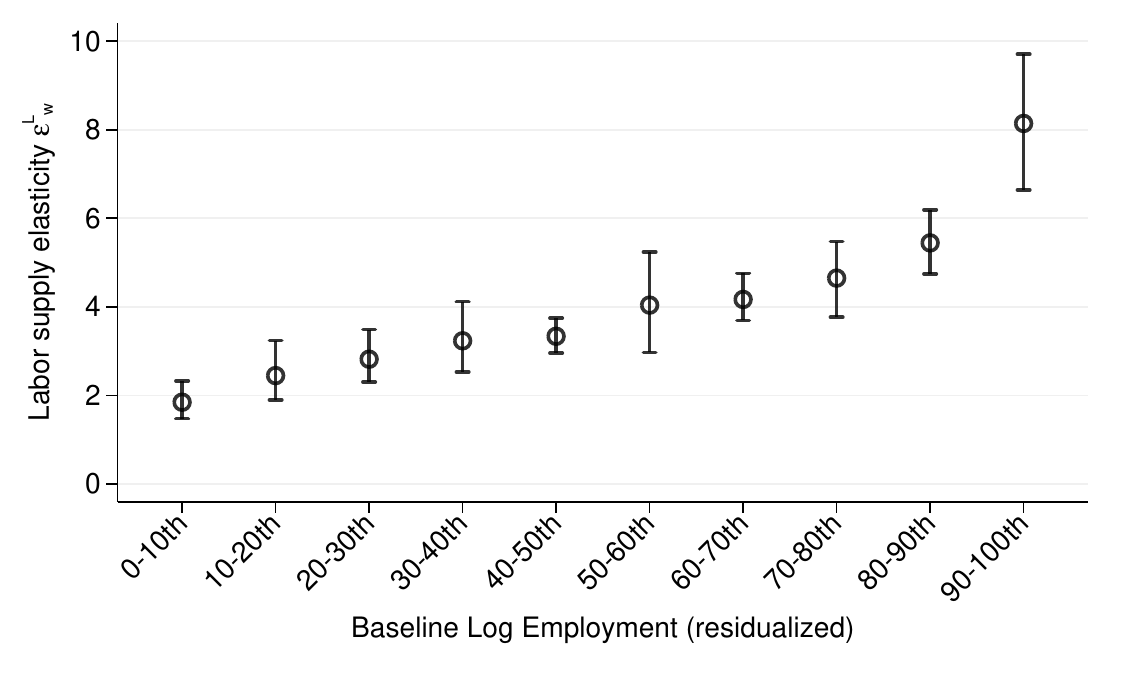}
    \caption{Ranked by Baseline Employment}
\end{subfigure}
\begin{minipage}{\textwidth}
\scriptsize
\emph{Notes:} Elasticities are constructed from the post-shock average
effects $\bar{\beta}_Y$ following the dynamic specification in
Eq.~\eqref{eq:dynamic_did}: the labor supply elasticity $\varepsilon^L_w$
is the ratio of the employment and wage coefficients. Confidence bands are 95\% percentile intervals; their asymmetry reflects the skewed sampling distribution of a  coefficient ratio, most pronounced where the wage response is least precisely estimated. Firms are ranked by baseline log value added (panel (a)) or baseline log employment (panel (b)), residualized by two-digit industry and commuting zone. We estimate a separate dynamic specification within each decile of residualized baseline log value added and baseline log employment. We use  internal shock design for cohorts 2005--2019.
\end{minipage}
\end{figure}

\paragraph{Elasticity of effective labor supply.}
Figure~\ref{fig:elasticities_deciles_labor} shows that the effective wage-elasticity of labor supply $\varepsilon_w^L(L_j)$ rises steeply across the firm distribution, from about 2 at the bottom decile to about 8 at the top decile, suggesting that firms with higher value added (panel a) and firms that are larger in size (panel b) face a more elastic effective labor supply. Focusing on the average labor supply elasticity estimate would thus mask this apparent heterogeneity. In these calculations, we residualized pre-shock firm value added (panel a) and employment size (panel b) by two-digit industry and commuting zone, ranking firms into ten deciles of the baseline levels of value added and employment size, respectively, and estimate internal-shock specifications separately for each decile to recover the decile-specific elasticities. Ranking firms by residualized baseline value added or by employment gives similar patterns.

\paragraph{Structural parameters.}
We recover $\rho$ and $\eta$ from the decile-specific variation in $\varepsilon_w^L$ and $\varepsilon_L^{\mathcal{A}}$ using the relationship~\eqref{eq:eos_link}, with the augmented multiplier's elasticity $\varepsilon_L^{\AMP}$ in~\eqref{eq:AP_resp} built from the monetized amenity cost multiplier's elasticity $\varepsilon_L^{\AM}$ and the non-wage job attribute responses at each decile. Using the non-linear estimation procedure described in Appendix~\ref{app:identification}, we obtain $\hat\rho \approx \paRhoHat$ with 95\% CI $[\paRhoLo,\paRhoHi]$ and $\hat\eta \approx \paEtaHat$ with 95\% CI $[\paEtaLo,\paEtaHi]$. Using the monetized amenity cost multiplier alone, we get $\hat\rho\approx\paRhoAcc$ and $\hat\eta\approx\paEtaAcc$. Our estimates of the structural parameters thus suggest (i) a comparatively modest dispersion of idiosyncratic tastes, with $\hat\eta$ above the average effective labor supply elasticity of $\elasw$, as convex amenity cost structure lowers the effective elasticity below $\hat\eta$, and (ii) that wages and amenities are gross substitutes with  $\hat\rho>0$, though close to the Cobb--Douglas benchmark of $\rho =0$.

\begin{figure}[t!]
\centering
\caption{The Distribution of Amenity Cost Multiplier Elasticities.}
\label{fig:elasticities_deciles_multiplier}
\begin{subfigure}[t]{0.48\textwidth}
    \includegraphics[width=\linewidth]{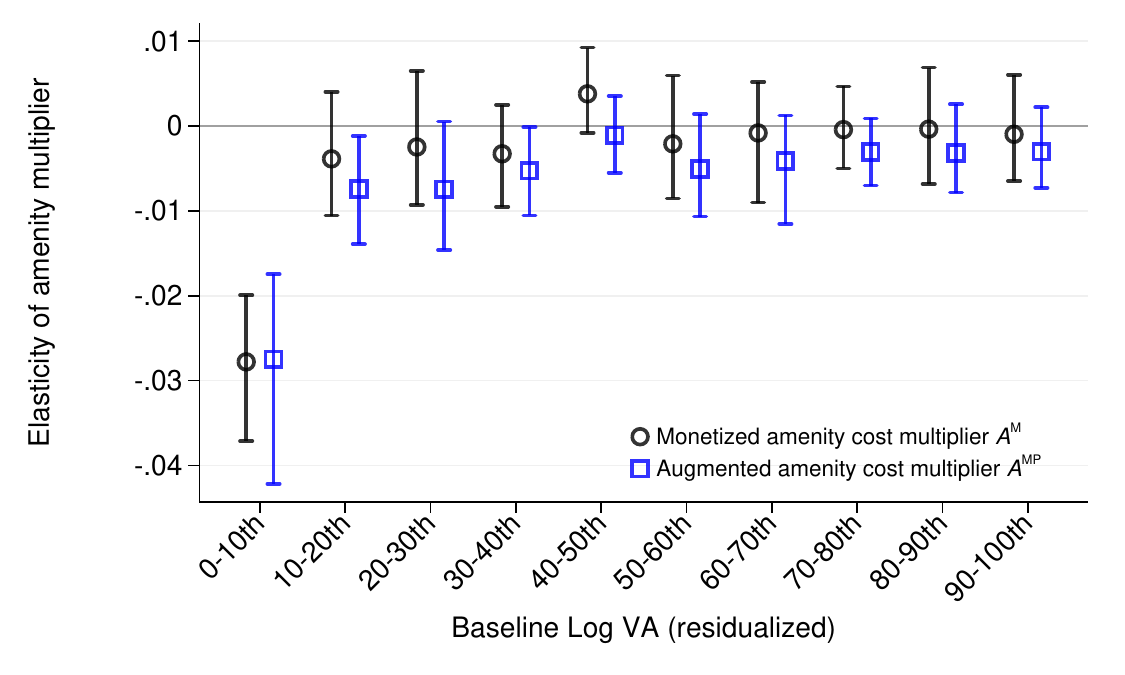}
    \caption{Ranked by Baseline Value Added}
\end{subfigure}\hfill
\begin{subfigure}[t]{0.48\textwidth}
    \includegraphics[width=\linewidth]{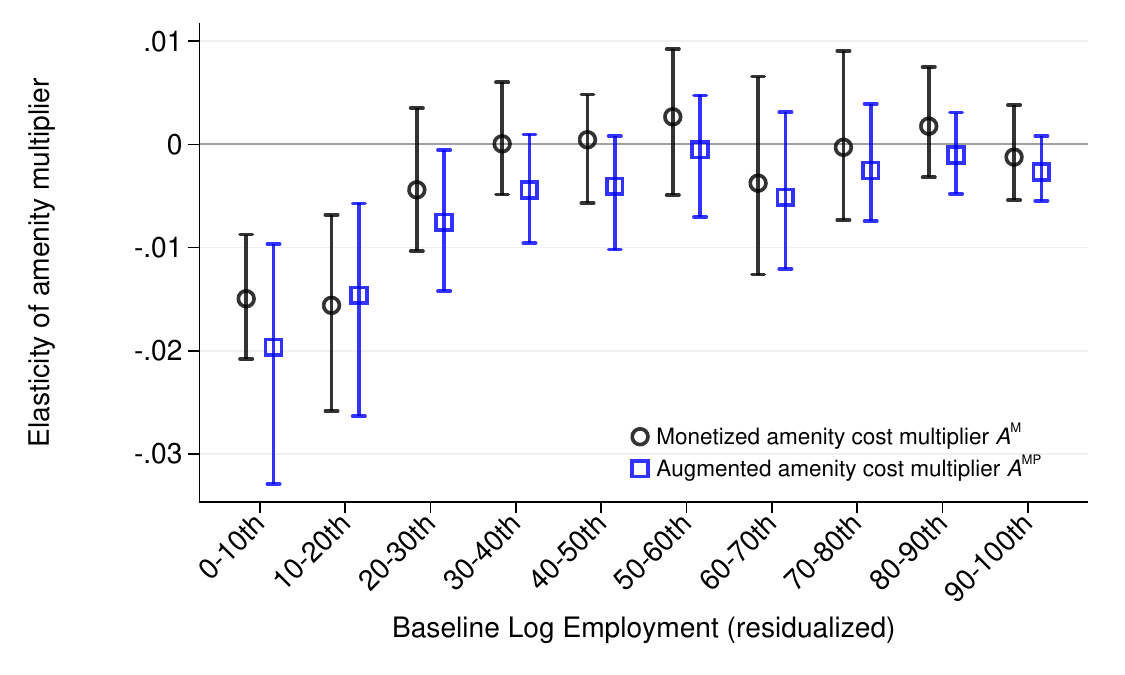}
    \caption{Ranked by Baseline Employment}
\end{subfigure}
\begin{minipage}{\textwidth}
\scriptsize
\emph{Notes:} Elasticities are constructed from the post-shock average
effects $\bar{\beta}_Y$ following the dynamic specification in
Eq.~\eqref{eq:dynamic_did}: the elasticity of the amenity cost multiplier
with respect to employment is shown for the monetized amenity cost multiplier, $\varepsilon^{\AM}_L$ (circles), and for the augmented amenity cost multiplier of Section~\ref{subsec:identification_structural}, which adds the attributes paid in output, $\varepsilon^{\AMP}_L$ (squares). Confidence bands are 95\% percentile intervals
from a parametric bootstrap ($B=1{,}000$) that draws each decile's
coefficient vector from $N(\hat\beta^d,\hat\Omega_d)$, and for the squares also the attribute responses and valuations, and recomputes the
ratios; their asymmetry reflects the skewed sampling distribution of a
coefficient ratio, most pronounced where the wage response is least
precisely estimated. Panel (a) ranks firms by baseline log value added and panel (b) by baseline log employment, both residualized by two-digit industry and commuting zone; in panel (b), the attribute responses are estimated by employment decile. We estimate a separate dynamic specification within each decile.  We use  internal shock design for cohorts 2005--2019.
\end{minipage}
\end{figure}

\begin{figure}[t!]
\centering
\caption{The Distribution of Total Compensation Markdowns and Cost Structure.}
\label{fig:markdown_deciles}
{\small\itshape Ranked by Baseline Value Added}\par\smallskip
\begin{subfigure}[t]{0.48\textwidth}
    \includegraphics[width=\linewidth]{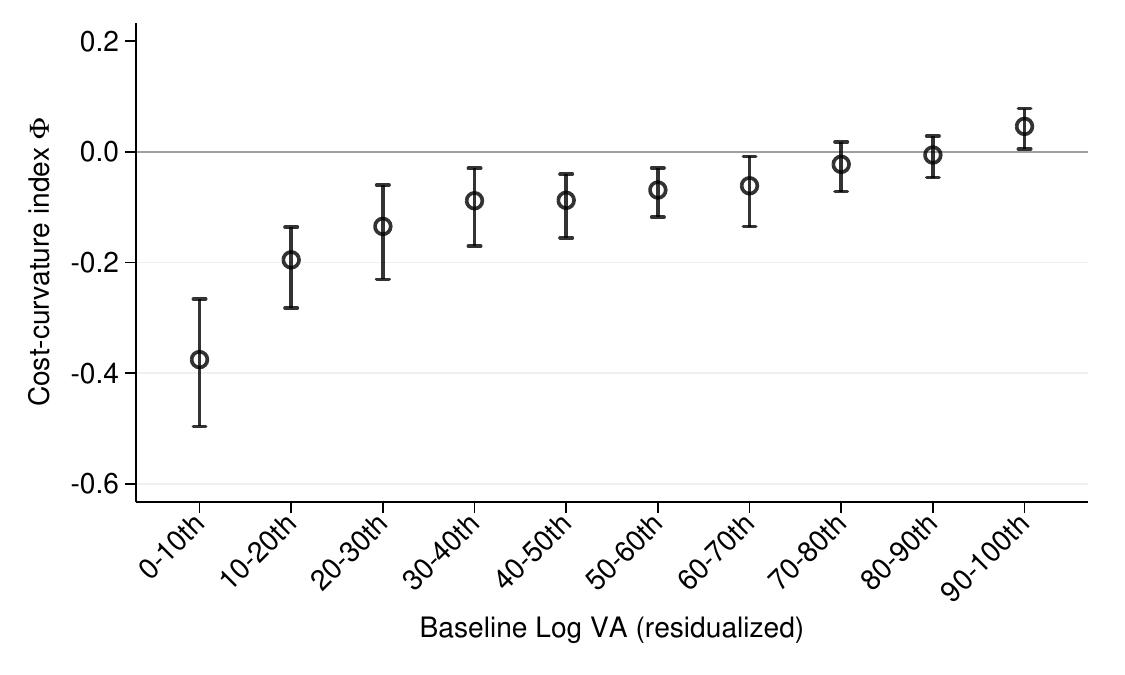}
    \caption{Amenity Cost-Curvature Index ($\Phi(L_j)$)}
\end{subfigure}\hfill
\begin{subfigure}[t]{0.48\textwidth}
    \includegraphics[width=\linewidth]{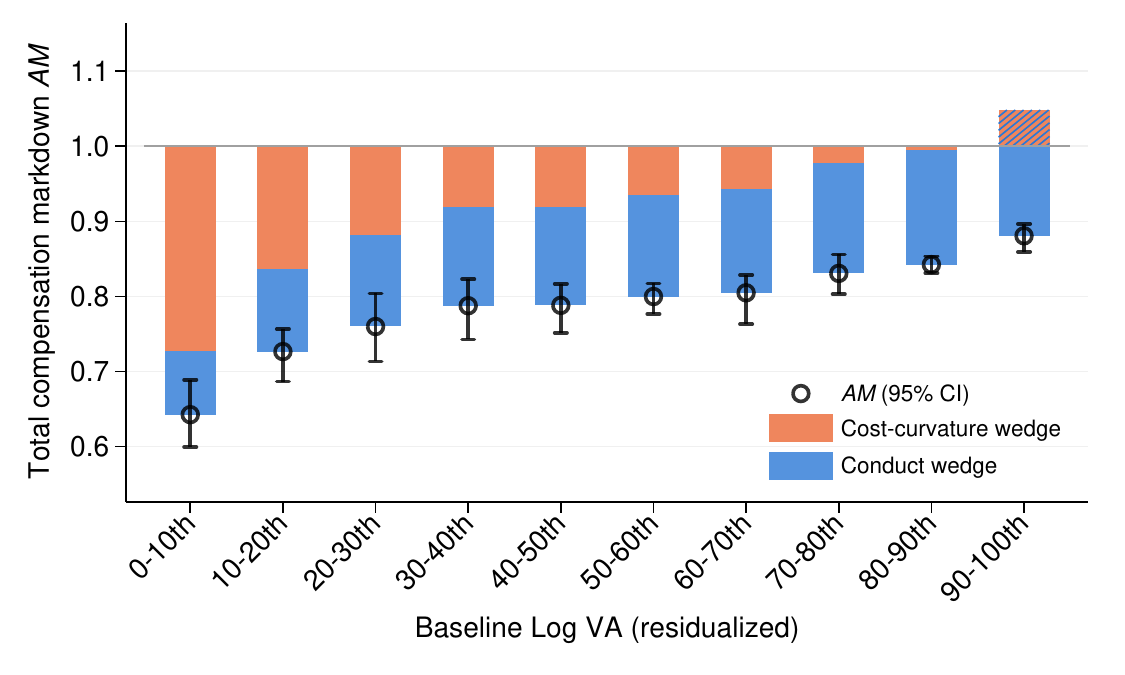}
    \caption{Total Compensation Markdown ($\mathcal{A}\mathcal{M}$)}
\end{subfigure}

\medskip
{\small\itshape Ranked by Baseline Employment}\par\smallskip
\begin{subfigure}[t]{0.48\textwidth}
    \includegraphics[width=\linewidth]{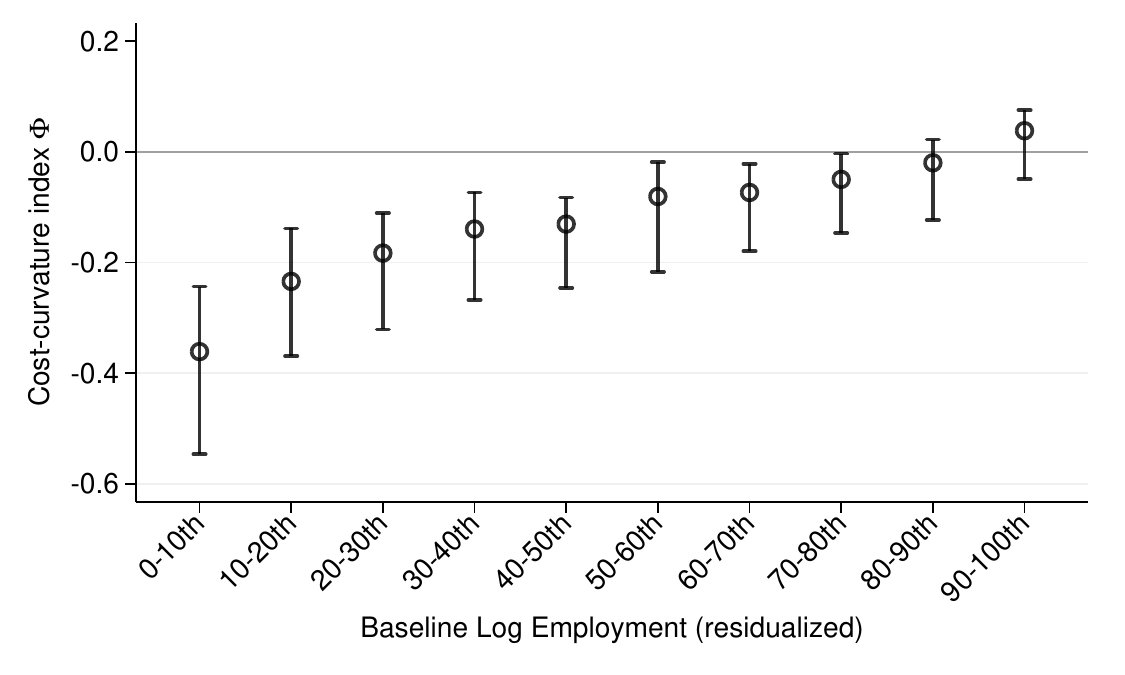}
    \caption{Amenity Cost-Curvature Index ($\Phi(L_j)$)}
\end{subfigure}\hfill
\begin{subfigure}[t]{0.48\textwidth}
    \includegraphics[width=\linewidth]{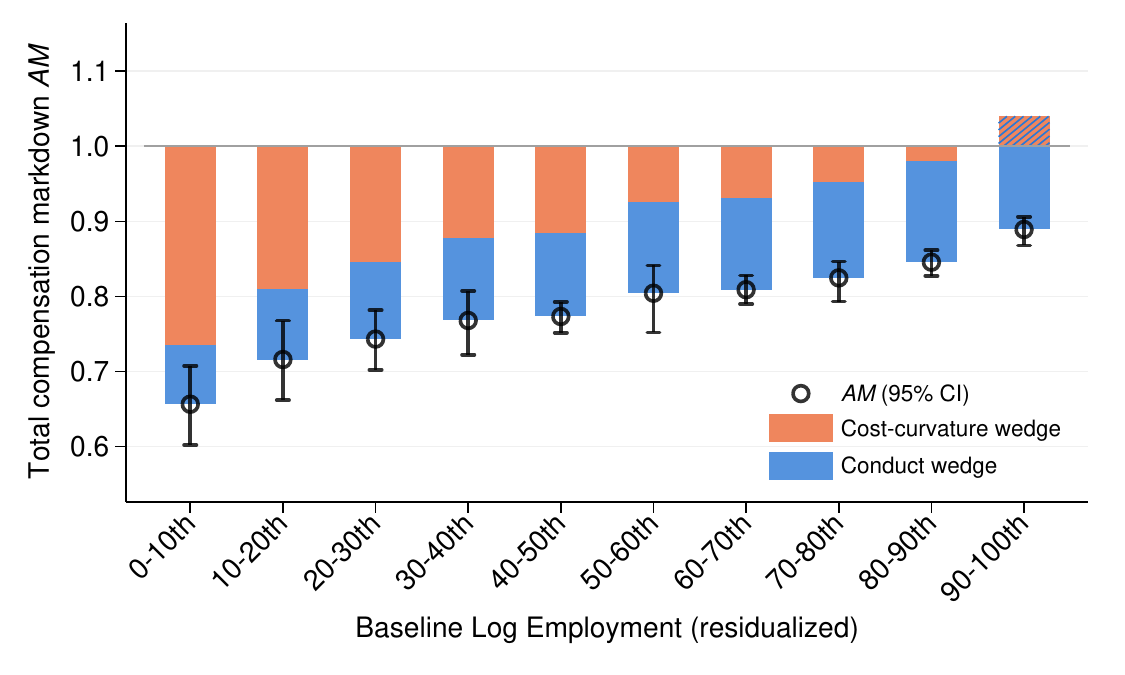}
    \caption{Total Compensation Markdown ($\mathcal{A}\mathcal{M}$)}
\end{subfigure}
\begin{minipage}{\textwidth}
\scriptsize
\emph{Notes:} Both objects are implied by the structural estimates
$(\hat\eta,\hat\rho)$ recovered in Appendix~\ref{app:identification}
(see Eq.~\eqref{eq:eos_link}), evaluated decile by decile at the
estimated effective labor supply elasticity $\varepsilon_w^L$. The
total compensation markdown is
$\mathcal{A}\mathcal{M} = \eta/(\eta + 1 - \eta\Phi)$ and the
cost-curvature index is $\Phi(L_j) = (1-\rho)\,(1/\eta - 1/\varepsilon_w^L)$, both over the amenities we observe, in money and in output, with $(\hat\eta,\hat\rho)$ estimated on the augmented multiplier (Appendix~\ref{app:identification}).
Negative $\Phi$ indicates convex aggregate amenity costs. Confidence
intervals are 95\% parametric-bootstrap percentile bands. Panels (b) and (d) split the gap $1-\mathcal{A}\mathcal{M}$
into the cost-curvature wedge, from one down to $1/(1-\Phi)$, and the conduct wedge of Section~\ref{subsec:defs}; where $\Phi>0$ the two overlap on the hatched band. Panels (a) and (b) rank firms by residualized baseline log value added by two-digit industry and commuting zone; panels (c) and (d) rank firms by baseline log employment, residualized in the same way, with the attribute responses estimated by employment decile. We estimate a separate dynamic specification within each decile.  We use internal shock design for cohorts 2005--2019.
\end{minipage}
\end{figure}

\paragraph{Elasticity of amenity cost multiplier.}
Next, we provide evidence on the size-elasticities of the amenity cost multipliers in Figure~\ref{fig:elasticities_deciles_multiplier}. For the monetized amenity cost multiplier, we rely on the natural empirical counterpart, which we measure using information on both the total non-wage personnel costs and the wage bill reported in firms' cost accounts. We find that the monetized amenity cost multiplier elasticity $\varepsilon_L^{\AM}(L_j)$, shown in black circles, is also heterogeneous, being significantly negative in the bottom deciles and indistinguishable from zero in the rest of the distribution. The elasticity of the augmented multiplier $\varepsilon_L^{\AMP}(L_j)$, shown in blue squares and computed from~\eqref{eq:AP_resp}, lies slightly below that of the monetized amenity cost multiplier. As firms in the bottom part of the firm distribution grow, wages rise by more than valued non-wage job attributes, so amenity provision costs fall relative to the wage. Again, we find similar results depending on whether we rank firms into ten deciles of residualized baseline log value added (panel a) or baseline log employment (panel b).

In a similar manner as the reduced-form shock responses illustrated in Section~\ref{sec:demand_shocks}, we show the event-study estimates in Appendix Figure~\ref{fig:external_amen_multiplier} for the monetized amenity cost multiplier, which is directly measured in firms' cost accounts. Using the internal demand shock (panel~a), we find that on average the monetized amenity cost multiplier adjusts by $\pctamenmultiplierwl$ in percentage terms among the shock exposed firms, indicating that wages absorb a slightly larger fraction of the shock than non-wage costs.\footnote{Using external shocks in Appendix Figure~\ref{fig:external_amen_multiplier}, panels~(c) and~(e), we also find declines in the multiplier by $\pctprocamenmultiplierwl$ for the procurement shock and $\pctexpamenmultiplierwl$ for the export shock, although with less precision. Panels~(b),~(d) and~(f) of Appendix Figure~\ref{fig:external_amen_multiplier} decompose these responses by cost component, expressed as shares $\omega_k=C_k/WL$. Other non-wage personnel costs are the driver of these declines.} Again, we find no evidence of differential pre-trends. Notably, the size-elasticities for the monetized amenity cost multipliers, shown in black circles in Figure~\ref{fig:elasticities_deciles_multiplier}, are the decile-specific ratios of these reduced-form multiplier responses to the corresponding employment responses, averaged over the post-shock period.

\paragraph{Total compensation markdowns and cost structure.}
Using the structural estimates of $\hat{\rho}$ and $\hat{\eta}$, we recover the implied distributions of the cost-curvature index $\Phi(L_j)$ and the total compensation markdowns $\mathcal{A}\mathcal{M}$ across the firm distribution, as illustrated in Figure~\ref{fig:markdown_deciles}. The cost-curvature index is negative across most of the firm distribution, implying a convex amenity cost structure and effective labor supply elasticities below the structural taste dispersion parameter $\eta$, with confidence intervals that exclude zero in the bottom \paExclBot{} deciles. The implied total compensation markdowns rise from around $\paAMBotPct$ percent of the marginal product in the bottom decile to around $\paAMTopPct$ percent in the top decile, so the gap between total compensation and marginal product of labor is larger in less productive firms.\footnote{For brevity, the measures of the cost-curvature index $\Phi(L_j)$ and the total compensation markdowns $\mathcal{A}\mathcal{M}$ presented in Figure~\ref{fig:markdown_deciles} are derived from the augmented amenity multiplier elasticity $\varepsilon_L^{\AMP}$, while Appendix Table~\ref{tab:rho_decile_gradient} reports the corresponding measures computed based on the monetized amenity cost multiplier elasticity $\varepsilon_L^{\AM}$. The total compensation markdowns differ marginally across these two model estimations.}

We further decompose the total compensation markdowns into cost-curvature and conduct wedges, as in \eqref{eq:AM_decomp}.  The right panels of Figure~\ref{fig:markdown_deciles} stack the two wedges, $1-1/(1-\Phi(L_j))$ and $1/(1-\Phi(L_j))-\mathcal{A}\mathcal{M}$, at each firm's observed employment.
In the bottom decile, about three quarters of the gap between compensation and marginal product reflects cost curvature. At the top, the cost-curvature factor is above one, and conduct accounts for the whole gap. Thus, for the smallest firms, most of the gap between compensation and marginal product is attributable to the amenity provision cost structure, not employer market power.  Conduct lowers compensation by $\paCondPctLo$ to $\paCondPctHi$ percent below what the cost structure alone would imply. Ranking firms by baseline value added or employment  gives the same picture.

\paragraph{Amenity cost multipliers and wage markdowns.}
As described in Section~\ref{subsec:identification_structural}, our identification of the total compensation markdowns uses the effective labor supply and amenity cost multiplier elasticities. Whether we consider only the monetized amenity cost multiplier or the augmented multiplier including production-relevant non-wage job attributes, the estimated multiplier elasticities in  Figure~\ref{fig:elasticities_deciles_multiplier} are relatively similar, and thus the total compensation markdowns we recover are robust to these choices. However, further decomposing the total compensation markdowns $\mathcal{A}\mathcal{M}$ into their two components: the amenity cost multipliers  $\mathcal{A}$ and the wage markdowns $\mathcal{M}$, hinges crucially on the extent to which the researcher is able to measure the relevant amenity bundle that employers determine endogenously. This admittedly requires an assumption stronger than Assumption~\ref{ass:prop} maintained thus far.

Using information on non-wage personnel costs from the firm accounts, we calculate an average value of the monetized amenity cost multiplier ${\mathcal{A}}^{\sM} \approx \Abar$, ranging between $\paAmBot$ and $\paAmTop$ across the deciles, for the extended sample used in the internal shock design. By comparison, our calculations suggest that production-relevant amenities are worth about \paAttrPct{} percent of the wage, where about three quarters of this amenity value reflects permanent jobs and regular work schedules. As a result, we find that including non-wage job attributes in our calculations raises the amenity cost multiplier by about \paThetaPct{} percent in every decile, such that the augmented multiplier $\AMP$ varies between $\paAdBot$ and $\paAdTop$. While the amenity value of non-wage job attributes is found to be large, it is in line with evidence from survey experiments: for instance, \citet{maestas2023value} estimate that a move from a job with none of the preferred amenities to one with all of them is equivalent to a 55 percent wage increase.%

Based on the comprehensive set of amenities observed in our data, combining both monetized amenities found in firms' cost accounts and the set of valued non-wage job attributes, we estimate that wage markdowns, $\mathcal{M}=\mathcal{A}\mathcal{M}/\AMP$, are $\paGapLo$ to $\paGapHi$ percentage points below the corresponding total compensation markdowns (Appendix Table~\ref{tab:rho_decile_gradient}). However, if unobserved endogenous amenities are present, these wage markdown measures can be interpreted as upper bounds. We also provide an alternative in which both the total compensation markdown and the wage markdown are computed with the monetized amenity cost multiplier $\AM$ alone. Even with this smaller multiplier, the gap remains substantial: across deciles, the wage markdown lies $\paGapMLo$ to $\paGapMHi$ percentage points below the total compensation markdown.

\paragraph{Discussion.} Our evidence shows that effective labor supply elasticities increase and total compensation markdowns are compressed as we move along the firm distribution. Decomposing the total compensation markdowns, we find that employer conduct accounts for about a quarter of the gap at the bottom of the firm distribution and for all of the gap at the top. These conclusions rest on three modeling assumptions maintained in our analysis: (i) labor is untaxed, (ii) firms do not interact strategically, and (iii) workers are homogeneous. We now consider each assumption in turn. First, we can extend our baseline model with progressive wage taxation, leaving amenities untaxed (see details in Appendix~\ref{app:taxation}). With Norwegian tax progressivity, the wage markdown is unchanged, the markdown on compensation as workers value it falls by at most $\paTaxAMDrop$ percentage points, and the gradient of $\Phi(L_j)$ flattens only slightly. Progressive labor taxation therefore leaves our interpretation unaltered.\footnote{The firm's labor costs also include payroll taxes, which enter neither the wage bill nor our measure of non-wage compensation. The payroll tax is proportional to the wage, so the total compensation markdown is unaffected (Appendix~\ref{app:tax_application}). Wage markdowns are lower once the payroll tax is counted as a labor cost. At the rate $t=0.141$, they range from $\paPayMBot$ to $\paPayMTop$ across deciles.} Second, in the calibrated model of \citet{berger2026labor}, oligopsonistic competition accounts for little of the variation in wage markdowns across firms in the United States, so we expect its role to be limited here as well, although this aspect could be evaluated more closely in our context. Third, worker heterogeneity offers an alternative interpretation of our evidence. In related work, \citet{volpe2024job} finds that high-wage firms in Norway face more elastic labor supply, and attributes it to worker composition: wage-sensitive workers concentrate in high-paying firms, while workers valuing amenities relatively more concentrate at the bottom of the firm distribution.\footnote{In \citet{berger2026labor}, heterogeneous marginal value of wealth works the other way: richer workers are less wage-sensitive, so high-wage firms face less elastic supply (as in the US and Brazil).} We provide an alternative, complementary explanation: the cost curvature of amenity provision varies across firms and is most convex at the bottom of the firm distribution. Estimating labor supply elasticities within worker types could in principle allow separating worker sorting from cost curvature, since sorting predicts the same elasticity at every firm for a given type while cost curvature does not. We leave this for future work.

\begin{table}[t!]
  \centering
  \caption{Comparisons to Conventional Measures of Employer Market Power.}
  \label{tab:elasticities_markdowns}
  \renewcommand{\arraystretch}{1.1}
  \small
  \begin{tabular}{l c c c c c c}
  \toprule
   & \multicolumn{2}{c}{Internal Shock} & \multicolumn{2}{c}{Procurement Sample}
   & \multicolumn{2}{c}{Export Sample} \\
   & \footnotesize{Extended} & \footnotesize{Contracts} &
  \footnotesize{Internal} & \footnotesize{Procurement} & \footnotesize{Internal}
   & \footnotesize{Export} \\
   & \footnotesize{Sample} & \footnotesize{Sample} & \footnotesize{Shock} &
  \footnotesize{Shock} & \footnotesize{Shock} & \footnotesize{Shock} \\
   & (1) & (2) & (3) & (4) & (5) & (6) \\
  \midrule
  \addlinespace[2pt]
  \multicolumn{7}{l}{\textit{Estimates of Effective Elasticities}} \\[3pt]
  \quad $\varepsilon^L_w = \bar{\beta}_L / \bar{\beta}_w$ & \elasw & \elaswredu
  & \elaswprocint & \elaswproc & \elaswexpint & \elaswexp \\
  \quad $\varepsilon^{w}_{VA} = \bar{\beta}_w / \bar{\beta}_{VA}$ & \elasrs & \elasrsredu & \elasrsprocint & \elasrsproc & \elasrsexpint & \elasrsexp \\
  \addlinespace[6pt]
  \multicolumn{7}{l}{\textit{Conventional Markdown}} \\[3pt]
  \quad $\varepsilon^L_w /(\varepsilon^L_w+1)$ & \markdownAMtextbook & \markdownAMtextbookredu & \markdownAMtextbookprocint & \markdownAMtextbookproc & \markdownAMtextbookexpint & \markdownAMtextbookexp \\
  \addlinespace[6pt]
  \multicolumn{7}{l}{\textit{Markdowns with the Monetized Amenity Cost Multiplier $\AM$}} \\[3pt]
  \quad Total Compensation $\mathcal{A}\mathcal{M}$ & \paTabAMM & \paTabAMMRedu & \paTabAMMProcint & \paTabAMMProc & \paTabAMMExpint & \paTabAMMExp \\
  \quad Wage $\mathcal{M}=\mathcal{A}\mathcal{M}/\bar{\mathcal{A}}^{\sM}$ & \paTabMM & \paTabMMRedu & \paTabMMProcint & \paTabMMProc & \paTabMMExpint & \paTabMMExp \\
  \quad $\mathcal{A}\mathcal{M}-\mathcal{M}$ (pp) & \paTabGapMM & \paTabGapMMRedu & \paTabGapMMProcint & \paTabGapMMProc & \paTabGapMMExpint & \paTabGapMMExp \\
  \quad $\mathcal{A}\mathcal{M}$ $-$ Conventional (pp) & $\paTabCorrM$ & $\paTabCorrMRedu$ & $\paTabCorrMProcint$ & $\paTabCorrMProc$ & $\paTabCorrMExpint$ & $\paTabCorrMExp$ \\
  \addlinespace[6pt]
  \multicolumn{7}{l}{\textit{Markdowns with the Augmented Amenity Cost Multiplier $\AMP$}} \\[3pt]
  \quad Total Compensation $\mathcal{A}\mathcal{M}$ & \paTabAM & \paTabAMRedu & \paTabAMProcint & \paTabAMProc & \paTabAMExpint & \paTabAMExp \\
  \quad Wage $\mathcal{M}=\mathcal{A}\mathcal{M}/\bar{\mathcal{A}}^{\sMP}$ & \paTabM & \paTabMRedu & \paTabMProcint & \paTabMProc & \paTabMExpint & \paTabMExp \\
  \quad $\mathcal{A}\mathcal{M}-\mathcal{M}$ (pp) & \paTabGap & \paTabGapRedu & \paTabGapProcint & \paTabGapProc & \paTabGapExpint & \paTabGapExp \\
  \quad $\mathcal{A}\mathcal{M}$ $-$ Conventional (pp) & $\paTabCorr$ & $\paTabCorrRedu$ & $\paTabCorrProcint$ & $\paTabCorrProc$ & $\paTabCorrExpint$ & $\paTabCorrExp$ \\
  \addlinespace[2pt]
  \bottomrule
  \end{tabular}
  \begin{minipage}{\textwidth}
  \scriptsize
  \vspace{0.25cm}
  \emph{Notes:} Elasticities are ratios of the post-shock average effects $\bar{\beta}_Y$. The formula for the conventional markdowns uses the average effective labor supply elasticity $\varepsilon^L_w = \bar{\beta}_L / \bar{\beta}_w$, estimated separately for each shock design and estimation sample. The total compensation markdown follows Eq.~\eqref{eq:foc_AwMRP} and the wage markdown divides it by the pre-shock multiplier mean value. The first set of markdowns uses the monetized amenity cost multiplier, and the second the augmented multiplier, which adds the attributes paid in output (Appendix~\ref{app:structural_estimation}). Columns (1)--(2) use the extended sample (cohorts 2005--2019) and the contracts sample (cohorts 2018--2019), columns (3)--(4) the procurement-shock sample, and columns (5)--(6) the export-shock sample.
  \end{minipage}
  \end{table}

\paragraph{Implications.} What are the implications of our findings for \textit{conventional measures} of employer market power? To answer this, we abstract from firm heterogeneity and focus on the average effective labor supply elasticity, $\varepsilon^L_w = \bar{\beta}_L / \bar{\beta}_w$, and compare our estimates of total compensation markdowns with those implied by conventional measures. As earlier, using the expression \eqref{eq:foc_AwMRP} for the total compensation markdown in our model, we can construct alternative measures of this markdown from different estimates of the average effective labor supply and amenity cost multiplier elasticities. To quantify the wage markdown, we must further quantify the average amenity cost multiplier, which, in our comprehensive internal-shock extended sample, is $\bar{\mathcal{A}}^{\sM} \approx \Abar$ based on the monetized amenities and $\bar{\mathcal{A}}^{\sMP} \approx \paTabAMP$ once the non-wage job attributes are also included. We can then compare the markdown measures derived from our model to the conventional markdown measure, which equals $\varepsilon^L_w/(\varepsilon^L_w+1)$.

Table \ref{tab:elasticities_markdowns} presents our quantification and decomposition of total compensation markdowns into its components across the three shock designs and samples, first using the monetized amenity cost multiplier and then using the augmented multiplier. Compared to the conventional markdown, our total compensation markdown implies a correction of between $\paTabCorr$ and $\paTabCorrExp$ percentage points with the augmented multiplier in our main samples (Columns 1, 4, and 6), and of between $\paTabCorrM$ and $\paTabCorrMExp$ percentage points with the monetized amenity cost multiplier. This correction is small because the amenity cost multiplier responds little to firm size: its size-elasticity is $\varepsilon^{\mathcal A}_L=\rho\,\Phi/(1-\rho)$, and with $\hat\rho$ close to zero it stays small even when costs are far from per-worker. The cost curvature itself is not small. At the average elasticity, $\Phi\approx\paPhiTab$, which pushes the effective labor supply elasticity below the structural $\hat\eta$. The conventional measure captures this, since it uses the measured elasticity, but it attributes the whole gap to employer conduct. Our decomposition shows that a substantial part of it reflects the cost structure, and Figure~\ref{fig:markdown_deciles} shows that this part varies widely across the firm distribution. Further, we find wage markdowns that are $\paTabGapIntExp$ to $\paTabGapIntProc$ percentage points below total compensation markdowns with the augmented multiplier, and $\paTabGapMMInt$ to $\paTabGapMMIntExpint$ percentage points below with the monetized amenity cost multiplier. This gap is the main difference between our markdown measures and the conventional measures reported in the literature.

\section{Conclusion}
\label{sec:conclusion}

This paper has examined how firms adjust workplace amenities in response to demand shocks and what those adjustments imply for measuring employer market power. Our theoretical framework embeds endogenous, multi-dimensional amenity provision into a tractable monopsony model in which the per-worker cost of each amenity can scale flexibly with firm size. From this structure, we derive a compact sufficient statistic---the amenity cost-curvature index---that summarizes how the amenity bundle shapes the firm's effective labor supply elasticity and the total compensation markdown. The model clarifies that the level of amenity provision, per-worker amenity cost, and the amenity cost multiplier convey distinct information about preferences and cost structure, and that conventional wage-based markdowns differ from the welfare-relevant total compensation markdowns by a measurable wedge.

Bringing these predictions to Norwegian administrative data that combine matched employer--employee records, firm cost accounts, and the near-universe of publicly posted job ads, we document that firms respond to positive demand shocks across essentially every margin of compensation we can measure. Three complementary identification strategies---abrupt within-firm value-added changes, first-time public procurement contract wins, and shift-share export demand shocks---yield broadly consistent patterns: employment responds more than wages, and non-wage compensation rises along with the wage bill. Using the internal shock, we further find that non-wage compensation and in-kind benefits rise per worker, the share of full-time contracts increases, and the content of job postings tilts toward training and workplace-quality amenities. At the same time, the amenity cost multiplier changes only modestly, because workers' preferences over wages and amenities are close to Cobb--Douglas, even though amenity provision costs are convex at most firms.

The implications for the measurement of employer market power are twofold. First, the gap between total compensation and marginal product is not all conduct. The estimated total compensation markdowns rise from roughly $\paAMBotPct$ to $\paAMTopPct$ percent of marginal product across the firm distribution, and decomposing them into a cost-curvature wedge due to the structure of provision costs and a conduct wedge due to employer market power shows that most of the gap at the least productive firms reflects the cost structure, while conduct lowers compensation by $\paCondPctLo$ to $\paCondPctHi$ percent relative to what that cost structure alone would imply. Second, the gap between wage markdowns and total compensation markdowns is substantial: our evidence shows that, across the firm distribution, wage markdowns are $\paGapLo$ to $\paGapHi$ percentage points below the corresponding total compensation markdowns. When markdowns are measured as the gap between wages and the marginal product (e.g., in a production function estimation), a meaningful fraction of what is attributed to employer rent extraction may be compensation workers receive in non-wage benefits. By comparison, the measurement challenge is less severe when markdowns are inferred from the inverse labor supply elasticity: our evidence shows that the correction arising from endogenous amenity feedback is small for the average firm, reassuring that studies using firm-level demand shocks to estimate elasticities are informative about employer market power even with endogenous amenity provision.

Several limitations in our work point to directions for further research. Our framework treats firms as atomistic and workers as homogeneous in their preferences. Heterogeneity in tastes across workers---and the sorting it induces across firms with different amenity profiles---is a natural next step, with implications for the distributional incidence of changes in employer market power \citep{volpe2024job,berger2026labor}. Moreover, our evidence focuses on private-sector firms in Norway, where the regulatory environment and collective bargaining shape wage-setting and the menu of available amenities \citep{bhuller2022jep,lagos2025union,arold2025worker}. Evidence from other settings and frameworks encompassing such constraints are thus valuable. Our analysis also abstracts from informational frictions about amenities on either side of the market---whether firms accurately perceive workers' valuations and how workers learn about non-wage job attributes over their careers---aspects that have begun to receive attention in recent work in this area \citep{caldwell2025firm,bhuller2026beliefs,adrjan2026jobads,cordes2026firms}. We leave this to future work.%

\printbibliography[heading=bibintoc,title={References}]
\end{refsection}

\clearpage
\appendix

{\centering \section*{\Huge{Appendix}}}

\setcounter{page}{1}
\global\long\def\thepage{[Appendix-\arabic{page}]}

\setcounter{footnote}{0}
\setcounter{figure}{0}
\setcounter{table}{0}
\setcounter{equation}{0}

\numberwithin{equation}{section}
\renewcommand{\thefigure}{\thesection.\arabic{figure}}
\renewcommand{\thetable}{\thesection.\arabic{table}}
\renewcommand{\theresult}{\thesection.\arabic{result}}
\renewcommand{\tablename}{Appendix Table}
\renewcommand{\figurename}{Appendix Figure}
\renewcommand{\theequation}{\Alph{section}.\arabic{equation}}

\begin{refsection}
\section{Theoretical Appendix}
\label{app:theory}
\setcounter{figure}{0}
\setcounter{table}{0}
\setcounter{equation}{0}
\setcounter{result}{0}

This appendix provides mathematical derivations for the baseline model of Section~\ref{sec:model} (Appendix~\ref{app:foc}), compares the monopsonist's wage-amenity bundle to that of a value-taking firm (Appendix~\ref{app:value_taker}), extends our model to allow for progressive earnings taxation (Appendix~\ref{app:taxation}), reviews existing models of imperfect competition with endogenous amenities (Appendix~\ref{app:literature}), and provides numerical illustrations of our model (Appendix~\ref{app:psi_figures}).

\subsection{Baseline Model: Mathematical Derivations}
\label{app:foc}

\subsubsection{Labor Supply}
\label{app:microfound}

Given the structure described in Section \ref{subsec:workers}, we let idiosyncratic tastes be independent draws $\xi_{ij}\sim\text{Fréchet}(\eta,\bar{a}_j)$ with CDF $\exp(-\bar{a}_j\,\xi^{-\eta})$, and maintain that
each worker selects the firm offering the highest utility.  By standard discrete choice results \citep{mcfadden1974conditional}, the probability of choosing firm $j$ is $P_j=\bar{a}_j V_j^\eta/\sum_{l\in\mathcal{J}}\bar{a}_l V_l^\eta$, and with $N$ workers the law of large numbers gives
\[
L_j = N\cdot P_j = \frac{N\,\bar{a}_j}{\mathbb{V}^\eta}\,V_j^\eta,\qquad \mathbb{V}^\eta \equiv \sum_{l \in \mathcal{J}} \bar{a}_l\,V_l^\eta,
\]
which is Equation~\eqref{eq:L_j}.  In the continuum limit $L_j/N\to 0$, firms treat $\mathbb{V}$ as fixed and the residual labor supply elasticity with respect to $V_j$ collapses from the exact $\eta(1-L_j/N)$ to $\eta$.

\subsubsection{Firm's Problem}
\label{app:firm_problem}

Firm $j$ is atomistic and takes $\bar{a}_j$ and $\mathbb{V}$ as given and chooses $(w_j, a_{j1}, \ldots, a_{jn})$ to maximize
\[
\pi_j = \psi_j\,L_j^\alpha - w_j L_j - \sum_k c_{jk}\,a_{jk}\,h_k(L_j) \qquad\text{subject to}\qquad L_j = (N\bar{a}_j/\mathbb{V}^\eta)\, V_j^\eta,
\]
with $V_j = (w_j^\rho + \sum_k\gamma_k a_{jk}^\rho)^{1/\rho}$.

\paragraph{FOC for $w_j$.}  Differentiating profit at given $\mathbf{a}_j$, $\partial\pi_j/\partial w_j=0$ yields
\begin{equation}
\mathrm{MRP}(L_j) - w_j - \sum_m c_{jm}\,a_{jm}\,h_m'(L_j) \;=\; \frac{L_j}{\partial L_j/\partial w_j}.
\label{eq:w_foc_bracket}
\end{equation}

\paragraph{FOC for $a_{jk}$.}  Differentiating profit at fixed $w_j$ and fixed $a_{jm}$, $m\neq k$, $\partial\pi_j/\partial a_{jk}=0$ gives
\begin{equation}
\mathrm{MRP}(L_j) - w_j - \textstyle\sum_m c_{jm}\,a_{jm}\,h_m'(L_j) \;=\; \frac{c_{jk}\,h_k(L_j)}{\partial L_j/\partial a_{jk}}.
\label{eq:a_foc_bracket}
\end{equation}

\paragraph{MRS condition.}  Dividing~\eqref{eq:a_foc_bracket} by~\eqref{eq:w_foc_bracket} gives
\[
\frac{\partial L_j/\partial a_{jk}}{\partial L_j/\partial w_j} \;=\; \frac{c_{jk}\,h_k(L_j)}{L_j}.
\]
Since $L_j$ depends on $a_{jk}$ and $w_j$ only through $V_j$, the left-hand side equals the ratio of the CES partials, $(\partial V_j/\partial a_{jk})/(\partial V_j/\partial w_j)=\gamma_k(a_{jk}/w_j)^{\rho-1}$, yielding the MRS condition
\begin{equation}
\gamma_k\!\left(\frac{a_{jk}}{w_j}\right)^{\!\rho-1} \;=\; \frac{c_{jk}\,h_k(L_j)}{L_j},
\label{eq:MRS_app}
\end{equation}
matching~\eqref{eq:MRS}.  Rearranging gives $a_{jk}^* = w_j\,\kappa_k(L_j)$ with $\kappa_k(L_j) \equiv \bigl(\gamma_k L_j/(c_{jk} h_k(L_j))\bigr)^{1/(1-\rho)}$.

\subsubsection{Derivation of the Wage Equation}
\label{app:wage_equation}

We simplify the two non-trivial terms in~\eqref{eq:w_foc_bracket}.  For the cost terms on the LHS, using $C_{jk} = c_{jk}\,a_{jk}\,h_k(L_j)$, $\varepsilon_L^{h_k} = L_j h_k'(L_j)/h_k(L_j)$, and $\omega_k(L_j) = C_{jk}/(w_jL_j)$:
\[
\sum_k c_{jk}\,a_{jk}\,h_k'(L_j) \;=\; \sum_k C_{jk}\,\frac{\varepsilon_L^{h_k}}{L_j} \;=\; w_j\sum_k\omega_k(L_j)\,\varepsilon_L^{h_k}.
\]
For the RHS, Equation~\eqref{eq:L_j} gives $\partial L_j/\partial V_j = \eta L_j/V_j$ and the CES aggregator at fixed $\{a_{jk}\}$ gives $\partial V_j/\partial w_j = (V_j/w_j)^{1-\rho}$.  Combining via chain rule and using $V_j^\rho = w_j^\rho\,\mathcal{A}(L_j)$ from~\eqref{eq:V_S}:
\[
\frac{L_j}{\partial L_j/\partial w_j} \;=\; \frac{L_j}{(\eta L_j/V_j)\,(V_j/w_j)^{1-\rho}} \;=\; \frac{V_j^\rho\,w_j^{1-\rho}}{\eta} \;=\; \frac{\mathcal{A}(L_j)\,w_j}{\eta}.
\]
Substituting both into~\eqref{eq:w_foc_bracket}, dividing by $w_j$, and using $\mathrm{MRP}(L_j) = \alpha\psi_j L_j^{\alpha-1}$ yields the wage equation~\eqref{eq:eq_demand}:
\begin{equation}
w_j \;=\; \frac{\alpha\psi_j\,L_j^{\alpha-1}}{1 + \mathcal{A}(L_j)/\eta + \sum_k\omega_k(L_j)\,\varepsilon_L^{h_k}}.
\end{equation}

\subsubsection{Elasticities of Effective Labor Supply and Amenity Cost Multiplier}
\label{app:wage_elasticity}

Log-differentiating the supply curve~\eqref{eq:L_j} gives $d\ln L_j = \eta\,d\ln V_j$.  Log-differentiating~\eqref{eq:V_S} gives $d\ln V_j = (1/\rho)\,d\ln\mathcal{A}(L_j) + d\ln w_j$.  Since $\mathcal{A}(L_j)$ depends on $L_j$ alone, $d\ln\mathcal{A}(L_j) = \varepsilon_L^\mathcal{A}\,d\ln L_j$.  Combining:
\[
d\ln L_j\,\bigl(1 - \eta\varepsilon_L^\mathcal{A}/\rho\bigr) \;=\; \eta\,d\ln w_j,
\]
yielding $\varepsilon_w^L = \eta/(1 - \eta\varepsilon_L^\mathcal{A}/\rho)$.  We now express $\varepsilon_L^\mathcal{A}$ explicitly. Log-differentiating $\omega_k(L_j) = \gamma_k\kappa_k(L_j)^\rho$ from~\eqref{eq:omega_explicit} with $\kappa_k(L_j) \propto (L_j/h_k(L_j))^{1/(1-\rho)}$:
\begin{equation}
\frac{d\ln\omega_k(L_j)}{d\ln L_j} \;=\; \frac{\rho(1-\varepsilon_L^{h_k})}{1-\rho}.
\label{eq:dlnomega}
\end{equation}
Since $\mathcal{A}(L_j) = 1+\sum_k\omega_k(L_j)$, $d\mathcal{A}(L_j) = \sum_k d\omega_k(L_j)$ and $d\ln\mathcal{A}(L_j) = d\mathcal{A}(L_j)/\mathcal{A}(L_j) = \sum_k(\omega_k(L_j)/\mathcal{A}(L_j))\,d\ln\omega_k(L_j) = \sum_k\sigma_k(L_j)\,d\ln\omega_k(L_j)$, recovering the multiplier elasticity~\eqref{eq:epsLA},
\[
\varepsilon_L^\mathcal{A} \;=\; \sum_k\sigma_k(L_j)\,\frac{\rho(1-\varepsilon_L^{h_k})}{1-\rho} \;=\; \frac{\rho\,\Phi(L_j)}{1-\rho}.
\]
Hence $\varepsilon_w^L = \eta/(1-\eta\,\Phi(L_j)/(1-\rho))$.

The index $\Phi(L_j)$ itself drifts along supply.  By the product rule,
\begin{equation}
\frac{d\Phi(L_j)}{d\ln L_j}\;=\;\sum_m (1-\varepsilon_L^{h_m})\,\frac{d\sigma_m}{d\ln L_j}\;-\;\sum_m \sigma_m\,\frac{d\varepsilon_L^{h_m}}{d\ln L_j},
\label{eq:dPhi}
\end{equation}
where the first term equals $\tfrac{\rho}{1-\rho}\mathrm{Var}_\sigma(1-\varepsilon_L^{h_m})$: since $\sigma_m=\omega_m/\mathcal{A}(L_j)$, \eqref{eq:dlnomega} and $\varepsilon_L^\mathcal{A}=\tfrac{\rho}{1-\rho}\Phi(L_j)$ give $d\sigma_m/d\ln L_j=\sigma_m\tfrac{\rho}{1-\rho}[(1-\varepsilon_L^{h_m})-\Phi(L_j)]$, and summing against $(1-\varepsilon_L^{h_m})$ gives $\tfrac{\rho}{1-\rho}[\sum_m\sigma_m(1-\varepsilon_L^{h_m})^2-\Phi(L_j)^2]$.  Since $d\ln L_j=\varepsilon_w^L\,d\ln w_j$ along supply, differentiating $\varepsilon_w^L=\eta/(1-\eta\Phi(L_j)/(1-\rho))$ gives $d\ln\varepsilon_w^L/d\ln w_j=\eta^2(1-\rho)\,(d\Phi(L_j)/d\ln L_j)/(1-\rho-\eta\Phi(L_j))^2$, which with~\eqref{eq:dPhi} gives:
\begin{equation}
\frac{d\ln\varepsilon_w^L}{d\ln w_j}\bigg|_{\text{supply}} \;=\; \frac{\eta^2}{\big(1-\rho-\eta\,\Phi(L_j)\big)^2}\left[\rho\mathrm{Var}_\sigma(1-\varepsilon_L^{h_k}) \;-\;(1-\rho) \sum_k\sigma_k(L_j)\,\frac{d\varepsilon_L^{h_k}}{d\ln L_j} \right],
\label{eq:super_elasticity}
\end{equation}
where $\mathrm{Var}_\sigma(1-\varepsilon_L^{h_k}) \equiv \sum_k\sigma_k(L_j)\,(1-\varepsilon_L^{h_k})^2 - \Bigl(\sum_k\sigma_k(L_j)\,(1-\varepsilon_L^{h_k})\Bigr)^2$. We refer to \eqref{eq:super_elasticity} as the super-elasticity of effective labor supply \citep{kline2025labor}. The first term in the bracket is the \emph{curvature-dispersion}, scaled by $\rho$ and non-zero when cost-curvature elasticities differ across amenities; the second is the \emph{curvature-drift}, the share-weighted drift of the $\varepsilon_L^{h_k}$ along the supply curve, scaled by $1-\rho$ and non-zero whenever some $h_k$ is non-iso-elastic.

\subsubsection{Size-Elasticity of the Wage Markdown}
\label{app:wage_eq_slope}

Log-differentiating $w_j = \mathcal{M}(L_j)\,\alpha\psi_j\,L_j^{\alpha-1}$ yields the wage-equation elasticity
\[
\frac{d\ln w_j}{d\ln L_j}\bigg|_{\text{wage eq.}} \;=\; (\alpha-1) + \varepsilon_L^\mathcal{M}.
\]
It remains to compute $\varepsilon_L^\mathcal{M}$ explicitly. By~\eqref{eq:M_formula}, the wage markdown $\mathcal{M}(L_j)=w_j/\mathrm{MRP}(L_j)$ is the inverse of the wage-equation denominator in~\eqref{eq:eq_demand}, so
\[
\varepsilon_L^\mathcal{M}  \;\equiv\; \frac{d\ln\mathcal{M}(L_j)}{d\ln L_j} \;=\; -\mathcal{M}(L_j)\cdot\frac{d}{d\ln L_j}\!\left[1+\frac{\mathcal{A}(L_j)}{\eta}+\sum_k\omega_k(L_j)\,\varepsilon_L^{h_k}\right].
\]
Using $d\mathcal{A}/d\ln L_j = \sum_k d\omega_k/d\ln L_j$, the product rule on $\omega_k\,\varepsilon_L^{h_k}$, and substituting $d\omega_k/d\ln L_j = \omega_k\,\rho(1-\varepsilon_L^{h_k})/(1-\rho)$ from~\eqref{eq:dlnomega}:
\begin{align}
\varepsilon_L^\mathcal{M} &= -\mathcal{M}(L_j)\!\left[\frac{\rho}{1-\rho}\sum_k\omega_k(L_j)\,(1-\varepsilon_L^{h_k})\!\left(\frac{1}{\eta}+\varepsilon_L^{h_k}\right) + \sum_k\omega_k(L_j)\,\frac{d\varepsilon_L^{h_k}}{d\ln L_j}\right],
\label{eq:chi_def}
\end{align}
which we refer to as the size-elasticity of the wage markdown. The first term is a \emph{share-drift} channel: it vanishes at the per-worker benchmark ($\varepsilon_L^{h_k}=1$ for all $k$) and at Cobb--Douglas preferences ($\rho=0$).  The second is a \emph{curvature-drift} channel: it vanishes whenever individual cost technologies $h_k$ are iso-elastic, regardless of $\rho$.

\subsubsection{Conditions for Existence and Uniqueness}
\label{app:existence}

We impose two regularity assumptions.  The cost functions $h_k$ are twice continuously differentiable, with $\varepsilon_L^{h_k}$ bounded as $L_j\to0$.  Total compensation $\mathcal{A}(L_j)w_j(L_j)$ stays bounded as $L_j\to0$ and bounded away from zero as $L_j\to\infty$.\footnote{For $\rho>0$ both limit requirements are implied by (i) and (ii): compensation per worker has employment elasticity $1/\eta-\Phi(L_j)>\rho/\eta$ under (i), so it rises with firm size, and a diverging size elasticity of any $h_k$ with a nonvanishing share would make $\varepsilon_L^\mathcal{M}$ diverge, contradicting (ii).  For $\rho\leq0$ they add mild limits at small firms: no $h_k$ may have a size elasticity that diverges as $L_j\to0$, which excludes only functions like $e^{-1/L_j}$, and under complements $\Phi(L_j)$ may in addition not exceed $1/\eta$ as $L_j\to0$.}  We further maintain two conditions at every $L_j$, which deliver a unique interior equilibrium under the regularity assumptions.

\begin{enumerate}
\item[(i)] Supply slopes up: $1 - \rho - \eta\,\Phi(L_j) > 0$.  Automatic at $\Phi(L_j) \leq 0$; since $\varepsilon_L^{h_k}\geq0$ implies $\Phi(L_j)\leq1-1/\mathcal{A}(L_j)$, it is satisfied whenever the amenity share of total compensation is below $(1-\rho)/\eta$.
\item[(ii)] Wage-equation slopes down: $\varepsilon_L^\mathcal{M} - (1 - \alpha) < 0$.  Automatic at $\rho=0$ with iso-elastic costs (where $\varepsilon_L^\mathcal{M}=0$).
\end{enumerate}

Along its labor supply curve the firm effectively chooses only $L_j$: inverting~\eqref{eq:eq_supply} gives the wage that recruits $L_j$ workers, $w_j(L_j)=(L_j\mathbb{V}^\eta/N\bar a_j)^{1/\eta}\mathcal{A}(L_j)^{-1/\rho}$, the amenities follow from~\eqref{eq:astar_ratio}, and total compensation per worker is $\mathcal{A}(L_j)w_j(L_j)$ by~\eqref{eq:amultiplier}.  Profit is
\begin{equation}
\pi_j(L_j)\;=\;\psi_jL_j^{\alpha}-L_j\,\mathcal{A}(L_j)\,w_j(L_j).
\label{eq:FOC_V}
\end{equation}
Revenue has derivative $\mathrm{MRP}(L_j)=\alpha\psi_jL_j^{\alpha-1}$.  Log-differentiating the compensation bill along supply gives
\[
\frac{d\ln\bigl[L_j\mathcal{A}(L_j)w_j(L_j)\bigr]}{d\ln L_j}\;=\;1+\varepsilon_L^\mathcal{A}+\frac{1}{\varepsilon_w^L}\;=\;1+\frac{\rho\,\Phi(L_j)}{1-\rho}+\frac{1}{\eta}-\frac{\Phi(L_j)}{1-\rho}\;=\;1+\frac{1}{\eta}-\Phi(L_j),
\]
by~\eqref{eq:epsLA} and~\eqref{eq:eps_w_def}.  The derivative of the compensation bill is this elasticity times compensation per worker, so the marginal cost of a hire is $\mathcal{A}(L_j)w_j(L_j)\bigl[1+1/\eta-\Phi(L_j)\bigr]=w_j(L_j)\bigl[1+\mathcal{A}(L_j)/\eta+\sum_k\omega_k(L_j)\varepsilon_L^{h_k}\bigr]=w_j(L_j)/\mathcal{M}(L_j)$, the supply wage over the markdown~\eqref{eq:M_formula}.  Hence
\begin{equation}
\pi_j'(L_j)\;=\;\mathrm{MRP}(L_j)-\frac{w_j(L_j)}{\mathcal{M}(L_j)}.
\label{eq:pi_prime}
\end{equation}
Profit rises where the marginal revenue product exceeds the marginal cost of a hire and falls where it falls short.  Consider their ratio, $\mathcal{M}(L_j)\mathrm{MRP}(L_j)/w_j(L_j)$, which equals one at the first-order condition $\pi_j'(L_j)=0$.  Its log-derivative is $(\alpha-1)+\varepsilon_L^\mathcal{M}-1/\varepsilon_w^L$, with $\varepsilon_L^\mathcal{M}$ from Appendix~\ref{app:wage_eq_slope} and $1/\varepsilon_w^L$ from~\eqref{eq:eps_w_def}, which regroups into the slope of the wage equation minus the slope of supply:
\begin{equation}
\frac{d\ln\bigl[\mathcal{M}(L_j)\mathrm{MRP}(L_j)/w_j(L_j)\bigr]}{d\ln L_j}\;=\;\underbrace{\varepsilon_L^\mathcal{M}-(1-\alpha)}_{\text{wage-equation slope}}\;-\;\underbrace{\frac{1-\rho-\eta\,\Phi(L_j)}{\eta(1-\rho)}}_{\text{supply slope}}.
\label{eq:slope_gap}
\end{equation}
Under (ii) the first term is negative and under (i) the second is positive, so the log-ratio falls with employment.  Two consequences follow.  \emph{Uniqueness:} the ratio $\mathcal{M}(L_j)\mathrm{MRP}(L_j)/w_j(L_j)$ is strictly decreasing, so it equals one at most once.  \emph{Existence:} by the regularity assumptions and $\Phi(L_j)<1$, the marginal cost of a hire is bounded at small $L_j$ and bounded away from zero at large $L_j$, while $\alpha<1$ sends $\mathrm{MRP}(L_j)$ to infinity and to zero at the same ends.  The ratio thus starts above one, ends below it, and by continuity equals one at an interior $L_j^*$.  Writing~\eqref{eq:pi_prime} as $\pi_j'(L_j)=\frac{w_j(L_j)}{\mathcal{M}(L_j)}\bigl[\mathcal{M}(L_j)\mathrm{MRP}(L_j)/w_j(L_j)-1\bigr]$, the bracket vanishes at $L_j^*$, so only its derivative enters $\pi_j''(L_j^*)$, and by~\eqref{eq:slope_gap} the second-order condition holds,
\[
\pi_j''(L_j^*)\;=\;\frac{w_j(L_j^*)}{L_j^*\,\mathcal{M}(L_j^*)}\left[\varepsilon_L^\mathcal{M}-(1-\alpha)-\frac{1-\rho-\eta\,\Phi(L_j^*)}{\eta(1-\rho)}\right]\;<\;0.
\]
Profit therefore rises on $(0,L_j^*)$ and falls beyond, so $L_j^*$ is the global and interior optimum, and since the compensation bill vanishes with $L_j$, $\pi_j(L_j^*)>0$: the firm prefers operating to shutdown.

Conditions (i) and (ii) are sufficient rather than necessary.  The argument uses only that the log-ratio in~\eqref{eq:slope_gap} falls with employment, and (i) and (ii) sign its two pieces separately.  The unique interior equilibrium therefore survives even where the wage falls with firm size, as long as the ratio still falls.  We impose (i) for simplicity, since the comparative statics divide by $d\ln w_j>0$.

\paragraph{Value-taking benchmark.} Proposition~\ref{res:underprovision} in Section \ref{app:value_taker} below compares the monopsonist with a firm that takes the job value $V^c_j$ as given. The comparison requires that this firm also has a unique interior optimum. Its compensation bill is $L_j\mathcal{A}(L_j)w_j$ with $w_j$ set so that the job delivers $V^c_j$. By~\eqref{eq:epsLA}, the compensation bill per worker falls with employment at rate $\Phi(L_j)$, and the marginal cost is~\eqref{eq:MC_vt}. Profit is single-peaked, with an interior maximum, when this marginal cost falls more slowly with employment than the marginal revenue product, uniformly in $L_j$ and by a strictly positive margin:
\[
\text{(ii$'$)}\qquad \Phi(L_j)+\frac{1}{1-\Phi(L_j)}\frac{d\Phi(L_j)}{d\ln L_j}<1-\alpha.
\]
Condition (ii$'$) is the counterpart of (i) and (ii) for a firm that does not internalize the supply term $\mathcal{A}(L_j)w_j/\eta$ in~\eqref{eq:MC_mono}. Without that term, only diminishing returns in production work against scale economies in providing amenities, and (ii$'$) requires that they prevail. We maintain (ii$'$) wherever Proposition~\ref{res:underprovision} is invoked.

\subsection{Comparison to the Value-Taking Benchmark}
\label{app:value_taker}

To see how employer market power shapes the composition of the wage-amenity bundle, consider the benchmark of \emph{value-taking} conduct: the firm takes the job value $V^c_j$ as exogenously given, chooses employment $L_j$, and pays the wage $w_j=V^c_j\,\mathcal{A}(L_j)^{-1/\rho}$. The optimality conditions~\eqref{eq:MRS} remain unchanged, so the two conducts offer the same composition at the same scale and differ only in position on a common expansion path. By~\eqref{eq:V_S}, delivering $V^c_j$ costs $\mathcal{A}(L_j)\,w_j=V^c_j\,\mathcal{A}(L_j)^{(\rho-1)/\rho}$ per worker, which with $V^c_j$ fixed varies with employment only through $\mathcal{A}(L_j)$. Differentiating the bill $L_j\cdot V^c_j\,\mathcal{A}(L_j)^{(\rho-1)/\rho}$ by the product rule and using~\eqref{eq:epsLA} gives the value-taker's marginal cost of hiring an additional worker:
\begin{equation}
\mathrm{MC}^c(L_j)\;=\;\mathcal{A}(L_j)\,w_j\Bigl[1+\tfrac{\rho-1}{\rho}\,\varepsilon_L^{\mathcal{A}}\Bigr]
\;=\;\bigl[1-\Phi(L_j)\bigr]\,\mathcal{A}(L_j)\,w_j.
\label{eq:MC_vt}
\end{equation}
The value-taker's marginal cost can be decomposed into two terms: the first term $\mathcal{A}(L_j)w_j$ is the marginal hire's compensation, while the second term $-\Phi(L_j)\mathcal{A}(L_j)w_j$ captures the change in provision costs for all inframarginal workers as size increases, which lowers marginal cost under concave costs and raises it under convex costs. The value taker expands until $\mathrm{MRP}(L_j)=\mathrm{MC}^c(L_j)$, with employment $L^c_j$, wage $w^c_j$, and amenities $a^c_{jk}=a_{jk}(L^c_j)$; the value $V^c_j$ it takes as given is the one the supply curve~\eqref{eq:eq_supply} assigns to $L^c_j$, so that both conducts lie on the same supply curve.

While the monopsonist similarly expands until $\mathrm{MRP}(L_j)=\mathrm{MC}(L_j)$, its marginal cost has a different structure. Intuitively, the monopsonist doesn't hire at a fixed job value, since expanding means climbing its supply curve, so the marginal hire raises the value for every inframarginal worker. Thus, the value a monopsonist delivers is no longer fixed but rises along the supply curve, $d\ln V_j/d\ln L_j=1/\eta$. Differentiation by the product rule adds to~\eqref{eq:MC_vt} an additional term capturing this rising value, $\mathcal{A}(L_j)w_j/\eta$:
\begin{equation}
\mathrm{MC}(L_j)=\;\bigl[1-\Phi(L_j)\bigr]\,\mathcal{A}(L_j)\,w_j\;+\;\frac{\mathcal{A}(L_j)\,w_j}{\eta}.
\label{eq:MC_mono}
\end{equation}
As this term is positive, it lifts marginal cost above the value-taker's at every scale, so at $L^c_j$ the monopsonist's profit is already falling. With a unique optimum for each conduct (Appendix~\ref{app:existence}), it stops strictly short, $L^*_j<L^c_j$. At this smaller scale, the wage is lower ($w^*_j<w^c_j$), the job value is also lower ($V_j\propto L_j^{1/\eta}$), and the marginal revenue product higher ($\mathrm{MRP}(L^*_j)>\mathrm{MRP}(L^c_j)$). However, the provision of amenities can move either way. Both firms lie on the common path $a_{jk}=w_j\bigl(\gamma_k L_j/(c_{jk}h_k(L_j))\bigr)^{1/(1-\rho)}$ from~\eqref{eq:astar_ratio}; log-differentiating it along the supply curve, where $d\ln w_j/d\ln L_j=1/\varepsilon_w^L >0$, we have:
\begin{equation}
\frac{d\ln a_{jk}}{d\ln L_j}\bigg|_{\text{supply}} \;=\; \frac{1}{\varepsilon_w^L} \;+\; \frac{1}{1-\rho}(1-\varepsilon_L^{h_k}).
\label{eq:path_gradient}
\end{equation}
Writing $a^*_{jk}=a_{jk}(L^*_j)$ and $a^c_{jk}=a_{jk}(L^c_j)$, the gradient is positive and the monopsonist under-provides the amenity ($a^*_{jk}<a^c_{jk}$) unless provision costs are strongly convex, $\varepsilon_L^{h_k}-1>(1-\rho)/\varepsilon_w^L$, in which case it \textit{over}-provides it ($a^*_{jk}>a^c_{jk}$). Proposition~\ref{res:underprovision} summarizes.

\begin{result}[Provision of Wages and Amenities under Monopsony]
\label{res:underprovision}
\textit{Let $L^c_j$ be the optimal employment and let $(w^c_j,\{a^c_{jk}\})$ be the optimal wage-amenity bundle offered under a value-taking conduct. \textbf{(i)} Monopsony strictly under-provides employment, wages, and hence job value ($L^*_j<L^c_j$, $w^*_j<w^c_j$, $\mathrm{MRP}(L^*_j)>\mathrm{MRP}(L^c_j)$). \textbf{(ii)} Both conducts satisfy the same MRS condition~\eqref{eq:MRS} along an expansion path, so equilibrium amenities differ only due to the scale difference ($L^*_j<L^c_j$). \textbf{(iii)} Amenity $k$ is under-provided, $a^*_{jk}<a^c_{jk}$, if it rises with scale along the expansion path, $\varepsilon_L^{h_k}-1<(1-\rho)/\varepsilon_w^L$ throughout $[L^*_j,L^c_j]$ (per-worker, concave, or mildly convex costs), and over-provided if it falls with scale throughout that interval (strongly convex costs).}
\end{result}

\begin{figure}[H]
    \centering
    \caption{Wage--Amenity Determination under Monopsony.}
    \label{fig:amenity_markdown}
    \begin{minipage}[t]{0.40\textwidth}
        \centering
        \resizebox{\textwidth}{!}{\begin{tikzpicture}[scale=1.0, >=Stealth, font=\small]

\path[use as bounding box] (-0.6,-0.4) rectangle (7.2,6.5);

\draw[->, thick] (0,0) -- (6.0,0) node[right] {$a_{jk}$};
\draw[->, thick] (0,0) -- (0,6.0) node[above] {$w_j$};

\draw[gray!60, semithick, dotted] (1.8, 1.8) -- (1.8, -0.05);
\draw[gray!60, semithick, dotted] (1.8, 1.8) -- (-0.05, 1.8);
\draw[gray!60, semithick, dotted] (2.5, 2.5) -- (2.5, -0.05);
\draw[gray!60, semithick, dotted] (2.5, 2.5) -- (-0.05, 2.5);

\draw[thick, CoolRed] (0,3.6) -- (3.6,0);

\draw[thick, CoolRed, dashed] (0,5) -- (5,0);

\draw[->, very thick, CoolGreen] (0,0) -- (4.6,4.6);
\node[CoolGreen, font=\scriptsize, anchor=west, rotate=45] at (2.8, 3.24) {expansion path};

\draw[very thick, CoolBlue] plot[smooth] coordinates
  {(0.458,6.000) (0.504,5.570) (0.554,5.165) (0.609,4.783) (0.670,4.423) (0.737,4.084) (0.811,3.766) (0.892,3.468) (0.981,3.189) (1.079,2.928) (1.187,2.684) (1.306,2.456) (1.437,2.244) (1.580,2.047) (1.738,1.864) (1.912,1.694) (2.103,1.537) (2.314,1.391) (2.545,1.257) (2.799,1.134) (3.079,1.021) (3.387,0.917) (3.726,0.821) (4.098,0.734) (4.508,0.655) (4.959,0.583) (5.455,0.517) (6.000,0.458)};
\node[CoolBlue, anchor=west, font=\footnotesize] at (6.02, 0.46) {$V^*_j$};

\draw[very thick, CoolBlue] plot[smooth] coordinates
  {(0.958,6.000) (1.025,5.668) (1.097,5.351) (1.174,5.047) (1.257,4.757) (1.345,4.481) (1.440,4.217) (1.541,3.966) (1.649,3.727) (1.765,3.499) (1.890,3.282) (2.023,3.077) (2.165,2.881) (2.317,2.696) (2.480,2.520) (2.654,2.354) (2.841,2.196) (3.041,2.047) (3.255,1.907) (3.484,1.774) (3.729,1.649) (3.991,1.531) (4.271,1.420) (4.572,1.315) (4.893,1.217) (5.237,1.125) (5.606,1.039) (6.000,0.958)};
\node[CoolBlue, anchor=west, font=\footnotesize] at (6.02, 0.96) {$V^{c}_j$};

\fill (1.8, 1.8) circle (2pt);
\node[anchor=east, font=\scriptsize] at (1.60, 1.78) {$(a_{jk}^{*},\, w_j^{*})$};

\fill (2.5, 2.5) circle (2pt);
\node[anchor=west, font=\scriptsize] at (2.78, 2.48) {$(a_{jk}^{c},\, w_j^{c})$};

\end{tikzpicture}}\\[2pt]
        {\small \textbf{(a)}~Per-worker ($\varepsilon_L^{h_k}=1$)}
    \end{minipage}\hspace{0.1cm}
    \begin{minipage}[t]{0.40\textwidth}
        \centering
        \resizebox{\textwidth}{!}{\begin{tikzpicture}[scale=1.0, >=Stealth, font=\small]

\path[use as bounding box] (-0.6,-0.4) rectangle (7.2,6.5);

\draw[->, thick] (0,0) -- (6.0,0) node[right] {$a_{jk}$};
\draw[->, thick] (0,0) -- (0,6.0) node[above] {$w_j$};

\draw[thick, CoolRed, dashed] (0,4.187) -- (6,0.398);

\draw[thick, CoolRed] (0,3.029) -- (2.238,0);

\draw[gray!60, semithick, dotted] (1.1, 1.54) -- (1.1, -0.05);
\draw[gray!60, semithick, dotted] (1.1, 1.54) -- (-0.05, 1.54);
\draw[gray!60, semithick, dotted] (3.4, 2.04) -- (3.4, -0.05);
\draw[gray!60, semithick, dotted] (3.4, 2.04) -- (-0.05, 2.04);

\draw[->, very thick, CoolGreen] plot[smooth] coordinates
  {(0.018,0.552) (0.078,0.798) (0.166,0.961) (0.408,1.203) (0.721,1.387)
   (1.100,1.541) (1.530,1.674) (2.014,1.793) (2.548,1.900) (3.400,2.040)
   (4.071,2.138) (4.871,2.236) (5.865,2.334)};
\node[CoolGreen, font=\scriptsize, anchor=west, rotate=7] at (3.60, 2.42) {expansion path};

\draw[very thick, CoolBlue] plot[smooth] coordinates
  {(0.215,6.000) (0.243,5.487) (0.275,5.009) (0.311,4.563) (0.352,4.148) (0.398,3.763) (0.450,3.406) (0.509,3.076) (0.576,2.772) (0.652,2.492) (0.737,2.234) (0.834,1.998) (0.944,1.782) (1.068,1.585) (1.208,1.406) (1.366,1.243) (1.545,1.096) (1.748,0.963) (1.978,0.844) (2.237,0.736) (2.531,0.640) (2.863,0.555) (3.239,0.479) (3.664,0.412) (4.145,0.352) (4.689,0.300) (5.304,0.255) (6.000,0.215)};
\node[CoolBlue, anchor=west, font=\footnotesize] at (6.02, 0.21) {$V^*_j$};

\draw[very thick, CoolBlue] plot[smooth] coordinates
  {(1.081,6.000) (1.152,5.686) (1.228,5.385) (1.308,5.096) (1.394,4.820) (1.485,4.556) (1.582,4.303) (1.686,4.061) (1.797,3.831) (1.914,3.610) (2.040,3.400) (2.173,3.200) (2.316,3.009) (2.468,2.828) (2.629,2.655) (2.802,2.491) (2.985,2.335) (3.181,2.187) (3.389,2.047) (3.611,1.914) (3.848,1.788) (4.100,1.669) (4.369,1.556) (4.655,1.450) (4.960,1.349) (5.285,1.255) (5.631,1.165) (6.000,1.081)};
\node[CoolBlue, anchor=west, font=\footnotesize] at (6.02, 1.08) {$V^{c}_j$};

\fill (1.1, 1.54) circle (2pt);
\node[anchor=south west, font=\scriptsize] at (0.80, 1.72) {$(a_{jk}^{*},\, w_j^{*})$};

\fill (3.4, 2.04) circle (2pt);
\node[anchor=north, font=\scriptsize] at (3.15, 1.92) {$(a_{jk}^{c},\, w_j^{c})$};

\end{tikzpicture}}\\[2pt]
        {\small \textbf{(b)}~Concave ($\varepsilon_L^{h_k}<1$)}
    \end{minipage}
    \par\medskip
    \begin{minipage}[t]{0.40\textwidth}
        \centering
        \resizebox{\textwidth}{!}{\begin{tikzpicture}[scale=1.0, >=Stealth, font=\small]

\path[use as bounding box] (-0.6,-0.4) rectangle (7.2,6.5);

\draw[->, thick] (0,0) -- (6.0,0) node[right] {$a_{jk}$};
\draw[->, thick] (0,0) -- (0,6.0) node[above] {$w_j$};

\draw[gray!60, semithick, dotted] (1.6, 1.97) -- (1.6, -0.05);
\draw[gray!60, semithick, dotted] (1.6, 1.97) -- (-0.05, 1.97);
\draw[gray!60, semithick, dotted] (3.0, 4.86) -- (3.0, -0.05);
\draw[gray!60, semithick, dotted] (3.0, 4.86) -- (-0.05, 4.86);

\draw[thick, CoolRed] (0,3.899) -- (3.233,0);

\draw[thick, CoolRed, dashed] (2.456,5.70) -- (5.95,0.306);

\draw[->, very thick, CoolGreen] plot[smooth] coordinates
  {(0.50,0.77) (0.80,1.00) (1.10,1.31) (1.35,1.62) (1.60,1.97) (1.90,2.45)
   (2.20,3.01) (2.50,3.64) (2.75,4.22) (3.00,4.86) (3.15,5.27)};
\node[CoolGreen, font=\scriptsize, anchor=south, rotate=66] at (2.35, 3.40) {expansion path};

\draw[very thick, CoolBlue] plot[smooth] coordinates
  {(0.444,6.000) (0.489,5.567) (0.539,5.158) (0.593,4.773) (0.653,4.410) (0.719,4.070) (0.792,3.750) (0.872,3.451) (0.961,3.170) (1.058,2.908) (1.165,2.663) (1.283,2.435) (1.413,2.222) (1.556,2.025) (1.713,1.842) (1.886,1.672) (2.077,1.515) (2.288,1.370) (2.519,1.237) (2.774,1.114) (3.055,1.001) (3.364,0.898) (3.705,0.803) (4.080,0.717) (4.493,0.639) (4.948,0.567) (5.448,0.503) (6.000,0.444)};
\node[CoolBlue, anchor=west, font=\footnotesize] at (6.02, 0.44) {$V^*_j$};

\draw[very thick, CoolBlue] plot[smooth] coordinates
  {(2.393,6.000) (2.476,5.816) (2.562,5.636) (2.650,5.461) (2.742,5.290) (2.837,5.124) (2.935,4.962) (3.037,4.803) (3.142,4.649) (3.251,4.499) (3.364,4.353) (3.480,4.210) (3.601,4.071) (3.725,3.936) (3.854,3.805) (3.988,3.677) (4.126,3.552) (4.269,3.431) (4.417,3.314) (4.570,3.199) (4.728,3.088) (4.892,2.980) (5.061,2.875) (5.236,2.773) (5.417,2.673) (5.605,2.577) (5.799,2.484) (6.000,2.393)};
\node[CoolBlue, anchor=west, font=\footnotesize] at (6.02, 2.39) {$V^{c}_j$};

\fill (1.6, 1.97) circle (2pt);
\node[anchor=north west, font=\scriptsize] at (1.71, 2.26) {$(a_{jk}^{*},\, w_j^{*})$};

\fill (3.0, 4.86) circle (2pt);
\node[anchor=west, font=\scriptsize] at (3.20, 4.84) {$(a_{jk}^{c},\, w_j^{c})$};

\end{tikzpicture}}\\[2pt]
        {\small \textbf{(c)}~Mildly convex ($\varepsilon_L^{h_k}>1$)}
    \end{minipage}\hspace{0.1cm}
    \begin{minipage}[t]{0.40\textwidth}
        \centering
        \resizebox{\textwidth}{!}{\begin{tikzpicture}[scale=1.0, >=Stealth, font=\small]

\path[use as bounding box] (-0.6,-0.4) rectangle (7.2,6.5);

\draw[->, thick] (0,0) -- (6.0,0) node[right] {$a_{jk}$};
\draw[->, thick] (0,0) -- (0,6.0) node[above] {$w_j$};

\draw[gray!60, semithick, dotted] (2.5, 2.3) -- (2.5, -0.05);
\draw[gray!60, semithick, dotted] (2.5, 2.3) -- (-0.05, 2.3);
\draw[gray!60, semithick, dotted] (2.0, 5.0) -- (2.0, -0.05);
\draw[gray!60, semithick, dotted] (2.0, 5.0) -- (-0.05, 5.0);

\draw[thick, CoolRed] (0,4.619) -- (4.979,0);

\draw[thick, CoolRed, dashed] (1.693,5.70) -- (4.192,0);

\draw[->, very thick, CoolGreen] plot[smooth] coordinates
  {(0.48,0.70) (0.83,0.90) (1.53,1.30) (2.17,1.80) (2.43,2.15) (2.50,2.30)
   (2.58,2.60) (2.61,2.87) (2.60,3.05) (2.49,3.60) (2.37,4.00) (2.19,4.50)
   (2.00,5.00) (1.92,5.20)};
\node[CoolGreen, font=\scriptsize, anchor=south, rotate=32] at (1.2, 1.17) {expansion path};

\draw[very thick, CoolBlue] plot[smooth] coordinates
  {(0.874,6.000) (0.938,5.655) (1.008,5.325) (1.082,5.011) (1.162,4.712) (1.248,4.426) (1.341,4.155) (1.440,3.897) (1.546,3.651) (1.661,3.418) (1.784,3.197) (1.915,2.988) (2.057,2.789) (2.209,2.601) (2.373,2.424) (2.548,2.256) (2.737,2.098) (2.939,1.948) (3.157,1.808) (3.390,1.675) (3.641,1.551) (3.910,1.434) (4.199,1.325) (4.510,1.222) (4.844,1.126) (5.202,1.036) (5.587,0.952) (6.000,0.874)};
\node[CoolBlue, anchor=west, font=\footnotesize] at (6.02, 0.87) {$V^*_j$};
\draw[very thick, CoolBlue] plot[smooth] coordinates
  {(1.631,6.000) (1.712,5.750) (1.797,5.509) (1.885,5.275) (1.979,5.049) (2.076,4.831) (2.179,4.621) (2.287,4.418) (2.400,4.222) (2.518,4.033) (2.643,3.851) (2.773,3.675) (2.910,3.506) (3.054,3.343) (3.205,3.186) (3.363,3.035) (3.530,2.890) (3.704,2.751) (3.887,2.617) (4.079,2.488) (4.281,2.365) (4.492,2.246) (4.714,2.132) (4.947,2.023) (5.192,1.919) (5.448,1.819) (5.717,1.723) (6.000,1.631)};
\node[CoolBlue, anchor=west, font=\footnotesize] at (6.02, 1.63) {$V^{c}_j$};

\fill (2.5, 2.3) circle (2pt);
\node[anchor=east, font=\scriptsize] at (2.35, 2.3) {$(a_{jk}^{*},\, w_j^{*})$};

\fill (2.0, 5.0) circle (2pt);
\node[anchor=south west, font=\scriptsize] at (2.10, 4.66) {$(a_{jk}^{c},\, w_j^{c})$};

\end{tikzpicture}}\\[2pt]
        {\small \textbf{(d)}~Strongly convex ($\varepsilon_L^{h_k}>1$)}
    \end{minipage}
    \par\medskip
    \begin{minipage}{\textwidth}
    \scriptsize
    \emph{Notes:} The figure shows wage--amenity packages for a single amenity $k$ (with several amenities the figure applies amenity by amenity, since~\eqref{eq:MRS} holds for each $k$). Blue curves are iso-value loci of $V_j=(w_j^\rho+\gamma_k a_{jk}^\rho)^{1/\rho}$ at $\rho=0.1$, $\gamma_k=1$. The two marked bundles are the monopsony optimum $(a^*_{jk},w^*_j)$, delivering job value $V^*_j$, and the value-taking optimum $(a^c_{jk},w^c_j)$, delivering the higher $V^c_j$. Red iso-cost lines have slope $-c_{jk}h_k(L_j)/L_j$, solid at the monopsony bundle and dashed at the competitive one. The green expansion path traces the tangencies as $L_j$ rises along the supply curve. In (d) the per-worker cost rises steeply enough that the gradient~\eqref{eq:path_gradient} turns negative at large $L_j$, bending the path backward.
    \end{minipage}
\end{figure}

Figure~\ref{fig:amenity_markdown} illustrates Proposition~\ref{res:underprovision} in the $(w_j,a_{jk})$ plane. Both bundles lie on the common expansion path, with the monopsony optimum at the lower job value (parts (i) and (ii)), and cost curvature governs how the composition shifts along it: at per-worker costs the path is a straight ray and monopsony a radial contraction (panel a), concave costs tilt it toward the amenity (panel b), and convex costs tilt it toward wages (panels c--d). Under mild convexity the amenity still rises with scale (panel c), but once the gradient~\eqref{eq:path_gradient} turns negative the path bends backward (panel d), the over-provision case of part (iii).

\subsection{Progressive Taxation}
\label{app:taxation}

Our baseline model treats wages and amenities symmetrically. We now extend our model to allow that wages are taxed, using the two-parameter tax schedule of \citet{heathcote2017optimal}, $w-T(w)=\lambda w^{1-\tau}$, with progressivity $\tau\in[0,1)$ and level $\lambda>0$. Amenities are untaxed, and firm costs are unchanged. Workers then value the disposable wage $g(w_j)\equiv\lambda w_j^{1-\tau}$, and the job-value index becomes $V_j=(g(w_j)^\rho + \sum_k\gamma_k a_{jk}^\rho)^{1/\rho}$.
The derivation follows Appendix~\ref{app:foc}. The tax matters only through how much of the gross wage the worker keeps. Under progressivity, the marginal net-of-tax rate $g'(w_j)=(1-\tau)\lambda w_j^{-\tau}$ differs from the average net-of-tax rate $g(w_j)/w_j=g'(w_j)/(1-\tau)$, and both appear in the optimal amenity--wage ratio as follows:
\[
\frac{a_{jk}^*}{w_j^*} \;=\; \frac{\Theta}{1-\tau}\,\left[\left(\frac{\gamma_k}{c_{jk}}\right) \cdot \left(\frac{L_j}{h_k(L_j)}\right)\right]^{1/(1-\rho)}, \quad \Theta \;\equiv\; \big[g'(w_j^*)\big]^{-\rho/(1-\rho)}, \quad \omega_k^{\textit{t}} \;=\; \frac{\Theta}{1-\tau}\,\omega_k(L_j),
\]
with $\omega_k(L_j)$ the baseline share from~\eqref{eq:omega_explicit}. We call $\Theta$ the tilt and $1/(1-\tau)$ the wedge. The tilt depends on the marginal rate $g'(w_j^*)$ and, for $\rho\neq0$, differs from one under any tax, because firms substitute between wages and amenities when wages are taxed. The wedge is the ratio of the average to the marginal rate and differs from one only when the tax is progressive.

First, we consider a proportional tax, $\tau=0$. The wedge equals one and the tilt is a constant, $\Theta=\lambda^{-\rho/(1-\rho)}$. If wages and amenities are substitutes, every firm raises amenities relative to wages by the same factor $\Theta$. The cost shares become $\omega_k^{\textit{t}}=\Theta\,\omega_k$, and the model is otherwise the same as the baseline. All baseline results keep their form, with the curvature index $\Phi^{\textit{t}}=\sum_k\sigma_k^{\textit{t}}(1-\varepsilon_L^{h_k})$ and the shares $\sigma_k^{\textit{t}}=\Theta\,\omega_k(L_j)/\mathcal A^{\textit{t}}$.

Next, we consider a progressive tax, $\tau>0$, so the tilt depends on the firm's wage, $\Theta=[(1-\tau)\lambda]^{-\rho/(1-\rho)}\,(w_j^*)^{\tau\rho/(1-\rho)}$. If wages and amenities are substitutes, firms that pay higher wages provide more amenities relative to wages, and fewer if they are complements. The tilt also enters the multiplier, which becomes $\mathcal A^{\textit{t}}=1+\Theta\sum_k\omega_k(L_j)$, with value shares $\sigma_k^{\textit{t}}$ and cost-curvature index $\Phi^{\textit{t}}$. The multiplier $\mathcal A^{\textit{t}}$ measures what workers receive. The multiplier based on what amenities cost the firm is $\hat{\mathcal A}\equiv1+\sum_k\omega_k^{\textit{t}}$. The two are equal under a proportional tax. Under a progressive tax, a unit of wages is worth more to workers on average than at the margin, by the wedge $1/(1-\tau)$, while the firm chooses amenities at the margin. The cost of amenities relative to the wage therefore overstates what amenities add to workers' value, and $\mathcal A^{\textit{t}}=1+(1-\tau)(\hat{\mathcal A}-1)$ is below $\hat{\mathcal A}$.

The firm's problem is solved as in the baseline. The value index simplifies to $V_j=g(w_j^*)\bigl(\mathcal A^{\textit{t}}\bigr)^{1/\rho}$, and the wage FOC becomes
\[
w_j \;=\; \frac{\alpha\psi_j L_j^{\alpha-1}}{1 + \mathcal A^{\textit{t}}/[\eta(1-\tau)] + \sum_k\omega_k^{\textit{t}}\,\varepsilon_L^{h_k}}.
\]
Log-differentiating supply, using $d\ln g(w_j)=(1-\tau)\,d\ln w_j$ and $d\ln\Theta=\tfrac{\tau\rho}{1-\rho}\,d\ln w_j$, gives the following two elasticities:
\[
\varepsilon_w^L \;=\; \frac{\eta\,\Bigl(1-\tau\,\tfrac{1/\mathcal A^{\textit{t}}-\rho}{1-\rho}\Bigr)}{1-\eta\Phi^{\textit{t}}/(1-\rho)},
\qquad
\varepsilon_L^{\hat{\mathcal A}} \;=\; \rho\,\frac{\mathcal A^{\textit{t}}}{\hat{\mathcal A}}\biggl[\frac{1}{\eta(1-\tau)}-\frac{1}{\varepsilon_w^L}\biggr].
\]
In the first expression, a progressive tax multiplies the labor supply elasticity by $1-\tau(1/\mathcal A^{\textit{t}}-\rho)/(1-\rho)$. The factor is below one when $\rho<1/\mathcal A^{\textit{t}}$, which holds at our estimates, so the tax lowers the elasticity. The second expression writes the size-elasticity of the cost-based multiplier $\hat{\mathcal A}$ in terms of $\varepsilon_w^L$ instead of $\Phi^{\textit{t}}$, which is not observed. We use this relation for estimation in Section~\ref{app:tax_application}. Both expressions equal their baseline counterparts at $\tau=0$.

\begin{resultvariant}[Markdowns under Progressive Wage Taxation]{res:markdown_amenities}{$'$}
\label{res:markdown_tax}
Let wages be taxed by $w-T(w)=\lambda w^{1-\tau}$ with amenities untaxed. \textbf{(i)} The total compensation markdown $\mathcal A^{\textit{t}}\mathcal M$ keeps the form of~\eqref{eq:foc_AwMRP} with an adjusted curvature index and a numerator wedge,
\begin{equation}
\mathcal A^{\textit{t}}\,\mathcal M \;=\; \frac{\eta\,(1-\tau)}{\eta+1-\eta\,\widetilde\Phi^{\textit{t}}},\qquad \widetilde\Phi^{\textit{t}} \;\equiv\; \Phi^{\textit{t}} + \tau/\mathcal A^{\textit{t}},
\label{eq:AM_tax}
\end{equation}
\textbf{(ii)} The wage markdown stays below it by the factor $\mathcal A^{\textit{t}}>1$, which for $\tau>0$ lies below the cost-based multiplier $\hat{\mathcal A}$. \textbf{(iii)} Under $\tau>0$ the canonical formula no longer recovers $\mathcal A^{\textit{t}}\mathcal M$, even at $\rho=0$.  All parts reduce to Proposition~\ref{res:markdown_amenities} at $\tau=0$, $\lambda=1$.
\end{resultvariant}

Substituting the supply elasticity into~\eqref{eq:AM_tax} gives the total compensation markdown implied by a measured elasticity,
\[
\mathcal A^{\textit{t}}\,\mathcal M \;=\; \frac{1-\tau}{\,1+\rho/\eta-\tau/\mathcal A^{\textit{t}}+\bigl[(1-\rho)-\tau(1/\mathcal A^{\textit{t}}-\rho)\bigr]/\varepsilon_w^L\,},
\qquad
\mathcal A^{\textit{t}} \;=\; 1+(1-\tau)(\hat{\mathcal A}-1).
\]
At $\rho=0$, the formula is $\varepsilon_w^L/(1+\varepsilon_w^L)$ times $\mathcal A^{\textit{t}}/\hat{\mathcal A}$, which is below one under a progressive tax, so the canonical formula overstates the markdown.

\begin{resultvariant}[Productivity Shocks under Progressive Wage Taxation]{res:demand_shocks}{$'$}
\label{res:demand_tax}
Holding $(\tau,\lambda)$ fixed, after a positive productivity shock $d\psi_j>0$: \textbf{(i)} The amenity level $a^*_{jk}$ rises iff $\varepsilon_L^{h_k}-1<\bigl[1-\rho(1-\tau)\bigr]/\varepsilon_w^L$. \textbf{(ii)} Per-worker amenity cost $C^*_{jk}/L^*_j$ rises iff $\rho(\varepsilon_L^{h_k}-1)<\bigl[1-\rho(1-\tau)\bigr]/\varepsilon_w^L$. \textbf{(iii)} The multiplier $\mathcal A^{\textit{t}*}$ rises iff $\rho\bigl[\Phi^{\textit{t}}+\tau(1-1/\mathcal A^{\textit{t}})/\varepsilon_w^L\bigr]>0$, and is invariant at $\rho=0$.  All parts reduce to Proposition~\ref{res:demand_shocks} at $\tau=0$.
\end{resultvariant}

Progressivity changes the baseline thresholds in two ways. First, $1-\rho$ becomes $1-\rho(1-\tau)$ in parts (i) and (ii).\footnote{The condition of part~(i) also replaces the condition in Proposition~\ref{res:underprovision}(iii). With the tax, a monopsonist under-provides an amenity relative to the competitive benchmark when the condition of part~(i) holds.} With a progressive tax, the marginal tax rate rises with the wage. A positive shock raises the wage, so pay in wages is taxed more heavily, while amenities stay untaxed. The thresholds are therefore less restrictive when wages and amenities are substitutes ($\rho>0$) and more restrictive when they are complements ($\rho<0$). Second, the tax lowers the effective labor supply elasticity $\varepsilon_w^L$, which enters every threshold.

\subsubsection{Application: Structural Estimation with Tax Progressivity}
\label{app:tax_application}

How much do the tax progressivity corrections matter for our findings?  Rearranging the expression for $\varepsilon_L^{\hat{\mathcal A}}$ above gives the functional relationship~\eqref{eq:eos_link} with tax-adjusted coefficients,
\[
\frac{1}{\varepsilon_w^L} \;=\; \frac{1}{\eta(1-\tau)} \;-\; \frac{1}{\rho}\,\frac{\hat{\mathcal A}}{\mathcal A^{\textit{t}}}\,\varepsilon_L^{\hat{\mathcal A}},
\]
so we estimate it in the same way as before. The estimates $(\hat\eta,\hat\rho)$ then give the parameters under the tax as $\eta=\hat\eta/(1-\tau)$ and $\rho=\hat\rho\,\hat{\mathcal A}/\mathcal A^{\textit{t}}$, evaluated at the cross-decile mean of the ratio since $\mathcal A^{\textit{t}}_d$ varies slightly across deciles.  We use two values of $\tau$ for Norway: $0.054$, estimated from income taxes alone \citep{qiu2026income}, and $0.169$, which also accounts for transfers \citep{holter2019laffer}. As in Section~\ref{sec:discussion}, $\hat{\mathcal A}$ is the augmented multiplier $\AMP$ and $(\hat\eta,\hat\rho)$ are the baseline estimates. We compute $\mathcal{A}^{\textit{t}}$ from $\hat{\mathcal A}$ decile by decile. We treat all non-wage compensation as untaxed. Since some in-kind benefits are in fact taxable, our results give an upper bound on the effect of the tax exemption.  We show the results in Appendix Table~\ref{tab:tax_robustness}. The wage markdown, obtained by dividing~\eqref{eq:AM_tax} by $\mathcal A^{\textit{t}}$, does not depend on $\tau$. With these parameters, the wage markdown equals $\bigl[\hat{\mathcal A}\bigl(1+\hat\rho/\hat\eta+(1-\hat\rho)/\varepsilon_w^L\bigr)\bigr]^{-1}$, the same as without taxes in every decile. The total compensation markdown $\hat{\mathcal A}\mathcal M$, with compensation measured at its cost to the firm, is also the same. In contrast, when compensation is measured at its value to workers, the total compensation markdown $\mathcal A^{\textit{t}}\mathcal M$ is at most $\paTaxAMDrop$ percentage points lower, because workers value amenity costs at the net-of-tax rate.  The cost-curvature index is almost unchanged under the two schedules: $\Phi$ rises slightly in the bottom deciles and falls slightly in the top deciles, by at most $\paTaxPhiMax$. The index keeps its sign in every decile, so the cost structure we find is not an artifact of untaxed amenities.

\paragraph{Employer payroll taxes.} Our main analysis excludes employer payroll taxes from non-wage personnel costs. The payroll tax is proportional to the wage at rate $t$, which corresponds to $\tau=0$ and $\lambda=1/(1+t)$ in the model above. The exclusion therefore leaves $\rho$, $\eta$, $\Phi(L_j)$ and the total compensation markdown unchanged. In contrast, the wage markdown does depend on the payroll tax. The wage recorded in the firm accounts does not include the payroll tax, so a worker costs the firm $(\mathcal A+t)w_j$ instead of $\mathcal A w_j$. The total compensation markdown of Section~\ref{sec:discussion} is therefore a markdown on labor cost including the payroll tax, and the wage markdown is the total compensation markdown divided by $\mathcal A+t$. The final row of Appendix Table~\ref{tab:tax_robustness} shows how the wage markdowns change once we account for payroll taxes. At the standard rate $t=0.141$, wage markdowns fall from $\paTaxMBot$--$\paTaxMTop$ to $\paPayMBot$--$\paPayMTop$.

\begin{table}[t!]
  \centering
  \caption{Structural Estimates under Alternative Tax-Transfer Schedules.}
  \label{tab:tax_robustness}
  \small
  \begin{tabular}{lccccc}
\toprule
 & $\rho$ & $\eta$ & $\Phi^{\textit{t}}$ (d1--d10) & $\mathcal{A}^{\textit{t}}\mathcal{M}$ (d1--d10) & $\mathcal{M}$ (d1--d10) \\
\midrule
$\tau=0$ & 0.060 & 5.51 & $-0.375$--$0.046$ & 0.642--0.881 & 0.422--0.566 \\
$\tau=0.054$ & 0.061 & 5.82 & $-0.365$--$0.041$ & 0.630--0.864 & 0.422--0.566 \\
$\tau=0.169$ & 0.064 & 6.63 & $-0.340$--$0.031$ & 0.605--0.828 & 0.422--0.566 \\
$t=0.141$ & 0.060 & 5.51 & $-0.375$--$0.046$ & 0.642--0.881 & 0.386--0.519 \\
\bottomrule
\end{tabular}

  \begin{minipage}{\textwidth}
  \scriptsize
  \vspace{0.25cm}
  \emph{Notes:} The table reports estimates from the structural estimation, obtained using the augmented amenity cost multiplier $\AMP$, and evaluated under alternative tax-transfer schedules. The $\tau=0$ row shows our baseline estimates, $\tau=0.054$ allows tax-only progressivity \citep{qiu2026income}, $\tau=0.169$ adds transfers \citep{holter2019laffer}, and the last row applies the employer payroll tax rate $t=0.141$ to the wage markdown (maintaining $\tau=0$). Ranges span productivity deciles 1--10 (value-added ranking).
  \end{minipage}
\end{table}

\paragraph{Effects of tax progressivity.} Appendix Table~\ref{tab:tax_robustness} holds the measured elasticities fixed and varies only the assumed tax-transfer schedule, so the $\tau=0$ row shows what we would conclude if we ignored taxes, not an economy without taxes. To calculate the effects of tax progressivity itself, we compare the markdown with the schedule~\eqref{eq:AM_tax} and without~\eqref{eq:foc_AwMRP} at the median-decile estimates, holding $(\eta,\rho)$, amenity costs and the cost-curvature index fixed. At $\tau=0.169$, the labor supply elasticity is about $\paTaxElasDropPct$ percent lower. The total compensation markdown falls by $\paTaxWiden$ percentage points, from $\paTaxAMWithout$ to $\paTaxAMWith$ of marginal product. Valuing amenity costs at the net-of-tax rate accounts for $\paTaxReval$ percentage points, and the lower labor supply elasticity for the remaining $\paTaxElasPart$ percentage points.

\citet{berger2026labor} find a comparable effect of tax progressivity. In an oligopsony model with worker and firm heterogeneity and the same tax function, wages fall by nearly $3$ percentage points relative to marginal product when progressivity rises from $0$ to $0.20$. In both models, the marginal tax rate under a progressive tax is higher than the average rate, so a one percent increase in the wage raises the after-tax wage by only $1-\tau$ percent. Labor supply to the firm is therefore less elastic with respect to the wage, and the firm sets the wage further below marginal product. Workers thus bear more than the statutory burden of the tax, since their share of output is also smaller. Unlike in \citet{berger2026labor}, firms in our model can also shift pay toward untaxed amenities. More amenities cushion workers' value, but narrow the tax base and distort the mix of wages and amenities, as does any exemption of amenities from taxable income. At our small estimate of $\rho$, the fall in the labor supply elasticity is offset only slightly by the shift toward amenities. Meanwhile, in \citet{berger2026labor}, the elasticity a firm perceives has two margins: the extensive margin of which firm to work at \citep{mcfadden1974conditional} and an intensive margin of hours \citep{macurdy1981empirical}. They estimate the hours margin to be small (a Frisch elasticity of $0.09$). Our model has only the extensive margin: adding an hours elasticity of $0.1$ to the total compensation elasticity in~\eqref{eq:foc_AwMRP} would raise our total compensation markdown by roughly half a percentage point.

\subsection{Models of Imperfect Competition and Endogenous Amenities}
\label{app:literature}

Appendix Table~\ref{tab:literature} summarizes models of imperfect labor market competition with endogenous amenity provision. We provide an overview across studies of the structure of amenity provision costs, the specification of preferences and the amenity bundle, the source of employer market power and whether the derived markdown depends on amenities, and finally how each case compares to the cost structure specified in our framework. We depart from this literature by allowing multi-dimensional amenities with heterogeneous cost structures, and by focusing on the implications for the measurement of employer market power.

\newgeometry{margin=0.6in}
\begin{landscape}
\begingroup
\renewcommand{\arraystretch}{0.95}
\scriptsize
\setlength{\tabcolsep}{3pt}
\begin{longtable}{@{}>{\raggedright\arraybackslash}p{0.125\linewidth}>{\raggedright\arraybackslash}p{0.095\linewidth}>{\raggedright\arraybackslash}p{0.10\linewidth}>{\raggedright\arraybackslash}p{0.10\linewidth}>{\raggedright\arraybackslash}p{0.15\linewidth}>{\raggedright\arraybackslash}p{0.10\linewidth}>{\raggedright\arraybackslash}p{0.085\linewidth}>{\raggedright\arraybackslash}p{0.175\linewidth}@{}}
\caption{Overview of Models of Imperfect Labor Market Competition with Endogenous Amenity Provision.}\label{tab:literature}\\
\toprule
Paper & Amenity & Worker Preferences & Provision Cost & Mapping to Our Model & Variation Used & Source of Market Power & Do Markdowns Depend on Amenities? \\
\midrule
\endfirsthead
\multicolumn{8}{@{}l}{\textit{Appendix Table~\ref{tab:literature}, continued.}}\\[2pt]
\toprule
Paper & Amenity & Worker Preferences & Provision Cost & Mapping to Our Model & Variation Used & Source of Market Power & Do Markdowns Depend on Amenities? \\
\midrule
\endhead
\midrule\multicolumn{8}{r@{}}{\textit{continued on next page}}\\
\endfoot
\bottomrule
\multicolumn{8}{@{}p{0.98\linewidth}@{}}{\emph{Notes:} In our notation, a per-worker provision cost for amenity $k$ corresponds to a size elasticity $\varepsilon_L^{h_k}=1$, which implies $\Phi(L_j)=0$ when this cost structure holds for every $k$. For brevity, this overview omits some models of search frictions with endogenous employer-provided health insurance. For instance, in \citet{dey2005equilibrium}, insurance is bargained jointly with the wage at a premium per insured worker, while in \citet{dizioli2016health}, \citet{aizawa2019labor}, \citet{fang2019household}, and \citet{aizawa2020equilibrium}, firms post a wage and an insurance offer and bear a fixed cost per firm, alone or in addition to the medical costs of their workers. None of these papers provide a markdown formula. We also omit models without employer market power in which amenities respond to demand. For instance, in \citet{charles2022demand}, price-taking firms without wage-setting power choose safety (via training that costs output and rental price that falls with output demand) against a constant per-accident cost. The last column refers to the markdown each paper states; \emph{Not derived} means that the paper states none.}\\
\endlastfoot
This paper & $n$ endogenous amenities $\mathbf{a}$, plus exogenous $\bar a$ & CES in $(w,\mathbf{a})$, $\hat\rho\approx0.06$ & $C_{k}=c_{k}h_k(L)a_{k}$, any size elasticity $\varepsilon_L^{h_k}\geq0$ & --- & Firm-level demand shocks (Norway) & Idiosyncratic tastes & \textbf{Yes:} the wage and total compensation markdowns depend on amenity cost structure \\
\midrule
\citet{hwang1998hedonic} & Generic amenity $a$ & Additive, $w+h(a)$ & Per worker, $C$ convex in $a$ & One amenity, $\varepsilon_L^{h_k}=1$; $\rho\to1$ limit, but $C$ is convex in $a$ & --- & Search frictions, wage posting & \textbf{Not derived} \\
\citet{lang2004pricing} & Job characteristic $h$, e.g., hours & General $u(w,h)$ & Forgone output & One production-relevant amenity; preferences not CES & --- & Search frictions, wage posting & \textbf{Not derived:} profit per hire depends only on the utility offered \\
\citet{flabbi2012effect} & Hours flexibility, binary & Linear in wage and flexibility, worker-specific valuation & Share of match productivity & One binary amenity, $\varepsilon_L^{h_k}=1$; $\rho\to1$ limit; match-level bargaining & Structural estimation (US) & Search frictions, bargaining & \textbf{Not derived:} wages are Nash bargained with a fixed weight \\
\citet{dube2022power} & Vector of amenities (11 in the survey) & CES in the wage and a linear amenity index & Per worker & One composite amenity, $\varepsilon_L^{h_k}=1$ & Hypothetical job offers; Walmart's minimum wage (US) & Idiosyncratic tastes & \textbf{Yes:} the Lerner rule for the wage has an amenity term \\
\citet{arold2025worker}, Appendix~B & Worker rights in the contract, value $A(T)$ & CES in after-tax wage and $A(T)$ & Fixed, independent of firm size & One composite amenity, $\varepsilon_L^{h_k}=0$; same CES; firm size exogenous & Provincial income tax rates (Canada) & Firm's offer under a strike threat & \textbf{Not derived} \\
\citet{bagga2025amenity} & Remote work, binary & Linear, $w+x_ia$, two worker types & One-time random upgrade cost per job & One binary amenity; $\rho\to1$ limit; no firm size & Aggregate amenity valuation and productivity shocks (US) & Search frictions, sequential auctions & \textbf{Not derived} \\
\citet{colonna2025} & Working conditions, one amenity $a$ & Cobb--Douglas in $(w,a)$ & Per worker; mandated floor with fines & One amenity, $\varepsilon_L^{h_k}=1$; $\rho=0$ case & Random workplace inspections (Mexico) & Idiosyncratic tastes & \textbf{Not derived:} the amenity cost enters the wage first-order condition \\
\citet{corradini2025collective}, Appendix~D & Female-friendly contract clauses & Stone--Geary in $(w,a)$ & Per worker & One amenity, $\varepsilon_L^{h_k}=1$; $\rho=0$ case at zero reference levels & Union priorities (Brazil) & Idiosyncratic tastes & \textbf{Yes:} the markdown includes the amenity elasticity, as in \citet{lagos2025union} \\
\citet{kroft2025imperfect}, Supplemental Appendix~H & Firm-specific amenity; allowed to respond to the shock & $\ln w+\ln g$, with $g$ the value of amenities & Not modeled & One composite amenity, $\rho=0$; their assumed gap is our $\varepsilon_w^L/\eta-1$ & Winning procurement auctions (US) & Idiosyncratic tastes & \textbf{Yes:} the wage-based markdown is biased if compensation and wages respond differently \\
\citet{lavetti2025estimating} & Fatality risk $R$ & $\ln w-h(R)$ & Per worker, proportional to the wage & One amenity, $\varepsilon_L^{h_k}=1$, $\rho=0$; cost proportional to the wage & Job changers across industry--occupation cells (Brazil) & Idiosyncratic tastes, offer frictions & \textbf{Not derived} \\
\citet{sulka2025providing} & Benefits $b$, binary, unobserved before hiring & Linear, $w+b$ & Per worker, net of a tax advantage & One binary amenity, $\varepsilon_L^{h_k}=1$; $\rho\to1$ limit & --- & Search frictions, wage posting & \textbf{Not derived:} workers receive their outside option \\
\citet{adams2023flexible} & Schedule flexibility $f$, amenity or disamenity & General $u(w,f)$ & Profit term per worker, a gain or a cost & One production-relevant amenity, $\varepsilon_L^{h_k}=1$; general preferences & Employer concentration (UK) & Cournot oligopsony, assumed supply curve & \textbf{Yes:} the wage is a markdown on output plus the amenity's profit term \\
\citet{biro2026tightness} & Generic amenity $a$; injury risk in the data & Linear, $w+a$ & Per worker or fixed, convex in $a$ & One amenity; $\rho\to1$ limit; $\varepsilon_L^{h_k}=1$ or $0$, but $C$ is convex in $a$ & Local labor market tightness (Hungary) & Search frictions, wage posting & \textbf{Not derived} \\
\citet{diporto2026careers} & Low managerial delegation $d$ & $v(w,d)$, $v_d\le0$ & Forgone output & One production-relevant amenity; general preferences & Family vs.\ non-family firms (Italy) & Idiosyncratic tastes & \textbf{Yes:} the markdown depends on delegation through the supply elasticity \\
\citet{faberman2026job} & Generic amenity $a$ & Linear, $w+\nu_ga$ & Per job, $C$ convex in $a$ & One amenity, $\varepsilon_L^{h_k}=1$; $\rho\to1$ limit, but $C$ is convex in $a$ & Gender differences in amenity valuation (US) & Search frictions, sequential auctions & \textbf{Not derived:} wages are set by sequential auctions with a fixed bargaining weight \\
\citet{lachowska2026work} & Work hours $h$ & $wh-\varepsilon_ih^\mu$ & Forgone output & One production-relevant amenity; not nested & Hours differences across employers (US) & Exogenous firm surplus & \textbf{No:} the markdown is an exogenous firm-specific profit margin \\
\citet{lagos2025union} & Value index of all contract clauses & Stone--Geary in $(w,a)$ & Per worker & One index, $\varepsilon_L^{h_k}=1$; $\rho=0$ case at zero reference levels & Union bargaining power (Brazil) & Idiosyncratic tastes & \textbf{Yes:} the markdown on total compensation includes the amenity elasticity \\
\citet{lochner2026amenities} & Flexibility requirement $e$ & Linear, $w-e-\varepsilon_i$ & Forgone output & One production-relevant amenity; $\rho\to1$ limit & Job requirements over the business cycle (Germany) & Search frictions, bargaining & \textbf{Not derived:} wages are Nash bargained collectively with a fixed weight \\
\citet{morchio2026gender} & Generic amenity $a$ & Linear, $w+\beta_ga$ & Per worker, $C$ convex in $a$ & One amenity, $\varepsilon_L^{h_k}=1$; $\rho\to1$ limit, but $C$ is convex in $a$ & Gender differences in pay and worker flows (Brazil) & Search frictions, wage posting & \textbf{Not derived:} the amenity's net value is folded into firm productivity \\
\citet{scott2026workplace} & Schedule variability and work outside regular hours & $\alpha\ln w+\beta'a$, by gender & Forgone output & Two production-relevant amenities; $\rho=0$ case & Instruments from the firm's other jobs (France) & Idiosyncratic tastes & \textbf{No:} the markdown depends only on the wage elasticity \\
\end{longtable}
\endgroup
\end{landscape}
\restoregeometry

\normalsize

\newpage

\subsection{Numerical Illustrations for the Baseline Model}
\label{app:psi_figures}

\setcounter{equation}{0}

\begin{figure}[H]
    \centering
    \caption{Comparative Statics in Productivity $\psi_j$ at $\rho=0$ (Cobb--Douglas).}
    \label{fig:markdown_psi_rho0}
    \includegraphics[width=0.9\textwidth]{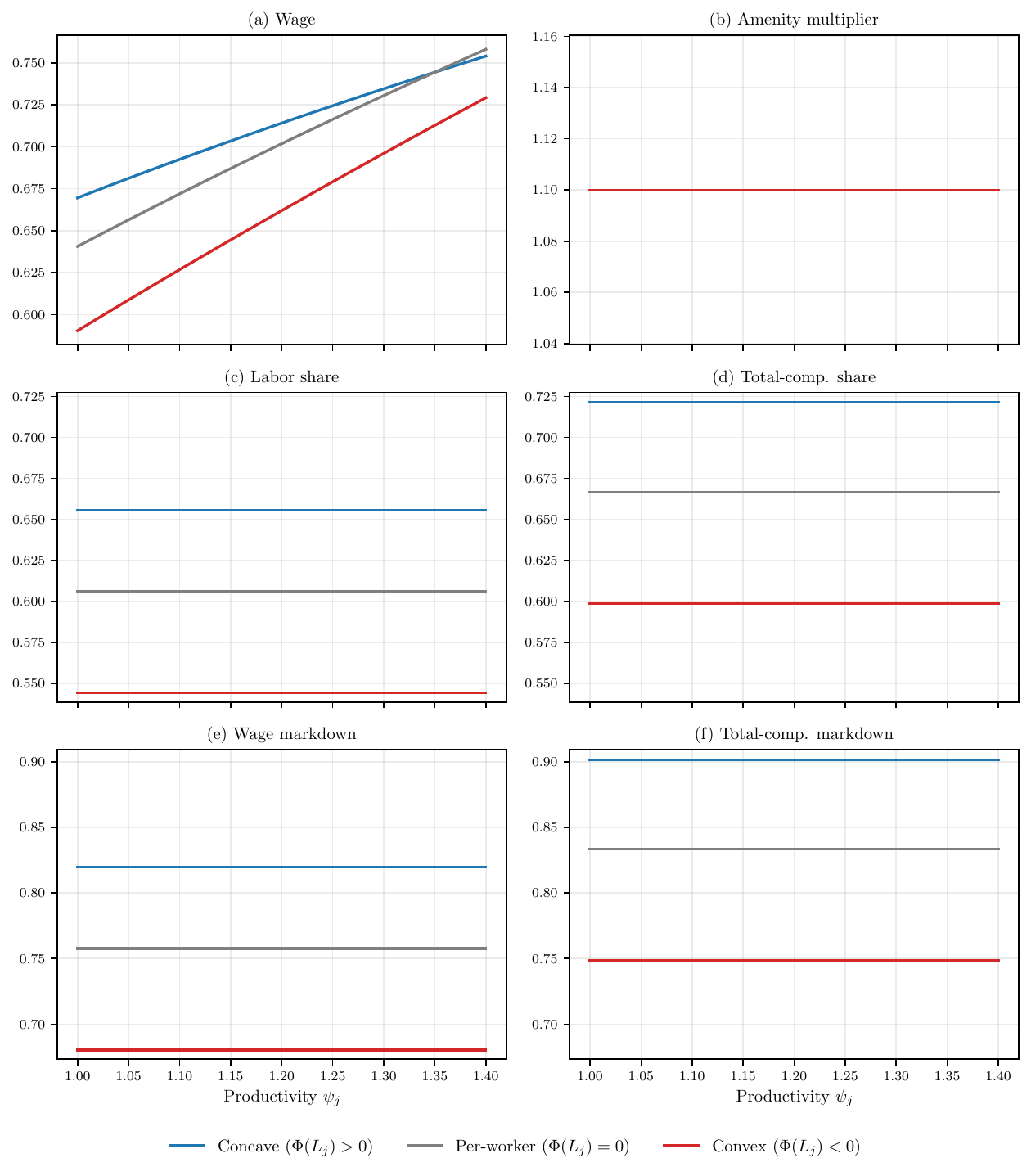}
    \par\medskip
    \begin{minipage}{\textwidth}
    \scriptsize
    \emph{Notes:} Within each panel, all lines are solid, and cost regimes are distinguished by color: concave ($\Phi(L_j)>0$, blue), per-worker ($\Phi(L_j)=0$, gray), convex ($\Phi(L_j)<0$, red). Panels show (a) the wage $w_j^*$; (b) the amenity multiplier $\mathcal{A}^*$; (c) the labor share $w_j^*L_j^*/R_j^*$ and (d) the total-compensation share $(w_j^*L_j^*+\sum_k C_{jk}^*)/R_j^*$ of revenue; and (e) the wage markdown $\mathcal{M}^*=w_j^*/\mathrm{MRP}^*$ and (f) the total-compensation markdown $\mathcal{A}^*\mathcal{M}^*=\mathcal{A}^*w_j^*/\mathrm{MRP}^*$.  The calibration has a single amenity with cost technology $h(L_j) = L_j^{\varepsilon_L^h}$ and parameters $\eta=5$, $\rho=0$, $\alpha=0.8$, $\gamma=0.10$, $c=1$, $N\bar{a}_j/\mathbb{V}^\eta=20$, and $\varepsilon_L^h \in \{0, 1, 2.5\}$.
    \end{minipage}
\end{figure}

\begin{figure}[H]
    \centering
    \caption{Comparative Statics in Productivity $\psi_j$: Substitutes ($\rho>0$).}
    \label{fig:markdown_psi_rho_pos}
    \includegraphics[width=0.9\textwidth]{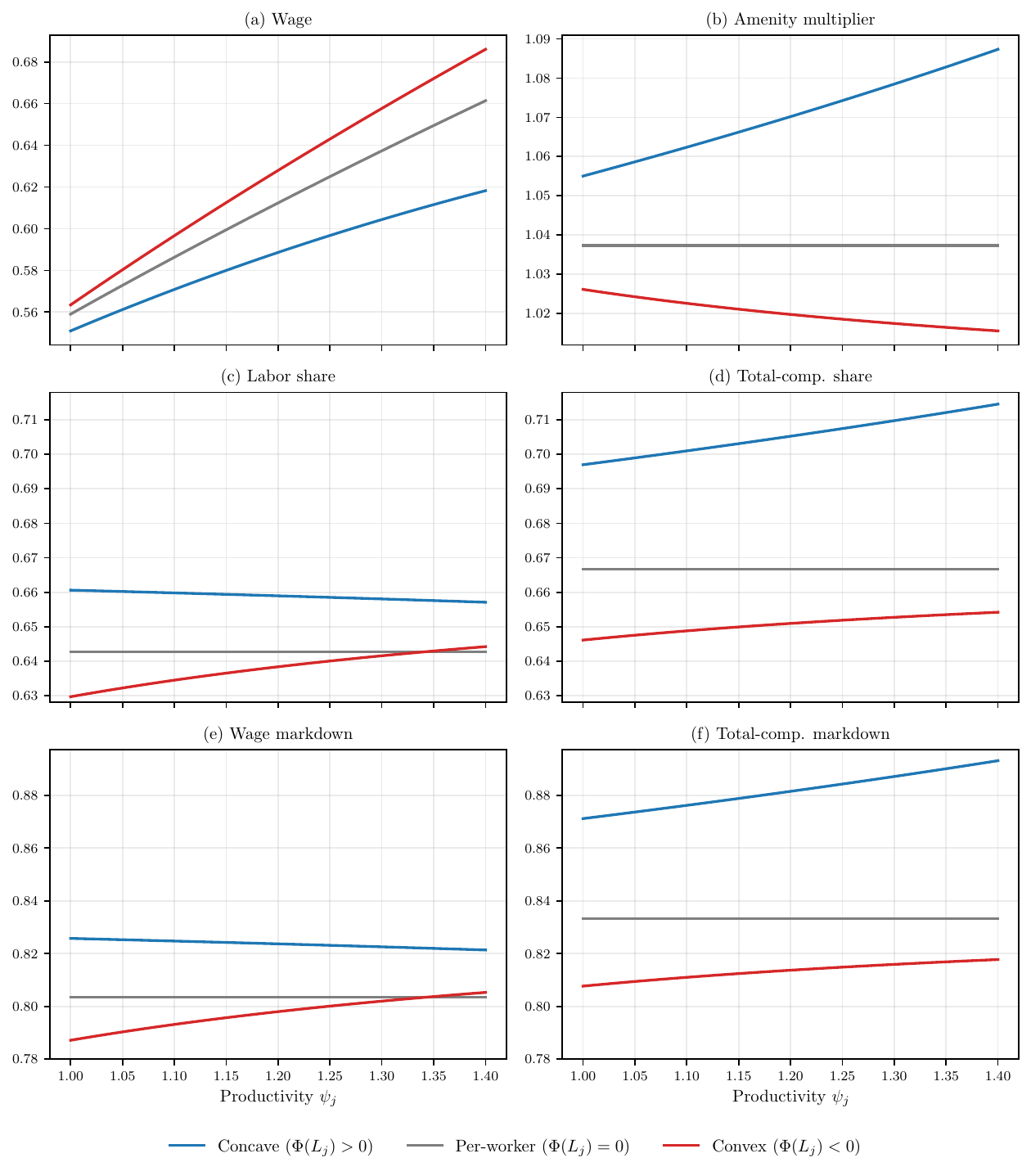}
    \par\medskip
    \begin{minipage}{\textwidth}
    \scriptsize
    \emph{Notes:} Within each panel, all lines are solid, and cost regimes are distinguished by color: concave ($\Phi(L_j)>0$, blue), per-worker ($\Phi(L_j)=0$, gray), convex ($\Phi(L_j)<0$, red). Panels show (a) the wage $w_j^*$; (b) the amenity multiplier $\mathcal{A}^*$; (c) the labor share $w_j^*L_j^*/R_j^*$ and (d) the total-compensation share $(w_j^*L_j^*+\sum_k C_{jk}^*)/R_j^*$ of revenue; and (e) the wage markdown $\mathcal{M}^*=w_j^*/\mathrm{MRP}^*$ and (f) the total-compensation markdown $\mathcal{A}^*\mathcal{M}^*=\mathcal{A}^*w_j^*/\mathrm{MRP}^*$.  The calibration has a single amenity with cost technology $h(L_j) = L_j^{\varepsilon_L^h}$ and parameters $\eta=5$, $\rho=0.3$, $\alpha=0.8$, $\gamma=0.10$, $c=1$, $N\bar{a}_j/\mathbb{V}^\eta=20$, and $\varepsilon_L^h \in \{0, 1, 2.5\}$.
    \end{minipage}
\end{figure}

\begin{figure}[H]
    \centering
    \caption{Comparative Statics in Productivity $\psi_j$: Complements ($\rho<0$).}
    \label{fig:markdown_psi_rho_neg}
    \includegraphics[width=0.9\textwidth]{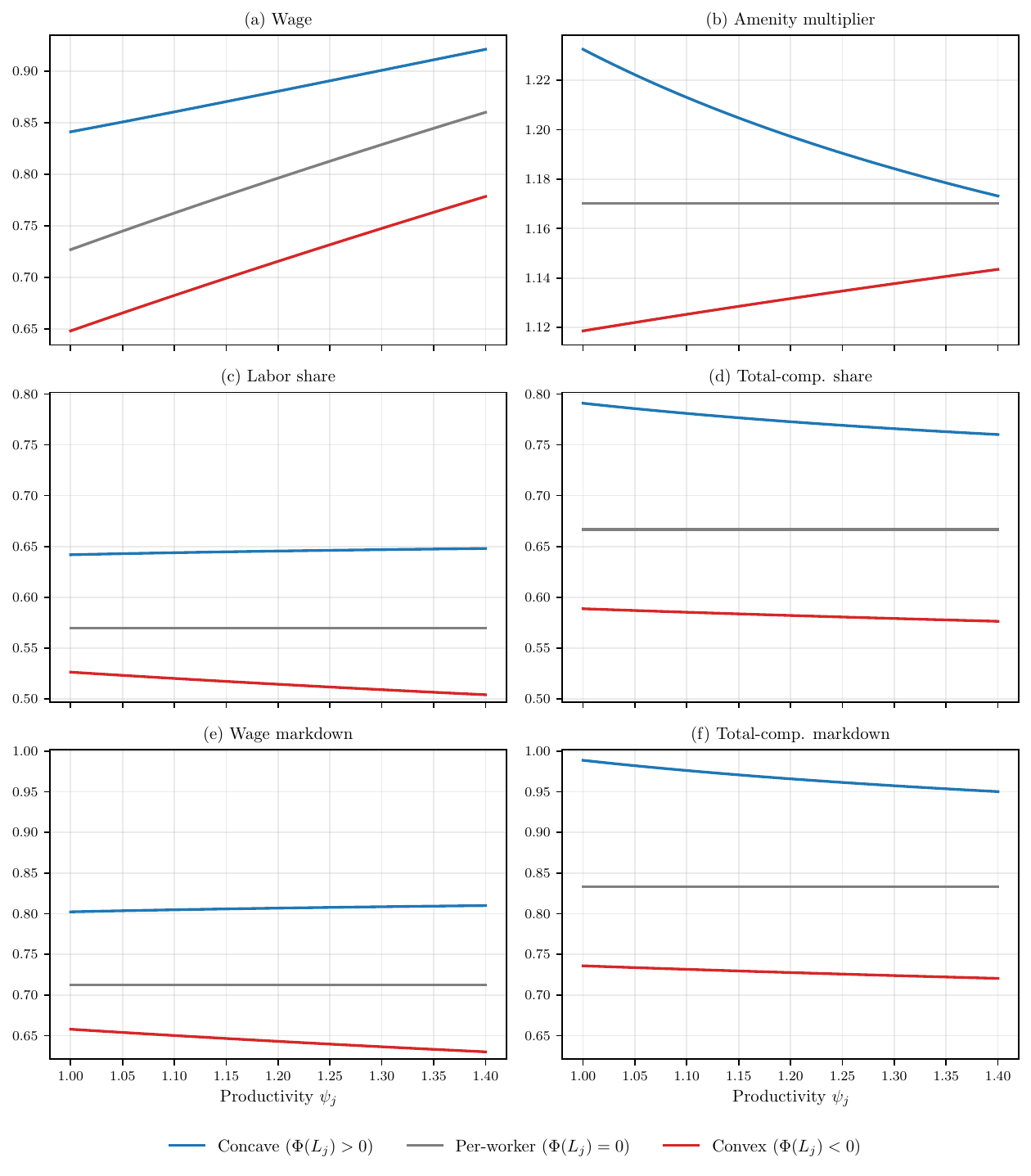}
    \par\medskip
    \begin{minipage}{\textwidth}
    \scriptsize
    \emph{Notes:} Within each panel, all lines are solid, and cost regimes are distinguished by color: concave ($\Phi(L_j)>0$, blue), per-worker ($\Phi(L_j)=0$, gray), convex ($\Phi(L_j)<0$, red). Panels show (a) the wage $w_j^*$; (b) the amenity multiplier $\mathcal{A}^*$; (c) the labor share $w_j^*L_j^*/R_j^*$ and (d) the total-compensation share $(w_j^*L_j^*+\sum_k C_{jk}^*)/R_j^*$ of revenue; and (e) the wage markdown $\mathcal{M}^*=w_j^*/\mathrm{MRP}^*$ and (f) the total-compensation markdown $\mathcal{A}^*\mathcal{M}^*=\mathcal{A}^*w_j^*/\mathrm{MRP}^*$.  The calibration has a single amenity with cost technology $h(L_j) = L_j^{\varepsilon_L^h}$ and parameters $\eta=5$, $\rho=-0.3$, $\alpha=0.8$, $\gamma=0.10$, $c=1$, $N\bar{a}_j/\mathbb{V}^\eta=20$, and $\varepsilon_L^h \in \{0, 1, 2.5\}$.
    \end{minipage}
\end{figure}

\newpage
\section{Identification and Structural Estimation}
\setcounter{figure}{0}
\setcounter{table}{0}
\label{app:identification}

This appendix describes the identification and estimation of structural parameters $(\eta,\rho)$.

\subsection{Point Identification}

The argument follows Section~\ref{subsec:identification_structural}. Under non-Cobb--Douglas preferences ($\rho\ne 0$) and a non-degenerate cost-curvature index ($\Phi(L_j)\ne\bar\Phi$), both $\varepsilon_w^L(L_j)$ and $\varepsilon_L^\mathcal{A}(L_j)$ vary across firms. Under a standard rank condition (at least two moments with sufficient variation in both), the cross-firm slope and intercept of~\eqref{eq:eos_link} then identify $(\eta,\rho)$. Given these parameters, $\Phi(L_j)$ follows from the wage elasticity $\varepsilon_w^L(L_j)$. This argument requires the homogeneity of $(\eta,\rho)$.

\subsection{Partial Identification}

The argument for point identification provided above breaks down if the cost-curvature index is degenerate or if preferences are Cobb--Douglas. We now characterize four configurations in which partial identification remains feasible despite the failure of point identification.

\begin{enumerate}

\item[(i)] \textit{Cobb--Douglas preferences ($\rho=0$) with non-degenerate cost-curvature ($\Phi(L_j)\ne\bar\Phi$).} At $\rho=0$, $\omega_k=\gamma_k$ regardless of $c_{jk}$, so $\bar{\mathcal A}=1+\sum_k\gamma_k$ is constant across firms. With $\Phi(L_j)$ varying, $\varepsilon_w^L$ varies across firms while $\varepsilon_L^\mathcal{A}=0$, and the markdown is $\bar{\mathcal A}\mathcal M^* = \varepsilon_w^L/(\varepsilon_w^L+1)$. Since $1/\varepsilon_w^L(L_j)=1/\eta-\Phi(L_j)$ at $\rho=0$, the profile of $\Phi(L_j)$ across firms is identified only up to the constant $1/\eta$.

\item[(ii)] \textit{Cobb--Douglas preferences ($\rho=0$) with degenerate cost-curvature ($\Phi(L_j)=\bar\Phi$).} As in case (i), $\bar{\mathcal A}$ is constant across firms, and now both elasticities are too, with $\varepsilon_L^\mathcal{A}=0$ and the markdown $\bar{\mathcal A}\mathcal M^*=\varepsilon_w^L/(\varepsilon_w^L+1)$. A constant multiplier $\bar{\mathcal A}$ across firms is consistent with $\rho=0$ but does not establish it, since per-worker costs with common unit costs produce the same pattern at any $\rho$. In addition, $\eta$ and $\bar\Phi$ cannot be separately identified under $\rho=0$. In the corner case $\bar\Phi=0$, $\varepsilon_w^L=\eta$, as in the canonical monopsony.

\item[(iii)] \textit{Non-Cobb--Douglas preferences ($\rho\ne 0$) with zero cost-curvature ($\Phi(L_j)=0$).} The multiplier $\bar{\mathcal A}_j$ varies across firms via $c_{jk}$ but not within firms; both elasticities are constant, $\varepsilon_w^L=\eta$ and $\varepsilon_L^\mathcal{A}=0$, and the markdown is canonical, $\bar{\mathcal A}_j\mathcal M_j^*=\eta/(\eta+1)$. Cross-firm variation in $\bar{\mathcal A}_j$, combined with $\varepsilon_L^\mathcal{A}=0$, pins down $\Phi(L_j)=0$ and $\eta=\varepsilon_w^L$; $\rho$ does not enter the elasticities and is not identified from them. If unit costs $c_{jk}$ are observable across firms (e.g.,\ from cost accounts), $\rho$ is identified from the cross-firm slope $d\ln\omega_k/d\ln c_{jk} = -\rho/(1-\rho)$, holding firm size fixed. A negative slope reveals substitutes ($\rho>0$) and a positive slope complements ($\rho<0$).

\item[(iv)] \textit{Non-Cobb--Douglas preferences ($\rho\ne 0$) with degenerate non-zero cost-curvature  ($\Phi(L_j)=\bar\Phi\ne 0$).} The multiplier $\bar{\mathcal A}_j$ varies across firms via $c_{jk}$ and within firms along the supply curve (since $\varepsilon_L^\mathcal{A}\ne 0$). This is the only case with constant $\varepsilon_w^L$ and $\varepsilon_L^\mathcal{A}\ne 0$ (distinguishing it from case (ii)). The triplet $(\eta,\rho,\bar\Phi)$ is not identified, but the markdown is pinned down as $\mathcal A^*\mathcal M^*=\varepsilon_w^L/[(\varepsilon_w^L+1)+\varepsilon_w^L\,\varepsilon_L^\mathcal{A}]$. The sign of $\varepsilon_L^\mathcal{A}$ further sign-identifies the primitives: $\mathrm{sign}(\rho)=\mathrm{sign}(\bar\Phi)$ if $\varepsilon_L^\mathcal{A}>0$, and opposite signs if $\varepsilon_L^\mathcal{A}<0$; the supply-stability requirement $\varepsilon_w^L>0$ adds the bound $\bar\Phi<(1-\rho)/\eta$.
\end{enumerate}

In each case the markdown $\mathcal{AM}$ remains identified, whereas the data are consistent with $\rho=0$ in cases~(i) and~(ii), pin down only $\eta$ and $\Phi(L_j)=0$ in case~(iii), and none of $(\eta,\rho,\bar\Phi)$ in case~(iv). A non-zero $\varepsilon_L^\mathcal{A}$ at any firm rules out cases~(i)--(iii), leaving case~(iv) or the non-degenerate case, and in either its sign sign-identifies $(\rho,\Phi)$ as in case~(iv). With the augmented multiplier, $\varepsilon_L^{\mathcal A}$ depends on $\rho$ through~\eqref{eq:AP_resp}, so in cases~(iii) and~(iv) the markdown is pinned down for a given $\rho$.

\subsection{Structural Estimation}
\label{app:structural_estimation}

We use the internal shock design, with firms split into deciles $d$ of residualized baseline value added and the shock responses estimated within each decile. For every decile, this yields the post-shock average coefficients on log employment, the log wage and the log monetized amenity cost multiplier, $(\hat\beta_L^d,\hat\beta_w^d,\hat\beta_{\AM}^d)$, with their covariance $\hat\Omega_d$. For each production-relevant job attribute $k\in\sP$, we measure the pre-period share of ads mentioning the attribute among the posting firms of the decile $\pi_{k,d}$, and estimate its post-shock average change $\Delta\pi_{k,d}$. Posting firms' shares and responses are used for all firms in their respective decile.

\paragraph{Valuing non-wage job attributes.} The provision of attribute $k\in\sP$ in decile $d$ is the share of ads that mention the attribute, i.e., $a_{k,d}=\pi_{k,d}$. As in Section~\ref{subsec:identification_structural}, the cost share of the attribute is then $\omega_{k,d}=\gamma_k(\pi_{k,d}/w_d)^\rho$. We obtain the preference weight $\gamma_k$ from survey-based estimates of willingness to pay. The survey gives $\nu_k$, the log wage difference that leaves a worker indifferent between a job with the attribute and an otherwise identical job without the attribute. We assume that raising the share $\pi_{k,d}$ by one percentage point is worth $\nu_k$ percent of the wage to workers. Holding job value constant, the wage then falls with the share at the rate $d\ln w_d/d\pi_{k,d}=-\nu_k$. In the model, the value of an additional percentage point of the share depends on the level of $\pi_{k,d}$ and on the wage $w_d$. We therefore require the model to match $\nu_k$ at the means across deciles, $\bar\pi_k$ and $\bar w$. The left side of~\eqref{eq:MRS} is the marginal rate of substitution between the attribute and the wage, which equals $\nu_k$ times the wage, so $\gamma_k(\bar\pi_k/\bar w)^{\rho-1}=\nu_k\bar w$. Multiplying by $\bar\pi_k/\bar w$ and solving for the preference weight, which is common to all deciles, gives $\gamma_k=\nu_k\bar\pi_k(\bar w/\bar\pi_k)^{\rho}$. The cost share is then
\begin{equation}
\omega_{k,d}=\nu_k\bar\pi_k\Bigl(\frac{\pi_{k,d}}{\bar\pi_k}\Bigr)^{\rho}\Bigl(\frac{\bar w}{w_d}\Bigr)^{\rho},
\label{eq:omega_d}
\end{equation}
where $\ln w_d$ is the pre-shock mean log hourly wage of stayers in the decile and $\ln\bar w$ its mean across deciles. The cost shares give the augmented multiplier $\AMP_d=\AM_d+\sum_{k\in\sP}\omega_{k,d}$ and the weights in~\eqref{eq:AP_resp},
\[
\sigma^{\sM}_d=\frac{\AM_d}{\AMP_d}=\frac{\AM_d}{\AM_d+\sum_{k\in\sP}\omega_{k,d}}
\qquad\text{and}\qquad
\sigma^{\sP}_{k,d}=\frac{\omega_{k,d}}{\AMP_d}=\frac{\omega_{k,d}}{\AM_d+\sum_{k'\in\sP}\omega_{k',d}},
\]
which sum to one within each decile. Appendix Table~\ref{tab:attr} reports the attributes' valuations and the means of their shares and responses.

\begin{table}[!htb]
\centering\scriptsize\setlength{\tabcolsep}{2.5pt}
\caption{Production-Relevant Amenities: Valuations, Shares, and Responses.}
\label{tab:attr}
\resizebox{\textwidth}{!}{\begin{tabular}{llccc}
\toprule
Attribute & Source of $\nu_k$ & $\nu_k$ & $\bar\pi_k$ & $\Delta\bar\pi_k$ \\
\midrule
Good career path & Audoly et al.\ (2024) & $0.084$ $(0.043)$ & $0.053$ & $0.005$ $(0.002)$ \\
On-the-job training & Maestas et al.\ (2023) & $0.054$ $(0.007)$ & $0.200$ & $0.006$ $(0.004)$ \\
Good workplace & Audoly et al.\ (2024), three-item mean & $0.111$ $(0.030)$ & $0.223$ & $0.010$ $(0.004)$ \\
Flexible hours & Andresen et al.\ (2026) & $0.251$ $(0.014)$ & $0.289$ & $-0.016$ $(0.004)$ \\
Regular work schedule & Andresen et al.\ (2026) & $0.171$ $(0.013)$ & $0.579$ & $0.012$ $(0.005)$ \\
Permanent job & Audoly et al.\ (2024) & $0.254$ $(0.043)$ & $0.826$ & $0.010$ $(0.003)$ \\
\bottomrule
\end{tabular}
}
\begin{minipage}{\textwidth}\scriptsize\vspace{3pt}
\emph{Notes:} $\nu_k$ is the valuation of the attribute as a share of the wage, and $\bar\pi_k$ and $\Delta\bar\pi_k$ are the pre-period share and its post-shock change, averaged across deciles. Standard errors are in parentheses. The three items for good workplace are good colleagues ($\paWtpColl$), work from home ($\paWtpHome$) and the inclusive work-life scheme ($\paWtpIA$). Relative to regular work schedule, \citet{bhuller2026beliefs} value flexible hours at $\paAblFlex$ and inconvenient hours at \paAblShift. We use inconvenient hours (e.g., shift work or overtime work required) as the reference category. The shares and responses are from the extended sample, cohorts 2005--2019.
\end{minipage}
\end{table}

\paragraph{Estimator.} The elasticities are $\varepsilon_w^L=\beta_L^d/\beta_w^d$ and $\varepsilon_L^{\mathcal{A}}$ from~\eqref{eq:AP_resp}, with $\varepsilon_L^{\AM}=\beta_{\AM}^d/\beta_L^d$ and $d\ln a_{jk}/d\ln L_j=(\Delta\pi_{k,d}/\pi_{k,d})/\beta_L^d$. Substituting into~\eqref{eq:eos_link} and multiplying through by $\beta_L^d$ gives the per-decile residual
\begin{equation*}
\hat r_d(\eta,\rho)=\hat\beta_w^d-\frac{1}{\eta}\hat\beta_L^d+\frac{\sigma^{\sM}_d}{\rho}\hat\beta_{\AM}^d+\sum_{k\in\sP}\sigma^{\sP}_{k,d}\Bigl(\frac{\Delta\pi_{k,d}}{\pi_{k,d}}-\hat\beta_w^d\Bigr).
\end{equation*}
The attribute term carries no $1/\rho$, because the cost shares move by $\rho$ times the change in provision relative to the wage. As the shares $\sigma^{\sM}_d$ and $\sigma^{\sP}_{k,d}$ depend on $\rho$ only weakly, $\rho$ is identified mainly from the cross-decile variation in the multiplier term. We fit $(\eta,\rho)$ by nonlinear least squares with parameter-dependent weights,
\begin{equation}
(\hat\eta,\hat\rho) \;=\; \arg\min_{(\eta,\rho)}\; \sum_{d} \frac{\hat r_d(\eta,\rho)^2}{\hat v_d(\eta,\rho)},
\label{eq:rho_estimator}
\end{equation}
where $\hat v_d = g_d'\hat\Omega_d g_d$ and $g_d=(-1/\eta,\,\sigma^{\sM}_d,\,\sigma^{\sM}_d/\rho)$ is the gradient of $\hat r_d$ in $(\hat\beta_L^d,\hat\beta_w^d,\hat\beta_{\AM}^d)$. Confidence intervals come from a parametric bootstrap. Each draw takes the decile coefficients from $N(\hat\beta^d,\hat\Omega_d)$. The attribute responses $\Delta\pi_{k,d}$ and the valuations $\nu_k$ are drawn independently from normal distributions given by their estimates and standard errors. For each draw, we re-estimate $(\eta,\rho)$ from~\eqref{eq:rho_estimator} and recompute $\AMP_d$, the cost-curvature index, the total compensation markdown and the wage markdown. The confidence intervals are the $2.5$th and $97.5$th percentiles across draws.

\subsection{Results: Preferences, Cost-Curvature and Markdowns}

With firms ranked by residualized baseline value added, we obtain $\hat\rho \approx \paRhoHat$ with 95\% bootstrap CI $[\paRhoLo,\paRhoHi]$ and $\hat\eta \approx \paEtaHat$ with 95\% CI $[\paEtaLo,\paEtaHi]$. Appendix Table~\ref{tab:rho_decile_gradient} reports the implied amenity cost multipliers, cost-curvature index and markdowns. Taken together, the non-wage job attributes are worth $\paAttrVal$ of the wage, where permanent jobs ($\paValPerm$) and regular work schedules ($\paValReg$) feature prominently in this valuation. In every decile, the augmented multiplier $\AMP_d$ is about \paThetaPct{} percent above the monetized amenity cost multiplier $\AM_d$. The difference is similar across deciles because $\rho$ is small, so the cost shares in~\eqref{eq:omega_d} vary little with provision and the wage. The cost-curvature index moves from $\paPhiBot$ in the bottom decile to $\paPhiTop$ in the top, with confidence intervals shown in Figure~\ref{fig:markdown_deciles}. The total compensation markdown rises from $\paAMBot$ to $\paAMTop$ and the wage markdown $\mathcal{A}\mathcal{M}_d/\AMP_d$ from $\paMBot$ to $\paMTop$.

\begin{table}[!htbp]
  \centering
  \caption{Amenity Cost Multipliers, Cost-Curvature and Markdowns.}
  \label{tab:rho_decile_gradient}
  \small\setlength{\tabcolsep}{4pt}
  \begin{tabular*}{\textwidth}{@{\extracolsep{\fill}} c *{8}{>{\centering\arraybackslash}p{1.4cm}} @{}}
\toprule
 & \multicolumn{4}{c}{Monetized amenity cost multiplier $\AM$} & \multicolumn{4}{c}{Augmented amenity cost multiplier $\AMP$} \\
\cmidrule(lr){2-5} \cmidrule(lr){6-9}
Decile & $\Phi_d$ & $\mathcal{A}\mathcal{M}_d$ & $\AM_d$ & $\mathcal{M}_d$ & $\Phi_d$ & $\mathcal{A}\mathcal{M}_d$ & $\AMP_d$ & $\mathcal{M}_d$ \\
\midrule
\multicolumn{9}{l}{\textit{Panel A: Firms ranked by residualized baseline value added}} \\[2pt]
1 & $-0.321$ & 0.642 & 1.097 & 0.585 & $-0.375$ & 0.642 & 1.522 & 0.422 \\
2 & $-0.142$ & 0.725 & 1.084 & 0.669 & $-0.195$ & 0.726 & 1.506 & 0.482 \\
3 & $-0.082$ & 0.758 & 1.081 & 0.701 & $-0.135$ & 0.760 & 1.503 & 0.506 \\
4 & $-0.036$ & 0.786 & 1.081 & 0.727 & $-0.088$ & 0.788 & 1.502 & 0.524 \\
5 & $-0.035$ & 0.786 & 1.083 & 0.726 & $-0.087$ & 0.788 & 1.504 & 0.524 \\
6 & $-0.017$ & 0.798 & 1.086 & 0.735 & $-0.069$ & 0.800 & 1.507 & 0.531 \\
7 & $-0.009$ & 0.803 & 1.090 & 0.736 & $-0.061$ & 0.805 & 1.511 & 0.533 \\
8 & $\phantom{-}0.029$ & 0.828 & 1.097 & 0.755 & $-0.022$ & 0.831 & 1.518 & 0.547 \\
9 & $\phantom{-}0.046$ & 0.840 & 1.111 & 0.756 & $-0.006$ & 0.842 & 1.531 & 0.550 \\
10 & $\phantom{-}0.098$ & 0.878 & 1.137 & 0.772 & $\phantom{-}0.046$ & 0.881 & 1.555 & 0.566 \\
\addlinespace[6pt]
\multicolumn{9}{l}{\textit{Panel B: Firms ranked by residualized baseline employment}} \\[2pt]
1 & $-0.313$ & 0.655 & 1.105 & 0.592 & $-0.361$ & 0.656 & 1.527 & 0.430 \\
2 & $-0.186$ & 0.714 & 1.093 & 0.653 & $-0.234$ & 0.716 & 1.516 & 0.472 \\
3 & $-0.134$ & 0.741 & 1.085 & 0.683 & $-0.183$ & 0.743 & 1.508 & 0.493 \\
4 & $-0.091$ & 0.766 & 1.085 & 0.706 & $-0.139$ & 0.768 & 1.508 & 0.509 \\
5 & $-0.082$ & 0.771 & 1.085 & 0.711 & $-0.130$ & 0.773 & 1.507 & 0.513 \\
6 & $-0.032$ & 0.802 & 1.085 & 0.739 & $-0.081$ & 0.804 & 1.508 & 0.533 \\
7 & $-0.024$ & 0.807 & 1.088 & 0.742 & $-0.073$ & 0.809 & 1.511 & 0.535 \\
8 & $\phantom{-}0.000$ & 0.823 & 1.092 & 0.753 & $-0.050$ & 0.825 & 1.515 & 0.544 \\
9 & $\phantom{-}0.030$ & 0.844 & 1.104 & 0.765 & $-0.020$ & 0.846 & 1.526 & 0.554 \\
10 & $\phantom{-}0.088$ & 0.888 & 1.124 & 0.790 & $\phantom{-}0.038$ & 0.889 & 1.544 & 0.576 \\
\bottomrule
\end{tabular*}

  \begin{minipage}{1\textwidth}
  \scriptsize
  \vspace{0.25cm}
  \emph{Notes:} Panel A ranks firms by residualized baseline value added and Panel B by residualized baseline employment. The first set of columns uses the monetized amenity cost multiplier $\AM_d$ and the second the augmented multiplier $\AMP_d$. Each set uses the estimates of $(\eta,\rho)$ obtained with its own multiplier: $(\paEtaAccTwo,\paRhoAcc)$ and $(\paEtaHatTwo,\paRhoHat)$ in Panel A, and $(\paEtaAccEmpTwo,\paRhoAccEmp)$ and $(\paEtaHatEmpTwo,\paRhoHatEmp)$ in Panel B. The wage markdown is calculated as $\mathcal{M}_d=\mathcal{A}\mathcal{M}_d/\AM_d$ in the first set, which counts only the monetized amenities as non-wage compensation, and as $\mathcal{M}_d=\mathcal{A}\mathcal{M}_d/\AMP_d$ in the second, which also counts the attributes paid in output. Extended sample, cohorts 2005--2019.
  \end{minipage}
\end{table}

Ranking firms by residualized baseline log employment instead gives $\hat\rho\approx\paRhoHatEmp$ with 95\% CI $[\paRhoLoEmp,\paRhoHiEmp]$ and $\hat\eta\approx\paEtaHatEmp$ with 95\% CI $[\paEtaLoEmp,\paEtaHiEmp]$. In Panel B of Appendix Table~\ref{tab:rho_decile_gradient}, the cost-curvature index rises from $\paPhiBotEmp$ in the smallest decile to $\paPhiTopEmp$ in the largest, the total compensation markdown from $\paAMBotEmp$ to $\paAMTopEmp$ and the wage markdown from $\paMBotEmp$ to $\paMTopEmp$. The two rankings give very similar results, both qualitatively and quantitatively. In both, amenity costs are convex in the bottom deciles, and the markdowns rise from the bottom to the top decile by nearly the same amount.

\section{Appendix Figures and Tables}
\label{app:empirical}
\setcounter{figure}{0}
\setcounter{table}{0}

\begin{figure}[H]
\centering
\caption{The Effects of Demand Shocks on Total Value Added.}
\label{fig:value_added_shocks}

\begin{subfigure}[t]{0.48\textwidth}
    \includegraphics[width=\linewidth]{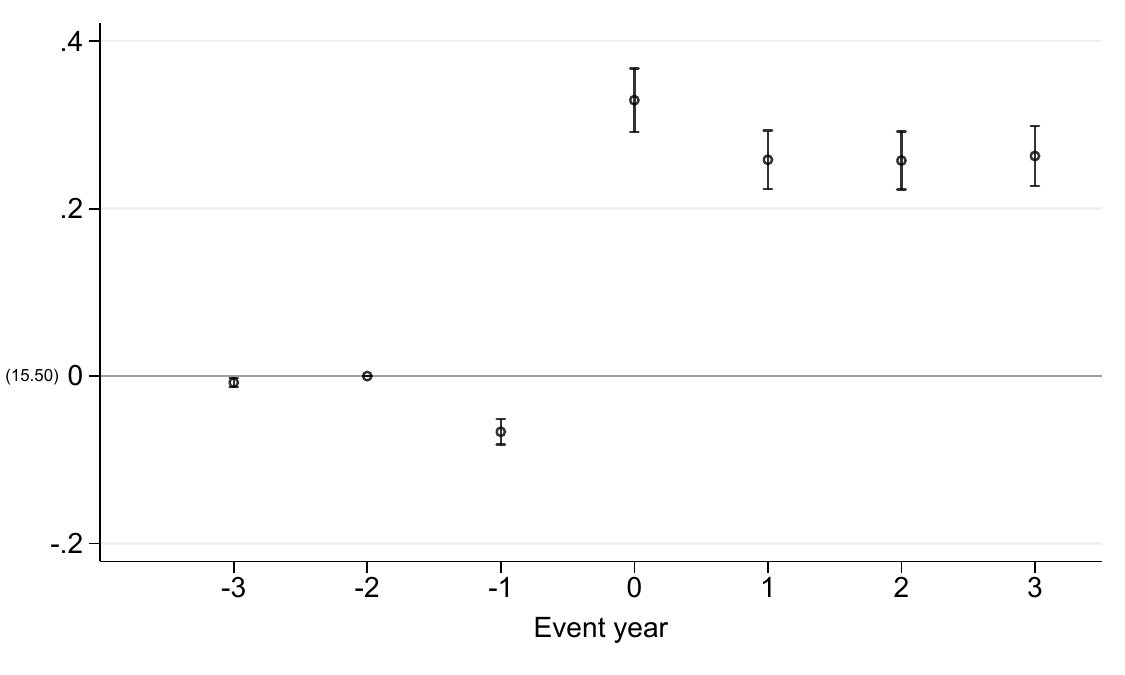}
    \caption{Internal Shock}
\end{subfigure}\hfill
\begin{subfigure}[t]{0.48\textwidth}
    \includegraphics[width=\linewidth]{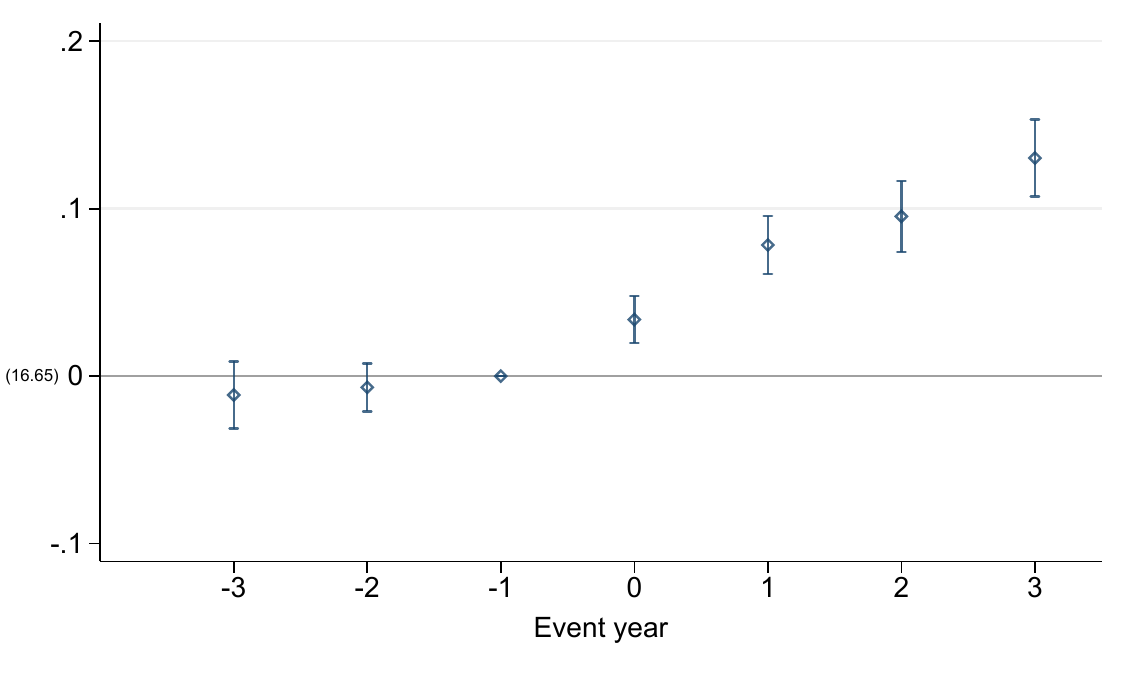}
    \caption{Procurement Shock}
\end{subfigure}

\vspace{0.5em}

\begin{subfigure}[t]{0.48\textwidth}
    \includegraphics[width=\linewidth]{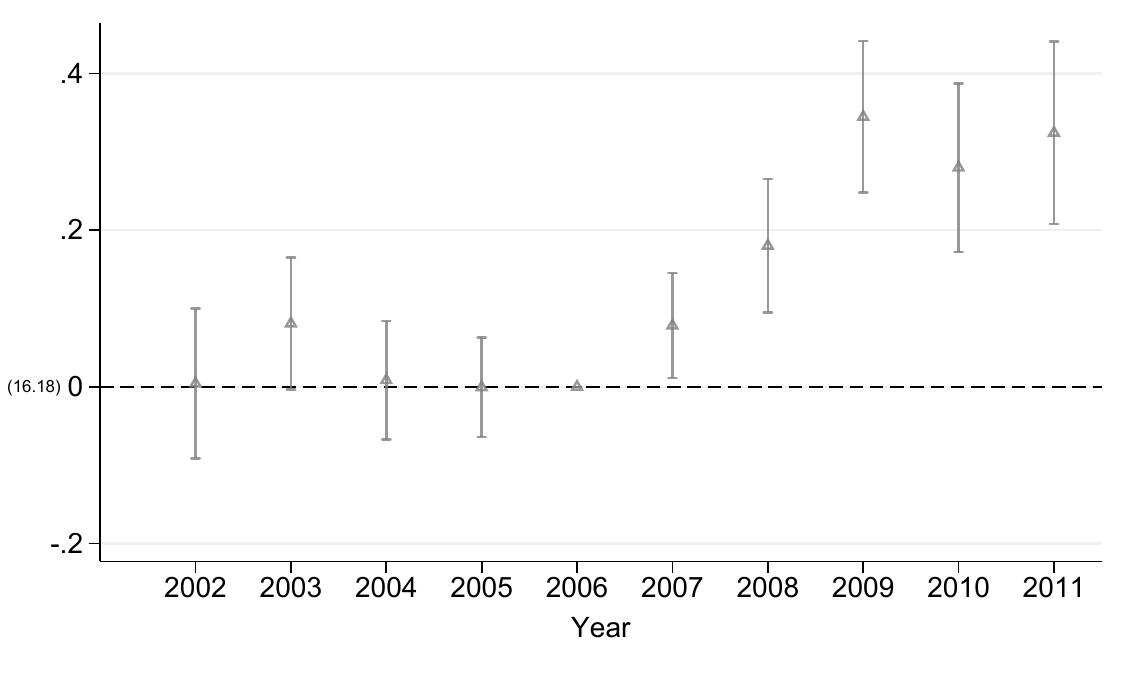}
    \caption{Export Shock}
\end{subfigure}

\begin{minipage}{\textwidth}
\scriptsize
\emph{Notes:} The main datasets used to construct outcomes are firm balance sheets. Coefficients are event-time estimates with 95\% confidence intervals from the dynamic specification in Eq.~(\ref{eq:dynamic_did}). Value-added outcomes are in logs. The internal shock (panel~a) uses a balanced $[-3,+3]$ event window for cohorts 2005--2019 with standard errors clustered at the industry and commuting zone level. The procurement shock (panel~b) uses a balanced event window for cohorts 2012--2019 with standard errors clustered at the firm level. For these two shocks, we restrict to firms observed in all years of the balanced $[-3,+3]$ event window and having at least two stayers. The export shock (panel~c) uses 2006 as the baseline period, and standard errors are clustered at the firm level. Stayers in this shock are defined as workers who are present at the firm for at least 8 years during the estimation window. The sample is restricted to firms with an export-to-revenue ratio of at least 1\% in 2005--2007.
\end{minipage}
\end{figure}

\begin{figure}[H]
\centering
\caption{The Effects of Demand Shock on Employment and Wage Margins.}
\label{fig:app_emp_wages}

\begin{subfigure}[t]{0.48\textwidth}
    \includegraphics[width=\linewidth]{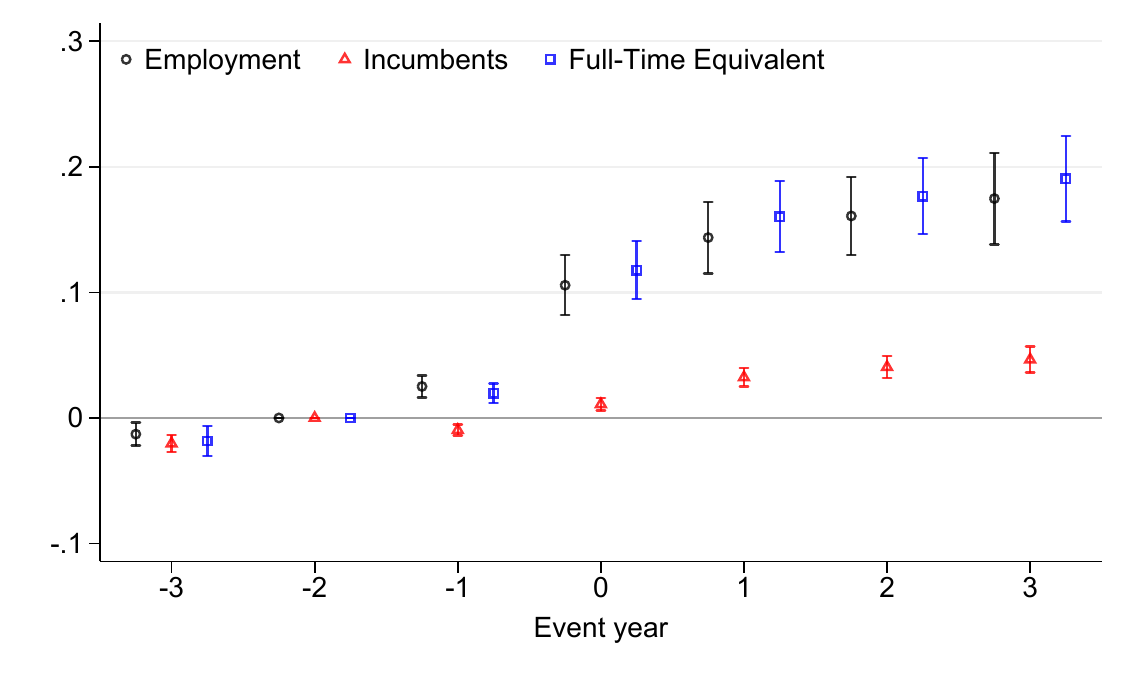}
    \caption{Employment Margins by Type}
\end{subfigure}\hfill
\begin{subfigure}[t]{0.48\textwidth}
    \includegraphics[width=\linewidth]{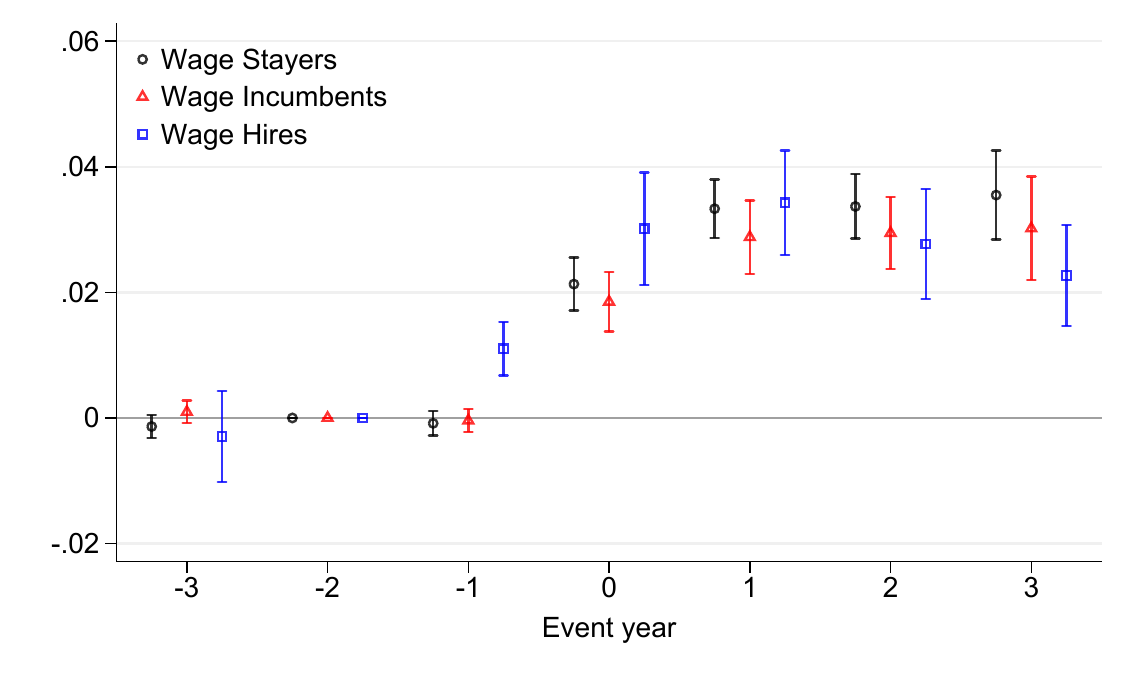}
    \caption{Wage Margins by Type}
\end{subfigure}

\begin{minipage}{\textwidth}
\scriptsize
\emph{Notes:} The dataset used to construct outcomes is the matched employer--employee data (A-meldingen).
Coefficients are event-time estimates with 95\% confidence intervals from the
dynamic specification in Eq.~(\ref{eq:dynamic_did}). All outcomes are in
logs. Wage outcomes are in hourly terms. We use the contracts sample (cohorts 2018--2019) for this internal shock and restrict this sample to firms observed in all years of the balanced
$[-3,+3]$ event window, and with at least two stayers throughout the full
estimation period. Estimates stack the 2018 and 2019 cohorts. Standard
errors are clustered at the industry and commuting zone level.
\end{minipage}
\end{figure}

\begin{figure}[H]
\centering
\caption{Validation: Employment and Wages Impacts Across Samples.}
\label{fig:bothshocks_emp_wage}

\begin{subfigure}[t]{0.43\textwidth}
    \includegraphics[width=\linewidth]{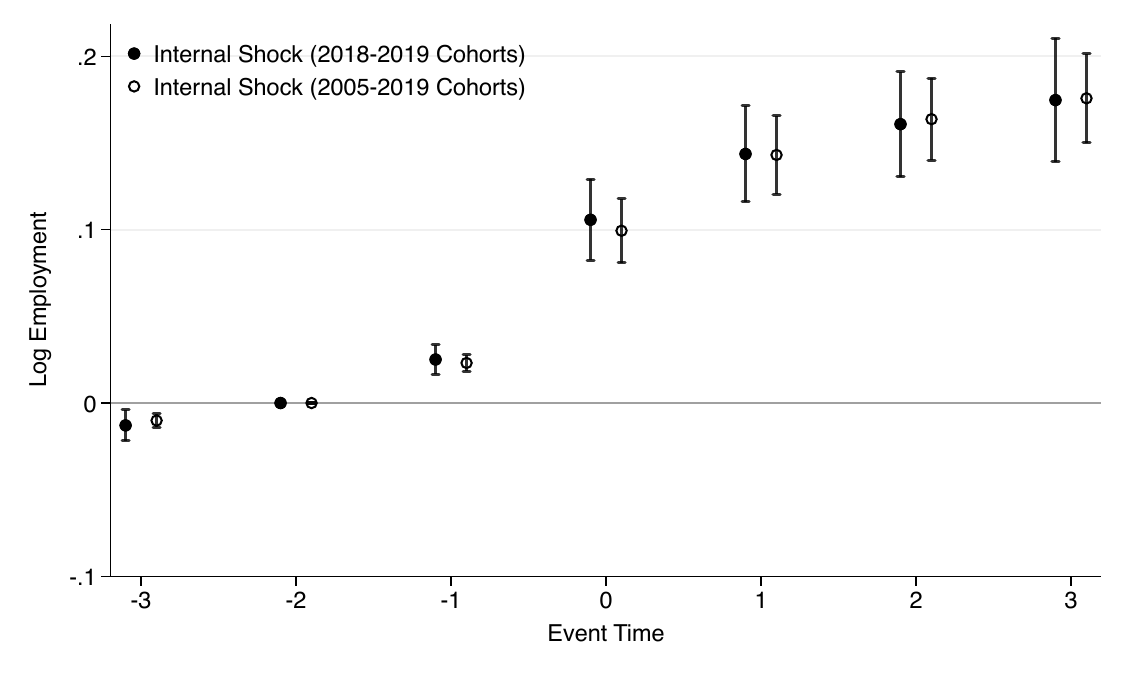}
    \caption{Employment: Internal Shock}
\end{subfigure}
\begin{subfigure}[t]{0.43\textwidth}
    \includegraphics[width=\linewidth]{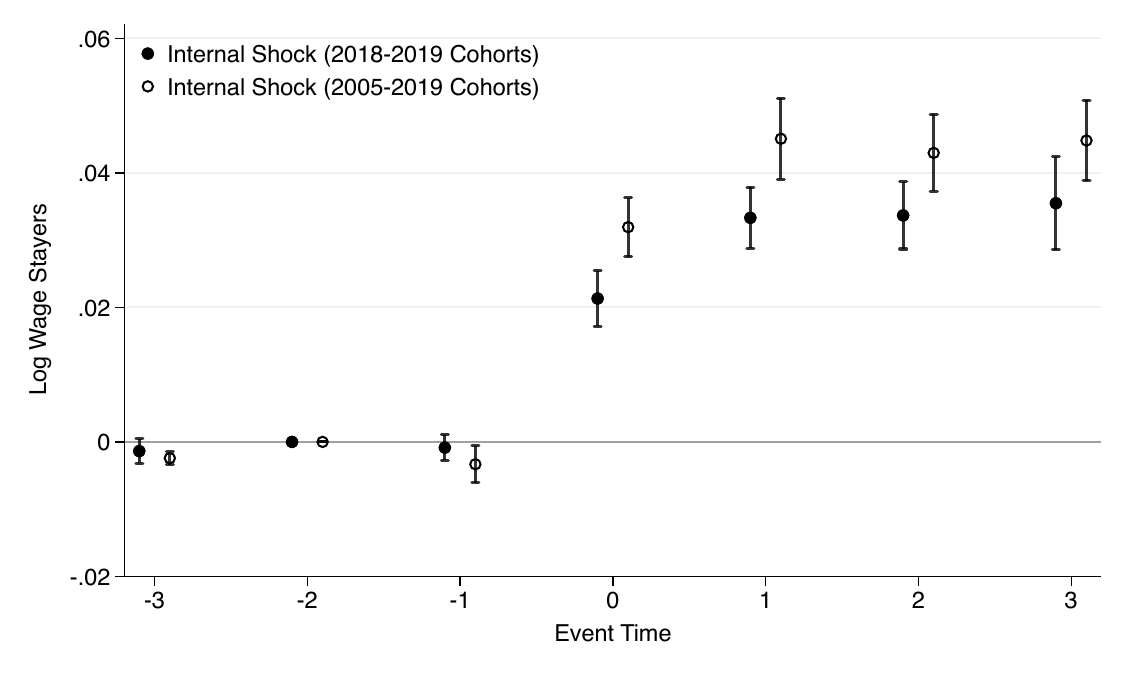}
    \caption{Wages: Internal Shock}
\end{subfigure}

\vspace{0.2cm}

\begin{subfigure}[t]{0.43\textwidth}
    \includegraphics[width=\linewidth]{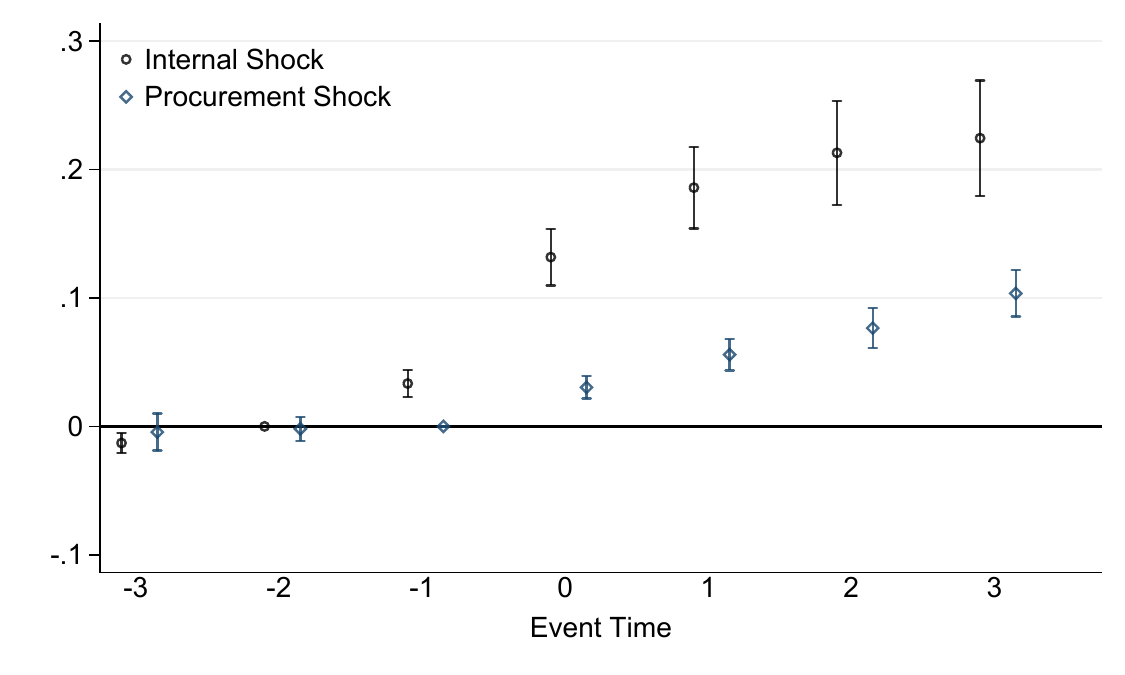}
    \caption{Employment: Procurement Sample}
\end{subfigure}
\begin{subfigure}[t]{0.43\textwidth}
    \includegraphics[width=\linewidth]{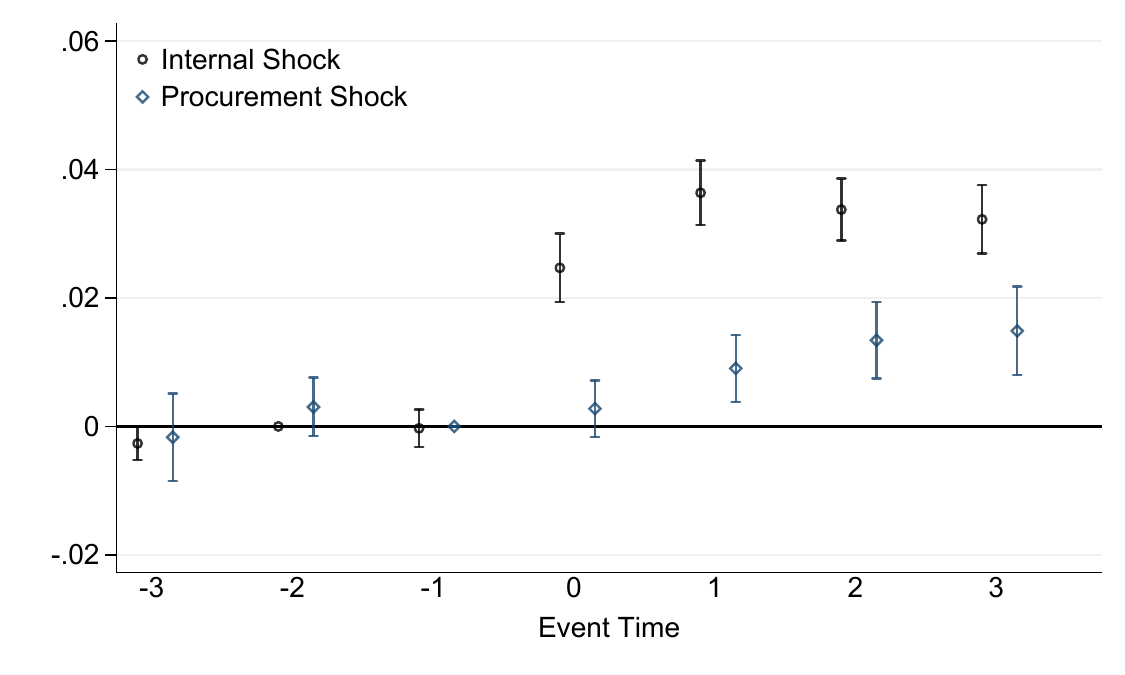}
    \caption{Wages: Procurement Sample}
\end{subfigure}

\vspace{0.2cm}

\begin{subfigure}[t]{0.43\textwidth}
    \includegraphics[width=\linewidth]{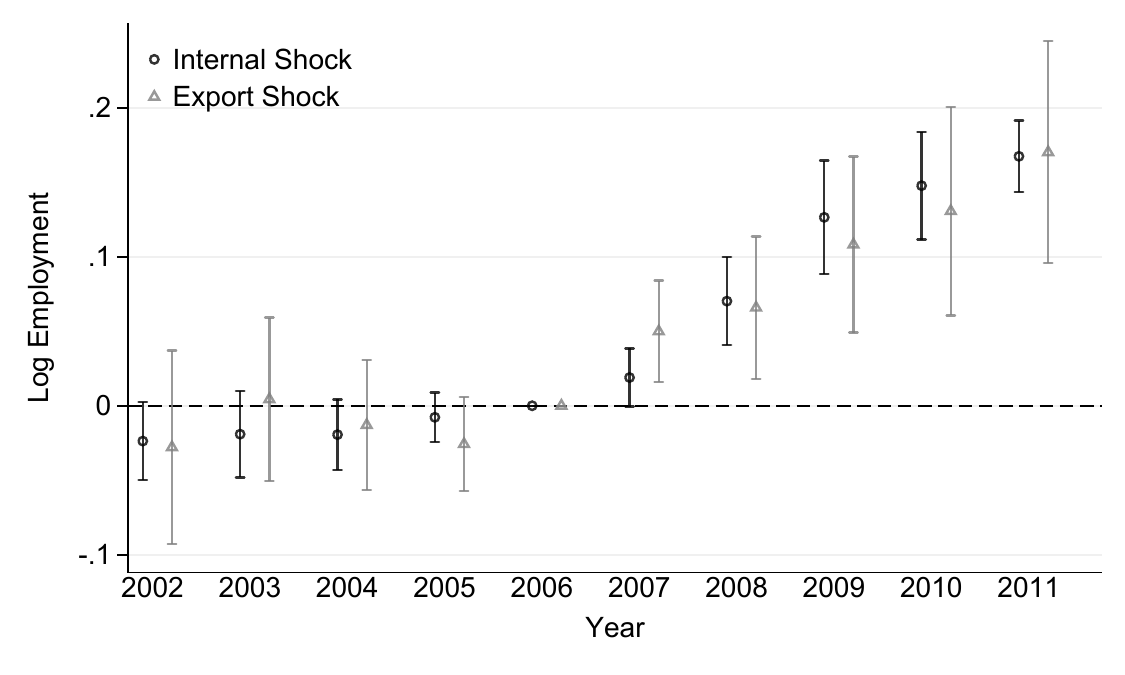}
    \caption{Employment: Export Sample}
\end{subfigure}
\begin{subfigure}[t]{0.43\textwidth}
    \includegraphics[width=\linewidth]{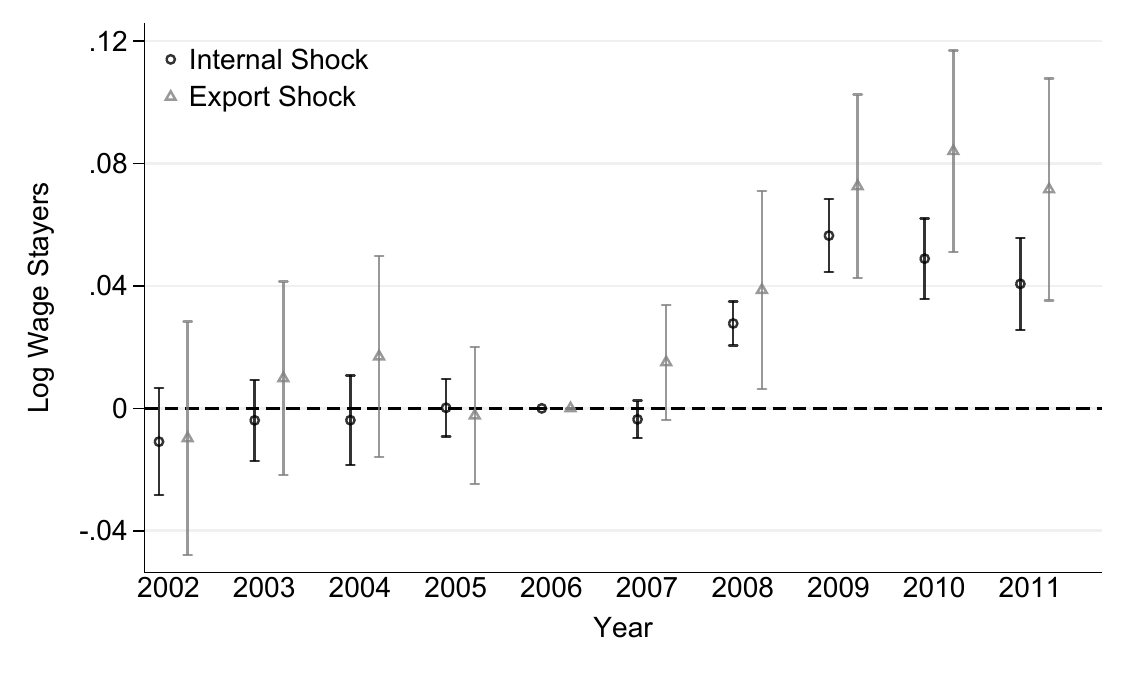}
    \caption{Wages: Export Sample}
\end{subfigure}

\begin{minipage}{\textwidth}
\scriptsize
\emph{Notes:} The dataset used to construct outcomes is the matched employer--employee data. Coefficients are event-time estimates with 95\% confidence intervals from the dynamic specification in Eq.~(\ref{eq:dynamic_did}). Panels (a) and (b) compare internal-shock estimates across the extended sample (cohorts 2005--2019) and the contracts sample (cohorts 2018--2019). Panels (c) and (d) re-estimate the internal shock restricted to the procurement-shock firms over the cohorts 2012--2019. Panels (e) and (f) re-estimate the internal shock using only the 2008 treatment cohort, restricted to export-shock firms in the 2002--2011 event window. Wages are defined for stayers. Standard errors are clustered at the firm level for the export shock, at the firm level for the procurement shock, and at the industry and commuting zone level for the internal shock.
\end{minipage}
\end{figure}

\begin{figure}[H]
\centering
\caption{Non-Wage Compensation and Non-Pecuniary Job Attributes.}
\label{fig:descr_contracts_jobads}

\begin{subfigure}[t]{0.32\textwidth}
    \includegraphics[width=\linewidth]{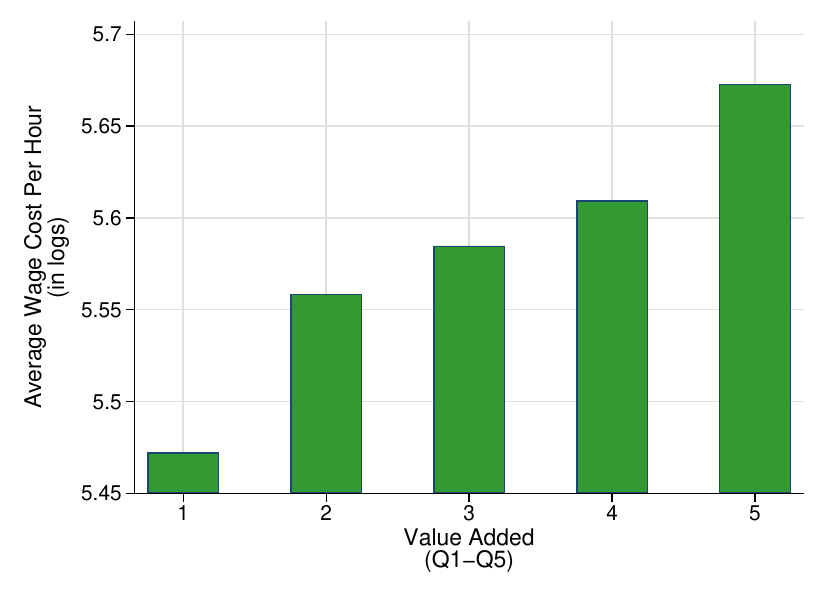}
    \caption{Hourly Wage Costs}
\end{subfigure}\hfill
\begin{subfigure}[t]{0.32\textwidth}
    \includegraphics[width=\linewidth]{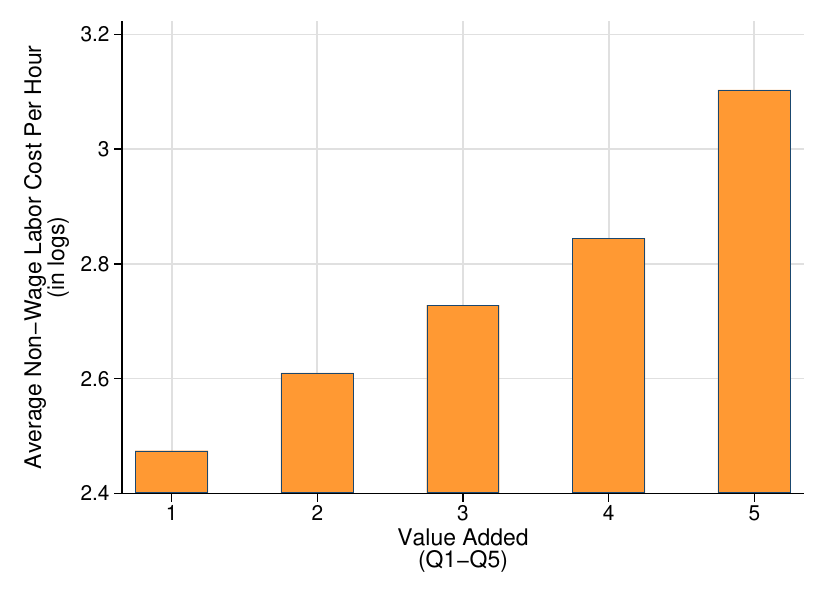}
    \caption{Non-Wage Compensation}
\end{subfigure}
\vspace{0.8em}
\begin{subfigure}[t]{0.32\textwidth}
    \includegraphics[width=\linewidth]{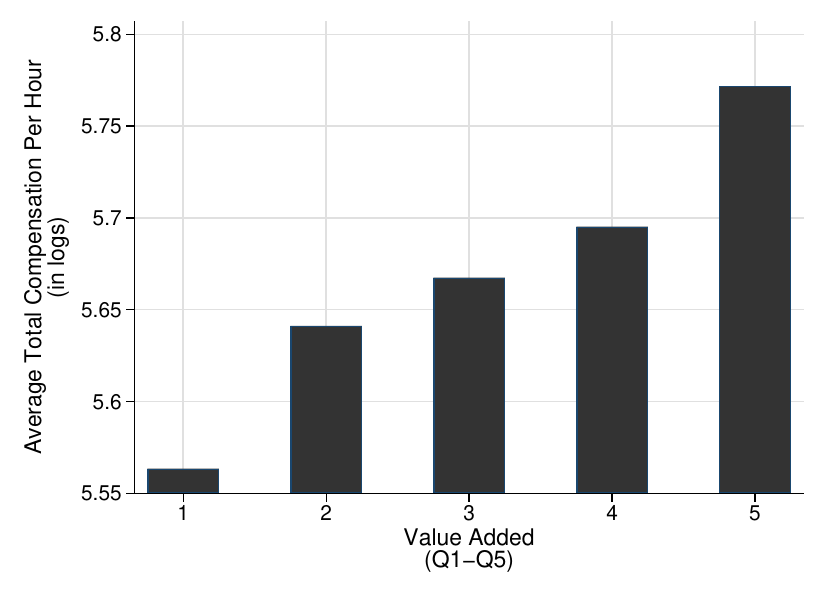}
    \caption{Total Compensation}
\end{subfigure}

\begin{subfigure}[t]{0.32\textwidth}
    \includegraphics[width=\linewidth]{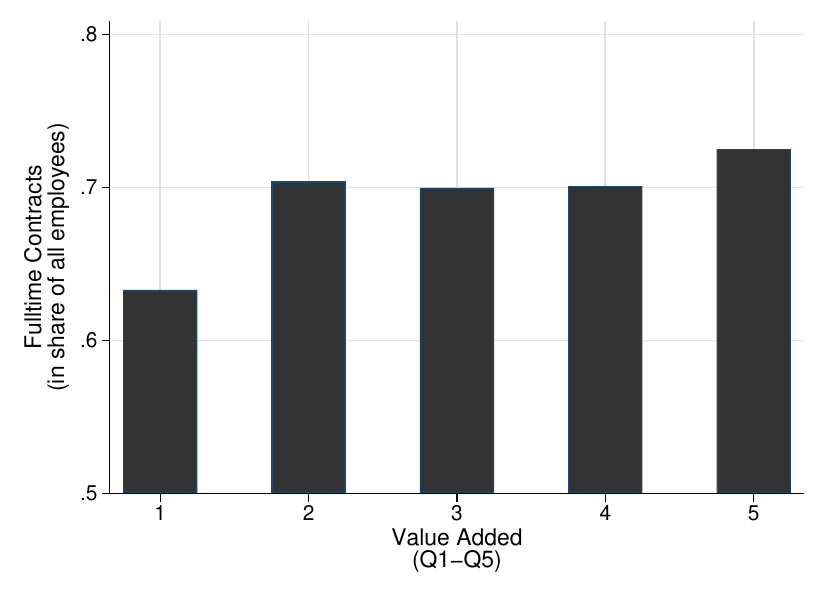}
    \caption{Full-Time Contracts}
\end{subfigure}\hfill
\begin{subfigure}[t]{0.32\textwidth}
    \includegraphics[width=\linewidth]{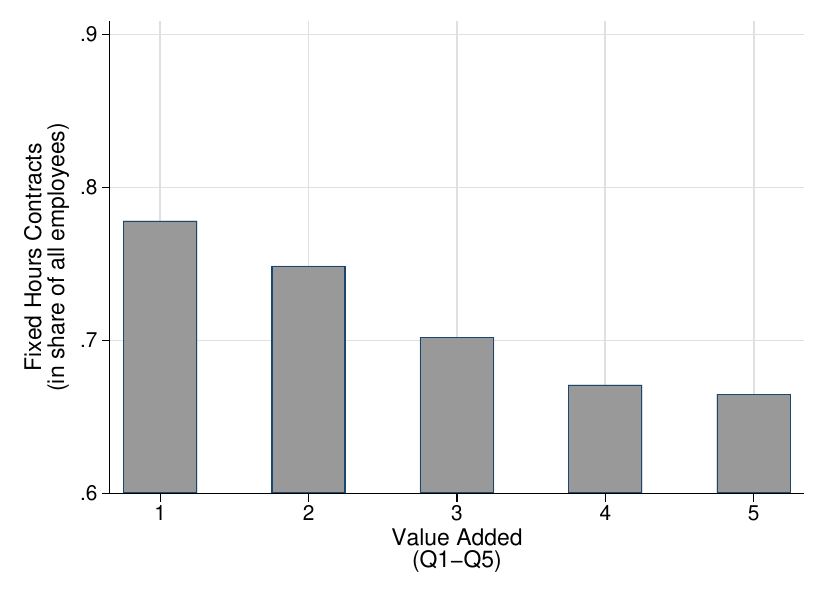}
    \caption{Fixed Hours Contracts}
\end{subfigure}\hfill
\begin{subfigure}[t]{0.32\textwidth}
    \includegraphics[width=\linewidth]{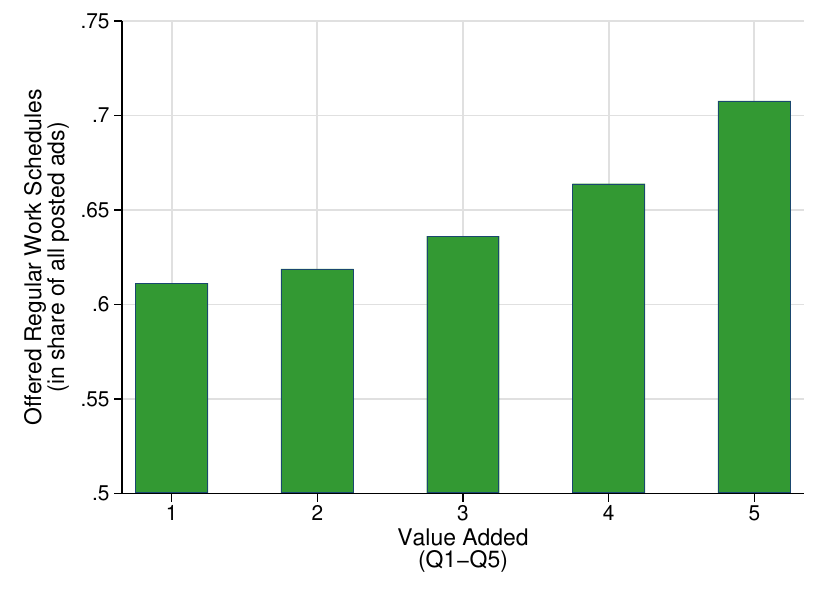}
    \caption{Regular Work Schedules}
\end{subfigure}

\begin{subfigure}[t]{0.32\textwidth}
    \includegraphics[width=\linewidth]{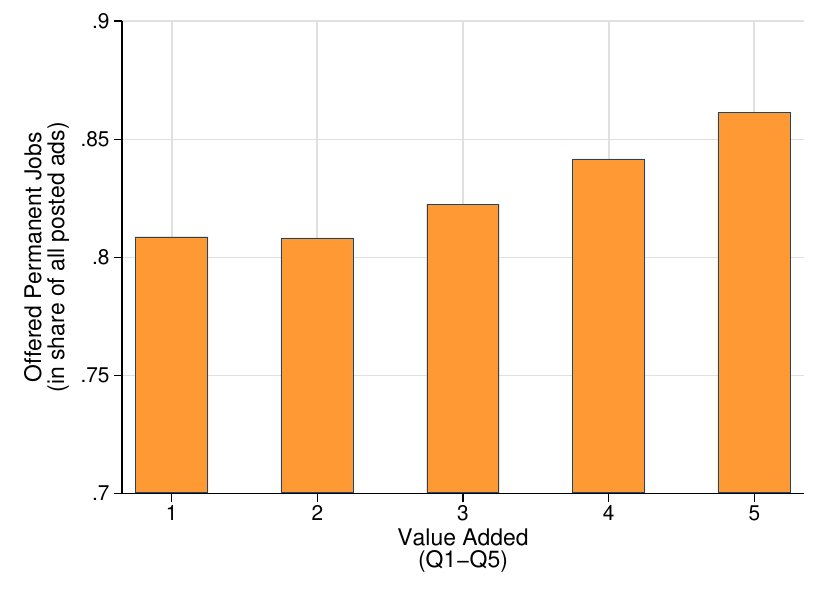}
    \caption{Permanent Jobs}
\end{subfigure}\hfill
\begin{subfigure}[t]{0.32\textwidth}
    \includegraphics[width=\linewidth]{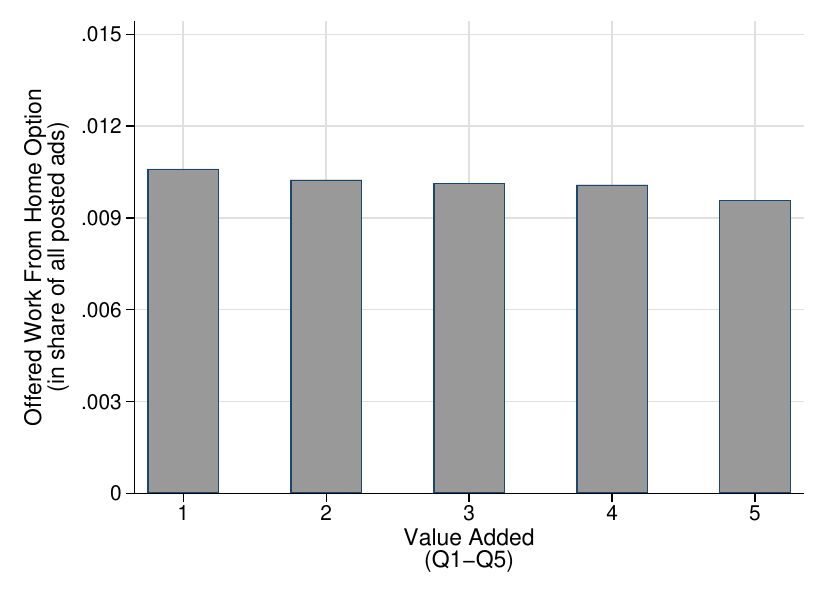}
    \caption{Work from Home}
\end{subfigure}\hfill
\begin{subfigure}[t]{0.32\textwidth}
    \includegraphics[width=\linewidth]{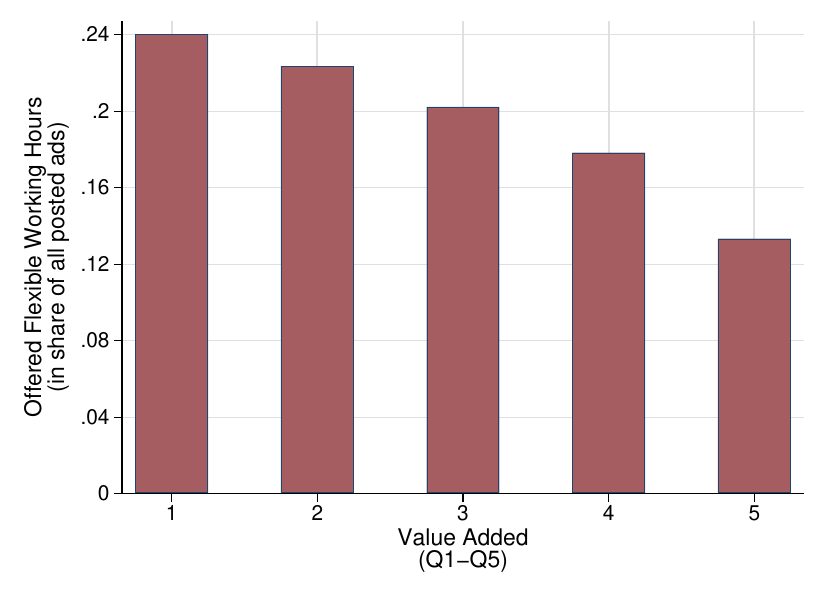}
    \caption{Flexible Work Hours}
\end{subfigure}

\begin{subfigure}[t]{0.32\textwidth}
    \includegraphics[width=\linewidth]{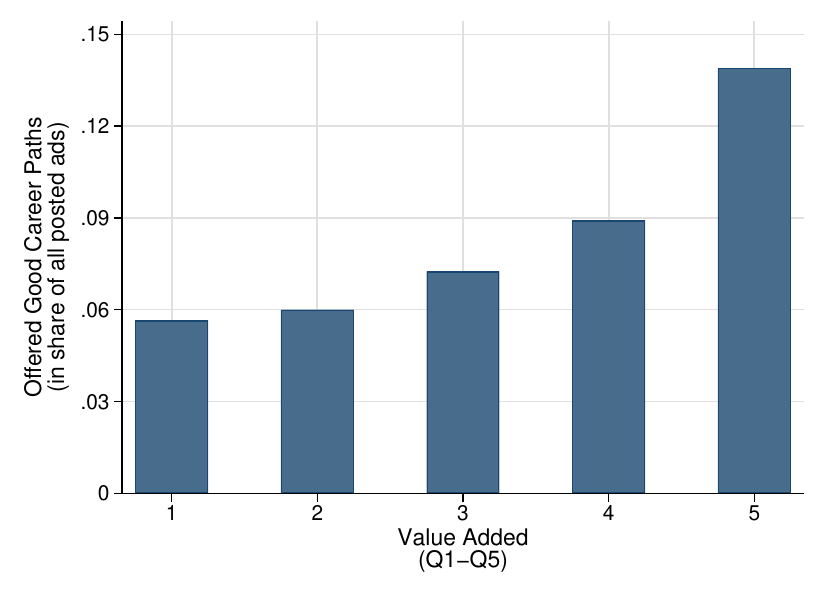}
    \caption{Good Career Paths}
\end{subfigure}\hfill
\begin{subfigure}[t]{0.32\textwidth}
    \includegraphics[width=\linewidth]{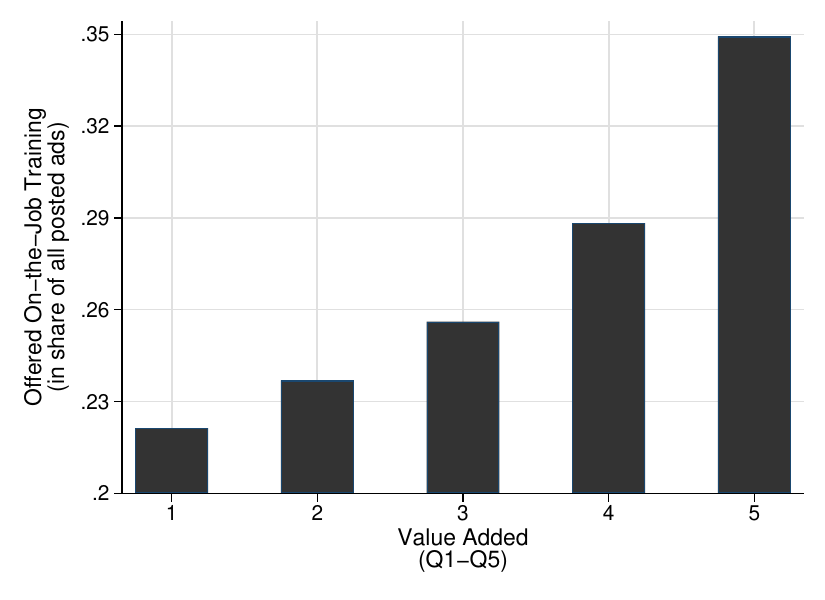}
    \caption{On-the-Job Training}
\end{subfigure}\hfill
\begin{subfigure}[t]{0.32\textwidth}
    \includegraphics[width=\linewidth]{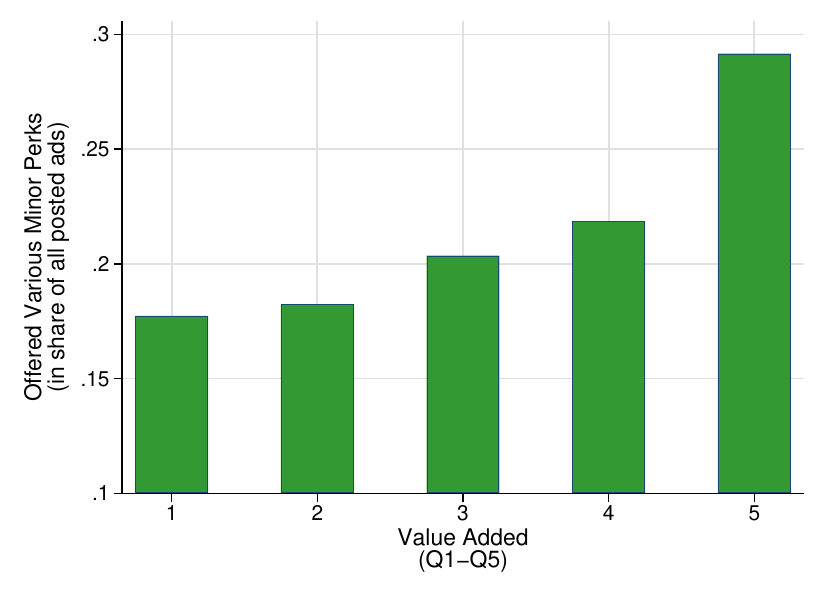}
    \caption{Additional Perks}
\end{subfigure}

\begin{minipage}{\textwidth}
\scriptsize
\emph{Notes:} Information on non-wage compensation (panels a--c) from firm cost accounts, 2015--2022. Information on contract types (panels d--e) from matched employer--employee data (A-meldingen), 2015--2022, and on advertised amenities (panels f--l) from job ads (NAV/FINN), 2015--2022, extracted using the text analysis methodology of \citet{bhuller2024ads}. Bars report average rescaled residual values (panels a--c) and average shares among all employees (panels d--e) or all posted ads (panels f--l), respectively, within firm value-added quintiles (Q1--Q5), constructed as residuals from regressions on two-digit industry, commuting zone, and year fixed effects, rescaled to have a mean equal to the sample average of the raw outcomes. All monetary values are reported in logs and CPI-adjusted to 2015 Norwegian Kroner (NOK).
\end{minipage}
\end{figure}

\begin{figure}[H]
\centering
\caption{The Effects of Demand Shocks on Wage and Non-Wage Compensation.}
\label{fig:external_nonwage_types}
\begin{subfigure}[t]{0.48\textwidth}
    \includegraphics[width=\linewidth]{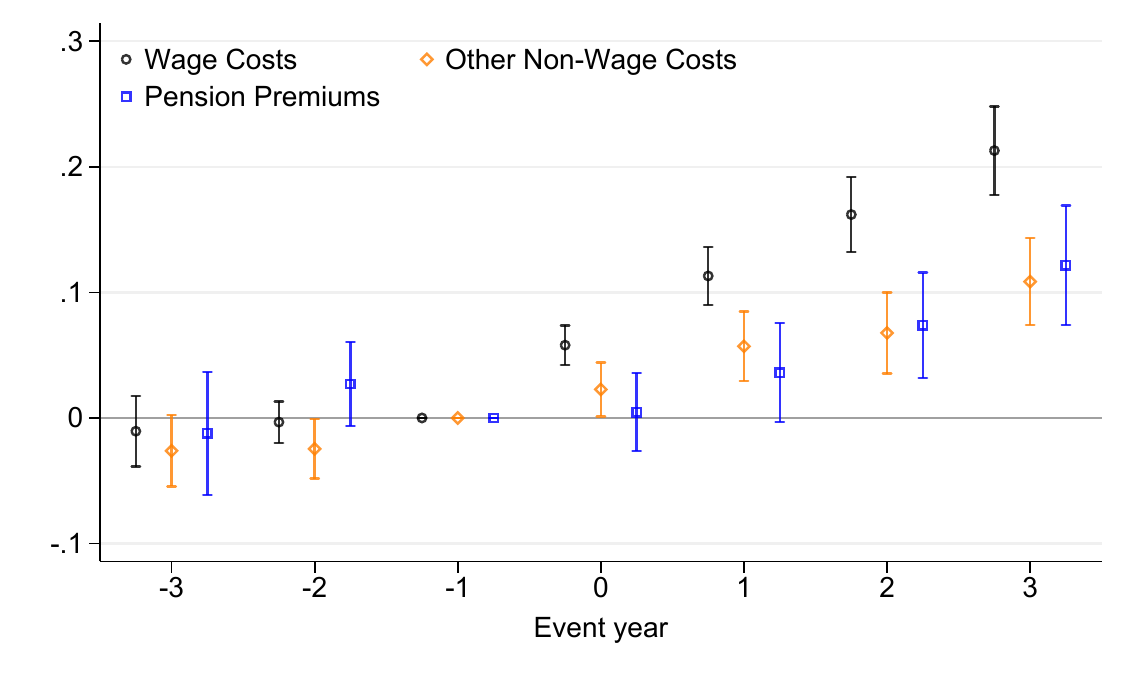}
    \caption{Procurement Shock: Total Compensation}
\end{subfigure}\hfill
\begin{subfigure}[t]{0.48\textwidth}
    \includegraphics[width=\linewidth]{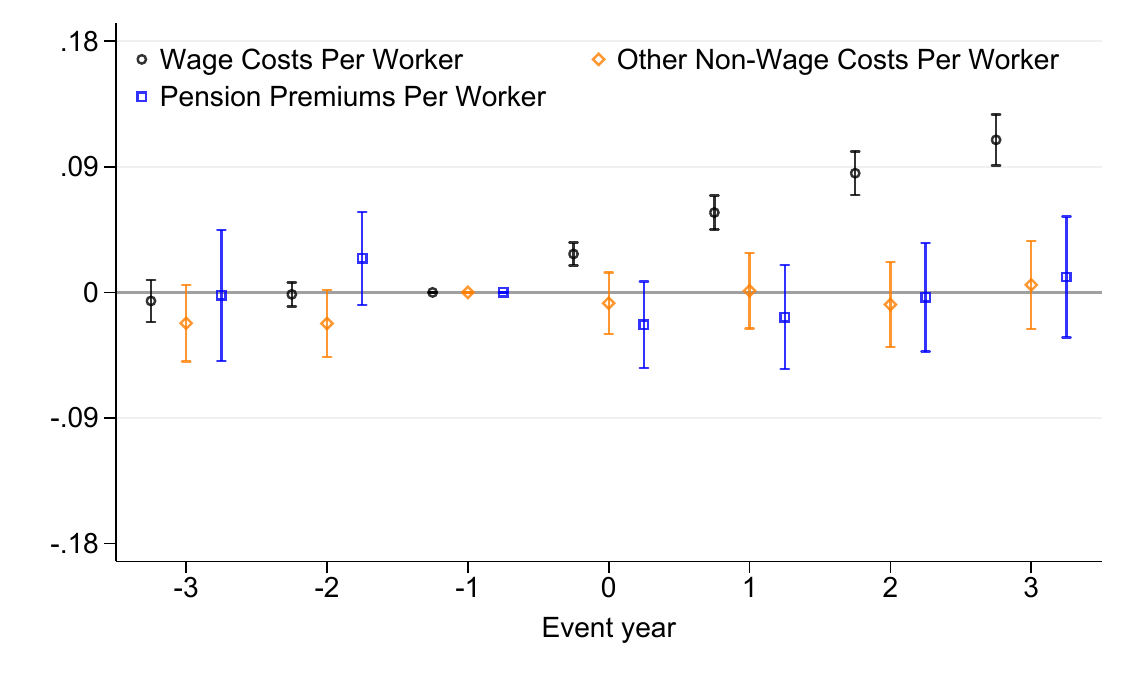}
    \caption{Procurement Shock: Per-Worker}
\end{subfigure}
\medskip
\begin{subfigure}[t]{0.48\textwidth}
    \includegraphics[width=\linewidth]{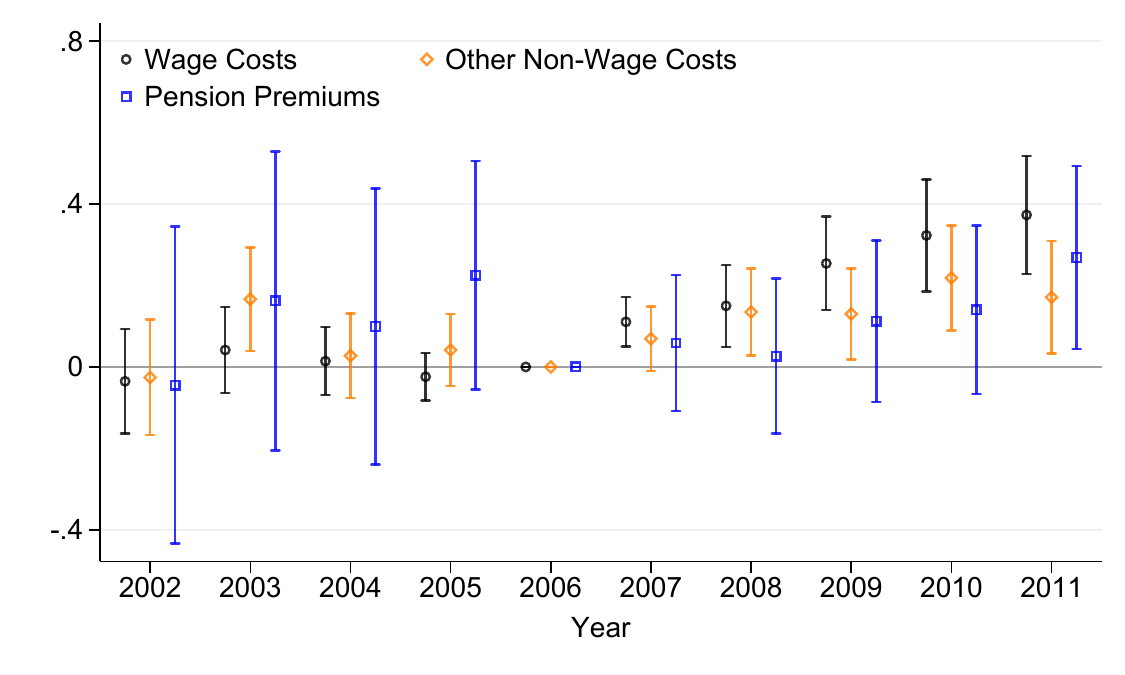}
    \caption{Export Shock: Total Compensation}
\end{subfigure}\hfill
\begin{subfigure}[t]{0.48\textwidth}
    \includegraphics[width=\linewidth]{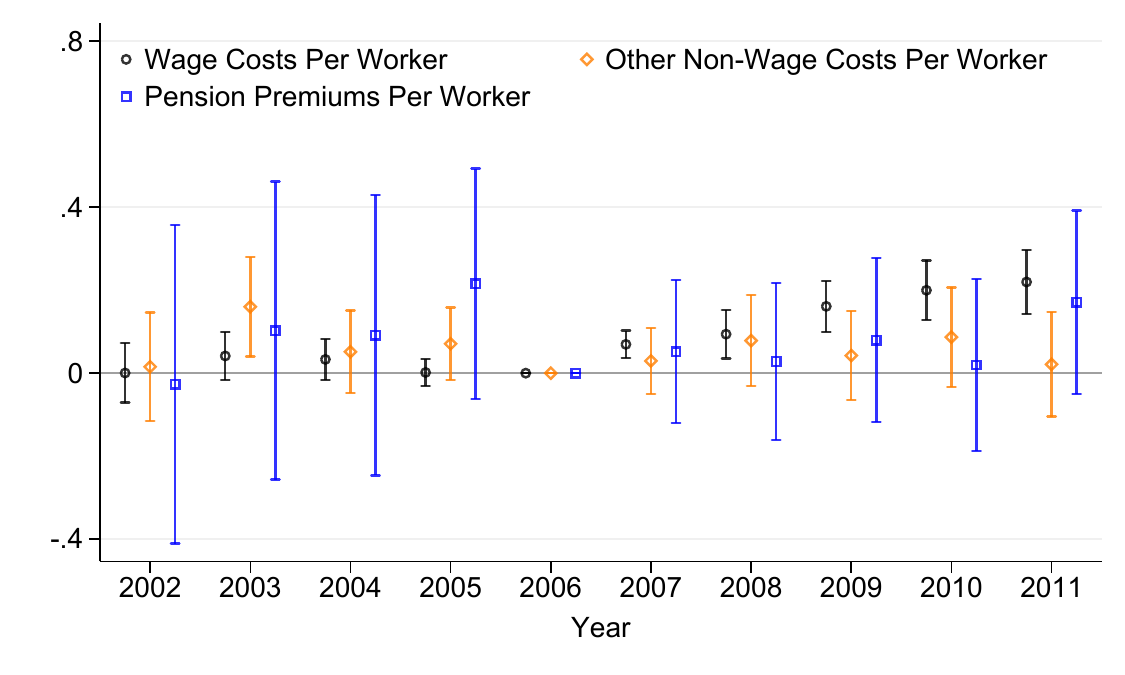}
    \caption{Export Shock: Per-Worker}
\end{subfigure}
\begin{minipage}{\textwidth}
\scriptsize
\emph{Notes:} Outcomes are constructed from firm cost accounts and matched employer--employee data. Panels~(a) and~(c) report total compensation costs by type, and panels~(b) and~(d) report per-worker compensation costs. All outcomes are in logs. Coefficients are event-time estimates with 95\% confidence intervals from Eq.~(\ref{eq:dynamic_did}). Panels~(a)--(b) show the procurement shock for the 2012--2019 cohorts, with event year $-1$ as the baseline and standard errors clustered at the firm level. These firms are observed throughout the balanced $[-3,+3]$ event window and have at least two stayers. Panels~(c)--(d) show the export shock, with 2006 as the baseline and standard errors clustered at the firm level. Export firms have an export-to-revenue ratio of at least 1\% in 2005--2007. Stayers in the export sample are workers present at the firm for at least eight years during the estimation window.
\end{minipage}
\end{figure}

\begin{figure}[H]
\centering
\caption{The Effects of Demand Shocks on the Amenity Cost Multiplier.}
\label{fig:external_amen_multiplier}
\begin{subfigure}[t]{0.48\textwidth}
    \includegraphics[width=\linewidth]{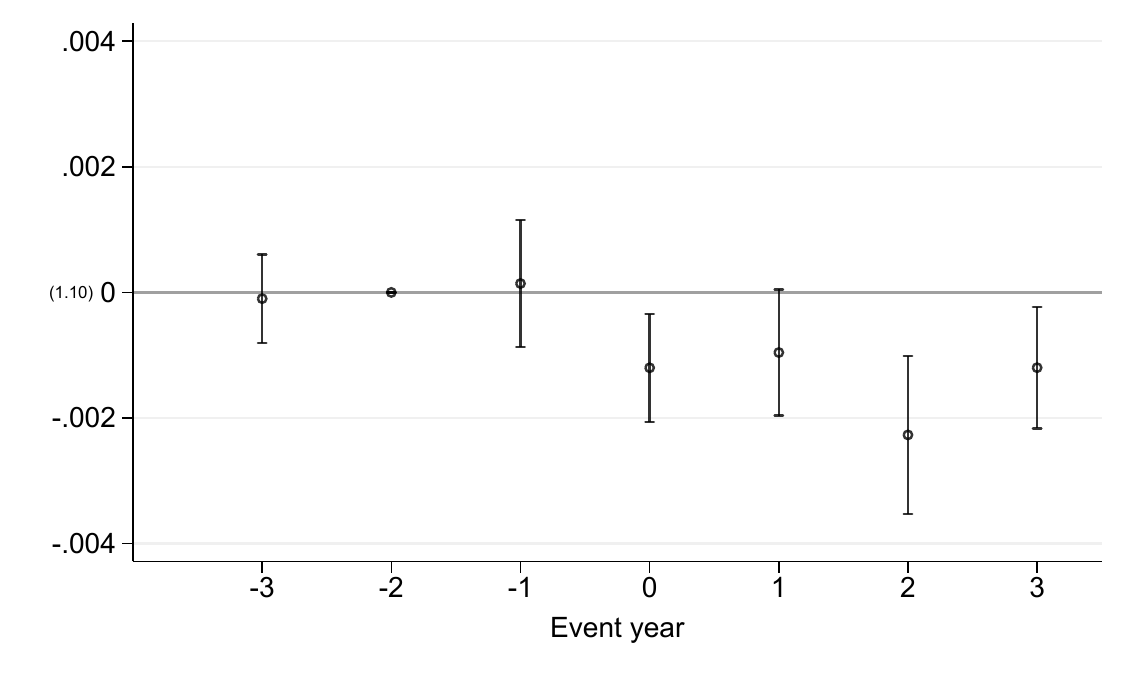}
    \caption{Internal Shock}
\end{subfigure}\hfill
\begin{subfigure}[t]{0.48\textwidth}
    \includegraphics[width=\linewidth]{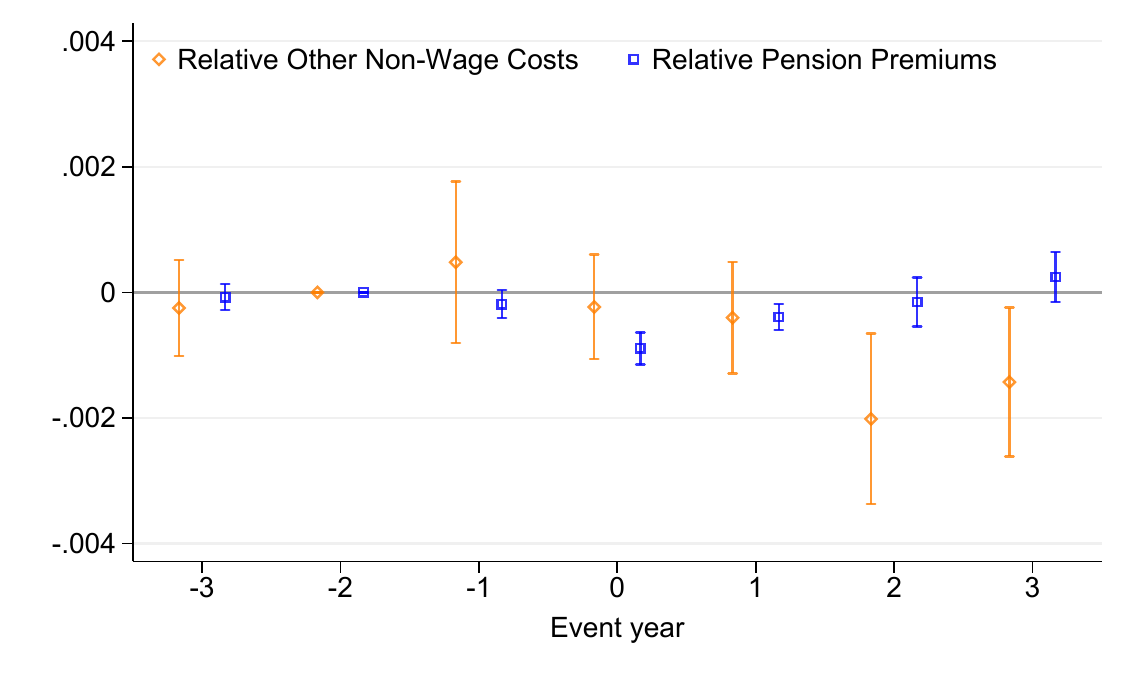}
    \caption{Internal Shock: by Cost Component}
\end{subfigure}
\begin{subfigure}[t]{0.48\textwidth}
    \includegraphics[width=\linewidth]{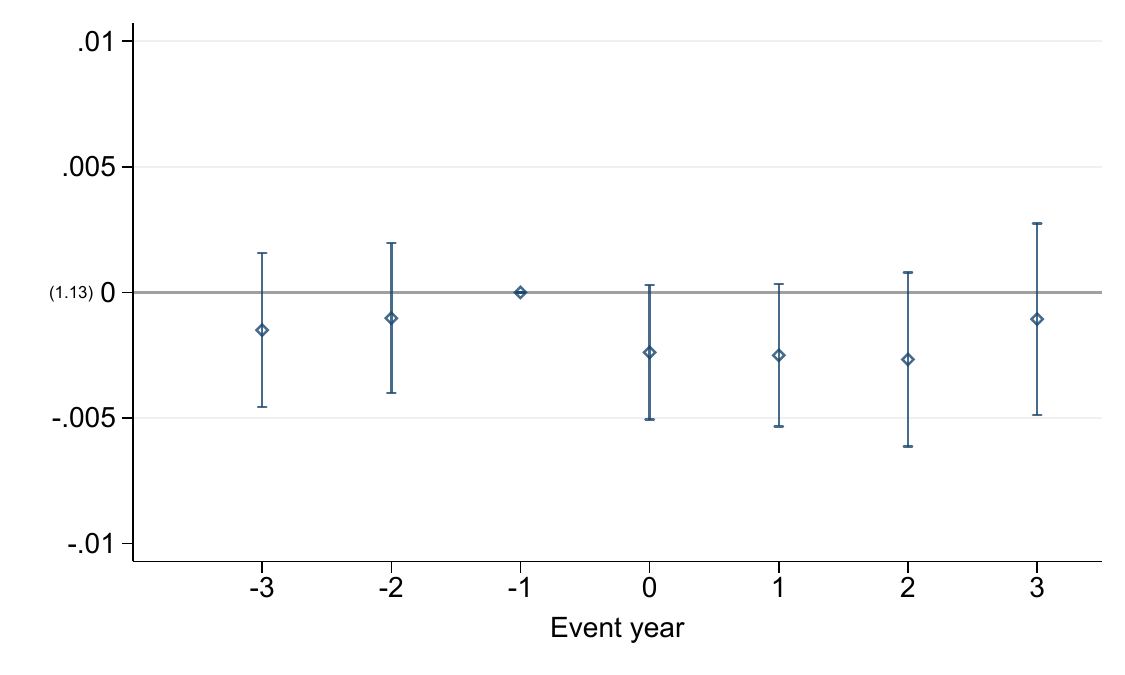}
    \caption{Procurement Shock}
\end{subfigure}\hfill
\begin{subfigure}[t]{0.48\textwidth}
    \includegraphics[width=\linewidth]{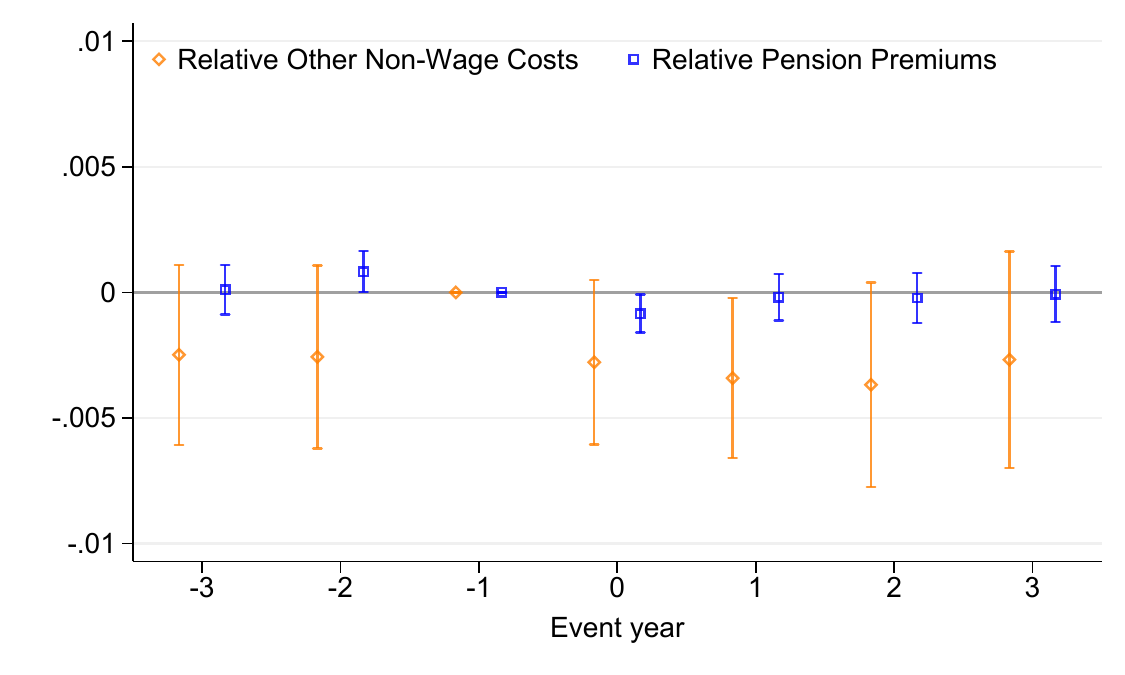}
    \caption{Procurement Shock: by Cost Component}
\end{subfigure}
\medskip
\begin{subfigure}[t]{0.48\textwidth}
    \includegraphics[width=\linewidth]{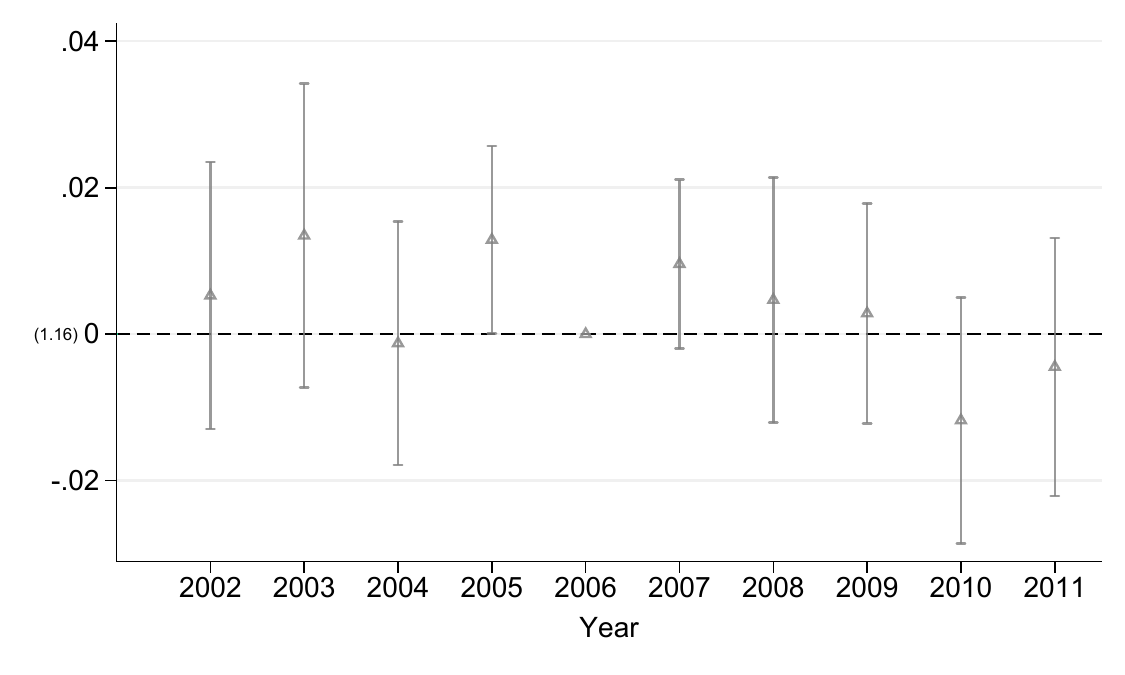}
    \caption{Export Shock}
\end{subfigure}\hfill
\begin{subfigure}[t]{0.48\textwidth}
    \includegraphics[width=\linewidth]{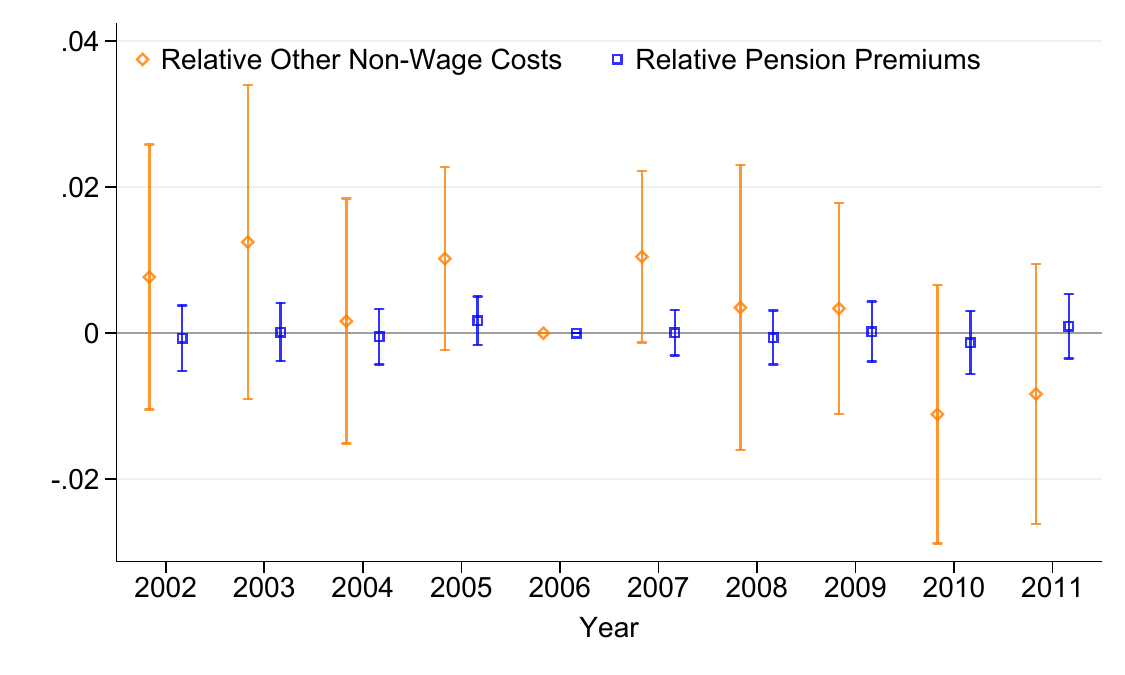}
    \caption{Export Shock: by Cost Component}
\end{subfigure}
\begin{minipage}{\textwidth}
\scriptsize
\emph{Notes:} Outcomes are constructed from firm cost accounts. Panels~(a), (c), and~(e) report the monetized amenity cost multiplier $\AM$ in levels, the counterpart of Eq.~\eqref{eq:amultiplier} over the amenities paid in money. Panels~(b), (d), and~(f) decompose it into cost shares $\omega_k=C_k/WL$, also in levels. Coefficients are event-time estimates with 95\% confidence intervals from Eq.~(\ref{eq:dynamic_did}). Panels~(a)--(b) show the internal shock for cohorts 2018--2019 with standard errors clustered at the industry and commuting zone level. Panels~(c)--(d) show the procurement shock for the 2012--2019 cohorts, with event year $-1$ as the baseline and standard errors clustered at the firm level. These firms are observed throughout the balanced $[-3,+3]$ event window and have at least two stayers. Panels~(e)--(f) show the export shock, with 2006 as the baseline and standard errors clustered at the firm level. Export firms have an export-to-revenue ratio of at least 1\% in 2005--2007. Stayers in the export sample are workers present at the firm for at least eight years during the estimation window.
\end{minipage}
\end{figure}

\newpage
\begingroup\footnotesize
\setlength{\LTcapwidth}{\linewidth}
\begin{longtable}{p{3.8cm} p{9.8cm}}

\caption{Variable Definitions and Construction.}
\label{tab:variable_construction} \\

\toprule
Variable & Definition and Construction \\
\midrule
\endfirsthead

\multicolumn{2}{l}{\textit{\tablename~\thetable{} -- continued from previous page}} \\[2pt]
\toprule
Variable & Definition and Construction \\
\midrule
\endhead

\midrule
\multicolumn{2}{r}{\textit{Continued on next page}} \\
\endfoot

\bottomrule
\multicolumn{2}{p{15cm}}{{\scriptsize
\emph{Notes:} All monetary variables are deflated to 2015~NOK (CPI, base year $2015=100$), and negative values are set to missing before taking logs. Matched employer--employee variables are firm-level outcomes from annual measures of \textit{A-meldingen}, and contract-type variables are measured from a reference month, November. Firm labor cost account variables are from \textit{N\ae{}ringsoppgaven}, and each can enter our analysis in three forms: the log of the total firm-level value, the log per worker obtained by dividing by the number of workers reported in the matched employer--employee data, and the log as a share of the total wage bill. Job-ad amenity shares follow the text-analysis methodology of \citet{bhuller2024ads} applied to NAV and FINN postings, excluding staffing agencies, and are defined only for firms with at least one posting in year $t$.}} \\
\endlastfoot

\addlinespace[2pt]

\multicolumn{2}{l}{\textit{Firm Balance Sheets}} \\[3pt]

\quad Value Added
  & Log of value added from annual firm balance sheets (\textit{regnskapsregisteret}), constructed as gross revenues minus operating costs plus labor costs and depreciation, restricted to firms with positive values,  in year $t$. \\[6pt]

\addlinespace[4pt]
\multicolumn{2}{l}{\textit{Matched Employee--Employer Data}} \\[3pt]

\quad Employment
  & Log of the number of workers employed at the firm in year $t$, drawn from annual matched employer--employee records, excluding workers with fewer than $52\!\times\!4$ annual contracted hours. \\[6pt]

\quad Hourly Wages
  & Log of the average real hourly wage across all workers or stayers in the firm, constructed as each worker's annual cash compensation divided by contracted hours, in year $t$. \\[6pt]

\quad In-Kind Compensation (\textit{naturalytelser beløp})
  & Log of the average monetary value of in-kind compensation across all workers or stayers in the firm in year $t$, covering non-cash benefits such as free car access, company phone, lodging or meal coverage, and compensation value of employee loans, fitness-center access, and travel allowances, as reported by the employer to the tax authorities via \textit{A-meldingen}. \\[6pt]

\quad In-Kind Contract (\textit{andel som får naturalytelser})
  & Firm-level share of workers or stayers in year $t$ receiving positive in-kind compensation as part of the contract from \textit{A-meldingen}, defined as the count of stayers with in-kind pay above zero divided by the total number of workers or stayers. \\[6pt]

\quad Full-Time Contract (\textit{heltidsarbeid})
  & Firm-level share of workers or stayers in year $t$ working more than 20 contracted hours per week, where total monthly hours sum contracted, variable, and overtime components from \textit{A-meldingen}. \\[6pt]

\quad Fixed Hours Contract (\textit{fast arbeidstid})
  & Firm-level share of workers or stayers in year $t$ without variable hours-based pay contracts (\textit{timelønnsavtale}) or contracts involving overtime hours (\textit{overtidsarbeid}) recorded in \textit{A-meldingen}. \\[6pt]

\addlinespace[4pt]
\multicolumn{2}{l}{\textit{Firm Labor Cost Accounts}} \\[3pt]

\quad Wage Bill
  & Total wage payments to employees in year $t$, recorded under entry P5000 of firms' annual business tax returns (\textit{n\ae{}ringsoppgaven}), which include salaries, bonuses, and commissions.
    Used as the log of the total value and the per-worker value, and as the denominator in the amenity multiplier. \\[6pt]

\quad Pension Premiums (\textit{pensjonskostnader})
  & Employer-sponsored pension premiums above the statutory minimum in year $t$, recorded under entry P5420 of firms' annual business tax returns (\textit{n\ae{}ringsoppgaven}). Used as a log of the total value, a log per worker, and a log as a share of the wage bill. \\[6pt]

    \quad Other Non-Wage Personnel Costs  (\textit{andre personalkostnader})
&
Non-wage personnel costs in year $t$ recorded under entry P5900 of firms' annual business tax returns (\textit{n\ae{}ringsoppgaven}), covering staff gifts, subsidized canteen meals, courses, company cabins, housing, sports-club memberships, occupational and travel insurance, and mandatory pension contributions. The measure also includes employee travel and food reimbursements, coming from P7155, P7165, and P7370, which cover travel, per diem and mileage allowances, subsistence, and representation expenses. Used as a log of total costs, a log per worker, and a log as a share of the wage bill. \\[6pt]

\quad Monetized Amenity Cost Multiplier \newline
  & The monetized amenity cost multiplier equals the ratio of total monetized labor costs to the wage bill in year $t$ from firms' annual business tax returns (\textit{n\ae{}ringsoppgaven}), with $\mathcal{A}^{\sM} \equiv (\text{wage bill} + \text{total monetized non-wage costs}) / \text{wage bill}$, where the total monetized non-wage costs include pension premiums \textit{P5420} and other non-wage personnel costs (\textit{P5900, P7155, P7165, and  P7370}). We winsorized at the 99th percentile. By construction, $\mathcal{A}^{\sM} =1$ if total monetized non-wage costs are zero. \\[6pt]

\addlinespace[4pt]
\multicolumn{2}{l}{\textit{Job Advertisements}} \\[3pt]

\quad Flexible Hours
  & The share of the firm's postings in year $t$ from NAV and FINN that mention convenient hours or flexible work hours, including flextime, standard hours, or no hour regulation, defined only for firms with at least one posting in year $t$. \\[6pt]

\quad Regular Work Schedule
  & The share of the firm's postings in year $t$ from NAV and FINN that do \textit{not} mention non-regular hours, including shift work, non-standard hours, zero-hour contracts, or overtime required, defined only for firms with at least one posting in year $t$. \\[6pt]

\quad Permanent Job
  & The share of the firm's postings in year $t$ from NAV and FINN that do \textit{not} advertise the position as temporary, defined only for firms with at least one posting in year $t$. \\[6pt]

\quad Good Workplace
  & The share of the firm's postings in year $t$ from NAV and FINN that mention at least one good-workplace attribute, including social environment, interesting tasks, good colleagues, work from home, open office, or employee involvement, defined only for firms with at least one posting in year $t$.
    \\[6pt]

\quad On-the-Job Training
  & The share of the firm's postings in year $t$ from NAV and FINN that mention on-the-job training, apprenticeship or trainee program, defined only for firms with at least one posting in year $t$. \\[6pt]

\quad Good Career Path
  & The share of the firm's postings in year $t$ from NAV and FINN that mention career development opportunities, defined only for firms with at least one posting in year $t$. \\[6pt]

\quad Financial Benefits
  & The share of the firm's postings in year $t$ from NAV and FINN that mention good financial benefits, including pension, insurance, or mortgage support, defined only for firms with at least one posting in year $t$. \\[6pt]

\quad Additional Perks
  & The share of the firm's postings in year $t$ from NAV and FINN that mention at least one additional perk, including proximity to nature, central location, exercise facilities, parking, company car, welfare benefits, company cabin, health benefits, subsidized food, or holiday benefits, defined only for firms with at least one posting in year $t$. \\[6pt]

\addlinespace[2pt]

\end{longtable}
\endgroup

\begin{table}[htbp]
\centering
\caption{Sample Sizes and Pre-Treatment Means Across Shocks.}
\label{tab:main_outcomes_sample_shocks}
\renewcommand{\arraystretch}{0.9}
\footnotesize
\begin{tabular}{l c c c}
\toprule
Outcome & Internal Shock & Procurement Shock & Export Shock \\
\midrule
\addlinespace[2pt]

\multicolumn{4}{l}{\textit{Firm Balance Sheets}} \\[1pt]

\quad Log Value Added
& \meanextenvalueadded
& \meanprocvalueadded
& \meanexpvalueadded \\
\quad Number of Observations
& \obsextenvalueadded
& \obsprocvalueadded
& \obsexpvalueadded \\
\quad Number of Firms
& \firmsextenvalueadded
& \firmsprocvalueadded
& \firmsexpvalueadded \\

\addlinespace[1pt]

\multicolumn{4}{l}{\textit{Matched Employee--Employer Data}} \\[1pt]

\quad Log Employment
& \meanextenemployment
& \meanprocemployment
& \meanexpemployment \\
\quad Log Hourly Wages
& \meanextenwagestayers
& \meanprocwagestayers
& \meanexpwagestayers \\
\quad Number of Observations
& \obsextenemployment
& \obsprocemployment
& \obsexpemployment \\
\quad Number of Firms
& \firmsextenemployment
& \firmsprocemployment
& \firmsexpemployment \\

\addlinespace[1pt]

\multicolumn{4}{l}{\textit{Firm Cost Accounts}} \\[1pt]

\quad Log Monetized Amenity Cost Multiplier ($\ln \mathcal{A}^{\sM}$)
& \meanextenamenmultiplierwl
& \meanprocamenmultiplierwl
& \meanexpamenmultiplierwl \\
\quad Number of Observations
& \obsextenamenmultiplierwl
& \obsprocamenmultiplierwl
& \obsexpamenmultiplierwl \\
\quad Number of Firms
& \firmsextenamenmultiplierwl
& \firmsprocamenmultiplierwl
& \firmsexpamenmultiplierwl \\

\addlinespace[1pt]
\bottomrule
\end{tabular}

\begin{minipage}{\textwidth}
\scriptsize
\emph{Notes:} Pre-treatment means are computed over the pre-shock period ($t<c$). The table reports the number of firm-cohort-year observations and the number of unique firms used in each estimation. Wage outcomes are based on the stayers sample. Internal shock estimates use cohorts 2005--2019, procurement shock estimates use cohorts 2012--2019, and export shock estimates restrict the sample to firms with an export-to-revenue ratio of at least 1\% in 2005--2007.
\end{minipage}
\end{table}

\begin{table}[htbp]
\centering
\caption{Industry Composition Across Estimation Samples.}
\label{tab:industry_shares_shocks}
\renewcommand{\arraystretch}{0.82}
\small
\begin{tabular}{l c c c}
\toprule
Industry & Internal Shock & Procurement Shock & Export Shock \\
\midrule
\addlinespace[2pt]
\quad Agriculture              & \shareintagri\%    & \shareprocagri\%    & \shareexpagri\% \\
\quad Mining                   & \shareintmining\%  & \shareprocmining\%  & \shareexpmining\% \\
\quad Manufacturing            & \shareintmanuf\%   & \shareprocmanuf\%   & \shareexpmanuf\% \\
\quad Electricity               & \shareintelec\%    & \shareprocelec\%    & \shareexpelec\% \\
\quad Water And Waste          & \shareintwater\%   & \shareprocwater\%   & \shareexpwater\% \\
\quad Construction             & \shareintconstr\%  & \shareprocconstr\%  & \shareexpconstr\% \\
\quad Trade                    & \shareinttrade\%   & \shareproctrade\%   & \shareexptrade\% \\
\quad Transport                & \shareinttransp\%  & \shareproctransp\%  & \shareexptransp\% \\
\quad Accommodation            & \shareintaccom\%   & \shareprocaccom\%   & 0.0\% \\
\quad Information              & \shareintinfo\%    & \shareprocinfo\%    & \shareexpinfo\% \\
\quad Finance                  & \shareintfinance\% & 0.0\%                 & 0.0\% \\
\quad Real Estate              & \shareintrealestate\% & \shareprocrealestate\% & \shareexprealestate\% \\
\quad Professional             & \shareintprof\%    & \shareprocprof\%    & \shareexpprof\% \\
\quad Other Industries          & \shareintrest\%   & \shareprocrest\%    & \shareexprest\% \\
\midrule
\quad Number of Firms & \firmsextenvalueadded & \firmsprocvalueadded & \firmsexpvalueadded \\
\addlinespace[2pt]
\bottomrule
\end{tabular}
\begin{minipage}{\textwidth}
\scriptsize
\emph{Notes:} This table reports the percentage share of treatment and control firms in the estimation
sample belonging to that industry, classified according to the Norwegian
Standard Industrial Classification (SN2007), which follows the European NACE
Rev.\ 2 classification. The internal shock sample uses cohorts 2005--2019,
procurement shock sample uses cohorts 2012--2019, and export shock sample
restricts the sample to firms with an export-to-revenue ratio of at least 1\%
in 2005--2007. ``Other Industries'' bundle NACE sections N through S.
\end{minipage}
\end{table}

\begin{table}[H]
\centering
\caption{Post-Treatment Average Effects Across Shocks.}
\label{tab:main_outcomes_postavg_shocks}
\renewcommand{\arraystretch}{0.8}
\small
\begin{tabular}{l c c c}
\toprule
 & Internal  & Procurement  & Export \\
Outcome: &  Shock & Shock & Shock \\
 & (1) & (2) & (3) \\
\midrule
\addlinespace[2pt]
\multicolumn{4}{l}{\textit{Firm Balance Sheets}} \\[3pt]
\quad Log Value Added
& \coefextenvalueadded
& \coefprocvalueadded
& \coefexpvalueadded \\
& \seextenvalueadded
& \seprocvalueadded
& \seexpvalueadded \\
\multicolumn{4}{l}{\textit{Matched Employee--Employer Data}} \\[3pt]
\quad Log Employment
& \coefextenemployment
& \coefprocemployment
& \coefexpemployment \\
& \seextenemployment
& \seprocemployment
& \seexpemployment \\
\addlinespace[2pt]
\quad Log Hourly Wages
& \coefextenwagestayers
& \coefprocwagestayers
& \coefexpwagestayers \\
& \seextenwagestayers
& \seprocwagestayers
& \seexpwagestayers \\
\multicolumn{4}{l}{\textit{Firm Cost Accounts}} \\[3pt]
\quad Log Monetized Amenity Cost Multiplier ($\ln \mathcal{A}^{\sM}$)
& \coefextenamenmultiplierwl
& \coefprocamenmultiplierwl
& \coefexpamenmultiplierwl \\
& \seextenamenmultiplierwl
& \seprocamenmultiplierwl
& \seexpamenmultiplierwl \\
\addlinespace[2pt]
\bottomrule
\end{tabular}
\begin{minipage}{\textwidth}
\scriptsize
\emph{Notes:} Reported coefficients correspond to the post-shock average effect
$\bar{\beta}_Y=(\beta_{Y,1}+\beta_{Y,2}+\beta_{Y,3})/3$, and for the export shock as $\bar{\beta}_Y=(\beta_{Y,2010}+\beta_{Y,2011})/2$, from the dynamic specification in Eq.~(\ref{eq:dynamic_did}). Standard errors are reported in parentheses. Wage outcomes are based on the stayers sample. Internal shock estimates use cohorts 2005--2019, procurement shock estimates use cohorts 2012--2019, and export shock estimates restrict the sample to firms with an export-to-revenue ratio of at least 1\% in 2005--2007. Standard errors are clustered at the industry and commuting zone level for internal shocks, at the firm level for procurement shocks, and at the firm level for export shocks. $* \ p<0.10$, $** \ p<0.05$, $*** \ p<0.01$.
\end{minipage}
\end{table}

\begin{table}[H]
\centering
\caption{Internal Shock (Contracts Sample): Post-Effects and Pre-Means.}
\label{tab:main_outcomes_descriptives_combined}
\label{tab:main_outcomes_sample}
\label{tab:main_outcomes_postavg}
\footnotesize
\setstretch{1}
\renewcommand{\arraystretch}{0.95}
\setlength{\tabcolsep}{4pt}
\begin{tabular}{l c c r r}
\toprule
Outcome & $\bar{\beta}_Y$ & \shortstack{Pre-Treatment\\Mean} & Obs. & Firms \\
\midrule
\multicolumn{5}{l}{\textit{Firm Balance Sheets}} \\[2pt]
\quad Log Value Added & \coefvalueadded & \meanvalueadded & \obsvalueadded & \firmsvalueadded \\[-2pt]
 & \sevalueadded & & & \\[2pt]
\addlinespace[4pt]
\multicolumn{5}{l}{\textit{Matched Employee--Employer Data}} \\[2pt]
\quad Log Employment & \coefemployment & \meanemployment & \obsemployment & \firmsemployment \\[-2pt]
 & \seemployment & & & \\[2pt]
\quad Log Hourly Wages  & \coefwagestayers & \meanwagestayers & \obswagestayers & \firmswagestayers \\[-2pt]
 & \sewagestayers & & & \\[2pt]
\quad Log In-Kind Compensation & \coefpayinkindstayers & \meanpayinkindstayers & \obspayinkindstayers & \firmspayinkindstayers \\[-2pt]
 & \sepayinkindstayers & & & \\[2pt]
\quad Share In-Kind Contract & \coefinkindstayers & \meaninkindstayers & \obsinkindstayers & \firmsinkindstayers \\[-2pt]
 & \seinkindstayers & & & \\[2pt]
\quad Share Full-Time Contract & \coeffulltimestayers & \meanfulltimestayers & \obsfulltimestayers & \firmsfulltimestayers \\[-2pt]
 & \sefulltimestayers & & & \\[2pt]
\quad Share Fixed Hours Contract & \coeffixedstayers & \meanfixedstayers & \obsfixedstayers & \firmsfixedstayers \\[-2pt]
 & \sefixedstayers & & & \\[2pt]
\addlinespace[4pt]
\multicolumn{5}{l}{\textit{Firm Labor Cost Accounts -- Totals}} \\[2pt]
\quad Log Wage Costs & \coefwagecost & \meanwagecost & \obswagecost & \firmswagecost \\[-2pt]
 & \sewagecost & & & \\[2pt]
\quad Log Other Non-Wage Costs & \coefnonwagecost & \meannonwagecost & \obsnonwagecost & \firmsnonwagecost \\[-2pt]
 & \senonwagecost & & & \\[2pt]
\quad Log Pension Premiums & \coefpenspremcost & \meanpenspremcost & \obspenspremcost & \firmspenspremcost \\[-2pt]
 & \sepenspremcost & & & \\[2pt]
\quad Log Monetized Amenity Cost Multiplier ($\ln \mathcal{A}^{\sM}$) & \coefamenmultiplierwl & \meanamenmultiplierwl & \obsamenmultiplierwl & \firmsamenmultiplierwl \\[-2pt]
 & \seamenmultiplierwl & & & \\[2pt]
\addlinespace[4pt]
\multicolumn{5}{l}{\textit{Firm Labor Cost Accounts -- Per Worker}} \\[2pt]
\quad Log Wage Costs & \coefwagecostperworker & \meanwagecostperworker & \obswagecostperworker & \firmswagecostperworker \\[-2pt]
 & \sewagecostperworker & & & \\[2pt]
\quad Log Other Non-Wage Costs & \coefnonwageperworker & \meannonwageperworker & \obsnonwageperworker & \firmsnonwageperworker \\[-2pt]
 & \senonwageperworker & & & \\[2pt]
\quad Log Pension Premiums & \coefpenspremperworker & \meanpenspremperworker & \obspenspremperworker & \firmspenspremperworker \\[-2pt]
 & \sepenspremperworker & & & \\[2pt]
\bottomrule
\end{tabular}
\vspace{3pt}
\begin{minipage}{\textwidth}
\scriptsize
\emph{Notes:} Effects are post-shock averages $\bar{\beta}_Y=(\beta_{Y,1}+\beta_{Y,2}+\beta_{Y,3})/3$ from the dynamic specification in Eq.~(\ref{eq:dynamic_did}), with standard errors in parentheses below the estimates. Pre-treatment means are computed over the pre-shock period. Obs.\ counts firm-cohort-year observations, and Firms counts unique firms in each estimation sample. Matched employer--employee outcomes, apart from employment, refer to stayers. The contracts sample stacks the 2018 and 2019 cohorts and includes firms observed throughout the balanced $[-3,+3]$ event window with at least two stayers. Standard errors are clustered at the industry and commuting zone level. $*\ p<0.10$, $**\ p<0.05$, $***\ p<0.01$.
\end{minipage}
\end{table}

\begin{table}[H]
\centering
\caption{Internal Shock (Job Postings Sample): Post-Effects and Pre-Means of Advertised Amenities.}
\label{tab:jobads_exten}
\footnotesize
\setlength{\tabcolsep}{4pt}
\renewcommand{\arraystretch}{0.8}
\begin{tabular}{l c c r r}
\toprule
Advertised Amenity & $\bar{\beta}_Y$ & \shortstack{Pre-Treatment\\Mean} & Obs. & Firms \\
\midrule
On-the-Job Training & \coefextentraining & \meanextentraining & \obsextentraining & \firmsextentraining \\[-3pt]
 & \seextentraining & & & \\[2pt]
Good Workplace & \coefextengoodworkplace & \meanextengoodworkplace & \obsextengoodworkplace & \firmsextengoodworkplace \\[-3pt]
 & \seextengoodworkplace & & & \\[2pt]
Flexible Hours & \coefextenconvenient & \meanextenconvenient & \obsextenconvenient & \firmsextenconvenient \\[-3pt]
 & \seextenconvenient & & & \\[2pt]
Regular Work Schedules & \coefextenregular & \meanextenregular & \obsextenregular & \firmsextenregular \\[-3pt]
 & \seextenregular & & & \\[2pt]
Additional Perks & \coefextenaddedperks & \meanextenaddedperks & \obsextenaddedperks & \firmsextenaddedperks \\[-3pt]
 & \seextenaddedperks & & & \\[2pt]
Financial Benefits & \coefextengoodfinance & \meanextengoodfinance & \obsextengoodfinance & \firmsextengoodfinance \\[-3pt]
 & \seextengoodfinance & & & \\[2pt]
Permanent Job & \coefextenpermanent & \meanextenpermanent & \obsextenpermanent & \firmsextenpermanent \\[-3pt]
 & \seextenpermanent & & & \\[2pt]
Good Career Path & \coefextencareer & \meanextencareer & \obsextencareer & \firmsextencareer \\[-3pt]
 & \seextencareer & & & \\[2pt]
\bottomrule
\end{tabular}
\begin{minipage}{\textwidth}
\scriptsize
\emph{Notes:} Outcomes are firm-year shares of job ads mentioning each amenity, from FINN/NAV linked to the extended sample. Effects are post-shock averages $\bar{\beta}_Y=(\beta_{Y,1}+\beta_{Y,2}+\beta_{Y,3})/3$ from Eq.~(\ref{eq:dynamic_did}), with standard errors in parentheses below the estimates. Means are computed over event years $t-c\leq-2$. Effects and means are in share units. Obs.\ counts firm-cohort-year observations, and Firms counts unique firms in each estimation sample. The underlying sample pools the 2005--2019 cohorts, with firms observed throughout the balanced $[-3,+3]$ event window and at least two stayers. Job-ad outcomes are measured for posting firms. Standard errors are clustered at the industry and commuting zone level. $*\ p<0.10$, $**\ p<0.05$, $***\ p<0.01$.
\end{minipage}
\end{table}

\printbibliography[heading=bibintoc,title={Appendix References}]
\end{refsection}

\end{document}